\documentclass[a4paper,12pt]{book}
\usepackage[utf8]{inputenc}
\usepackage{xspace}			   
\usepackage[T1]{fontenc}
\usepackage{libertine}
\usepackage{libertinust1math}
\usepackage[ttdefault=true,scaled=0.85]{AnonymousPro}

\usepackage{subfig}
\usepackage[margin=28mm,includeheadfoot]{geometry}
\usepackage{setspace}
\usepackage{ragged2e}
\usepackage[all]{xy}
\usepackage{graphicx} 
\usepackage{color}
\usepackage{pifont}
\usepackage[svgnames]{xcolor} 
\usepackage{floatflt}
\usepackage{amsmath, mathtools}
\usepackage{amssymb,amsfonts,dsfont}	
\usepackage{bm}				
\usepackage{bbm,upgreek}		
\usepackage{esint}			
\usepackage{etoolbox}			
\usepackage{calc}				
\usepackage{eso-pic}			
\usepackage{datetime}			

\usepackage{tikz}
\usetikzlibrary{patterns, decorations.pathreplacing,decorations,decorations.markings}
\usepackage{slantsc}
\makeatletter
\patchcmd{\chaptermark}{\MakeUppercase}{\scshape\slshape}{}{}%
\patchcmd{\sectionmark}{\MakeUppercase}{\scshape\slshape}{}{}%
\makeatother
\numberwithin{equation}{section} 
\numberwithin{figure}{chapter}
\numberwithin{table}{chapter}
\mathtoolsset{showonlyrefs,showmanualtags}
\usepackage{titlesec}                        
\titleformat{\chapter}[display]{\Huge\sffamily\bfseries}%
	{\chaptertitlename~\thechapter}{1ex}{}
\titleformat{\section}[hang]{\Large\sffamily\bfseries}%
	{\rlap{\thesection}}{2em}{}
\titleformat{\subsection}[hang]{\large\sffamily\bfseries}%
	{\rlap{\thesubsection}}{3em}{}
\usepackage[numbers,sort&compress]{natbib}
\definecolor{blue}{RGB}{0, 0, 128}
\definecolor{darkgray}{RGB}{64, 64, 64}   
\definecolor{teal}{RGB}{0, 100, 100}
\definecolor{darkred}{RGB}{139, 0, 0}

\usepackage[backref=page]{hyperref}
\hypersetup{linktocpage,colorlinks,citecolor={blue},pdfdisplaydoctitle=true,pdfpagemode=UseOutlines,bookmarksnumbered=true,urlcolor={teal},linkcolor={darkred}}
\usepackage{mathrsfs}
\usepackage{amsthm}
\usepackage{epigraph}
\usepackage{floatrow}

\usepackage{tikz}
\usetikzlibrary{positioning}
\usetikzlibrary{calc}

\definecolor{mylightred}{RGB}{211,79,73}
\definecolor{mydarkred}{RGB}{199,44,38}
\definecolor{mylightgreen}{RGB}{78,153,67}
\definecolor{mydarkgreen}{RGB}{43,129,33}
\definecolor{mylightpurple}{RGB}{150,107,178}
\definecolor{mydarkpurple}{RGB}{126,78,160}
\definecolor{mylightblue}{RGB}{49,101,205}
\definecolor{mydarkblue}{RGB}{20,92,205}

\tikzset{
  juliadot/.style args={#1,#2}{shape=circle,line width=0.03ex,minimum width=0.4ex,fill=#1,draw=#2}
}

\newcommand{\tadpole}[0]{
	\begin{tikzpicture}[scale=0.85,baseline=-0.2em]
        \draw[] (0,0) -- (2,0);
        \draw[] (1,0) to[out=10, in=0, looseness=1.5] (1,1) to[out=-180, in=170, looseness=1.5] (1,0);
        \node[circle, fill=black, inner sep=1.5pt] at (1,0) {};
	\end{tikzpicture}
}

\newcommand{\blue}[1]{\textcolor{blue}{#1}}

\newcommand{\ora}[1]{\textcolor{orange}{#1}}

\theoremstyle{remark}
\newtheorem{remark}{Remark}

\theoremstyle{plain}

\theoremstyle{plain}

\theoremstyle{plain}

\newcommand{\bra}[1]{\langle #1 |} 
\newcommand{\ket}[1]{| #1 \rangle } 

\definecolor{cbl}{rgb}{0,0,1}

\definecolor{crd}{rgb}{1,0,0}
 
\newcommand{\upd}{\mathrm{d}}
\newcommand{\tr}{\mathrm{tr}}
\newcommand{\xb}{\mathbf{x}}
\newcommand{\yb}{\mathbf{y}}

\newcommand{\db}{\mathbf{d}}

\newcommand{\ie}[0]{\textit{i.e.} }
\newcommand{\eg}[0]{\textit{e.g.} }

\newcommand{\hpsi}{\hat{\psi}}

\newcommand\e{\mathrm{e}}

\newcommand{\braket}[1]{\left\langle #1 \right\rangle } 

\newcommand\Id{\mathrm{Id}}
\newcommand\id{\Id}
\newcommand\hphi{\hat{\phi}}

\newcommand{\Dmu}{{\mathcal{D}}_L}

\DeclareMathOperator*{\argmin}{argmin}

\csname endofdump\endcsname

\begin{document}

\begin{titlepage}
\centering

\vspace*{1.5cm}

{\huge\bfseries Some progress on the use of the variational method in quantum field theory\par}

\vspace{0.9em}
{\color{gray}\rule{0.55\textwidth}{0.4pt}}\par
\vspace{0.9em}

{\large\itshape A journey into relativistic continuous matrix product states\\[0.2em]
and recent extensions\par}

\vspace{1.5cm}

{\Large Antoine Tilloy\par}

\vspace{0.5em}

{\small\itshape
Laboratoire de Physique de l'\'{E}cole Normale Sup\'{e}rieure,
Mines Paris, Inria, CNRS, ENS-PSL,\\[0.2em]
Centre Automatique et Syst\`{e}mes (CAS),
Sorbonne Universit\'{e}, PSL Research University, Paris, France\par}

\vspace{1.5cm}

\begin{minipage}{0.82\textwidth}
\small\setlength{\parindent}{0pt}
Strongly coupled quantum field theories in $(1+1)$ dimensions are notoriously hard to
solve non-perturbatively. Variational methods, despite their success for quantum
many-body physics on the lattice, have long lacked a natural ansatz adapted to the relativistic
setting. This monograph explains the intuition behind \textit{relativistic continuous matrix product states} (RCMPS), a variational ansatz tailored to $(1+1)$-dimensional QFT, and reports on
several years of progress in developing and applying this approach. Using Riemannian
optimization on the manifold of RCMPS, we obtain competitive non-perturbative
approximations to ground state energies and local observables in the $\phi^4$,
Sine-Gordon, and Sinh-Gordon models, including in strongly coupled regimes where
perturbation theory fails. We then describe extensions to models with several interacting
fields. Beyond energy density and local observables, we show how the framework can be used to evaluate non-local observables (defects) and, through an original linear programming approach, to extract spectral data such as particle masses. We close by discussing the current limitations of the method and the most promising directions for future work.
\end{minipage}

\vspace*{\fill}

\end{titlepage}
\thispagestyle{empty}

\frontmatter

\chapter*{Acknowledgments}
\phantomsection
\addcontentsline{toc}{chapter}{Acknowledgments}
After writing this fairly long manuscript, my first thanks go to the referees who agreed to read it. Slava Rychkov, Nobert Schuch, and Xavier Waintal accepted this ungrateful task, and may forever remain the only readers of the present work. I am deeply grateful for the time they gave, for their technical comments, and for their encouragements. Next come Christoph Kopper, Adam Nahum, Anne Nielsen, and Jean-Bernard Zuber who honored me by accepting to be part of the Habilitation jury that evaluated an earlier version of this monograph.

This manuscript presents work done during a transition period from a time when I was working mostly alone, to a time when I had the chance of supervising (or rather collaborating with) bright students and postdocs. Karan Tiwana, Edoardo Lauria, Sophie Mutzel, and Molly Kaplan directly contributed to results that are presented here, and substantially shaped my views on these questions. It is hard to give justice to how much I learned from them. Without them, I would have certainly written this manuscript much faster, but it would have been far thinner. 

During this period, I also collaborated with other students and postdocs, on fascinating questions orthogonal to the subject of this manuscript. Pierre Guilmin, Roberto Negrin, Florent Goulette, and Gustave Robichon substantially impacted the way I think about quantum physics. Without them, I would also have written this manuscript much faster, it probably would not have been much thinner, but I would have been far less motivated about physics in general.

I am also indebted to my colleagues at Inria and Mines, who provided me with an exceptionally stimulating environment. In particular, I am grateful to Mazyar Mirrahimi and Pierre Rouchon, who trusted me from the beginning despite my peculiar research interests, and supported me practically and scientifically. Florent Di Meglio gave me precious advice to get this manuscript written in a reasonable amount of time, and provided a friendly pressure to ultimately get it all done.

Finally, I thank Izel Sari, my wife and lover, who makes life bearable. Her being gives me the peace that makes writing possible.

\vskip1.0cm
\noindent \emph{Most of the original research presented in this monograph has been possible thanks to funding from the European Research Council (ERC) from the QFT.zip project (grant agreement No.\ 101040260).}

\tableofcontents

\mainmatter
\chapter[Introduction]{Introduction}\label{ch:introduction}

\section{Quantum field theory}

Quantum field theory (QFT) is the most accurate prediction toolbox we have for problems involving quantum matter interacting with $3$ of the fundamental forces (gravity being so far left out). Following the Wilsonian revolution, it is also a framework that has found use almost everywhere else in theoretical physics, in particular in condensed matter theory, statistical physics, and even quantum information.

In this monograph, we are interested in QFT models that could in principle be valid all the way down (\emph{aka} ``UV complete'', without short-distance cutoff, non-perturbatively renormalizable, or just \emph{true} QFT, in contrast with \emph{effective} QFT). It is not known if this is the right way to think of QFT in fundamental physics. Indeed, the \emph{Standard Model} of particle physics, which is currently the instantiation of QFT fitting observations best, is not UV complete as a whole. Further, even its UV-complete subparts, like quantum chromodynamics (QCD), could also \emph{naturally} emerge from many different microscopic descriptions\footnote{Renormalizable QFT, because they are the \emph{only} theories that can be valid all the way down, are obvious candidates for fundamental descriptions of Nature. Renormalizable QFT, because they are the \emph{only} possible large distance behavior of any model (under mild constraints), are obvious candidate for \emph{effective} descriptions of Nature.}. The final completion of the laws of physics may or may not be a \emph{true} QFT.

Nonetheless, no matter their ultimate fate as a fundamental description, \emph{true} QFT models are still an absolutely crucial part of theoretical physics. In my opinion, they deserve to be studied seriously, without hiding behind their plausibly effective nature or waiting for the connection with gravity to be first understood. After all, they are the most reasonable models consistently putting together special relativity and quantum mechanics and it is interesting to know how far they can be pushed, even just in principle.

There are three problems one faces when working with such true QFT: defining it (what object is it mathematically?), interpreting it (what does it say about the world if true?), and computing with it (what quantitative predictions can one extract?).

The definition problem has been partly solved, thanks to heroic efforts by mathematical physicists. Starting in the late sixties, they first understood what properties a genuine QFT model should verify (with ``axiomatic field theory'') and then explicitly constructed non-trivial examples of models with interactions (with ``constructive field theory''). This program is sometimes unfairly presented as a failure, because constructive field theorists have not managed to construct non-trivial interacting models in $3+1$ space-time dimensions. One candidate for such true QFT in $3+1$ dimensions is non-Abelian Yang-Mills theory, and defining it properly remains one of the unsolved Millennium problems selected by the Clay Mathematics Institute. Yet in lower dimensions, the program has mostly succeeded, and led to the rigorous definition of several quantum field theories in $1+1$ and $2+1$ dimensions (starting with the self-interacting scalar $\phi^4$). Recently, fairly exotic QFT models like Liouville theory have been given a fully rigorous probabilistic construction. At least a fraction of models are well understood mathematically, and provide a sound basis one can build upon. This is making the results in this memoir possible.

The interpretation problem remains largely open, despite intense efforts in the quantum foundations community. Independently of its extra mathematical difficulty, QFT shares the same foundational problems as standard quantum mechanics. In particular, the \emph{measurement problem} is not in any way tamed by QFT. Worse, some solutions that have been constructed for standard quantum mechanics become far more convoluted and anaesthetic when extended to QFT. Logically, one should probably go the other way, and devise a solution to the measurement problem specifically tailored to QFT (its non-relativistic limit would then be allowed to be peculiar or counter-intuitive). In an ideal world, the ultimate solution to the measurement problem would work \emph{only} for QFT, and even only for the specific type instantiated by Nature (the Standard Model or its extensions). This would make the laws of physics far less contingent than they seem.  I believe making sense of QFT in this strong sense is absolutely crucial, and I have sunk unwise amounts of time into it~\cite{tilloy2017interactingquantumfieldtheories}, unfortunately without real progress. This monograph thus focuses on easier problems.

The computational problem is still very difficult, but has seen major progress in the recent years, with an explosion of new methods. This monograph presents a very modest contribution to this exciting effort. 
\section{Solving models}
\epigraph{QCD must be exactly soluble, or else I cannot imagine what the physics textbooks of the future will look like.}{Alexander Polyakov\footnotemark}

\footnotetext{According to John Preskill in the \href{https://quantumfrontiers.com/2012/12/11/fundamental-physics-prize-prediction-polyakov/}{Quantum Frontier blog}, Polyakov said this during a visit at Harvard in 1978.}

Most of our problems come from the fact that Polyakov's hopes, expressed in the quote above, have unfortunately not materialized. It seems that the QFT models we most naturally come up with, that appear in the Standard Model or as effective descriptions in condensed matter theory, are not exactly solvable. This is not surprising: QFT is a quintessential instance of the quantum many body problem (where the \emph{many} is actually \emph{infinitely many}). And the quantum many body problem is almost always not analytically solvable. Solvable models are at best points or lines in the huge space of reasonable theories. This limits the predictive power of theoretical physics in the quantum realm: most of the time we know (a good approximation to) the fundamental laws, but are just unable to solve them to predict anything.

\subsection{The state of the art: perturbation theory and lattice}

Historically, the main way to extract approximate predictions from QFT has been perturbation theory, aka the expansion in Feynman diagrams. This tool is so tied to QFT that many textbooks equate a QFT model with its corresponding diagrammatic rules. 

The problems of perturbation theory are well known. The first problem is mild: the expansion is term by term divergent. This is just the sign that we took the continuum limit a bit too naively, and is cured by standard renormalization theory. The second problem is more serious, and it is understood since Dyson that the expansion is only asymptotic, \ie its radius of convergence is zero. As a result, the expansion is rigorously valid only for infinitely small coupling. In principle, this problem can also be cured by the theory of Borel resummation and its resurgence theory refinements. In cases where the theory is well understood, it allows to (approximately) re-sum the perturbative series for finite coupling values, provided one can compute a very large number of terms. This brings us to the last and most serious practical issue with perturbative expansions in QFT: their cost increases brutally (in fact usually factorially) with the order, and thus one can rarely ever compute more than a few terms. This is usually too few to go to large values of the coupling, \emph{even if} the series had an infinite convergence radius. 

One can still obtain exceptionally precise results for simple models~\cite{serone2018symmetric,serone2019broken,heymans2021N8LO}, where the expansion is sufficiently simple that the factorial wall appears only fairly late. Perturbation theory is certainly not dead, but its main issue is really practical: we need to find ways to compute many terms efficiently on a computer\footnote{There are some conjectures for how to do so. In condensed matter theory, it is sometimes possible to tame the factorial explosition by grouping diagrams together into a determinant / permanent~\cite{rossi2017diagrammatic}, and similar strategies could be attempted in relativistic QFT. Even if the factorial growth is tamed, one still needs to compute integrals of divergent integrands over many variables. For smooth and bounded integrands, tensor cross interpolation shows promise, with fast convergence~\cite{fernandez2022tci}. For more general integrands, recent progress in tropical Monte Carlo algorithms allows to reach the convergence rate that the standard Monte Carlo method obtains for bounded integrands~\cite{borinsky2023tropical}. One may dream of one day combining all these approaches to make perturbation theory numerically efficient even at large coupling.}.

The other more recent but still well established approach to do calculations in QFT is lattice Monte-Carlo. In my opinion, lattice field theorists are the real heroes of the field, and have obtained by far the most impressive results in strongly coupled quantum field theory. They certainly did not get the recognition they deserve, at least in the general public\footnote{It is quite telling that, in the part of the young public that is interested in physics, almost every teenager knows Edward Witten, Leonard Susskind, or Nima Arkani-Hamed. But almost no one has heard of Martin Lüscher. Arguably, we have learned more quantitatively about the fundamental interactions of Nature thanks to Lüscher than thanks to any string theorist (or, for that matter, any tensor network specialist!).} but also in theoretical physics. Lattice Monte-Carlo is impressive in that physicists have managed to go the whole way, in 40 years, from toy models, to the complex field theories of the real world. This was made possible partly thanks to the improvement in computers, but mostly, thanks to an endless stream of smart tricks and ideas. I believe it is important to be clear that this memoir, and more generally the work done with tensor networks in field theory, goes nowhere near their achievements!

That said, in the long run, lattice Monte Carlo has objective limits, and a subjective taste of \emph{not enough}. The objective limits are twofold. The first problem is the mildest: even when it works, the precision of the Monte-Carlo method is difficult to increase by orders of magnitude. The precision is limited by statistical uncertainties, which decay only as $1/\sqrt{N}$ (where $N$ is the number of samples, proportional to the effort put in) and systematic lattice discretization errors. Both types of error are well understood, and reducing them is in principle only polynomially expensive. However, if the Moore law breaks down as expected in the coming decade, the precision of the Monte-Carlo method for the Standard Model will remain inevitably limited\footnote{For example, as far as lattice spacing is concerned, Lüscher himself wrote in his 1997 les Houches lecture notes~\cite{luscher1998advancedlatticeqcd} \textit{``In hadron mass calculations, for example, current lattice spacings are usually not much smaller than 0.1 fm. This will remain so for quite some time, because the simulation programs slow down proportionally to $a^5$ (or even a larger power of $a$) if all other parameters are held fixed.}''. Nowadays, the smallest lattice spacing used in calculations are not fundamentally different (\eg $a=0.025$fm is the absolute smallest considered in~\cite{petreczky2019qcdfinelattice}).}. Of course, as happened in the past, theory alone may gain us a few orders of magnitude (for example, with the recently introduced normalizing flows~\cite{cheng2026normalizingflows}). But I think it is fair to expect ultimate limits in precision for such a well understood approach.

The second objective problem, more crucial, is the \emph{sign problem}: for some observables, the Monte Carlo method simply does not work. When the quantity estimated $\langle \mathcal{O}\rangle$ is small (say order $1$), but obtained as a sum of very large terms $\mathcal{O}(\omega_i)$ (say of order $2^n$) of opposite signs, the Monte Carlo method is exponentially expensive. This happens typically when the action $S(\phi)$ appearing in the field pseudo-measure $\mathcal{D}\phi\, \e^{-S(\phi)}$ is not real.  This is the case for simulations at finite baryon density, for real-time dynamics, or when including chiral Fermions (which is technically needed to simulate the full Standard Model). The bottom line is that turning a quantum mechanical problem into a statistical mechanical problem is not always possible \emph{efficiently} (otherwise, quantum computers would be useless).

More subjectively, it is a pretty common view that there is something not fully satisfying with the Monte Carlo method, at least as sole probe of (some regime of) our most fundamental theories. I think part of this dislike is unwarranted (some physicists, for example, find these types of methods ``dirty''). But even for physicists who are absolutely sold to the beauty of numerics, as I am, there is a feeling that one should be able to do better. That the impressive achievements of lattice QCD notwithstanding, there must be a way to do it without harnessing randomness, and maybe even without discretizing space-time.

\subsection[Two philosophies forward]{Two philosophies forward -- skyscraper vs pile of dirt\texorpdfstring{\protect\footnotemark}{}}\label{sec:pileofdirt}

\footnotetext{This subsection is a distilled and revamped version of \href{https://atilloy.com/2020/06/09/the-skyscraper-and-pile-of-dirt-approaches-to-qft/}{an essay} that I posted on my blog in 2020.}

Aware of the limitations of perturbation theory and Monte Carlo, theoretical physicists have looked for other methods to understand quantum field theory. In a way that is perhaps insufficiently emphasized usually, this problem of QFT has been attacked from two completely different angles, by two different communities, and with a radically different philosophy. In a nutshell the philosophies are:
\begin{itemize}
\item    Make the problem simpler, hoping the tools you build to define and solve the simple instances are more or less generic, and work your way up to real QCD.
 \item   Make the problem more complex, so that it can miraculously be solved exactly thanks to the extra structure, and progressively reduce or deform the structure to get to real QCD.
\end{itemize}
In this monograph, we will present a very modest instance of the first approach. However, the second one deserves to be presented first because I think it is paradoxically the most mainstream (at least in terms of fame of those who defend it) and perhaps the most surprising. It is one that illuminates best the mindset of theoretical physicists and high-energy physicists in particular.

\subsubsection*{The appeal of the gargoyles}

At first sight it seems counter-intuitive that making things more complex makes the problem solvable. But I think it is a quite standard pattern in theoretical physics, at least in the last century. Typically you start from a simple non-trivial model. You realize that the things you would want to compute can be expressed as a sum of infinitely many complicated terms. It is hopeless, too generic, almost random. Then, you complexify the model to add “structure” in such a way that this infinite mess orders itself into a few simple terms, through magical cancellations, or because all the terms now make a geometric series. At the same time you can use additional structure to cure divergences in a neater way than with standard renormalization. The most famous extra structure is supersymmetry, but in lower dimensions (1+1 instead of 3+1) there is also an infinite family of quantum field theories precisely designed to make the difficulty go away (the so called “integrable” QFTs).

The idea of this approach is to reproduce the success of perturbation theory, but this time with new interacting candidates as solvable starting points. This goal has mesmerized physicists: the infinite richness of exactly solvable theories has provided nice qualitative understanding of some weird phenomena like quark confinement, or given insights into phase diagrams. However, I think it will not go beyond that, because we see in these models only the structure we put in. This ordering, this simplification of the mess is precisely what does not happen in most of the real quantum field theories we see in Nature.

\begin{figure}
  \centering
  \includegraphics[width=0.4\textwidth]{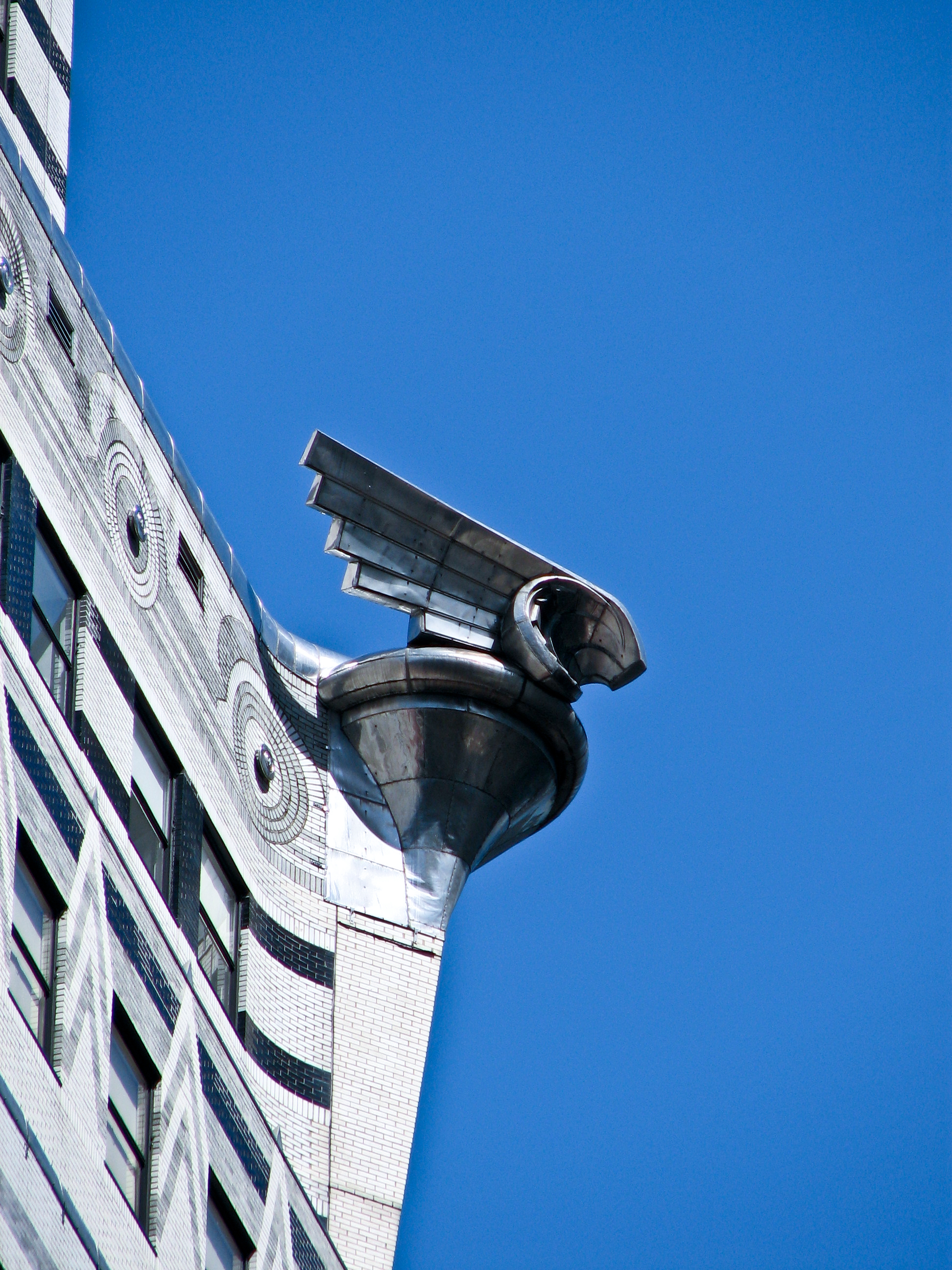}
  \caption{One of the ornaments of the Chrysler building – Norbert Nagel / Wikimedia Commons CC-BY-SA 3.0}
  \label{fig:gargoyle}
\end{figure}

It is a bit like trying to learn about mountains and volcanoes by studying the Eiffel tower or the Chrysler building. They are all pointy things going upward, but what we learn from one by studying the other is necessarily limited. Further the Chrysler building has beautiful metallic gargoyles (see Fig. \ref{fig:gargoyle}), which end up attracting all the attention. In the same way, the idiosyncrasies of the structured solvable models\footnote{Examples of idiosyncrasies are peculiar thermalization because of conserved charges, lack of renormalization/flow of certain quantities, absence of particle production, etc. . } become the center of attention, an interest on their own, and we end up forgetting what we came here for\footnote{In \emph{Le diner de cons}, François Pignon has to call Juste Leblanc to get the phone number of Brochant's mistress. As an excuse for the call, he has to pretend he is interested in purchasing the rights to turn Leblanc's book into a movie. After a long and confusing phone call, Pignon hangs up and, proud of himself, announces that he managed to successfully purchase the rights.}. We get too fascinated by the gargoyles.

A historical hope was that, deep down, Nature was like the Chrysler building, that there was structure all the way down. But there is not much empirical evidence for this belief. It seems the complex models that make things simpler are really man-made, and don’t describe what we see. Now I would say most people in the field no longer use this argument as a defense of the study of highly structured theories, but rather say that these well understood theories give us a window into what really happens in those, more realistic, that we cannot solve. I do not doubt that there is a qualitative window in some cases, but can it be turned into something quantitative?

\subsubsection*{The alternative pile-of-dirt approach}

The alternative is to go from simple to more complicated, to attack the mount Everest of QCD by starting from piles of dirt, onto hills, and then small mountains. This is the strategy that was followed by the warrior monks of constructive field theory, and by the hyper-pragmatic lattice Monte Carlo theorists. I think this is a strategy that can work again for new techniques. One should aim to find what makes QFT \emph{generically} difficult and subtle, but fight the difficulty in a much simpler context. 

Of course, the risk of this approach is that in some cases one cannot do the last steps. By simplifying, we may remove too much of the essence of the difficulty, and the methods we develop may not transfer to the more difficult examples. This is what happened with constructive field theory, where the just-renormalizable models have remained out of reach of the techniques invented for $\phi^4_2$ and $\phi^4_3$. This is also plausibly the case with the tensor network methods we will discuss, that certainly do not transfer \emph{as is} to $3+1$d QFT! At least in that case, we are left with an understanding the intermediate problems, that are arguably interesting in their own right, although not fundamental.

\begin{figure}\label{fig:pile_of_dirt}
\begin{center}
    \includegraphics[width=0.2\textwidth]{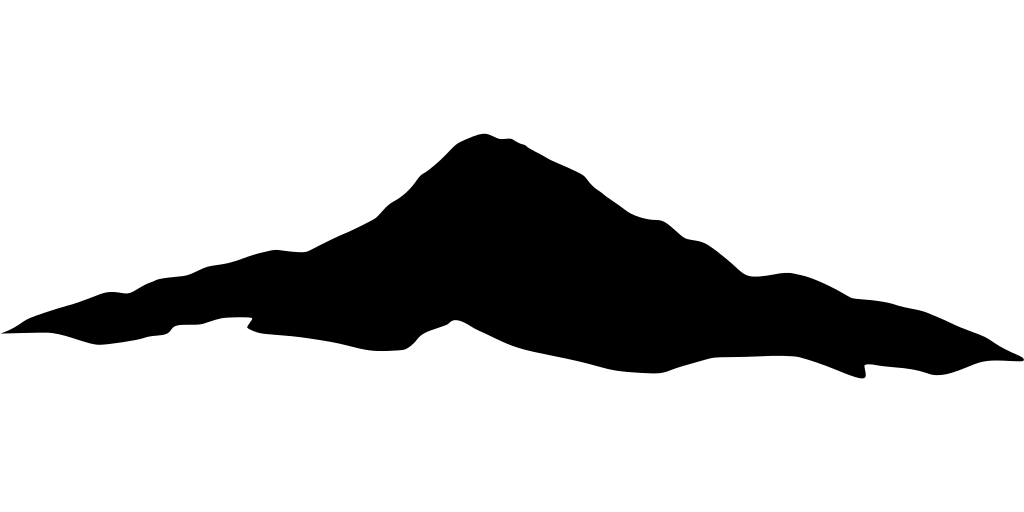} ~~~~~~
    \includegraphics[width=0.2\textwidth]{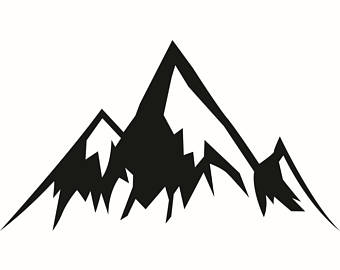} ~~~~~~~~~~~~
    \includegraphics[width=0.2\textwidth]{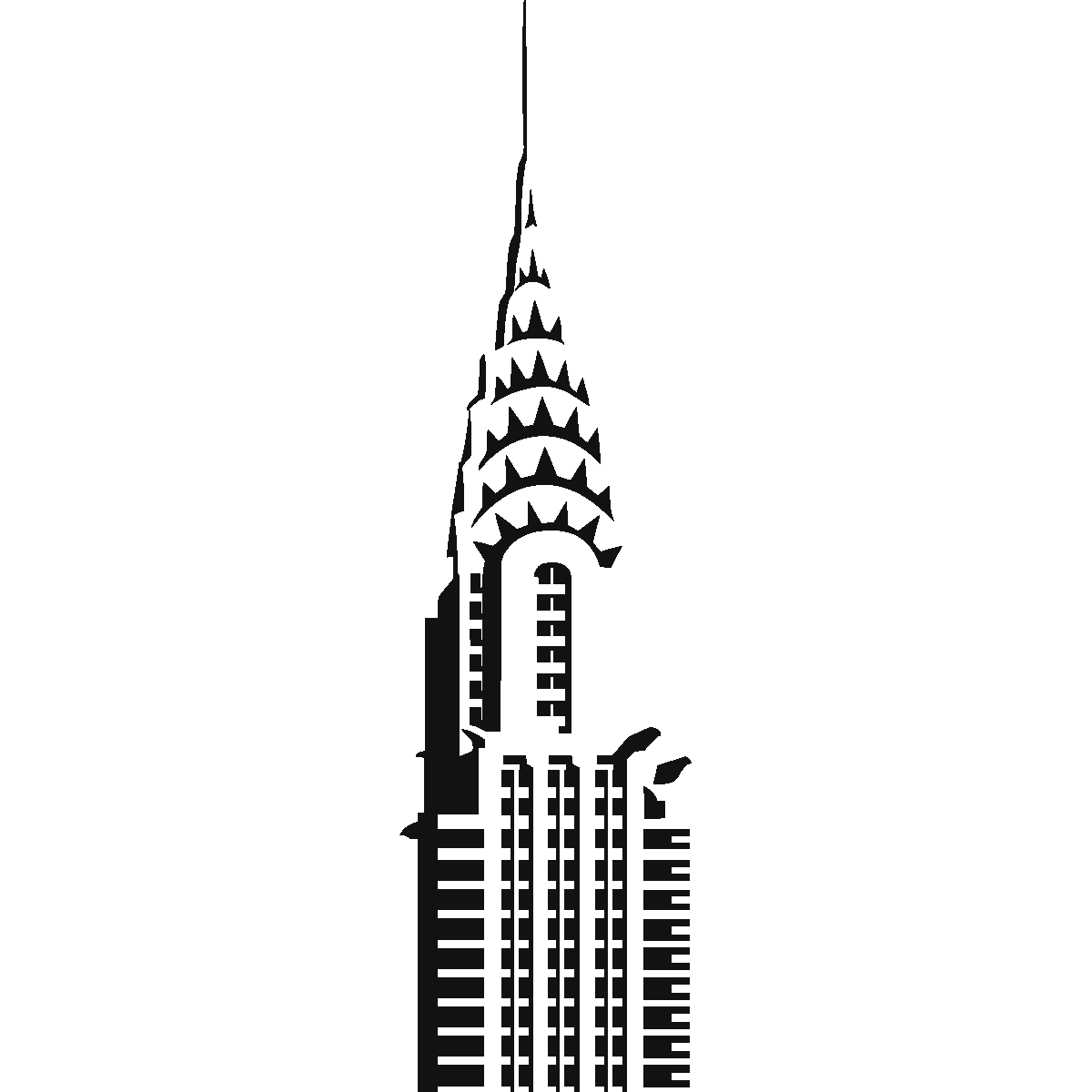}
    
    ~~~~~$\phi^4_2$ - pile of dirt ~~~~~~~~ ~~~~ $QCD$ - Everest ~~~~~$\mathcal{N}=4\,\, SYM$ - Chrysler building
\end{center}
\caption{To get closer to the Everest, we can start by climbing small hills, or by going up buildings. The latter may go higher at first, but it is not clear we learn much about alpinism.}
\end{figure}

\subsubsection*{The shape of theory space}

Apart from pure taste, our intuition about which approach should work best lie on expectations about the shape of theory space. The success of methods based on introducing extra structure depends on the density of these solvable points near physically motivated models. Further, we need to know how thick the well controlled zones around a solvable point are.

My expectation is that these solvable points are not close enough to interesting theories, or that they are too thin, to be useful quantitatively. For example in $1+1$ dimensions, integrable models without particle production cannot be close to theories where particle production is large! And it seems difficult to get closer to a strongly coupled theory with particle production by expanding with respect to a small parameter. We do not get anywhere close to a mountain by slightly perturbing a building.

The features that make a theory \emph{analytically} tractable are deeply not-generic, and this is why purely analytical methods are limited in what they can achieve. What we want, instead, is to exploit generic features of QFT, that make them \emph{numerically} manageable. As we will see, QFT ground states are locally entangled. This property is very special in the space of states, but generic for QFT ground states. This is the example of a generic feature that helps with numerical resolution, but there are, of course, possibly many others.

\begin{remark}[Where do bootstrap approaches stand?]
  In this crude classification of approaches, recent bootstrap methods (conformal bootstrap and S-matrix bootstrap) may seem like a peculiar in-between. Certainly, these approaches use structure, and in some cases extra structure, but I would claim this extra structure is generic enough. For example, the properties demanded of the S-matrix in the S-matrix bootstrap (unitarity, analyticity) are expected to hold for most relativistic QFT. In the conformal bootstrap, demanding conformal invariance is certainly asking for more than what a standard QFT comes with. However, one can still answer questions about generic QFT with the conformal bootstrap, because relativistic QFT are (almost always) conformally invariant at short distances. Further, as a referee of an earlier version of this manuscript wisely adds, conformal invariance can emerge at large distances, at a renormalization group fixed point. This is to be contrasted with other types of extra structure like supersymmetry, that typically need to be imposed at short distance to be present.
\end{remark}

\section{The variational method}
\setlength{\epigraphwidth}{0.55\textwidth}
\epigraph{I'd like to talk on some work I did on the variational principle in field theory. [...] I'm going to give away what I want to say, which is that I didn't get anywhere! I got very discouraged and I think I can see why the variational principle is not very useful. So I want to take, for the sake of argument, a very strong view - which is stronger than I really believe - and argue that it is no damn good at all!}{Richard Feynman\footnotemark}

\footnotetext{The quote can be found in the transcript of a talk given by Feynman in 1987 at a conference entitled ``\emph{Variational calculations in quantum field theory}''\cite{feynman1988}.}

\subsection{General idea and Feynman requirements\texorpdfstring{\protect\footnotemark}{}}
\footnotetext{This subsection and the next are a reworked version of an explanation that I first wrote in the B2 part of my ``\texttt{QFT.zip}'' ERC starting grant proposal.}

The idea of the variational method is to take seriously the fact that a Quantum Field Theory is a particular Quantum Theory, with a Hilbert space $\mathscr{H}$, a Hamiltonian operator $H$, etc.. Unfortunately, the Hilbert space of QFT is continuously infinite. Even if it is discretized with a lattice cutoff, it is still the Hilbert space of a quantum many-body system that grows exponentially with system size and thus quickly becomes unmanageable numerically. If we could \emph{compress} states in this Hilbert space down to a small (finite) number of parameters, then we would be in a much better situation. This is what we ultimately aim to do. In what follows I will restrict myself to the ground state search problem, but the method extends to the low lying spectrum~\cite{vanderstraeten2019spectrum,vanderstraeten2019tangentspace}, thermal states~\cite{verstraete2004thermal,kshetrimayum2019thermal}, and even dynamics~\cite{hackl2020}. 

For ground state problem, the variational method has two steps:
\vskip.3cm
\begin{minipage}{0.73\textwidth}
\begin{enumerate}
\item \textit{Compression} -- based on physical insights, \emph{guess} a finite dimensional submanifold $\mathcal{M} \subset \mathscr{H}$ of the Hilbert space (aka an ansatz wavefunction) that should contain (or be close to) the states of interest,
\item \textit{Minimization} -- minimize the energy\footnotemark (expectation value of the Hamiltonian $H$) over this submanifold, \eg through gradient descent, to find an approximation of the ground state
\begin{equation}
\ket{0} \simeq \argmin_{\ket{\psi}\in \mathcal{M}} \frac{\bra{\psi} H \ket{\psi}}{\bra{\psi}\psi\rangle} \, .
\end{equation}
\end{enumerate}
\end{minipage}
\begin{minipage}{0.26\textwidth}
\centering
\includegraphics[width=0.55\textwidth]{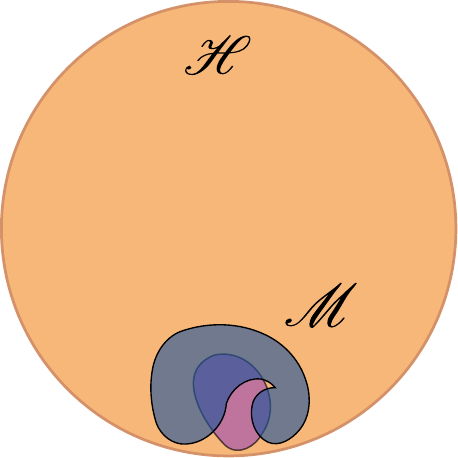}

~~

\includegraphics[width=0.65\textwidth]{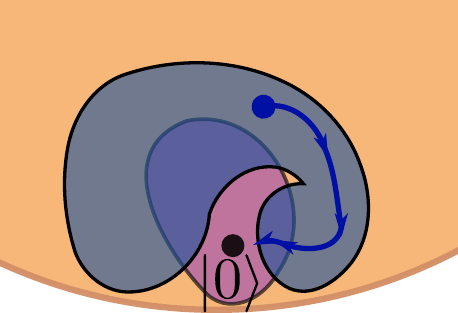}

~~
\end{minipage}
\footnotetext{Or, better, minimize the energy \emph{density}.} 

By increasing the dimension of $\mathcal{M}$, and provided the physical guess is good, one hopes to refine the approximation arbitrarily, and get as close as needed to the correct ground state. Note already that we allow in principle for a \emph{submanifold} $\mathcal{M}$ (not just a vector space $\mathscr{V}\subset \mathscr{H}$). This massively increases the expressiveness, but also makes the energy not quadratic as a function of the free parameters, and usually not even convex. As a result the optimization part is usually done with local techniques, without guarantee of global optimality. However, even if the minimization reaches a local minimum or gets stuck in a plateau, the quality of the state obtained can be rigorously evaluated: a state with lower energy is a better approximation to the ground state.

Although Feynman expressed his skepticism the variational method could be used in QFT~\cite{feynman1988}, he was constructive in his criticism. He put forward $3$ desirable properties for the submanifold $\mathcal{M}$. I reformulate them in a modern way here:
\begin{enumerate}
  \item \textit{Extensiveness} -- as the size $L$ of the system grows, or the number of particle $N$ grows, the size of the Hilbert space $\mathscr{H}$ grows exponentially. The submanifold $\mathcal{M}$ should however be parameterized extensively, that is its dimension should grow at most linearly as the system grows (or at worse polynomially)
\item \textit{Computability} -- for a state $\ket{\psi}\in\mathcal{M}$, there should be an efficient method to compute expectation values $\bra{\psi}\mathcal{O}\ket{\psi}$ that does not require explicitly summing over the (exponentially or infinitely many) states in the Hilbert space basis. This is required if only to compute the energy and minimize it.
\item \textit{Short distance regularity (for relativistic QFT)} -- the minimization problem should be well defined, and minimizing a (possibly divergent) energy density should not yield a runaway behavior where the approximation at physically relevant length-scales degrades through the optimization.
\end{enumerate}
The last requirement is the most technical and subtle, and we will discuss it more thoroughly in \ref{sec:relativistic_difficulties}. It is related to the singular short distance behavior of relativistic quantum field theories and is important only in this specific context.

Feynman's conditions should be seen as an ideal to thrive for, and are not strictly \emph{necessary} for the variational method to be usable at all. In fact, many popular ansatz wave-functions do not verify these requirements. Hamiltonian truncation~\cite{james2018hamiltonian_truncation,eliasmiro2016,eliasmiro2017-1,eliasmiro2017-2,rychkov2015}, which we will discuss later, uses as submanifold a \emph{vector space} which makes the minimization step trivial (it is a linear eigenvalue problem) but breaks extensiveness (the cost of the method is $\propto \e^{LE_T}$ where $E_T$ is a truncation energy). The recently introduced neural network wave-functions~\cite{carleo2017neural,choo2020fermionic} are sparsely parameterized, but break the strong form of the second requirement of computability: their expectation values have to be evaluated by carrying the sum over states approximately with Monte Carlo. Both methods can still be efficient for a wide class of problems, \eg because numerical prefactors are favorable despite prohibitive asymptotics or because importance sampling is efficient in many situations of interest. 

However, \emph{if it is possible for the problem considered}, it is clearly better to fit the $3$ requirements above: the method will then typically be better behaved and reliable (because of computability), not requiring a finite size cutoff (because of extensiveness) and not requiring a short  distance cutoff (because of regularity). As I will argue, at least in some cases (low dimension, sufficiently relevant operators), this seems feasible.

\subsection{Solving the IR: Tensor network states}

\subsubsection*{Intuition for compression}
Tensor network states provide a submanifold fitting precisely the first two requirements of Feynman on the lattice (where the third is irrelevant). Let us explain how a compression is possible. Certainly, generic states are not compressible, in the same way that a picture with random pixels cannot be efficiently compressed with JPEG. 

\begin{figure}[!h]
\floatbox[{\capbeside\thisfloatsetup{capbesideposition={right,top},capbesidewidth=0.4\textwidth}}]{figure}[0.57\textwidth]
{\caption{\small Images with random pixels cannot be compressed because they are generic. Images of cats can because they are not.}\label{fig:jpg}}
{\includegraphics[width=0.225\textwidth]{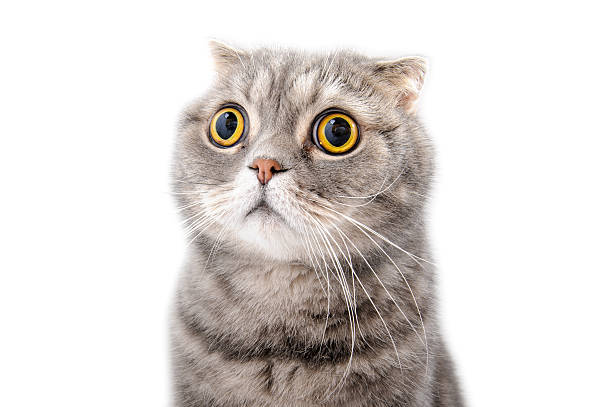}~~~~~~
    \includegraphics[width=0.145\textwidth]{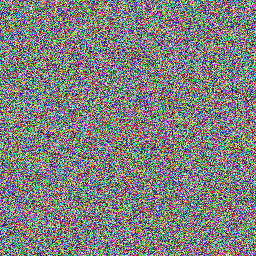}}
\end{figure}

There is a compression possible because the states we target are \emph{highly unusual}, they are the low energy states of local Hamiltonians. Such states are much more locally entangled than typical states in the Hilbert space, and verify the \emph{area law}. More precisely, for a low energy state of a local Hamiltonian, the entanglement entropy $S_\mathcal{R}$ of a regular connected subregion $\mathcal{R}$ of space with the rest grows proportionally to the border of $\mathcal{R}$, \ie $S_\mathcal{R}\propto |\partial \mathcal{R}|$ (with additional logarithmic factors in the gapless case)~\cite{srednicki1993,wolf2008,eisert2010}. Typical states in the Hilbert space (\eg sampled with uniform measure over rays) verify instead a volume law, \ie $S_\mathcal{R}\propto |\mathcal{R}|$~\cite{page1993volume,dahlsten2014volume}. This gives the heuristic picture in Fig. \ref{fig:area_law} below. 

\begin{figure}[!h]
\floatbox[{\capbeside\thisfloatsetup{capbesideposition={right,top},capbesidewidth=0.4\textwidth}}]{figure}[0.57\textwidth]
{\caption{\small Heuristic representation of entanglement between a closed subregion of the lattice and the rest, for a low energy state of a local gapped Hamiltonian (left) and a typical random state (right).}\label{fig:area_law}}
{\includegraphics[width=0.25\textwidth]{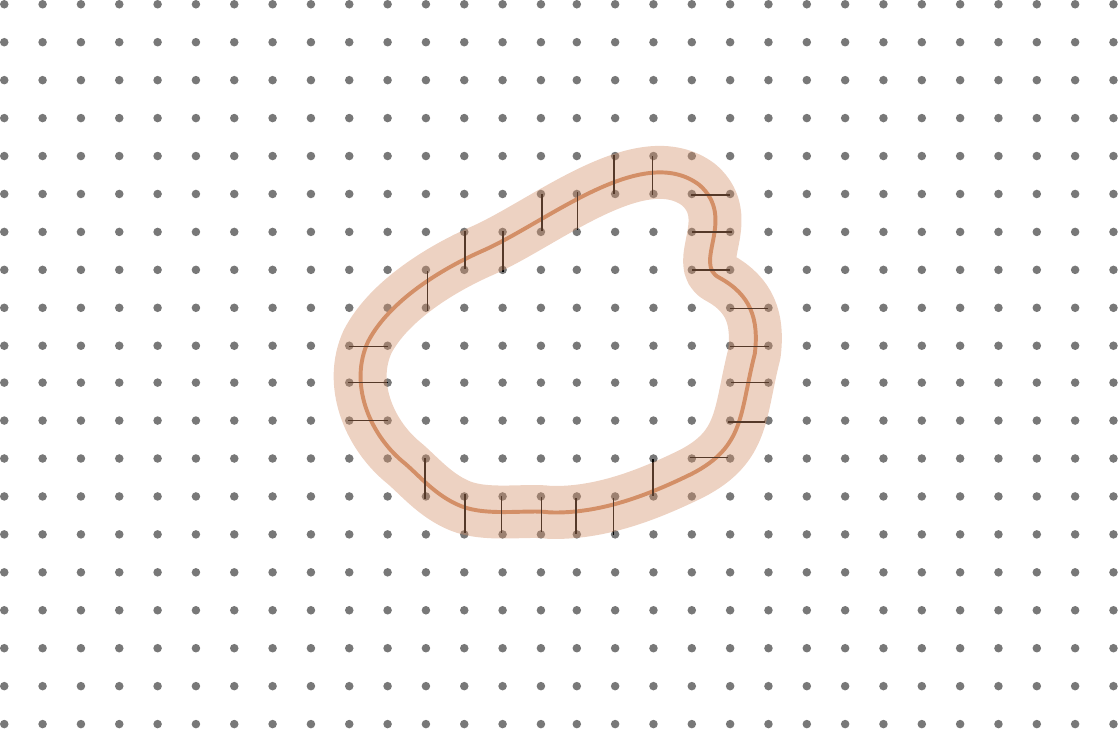}~~~~ \includegraphics[width=0.25\textwidth]{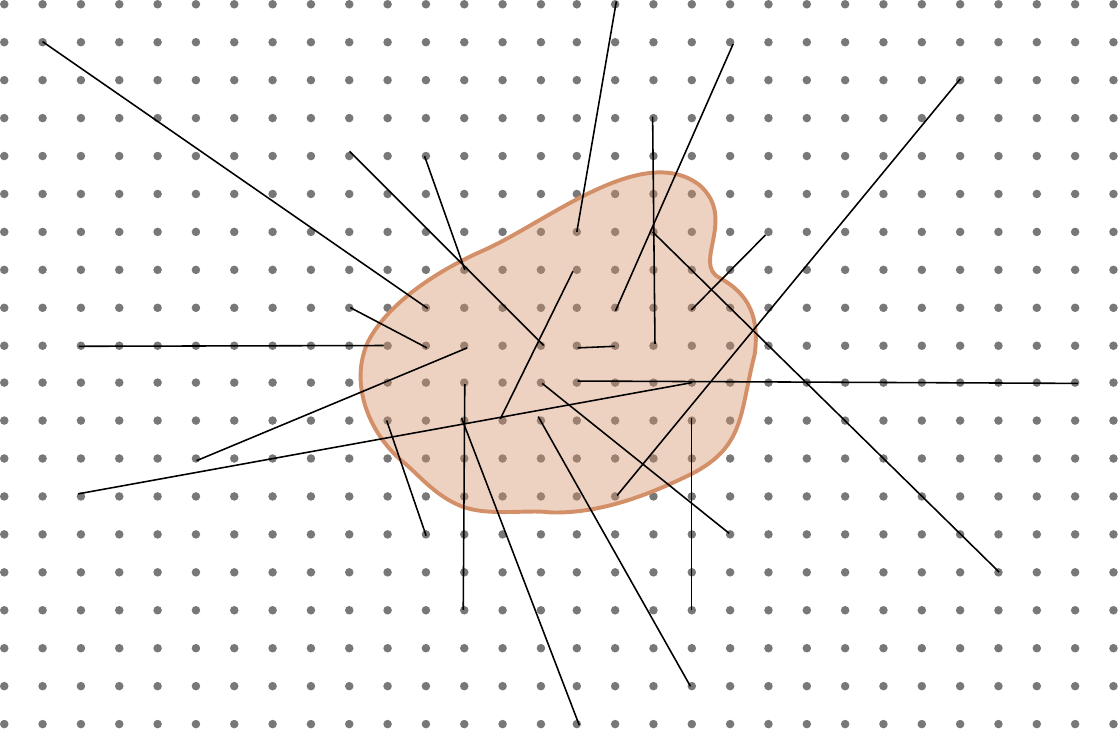}}
\end{figure}

Tensor network states leverage this crucial insight from quantum information theory to target directly this corner of the Hilbert space verifying the area law. Targeting this corner with a \emph{polynomial reduction in entanglement} enables a \emph{local} parameterization of states, and thus an \emph{exponential compression} of the number of parameters.

\subsubsection*{Matrix product states}
Concretely, tensor network states are obtained from a product of an extensive number of low rank tensors $A_{k_1 k_2 \dots k_r}^j$ with the indices $k$ contracted along the links of a given network~\cite{bridgeman2017review}. The contracted indices $k$ are called bond indices, and the number of values they can take is called the \emph{bond dimension}~$D$. The larger the bond dimension, the larger the submanifold dimension and number of adjustable parameters. The remaining free index $j$ is called physical, and corresponds the local Hilbert space basis states. 

One can illustrate how Feynman's requirements are met with a simple example on the line. Consider a chain of $N$ qudits (local Hilbert space $\mathbb{C}^d$). A generic quantum state for this chain is
\begin{equation}
\ket{\psi} = \sum_{j_1,j_2,\cdots, i_N = 1 }^d c_{j_1 j_2 \cdots j_N} \ket{j_1,j_2,\cdots,j_N} \, ,
\end{equation}
and thus requires $d^N$ complex coefficients $c_{j_1 j_2 \cdots j_N}$ to be specified (or equivalently a rank $N$ tensor $c$). An example of tensor network ansatz that is adapted if we are looking for low energy states of a gapped Hamiltonian is the matrix product states (MPS)~\cite{fannes1992}. They are obtained from the contraction of a rank $3$ tensor $A_{kl}^j$ (or equivalently, $d$ matrices $[A^1]_{kl}, [A^2]_{kl}$, \dots, $[A^d]_{kl}$):
\begin{equation}
\ket{A} = \sum_{j_1,j_2,\cdots, i_N=1}^d (L| A^{j_1} A^{j_2} \cdots A^{j_N} |R)\,  \ket{j_1,j_2,\cdots,j_N} \,
\end{equation}
where $(L|$ and $|R)$ are two arbitrary vectors $\in \mathbb{C}^D$. Here, the contraction of the lower indices of $A$ is given by the matrix product. At this stage it is more convenient to use a graphical notation where $A^{\blue{j}}_{\mathbf{\ora{k}}\mathbf{\ora{l}}}=\hbox{\includegraphics[scale=.65]{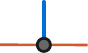}}$ and connected lines correspond to a contraction of indices, \ie $\ora{k}\, \vcenter{\hbox{\includegraphics[scale=.65]{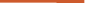}} } \, \ora{l} = \sum\delta_{\ora{k}\ora{l}}$. With this notation, the wave-function $c$, and by abuse of notation the state itself, read
\begin{equation}
\ket{A} \longleftrightarrow c_{\blue{j_1 j_2 \cdots j_N}} ={\hbox{\includegraphics[scale=.5]{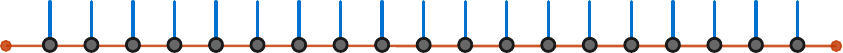}} }\, ,
\end{equation}
where the orange dots at the endpoints correspond to the final scalar product with the boundary vectors. The state parameterization is local. In the translation invariant case, the total number of parameters is independent of the system size $d\times D^2 \ll d^N$  (and grows only linearly $d\times D^2\times N$ in the general case). This solves Feynman's first requirement of extensive parameterization. One can easily show by a direct calculation that MPS verify the area law, and thus target the right corner of the Hilbert space. 

\subsubsection*{Computing observables}
Expectation values are also efficiently computable, for example the correlation function of two local observables $\mathcal{O}$ reads in graphical notation
\begin{equation}\label{eq:ladder}
\bra{A} \mathcal{O}(j_a)\mathcal{O}(j_b) \ket{A} = \vcenter{\hbox{\includegraphics[scale=.5]{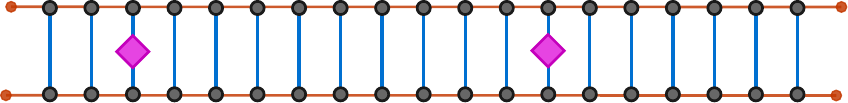}} } \, ,
\end{equation}
where the purple squares are the tensor notation for local $\mathcal{O}$ (matrices on physical indices, or rank 2 tensors). This ladder tensor network is contractible with a number of operations that does not grow exponentially with $N$, as it requires only the (iterated) multiplication of two different maps
\begin{equation}\label{eq:maps}
\Phi = \vcenter{\hbox{\includegraphics[scale=.5]{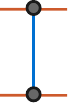}} } \hskip0.5cm \text{and} \hskip0.5cm \Phi_{\mathcal{O}} = \vcenter{\hbox{\includegraphics[scale=.5]{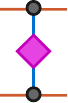}} } \, .
\end{equation}
This solves Feynman's second requirement of efficient computability. 

\subsubsection*{Expressive power beyond extensiveness}

One can further show that MPS give good approximations for ground states of local gapped Hamiltonians. More precisely, minimizing the energy density over all MPS of bond dimension $D$, one gets an approximation to the ground state with an error for local observables that decreases superpolynomially~\cite{verstraete2006groundstates,huang2015computing}, \ie that is $o(D^{-\alpha})$ for \emph{all} $\alpha>0$. Of course, the minimization itself may still be difficult, that is, finding this best MPS of a given bond dimension is not theoretically proved to be doable efficiently in general. However, existing heuristics like the density matrix renormalization group (DMRG, for finite systems with open boundaries)~\cite{white1993} or the variational uniform MPS algorithm (VUMPS, for infinite translation invariant systems) do converge for practical problems. Ultimately with MPS, one can usually solve ground state problems for gapped systems on the line to machine precision, only with a modest computational effort.

\subsubsection*{Other tensor network states}
In addition with MPS, the most important tensor network states that have been proposed are the projected entangled pair states (PEPS)~\cite{verstraete2004renormalization}, adapted to $2$ space dimensions and more, and the multiscale entanglement renormalization ansatz (MERA)~\cite{vidal2007,montangero2009,evenbly2013}, adapted to critical (gapless) systems in $1$ space dimension (see Fig. \ref{fig:mera_and_peps} below). 
\begin{figure}[!h]
\floatbox[{\capbeside\thisfloatsetup{capbesideposition={right,top},capbesidewidth=0.4\textwidth}}]{figure}[0.57\textwidth]
{\caption{\small PEPS on the left, adapted to gapped systems in $2$ space dimensions (or more), and MERA on the right, adapted to gapless systems in $1$ space dimensions. As before, the bond indices contracted over are shown in orange, while physical indices are shown in blue.}\label{fig:mera_and_peps}}
{\includegraphics[width=0.3\textwidth]{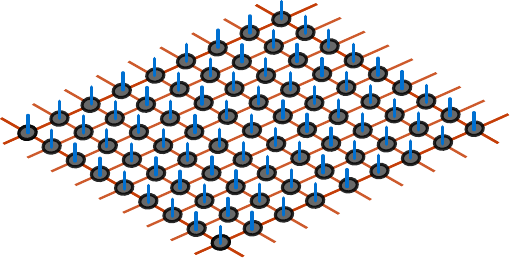}~~ \includegraphics[width=0.2\textwidth]{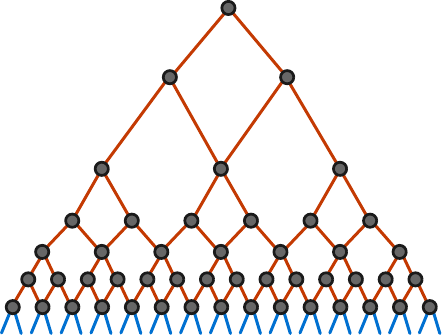}}
\end{figure}

PEPS also provide a local (extensive) parameterization of states, Feynman's first requirement is thus naturally satisfied. They also verify the area law. However computing expectation values is not as straightforward as for MPS with formula \eqref{eq:ladder}. The two dimensional equivalents of the maps $\Phi$ and $\Phi_\mathcal{O}$ in \eqref{eq:maps} have a size that still grows exponentially with $N$. Consequently, there exists no general purpose exact method of contraction~\cite{haferkamp2020pepscontraction}. We instead have efficient controllable approximation routines that converge fast for most problems~\cite{Ran2020}. Thus Feynman's second requirement is also satisfied for this ansatz, albeit in a weaker form.

\subsection{Solving the UV: Hamiltonian truncation}

Hamiltonian truncation attacks a problem orthogonal to that of tensor networks states: find the right Hilbert space on which to truncate a (typically singular) quantum field theory. In doing so, it solves Feynman's third requirement, and allows to use the variational approach for a continuum theory without excessive sensitivity to high frequencies. The approach has a long history, which, at least in QFT, starts with Brooks and Frautschi \cite{brooks1984}, and later Zamolochikov and Yurov~\cite{yurov1990truncated}.

\subsubsection*{Basics of truncation}
In Hamiltonian truncation, one starts from a decomposition of the total Hamiltonian into a free part (or at least exactly diagonalizable part, which could be a general CFT), and an interaction:
\begin{equation}
  H = H_0 + \lambda V \, .
\end{equation}
But unlike in perturbation theory, one does not assume $\lambda$ to be small. The idea is then to write $V$ as an infinite matrix in the basis that diagonalizes $H_0$, truncate it to some finite size, and finally find its lowest eigenvector.

The first difficulty is that in infinite space, the spectrum of $H$ is continuous and thus cannot be meaningfully discretized. One thus puts a large distance cutoff $L$, which makes the eigen-energies $E_0(i)$ of $H_0$ discrete. Writing $\ket{i}$ the corresponding eigenvectors, we can consider the Hilbert space spanned by all vectors of energy less than a cutoff $E_c$: $\mathscr{H}_{E_c} = \text{span} \{ \ket{i} ~ , ~ \bra{i} H_0 \ket{i} \leq E_c\}$.

To proceed with diagonalization in this finite dimensional Hilbert space, we need only that $V_{ij} := \bra{i} V \ket{j}$ is computable. This is generically the case, and allows to write the full truncated Hamiltonian $H^{E_c}_{ij} = E_0(i) \delta_{ij} + \lambda V_{ij}$. One ultimately finds the lowest eigenvector and corresponding energy either with a dense diagonalization, or with an iterative method like the Lanczos algorithm. As the cutoff $E_c$ is sent to $+\infty$, one expects to recover exact finite volume results (and then, later on, one should take the thermodynamic limit $L\rightarrow +\infty$).

\subsubsection*{Why it works: renormalization intuition}

The rigorous reason why Hamiltonian truncation works, which is a first step to understand how it can be systematically refined, is explained particularly clearly by Rychkov and Vitale with old fashioned perturbation theory arguments\cite{rychkov2015}. Intuitively, for a theory that is super-renormalizable, the interaction is strongly relevant, and thus has an infinitely small contribution when zooming \emph{in}. In an infinitely small volume, the physics is entirely dominated by $H_0$, and the total Hamiltonian must thus be almost diagonal in the basis of $H_0$. The interaction $V$ adds a contribution that is primarily at low energy, and can be captured by expanding in the low energy states of $H_0$.

As we will later see, there are subtleties, and Hamiltonian truncation works better the more strongly renormalizable the operator $V$ is.

\subsubsection*{Scalar field example}
In the example of a scalar field, the free Hamiltonan $H_0$ is (after expansion into normal modes and normal ordering)
\begin{equation}  
  H_0 = \frac{1}{2\pi} \int_\mathbb{R} \upd k \, \omega_k \; a_k^\dagger a_k\end{equation}
  where $[a_k,a_{k'}^\dagger] = 2\pi \delta(k-k')$ and $\omega_k = \sqrt{k^2+m^2}$. Putting the problem in finite volume $[0,L]$ (with periodic boundary conditions) instead of $\mathbb{R}$ makes $k$ discrete, with values $\e^{2i\pi n/L}, ~ n \in \mathbb{Z}$. The corresponding modes $a_n$ verify $[a_n,a_{n'}^\dagger] = \delta_{n,n'}$. The (still infinite dimensional) Hilbert space $\mathscr{H}_L$ is spanned by vectors of the form $\ket{n_1,m_1,\cdots, n_r,m_r} \propto a_{n_1}^{m_1\dagger} \cdots a_{n_r}^{m_r\dagger} \ket{0}$ where $\ket{0}$ is the Fock vacuum. There is a finite number of states with (free) energy $\langle n_1,m_1, \cdots n_r,m_r | H_0 |n_1,m_2, \cdots,n_r,m_r\rangle \leq E_c$. For a simple potential $V$, for example $\int \upd x :\! \phi^4(x) \!:$ or $- \int \upd x:\!\cos(\beta \phi(x))\! :$, one can systematically compute the matrix elements $V_{(n,m),(n',m')}$ using the canonical commutation relations. Finally diagonalizing the resulting $H_{(n,m),(n',m')}$ gives a quite accurate approximation of the finite volume spectrum, provided we did not take $L$ too large nor $E_c$ too small.

\begin{remark}[The cost of plain Hamiltonian truncation]
Taking larger $E_c$ increases the number of available states exponentially. However, one can show that, on the example of $V\propto\phi^4$, the error on local observables scales only like $1/E_c^2$. It can be improved to $1/E_c^3$ or $1/E_c^4$, with renormalization group refinements, but remains a power law. It is thus a method that is \emph{a priori} not efficient for asymptotic precision (it does not improve much if you use a lot more computing resources).

Taking larger $L$ while keeping the error approximately fixed also increases the number of states exponentially. This is because there are more modes around the same momenta when $L$ is larger, and also because $E_c$ needs to be increased linearly to keep a fixed density. In this respect, Hamiltonian truncation really behaves like exact diagonalization on the lattice. Studying the thermodynamic limit (and \eg phase transitions) is \emph{a priori} difficult.

Despite these two asymptotic difficulties, Hamiltonian truncation estimates are often very good, and are a tough benchmark to beat! Especially for massive models, one can take $L$ of the order of a few inverse masses only. In this regime, and with renormalization group refinements, the error $\propto 1/E_c^\ell$ can be already quite small for the reachable $E_c$ (although, again, it would be difficult to substantially improve by throwing more computing power at it).
\end{remark}

Hamiltonian truncation is a useful numerical tool, but above all it is a great starting point for other numerical methods. Its main value lies in showing how to bypass Feynman's third objection. The exponential cost it brings is only an artifact of its lack of extensiveness. But we already learned how to solve this problem on the lattice. In the next chapter, we will explain how to put the things we have learned together, and construct an ansatz that fits all of Feynman's requirements in $1+1$ dimensions.

\chapter{Relativistic continuous matrix product states}

\section[Continuous matrix product states]{Continuous matrix product states\protect\footnotemark}
\epigraph{I think I can safely say that nobody understands continuous matrix product states.}{Frank Verstraete -- (approximate wording)}

\footnotetext{This section is a shortened and corrected version of unpublished lecture notes I wrote for the \textit{Tensor21} summer school in Barcelona in 2021.}

To apply matrix product states to QFT, one first needs to take their continuum limit, so that we can at least define states in the right Hilbert space (which is a continuous Fock space). This step is absolutely crucial, but is essentially orthogonal to the UV difficulties of relativistic theories (for which we will need Hamiltonian truncation intuition). To solve problems step by step we thus consider continuous matrix product states in the context of non-relativistic QFT first. They were introduced in 2010 in the breakthrough paper of Cirac and Verstraete~\cite{verstraete2010}, further developed by Haegeman and collaborators~\cite{haegeman2013calculus,haegeman2010relativistic} who were then followed by many others.

\subsection{Non-relativistic bosonic QFT}

Before defining the CMPS, it is helpful to say what it is an ansatz for: non-relativistic quantum field theories in $1$ space dimension. To avoid complications, we will stick to the bosonic case. The corresponding Hilbert space $\mathscr{H}$ is the symmetric Fock space $\mathcal{F}[L^2(I)]$ where $L^2$ is the space of square integrable functions on $I$, which is the space interval on which the field theory is defined (\eg $[0,L]$ or $\mathbb{R}$). Formally, the Fock space is simply the direct sum of $0$, $1$, $2$, $3$, .. to arbitrarily many (symmetric) particle states
\begin{equation}
    \mathcal{F}[L^2(I)] = \bigoplus_{n=0}^{+\infty} S_+ L^2(I)^{\otimes n} \, ,
\end{equation}
where $S_+$ means taking the symmetric subspace. More concretely, a state in this Hilbert space can be represented in second quantized form as
\begin{equation}\label{eq:Fock-state}
    \ket{\Psi} = \sum_{n=0}^{+\infty} \int_{I^{n}} \upd x_1\cdots \upd x_n \; \varphi_n(x_1,\cdots,x_n)\, \hpsi^\dagger(x_1) \cdots \hpsi^\dagger(x_n) \ket{0}_\psi \,.
\end{equation}
In this expression, $\hpsi^\dagger(x)$ is the bosonic creation operator in $x$ which verifies the canonical commutation relations $[\hpsi(x),\hpsi^\dagger(y)]=\delta(x-y)$. The state $\ket{0}_\psi$ is the Fock vacuum associated to $\hpsi$, \ie the state without particles, verifying $\forall x,\, \hpsi(x) \ket{0}_\psi=0$. The functions $\varphi_n(x_1,\cdots,x_n)$ are the (symmetric) $n$-particle wavefunctions. Note that the sum in \eqref{eq:Fock-state} is allowed to be infinite, but a rather strong constraint is that the state has to be normalizable, \ie $\bra{\psi}\psi\rangle<+\infty$.

It helps to have in mind a prototypical Hamiltonian that naturally lives in such
a Hilbert space. A good example is the Lieb-Liniger Hamiltonian, aka Bose gas with delta interaction 
\begin{equation}\label{eq:H_LL}
    H_\text{LL} = \int_I \partial_x\hpsi^\dagger \partial_x \hpsi + 
    c\, \hpsi^\dagger \hpsi^\dagger \hpsi\hpsi  - \mu \hpsi^\dagger\hpsi\, ,
\end{equation}
which happens to be exactly solvable, and thus provides a convenient benchmark.
The first term is the kinetic term, the second is a repulsive contact interaction,
while the last one corresponds to a chemical potential forcing a non-zero particle 
density (otherwise the ground state would be empty).
More generally, one could consider higher order $3$ point interactions 
$\propto\hpsi^{\dagger^3}\hpsi^3$, interactions extending over some range 
$\propto\int \upd y\, V(x-y) \hpsi^\dagger(x)\hpsi^\dagger(y)\hpsi(x)\hpsi(y)$, or
even terms breaking particle conservation \eg $\propto\hpsi^{\dagger 2} + \hpsi^2$
which would no longer yield an integrable Hamiltonian
but for which the methods we will discuss should apply. However, in this section, one should keep the kinetic term $\partial_x\hpsi^\dagger \partial_x\hpsi$, which is the second quantized rewriting of the standard Schr\"odinger Hamiltonian (with $\hbar^2/(2m)=1$), and that is necessary to get a physically reasonable non-relativistic model.

\subsection{Definition}
Let us consider a translation invariant QFT model for simplicity. Let $Q$ and $R$ be two arbitrary $D\times D$ complex matrices. A continuous matrix product states (CMPS) is defined as
\begin{equation}\label{eq:cmps}
    \ket{Q,R} = \tr \left\{\mathcal{P} \exp\left[ \int_I \upd x \, Q \otimes \mathds{1} + R \otimes \hpsi^\dagger(x)\right] \right\} \ket{0}_\psi
\end{equation}
where
$\mathcal{P}$ is the path ordering operator, which puts largest arguments on the right\footnote{In this convention, the path-ordering operator $\mathcal{P}$ gives a reversed order compared to the one obtained with the standard time-ordering operator. It is however merely a matter of convention, which allows to have the expression of wave-functions and correlation functions ordered with arguments increasing from left to right, mimicking the physical ordering.} \ie
\begin{equation}
    \mathcal{P} \mathcal{O}(x_1) \mathcal{O}(x_2) = \left\{ \begin{array}{r}
         \mathcal{O}(x_1) \mathcal{O}(x_2) \; \text{if} \; x_2> x_1  \\
         \mathcal{O}(x_2) \mathcal{O}(x_1) \; \text{if} \; x_2< x_1
    \end{array} \right.
\end{equation}
The trace in \eqref{eq:cmps} is taken over the auxiliary $D\times D$ matrix space and $D$ is the bond dimension, which is \emph{really} the direct analog of the bond dimension of MPS. The two matrices $Q$ and $R$ contain the free parameters of the state, that need to be adjusted \eg to find the ground state of a given Hamiltonian. Note that the state produced at least formally belongs to the appropriate field theory Hilbert space: 
$\ket{Q,R} = \tr [ U_{0,L} ] \ket{0}_\psi$, where $U_{0,L}$ is an operator acting on $\mathcal{F}[L^2(I)]\otimes \mathbb{C}^D$. Taking a trace gives an operator acting on the Fock space, and thus acting on $\ket{0}_\psi$, we ultimately get a state in the Fock space. To make this more explicit, we can write the CMPS \eqref{eq:cmps} in the wave-function representation of \eqref{eq:Fock-state}. 
\begin{equation*}
  \ket{Q,R} = \sum_{n=0}^{+\infty} \int_{0< x_1<x_2<\cdots < x_n<L} \hskip-2cm\upd x_1\,\upd x_2\cdots \upd x_n \; \varphi_n(x_1,\cdots,x_n)\, \hpsi^\dagger(x_1) \cdots \hpsi^\dagger(x_n) \ket{0}_\psi
\end{equation*}
with
\begin{equation}\label{eq:cmps-wavefunction}
    \varphi_n(x_1,\cdots,x_n) = \tr\left[\e^{Q x_1} R \e^{Q(x_2-x_1)} R \cdots \e^{Q(x_n-x_{n-1})} R\e^{Q (L-x_n)}\right]
\end{equation}
This can be seen as an alternative definition of CMPS. This representation has the advantage of being more explicit in terms of the decomposition of $\mathcal{F}[L^2(I)]$ into a direct sum, but makes the connection with matrix product states less transparent.

A striking aspect of the CMPS ansatz is that it requires \emph{only two} $D\times D$ matrices (at least in the translation invariant case). For matrix product states, the number of matrices is given by the local physical dimension (\eg \, $2$ for spin $1/2$, $3$ for spin $1$, etc.), which is technically infinite for a chain of bosons. In the discrete, this requires artificially truncating at (potentially very large) local physical dimensions. This is the illustration of the drastic simplification (or compression) brought by the continuum limit for bosonic systems\footnote{In fact, in the continuum,  the bosonic Fock space is not bigger than the fermionic Fock space.}. In comparison with discrete Fock spaces, the continuum Fock space is almost empty. As a result, double occupation in an elementary volume $\upd x$ need not be considered. We will see this more clearly when considering the connection with the discrete.

Note finally that to simplify we considered a translation invariant state with periodic boundary conditions in \eqref{eq:cmps}. Both can be relaxed, exactly as with discrete MPS: one can make $Q$ and $R$ depend on $x$ and replace the trace by boundary vectors $\in \mathbb{C}^D$
\begin{equation}\label{eq:cmps-generalized}
  \ket{Q,R} = (\ell\, |\, \mathcal{P} \exp\left[ \int_I \upd x \, Q(x) \otimes \mathds{1} + R(x) \otimes \hpsi^\dagger(x)\right] \, |r)\; \ket{0}_\psi\,.
\end{equation}
Technically, the number of free parameters becomes infinite (a set $Q(x),R(x)$ per point of space $x$) and one needs to choose a functional ansatz for $Q(x)$ and $R(x)$ (\eg a finite sum of Hermite functions, a finite Fourier sum or a spline) to recover a finite number of parameters that fit in a computer.

\subsection{From CMPS back to MPS}

The easiest way to connect CMPS to their discrete MPS counterpart is simply to cut space into small intervals. Let us do it in a pedestrian yet reasonably rigorous way. Let $\varepsilon\ll L$ be the lattice spacing, $x_j=j\varepsilon$ for $0\leq j \leq L/\varepsilon=N$ the discretized positions. The path-ordered exponential can be approximated by a finite product when $\varepsilon$ is small enough:
\begin{equation}
    \ket{Q,R} = \tr\left\{\prod_{j=0}^{L/\varepsilon-1} \exp\left[\varepsilon Q + R \int_{j\varepsilon}^{(j+1)\varepsilon} \hpsi^\dagger(x)\right] \right\}\ket{0}_\psi+O(\varepsilon)\, .
\end{equation}
We now introduce a new discrete (or coarse grained) bosonic operator $b_j$:
\begin{equation}\label{eq:def_coarse_graining}
    b_j:= \frac{1}{\sqrt{\varepsilon}} \int_{j\varepsilon}^{(j+1)\varepsilon} \hpsi(x) \, .
\end{equation}
Let us verify its commutation relations. Clearly, $\forall j\neq k, \; [b_j,b_k^\dagger]=0$. For $k=j$,
\begin{align}
    [b_j,b_j^\dagger] &= \frac{1}{\varepsilon}\int_{j\varepsilon}^{(j+1)\varepsilon} \upd x \int_{j\varepsilon}^{(j+1)\varepsilon} \upd y \, [\hpsi(x),\hpsi^\dagger(y)]\\
    &= \frac{1}{\varepsilon} \int_{j\varepsilon}^{(j+1)\varepsilon} \upd x \int_{j\varepsilon}^{(j+1)\varepsilon} \upd y \, \delta(x-y)\\
    &= \frac{1}{\varepsilon}\times \varepsilon = 1 
\end{align}
Hence $[b_j,b_k^\dagger]=\delta_{j,k}$. Note the sneaky factor $\sqrt{\varepsilon}$ in \eqref{eq:def_coarse_graining} that is crucial to go from the continuum Dirac $\delta$ to the discrete Kronecker $\delta$. We now have
\begin{equation}\label{eq:cmps_bj}
    \ket{Q,R}=\tr\left\{\prod_{j=1}^{L/\varepsilon} \exp\left[\varepsilon Q + \sqrt{\varepsilon} R\, b^\dagger_j \right] \right\}\ket{0}_\psi + O(\varepsilon)
\end{equation}
This is exactly an MPS. Indeed in the (un-normalized) number basis $\ket{n_1,n_2,\cdots,n_{N}}  = b_1^{\dagger n_1} b_2^{\dagger n_2}\cdots b_{N}^{\dagger n_N} \ket{0}_\psi$ we have
\begin{equation}
    \ket{Q,R} = \sum_{n_1,n_2,\cdots,n_N \in \mathbb{N}} \tr[A_{n_1} A_{n_2}\cdots A_{n_n}] \ket{n_1,n_2,\cdots n_N} + O(\varepsilon)
\end{equation}
where the $D\times D$ complex matrices $A_n$ are obtained (at leading non-trivial order in $\varepsilon$) by expanding the exponential in \eqref{eq:cmps_bj} 
\begin{equation}\label{eq:MPS_from_CMPS}
    \begin{array}{l}
         A_0 = \mathds{1} + \varepsilon Q  \\
         A_1=\sqrt{\varepsilon} R\\
         A_2=\varepsilon R^2/2 \\
         A_n=o(\varepsilon) \; \text{for} \; n\geq 3
    \end{array}
\end{equation}
Although the $A_n$ for $n\geq 3$ can in principle be computed, they all give subleading contributions. In fact, our starting point \eqref{eq:def_coarse_graining} [the discretization of the path-ordered exponential] is accurate only to leading order in $\varepsilon$, and thus one would need to be more careful to get the higher order terms.

\subsection{Computing observables}
We saw previously in \eqref{eq:cmps-wavefunction} that we could very explicitly write down the CMPS ``wavefunction'', that is, its decomposition in the natural Fock space basis. However, in general, having an explicit expression for a state does not imply we can have compact expressions for expectation values of local operators. Indeed, the state lives in an infinite dimensional Hilbert space, and computing expectation values a priori requires summing over infinitely many basis elements. With matrix product states, we know there are efficient routines to evaluate expectation values that do not require the explicit full summation (finite in this context, but still exponential in the system size). Indeed, we just had to contract a ladder from left to right \eqref{eq:ladder}. In the continuum, the same fortunately remains true, and evaluating expectation values remains cheap (with a polynomial cost in $D$).

\subsubsection*{The norm}
Before computing expectation values of local observables, let us consider the simplest expectation value, that of the identity (aka. the norm of $\ket{Q,R}$). For a translation invariant CMPS on the interval $I=[0,L]$:
\begin{equation}\label{eq:normCMPS}
    \bra{Q,R}Q,R\rangle=\tr_{\mathbb{C}^D\otimes \mathbb{C}^D}\left[\exp (L \mathbb{T})\right]
\end{equation}
with the transfer operator $\mathbb{T}$
\begin{equation}\label{eq:transferOperator}
    \mathbb{T} = Q\otimes \mathds{1} + \mathds{1} \otimes Q^* + R\otimes R^*
\end{equation}
The generalization to space dependent $Q,R$ and a non-trivial open boundary conditions [\ie $\tr[\cdot]$ replaced with $(\ell\,| \cdot |r)$] is straightforward. 

This result is not difficult to prove, and I detail $3$ options in appendix \ref{app:cmps_expect}. The easiest approach is just to discretize, reuse the MPS machinery, and finally take $\varepsilon \rightarrow 0$. The map $\e^{\varepsilon \mathbb{T}}$ appears simply as the equivalent of the discrete map $\Phi$ of eq. \eqref{eq:maps}. As always with CMPS one can also completely forget the discrete origin of the state, and carry all computations in the continuum, which provides the $2$ other proofs.

From all the possible proofs, it is clear that the result generalizes immediately to the computation of arbitrary \emph{overlaps} of CMPS $\bra{Q_1,R_1} Q_2,R_2\rangle$. One simply needs to formally substitute $Q^*,R^* \rightarrow Q_1^*,R_1^*$ and $Q,R\rightarrow Q_2,R_2$ in \eqref{eq:transferOperator}.

\subsubsection*{Normal-ordered correlation functions}
Computing general correlation functions is easy once one understands the computation of the norm. The main tool is the generating functional of normal ordered correlation functions:
\begin{equation}
    \mathcal{Z}_{j',j} = \frac{1}{\bra{Q,R} Q,R\rangle} \bra{Q,R} \exp\left[\int_I j'(x)\hpsi^\dagger(x)\right] \exp\left[\int_I j(x) \hpsi(x)\right] \ket{Q,R}\,.
\end{equation}
Clearly, differentiating this function with respect to $j,j'$ gives us access to all normal-ordered correlation functions, \eg 
\begin{equation}
    \langle \hpsi^\dagger(x)\hpsi(y)\rangle_{Q,R} = \frac{\delta}{\delta j'(x)} \frac{\delta}{\delta j(y)} \mathcal{Z}_{j',j} \bigg|_{j,j'=0}
\end{equation}

The generating functional admits the exact expression
\begin{equation}\label{eq:exact_Z}
    \mathcal{Z}_{j',j} = \frac{1}{\tr[\e^{L\mathbb{T}}]}\; \tr \left[\mathcal{P}\exp \left(\int_I \mathbb{T}_{j',j}\right)\right]
\end{equation}
where
\begin{equation}
    \mathbb{T}_{j',j}(x) = \mathbb{T} + j(x) R\otimes \mathds{1} + j'(x) \mathds{1}\otimes R^*
\end{equation}

This expression is easy to derive using the formula for the overlap of two CMPS. The first step is to rewrite $\mathcal{Z}_{j',j}$ by commuting the two exponentials containing $j'$ and $j$. This is done easily using the Baker-Campbell-Hausdorff formula which gives, for $[X,Y]\propto \mathds{1}$, $\e^X\e^Y=\e^{[X,Y]}\e^Y\e^X$:
\begin{align}
    \mathcal{Z}_{j',j} &=\frac{1}{\bra{Q,R} Q,R\rangle} \bra{Q,R} \exp\left[\int_I j(x)\hpsi(x)\right] \exp\left[\int_I j'(x) \hpsi^\dagger(x)\right] \ket{Q,R} \exp\left(- \int_I j'j\right)\\
    &=\frac{\bra{Q,R+j^*} Q,R+j'\rangle}{\bra{Q,R} Q,R\rangle}\exp\left(- \int_I j'j\right)\\
    &=\frac{1}{\tr[\e^{L\mathbb{T}}]}\; \tr \left[\mathcal{P}\exp \left(\int_I \mathbb{T}_{j',j}\right)\right]\,, 
\end{align}
where the exponential of $j'j$ exactly cancels the quadratic term appearing in the overlap.

Now, from a straightforward functional differentiation of \eqref{eq:exact_Z}, we get that for $0<x_1<x_2<\cdots<x_n<L$:
\begin{equation}\label{eq:exact_correlations}
    \langle :\hpsi^{\delta_1}(x_1) \hpsi^{\delta_2}(x_2)\cdots \hpsi^{\delta_n}_n(x_n):\rangle_{Q,R}= \frac{1}{\tr[\e^{L\mathbb{T}}]} \tr \left[ \e^{x_1\mathbb{T}} R^{(\delta_1)}\e^{(x_2-x_1)\mathbb{T}}\cdots R^{(\delta_n)} \e^{(L-x_n)\mathbb{T}} \right]
\end{equation}
where $\delta_i$ can be either nothing or $\dagger$ and
\begin{align}
    R^{()} &= R\otimes \mathds{1}\\
    R^{(\dagger)} &= \mathds{1} \otimes R^*
\end{align}
For example, this formula gives the two point function explicitly:
\begin{equation}\label{eq:twopoint_cmps}
    \langle \hpsi^\dagger(x)\hpsi(y) \rangle_{Q,R} = \frac{1}{\tr[\e^{L\mathbb{T}}]} \tr \left[ \e^{x\mathbb{T}} (\mathds{1}\otimes R^*)\e^{(y-x)\mathbb{T}} (R\otimes\mathds{1}) \e^{(L-y)\mathbb{T}} \right]\,.
\end{equation}
Formula \eqref{eq:exact_correlations} shows that, in complete analogy with MPS, correlation functions for CMPS are always exponentially decreasing (at least as long as the matrices $Q,R$ are finite dimensional).

\subsubsection*{Simplifications: exploiting gauge freedom and taking the thermodynamic limit}

The first thing to notice is that the norm of the CMPS looks ill-behaved in the thermodynamic limit, and scales $\propto (\ell_1|r_1)\exp(\lambda_1 L)$ where $\lambda_1$, $|r_1)$, $|l_1)$ are the eigenvalue with largest real part of $\mathbb{T}$ and its associated right and left eigenvectors. We can cancel this behavior, without loss of generality, by simply substituting $Q\rightarrow Q-\lambda_1 \mathds{1}$. With this new choice, the leading eigenvalue of $\mathbb{T}$ is $0$ and all the other ones have strictly negative real part. Diagonalizing $\mathbb{T}$ one gets:
\begin{equation}
    \e^{L\mathbb{T}} = \sum_{j=1}^{D^2} \e^{\lambda_j} |r_j)(\ell_j| \underset{L\rightarrow \infty}{\longrightarrow} |r_1)(\ell_1|
\end{equation}
and thus in the thermodynamic limit, the exponential of the transfer matrix becomes a simple rank-$1$ projector.

The second step, analogous to what is done with MPS, is to go to a super-operator representation of the transfer operator. In a nutshell we are mapping the tensor-product vector space $\mathbb{C}^D\otimes \mathbb{C}^D$ to the space of matrices $\mathcal{M}_D(\mathbb{C})$. We now look at a vector $|v)$ on which $\mathbb{T}$ acts as a matrix
\begin{equation}
    |v) = \sum_{k,l} v_{k,l} \ket{k}\otimes \ket{l} \rightarrow v = \sum_{k,l} v_{k,l} \ket{k}\bra{l} \, .
\end{equation}
With this mapping, we can introduce the super-operator $\mathcal{L}$ which reproduces the action of $\mathbb{T}$ on $v$ now written as a matrix
\begin{equation}\label{eq:T_to_Lindblad}
\mathbb{T} |v) \rightarrow \mathcal{L}\cdot v = Qv +vQ^\dagger + R v R^\dagger \,.
\end{equation}
\begin{remark}[Scaling]An obvious reason why this rewriting is advantageous is that it reduces the complexity of the operations we have to carry. Indeed, multiplying operators acting on $\mathbb{C}^D\otimes \mathbb{C}^D$, which is what is needed to compute correlation functions, naively scales like $D^6$ whereas composing superoperators acting simply on the left and right of a $D\times D$ matrix scales like $D^3$.
\end{remark}

The third step, like in the discrete, is to note that there is a lot of redundancy in the parameterization of the CMPS. In particular, it is straightforward to see that conjugating the matrices $Q,R$ with an invertible matrix $U$ does not change the state
\begin{equation}
    \ket{U^{-1}QU,U^{-1}RU} = \ket{Q,R}\,.
\end{equation}
This can be seen either in the original definition of the CMPS \eqref{eq:cmps} or in its wave-function representation \eqref{eq:cmps-wavefunction}. This can be exploited to fix properties of $Q$ and $R$, to simplify computations without losing expressiveness. A particularly convenient choice is the so called \emph{left canonical form} which is obtained by taking $U=\ell_1$ where $(\ell_1|$ is the leading left eigenvector of $\mathbb{T}$. By definition, this matrix verifies
\begin{equation} 
\ell_1 Q +  Q^\dagger \ell_1 + R^\dagger \ell_1 R = 0
\end{equation}
Now taking $Q_\ell = C Q C^{-1}$ and $R_\ell = C R C^{-1}$ where $\ell_1=C^\dagger C$ we get
\begin{equation}\label{eq:left-canonical}
    Q_\ell + Q_\ell^\dagger + R_\ell^\dagger R_\ell = 0 \, .
\end{equation}
This implies that the identity matrix, once vectorized, is a left eigenvector of $\mathbb{T}$. Equivalently, this implies that $\mathcal{L}$ \eqref{eq:T_to_Lindblad} is of the Lindblad form. In practice, one can choose, without loss of generality, matrices verifying \eqref{eq:left-canonical} from the beginning. Writing $Q_\ell=-i K - R_\ell^\dagger R_\ell/2$, equation \eqref{eq:left-canonical} is equivalent to $K$ being self-adjoint. One can thus parameterize the CMPS directly with $K$ self-adjoint and $R_\ell$.

Finally, all these manipulations allow to rewrite correlation functions in the thermodynamic limit in a simpler way. For $-\infty<x_1<x_2<\cdots<x_n<+\infty$:
\begin{equation}
\langle :\hpsi^{\delta_1}(x_1) \hpsi^{\delta_2}(x_2)\cdots \hpsi^{\delta_n}_n(x_n):\rangle_{Q,R} = \tr \left[R^{(\delta_1)}\cdot \e^{(x_2-x_1)  \mathcal{L}}  \cdot R^{(\delta_2)}\cdots \e^{(x_{n}-x_{n-1}) \mathcal{L}} \cdot R^{(\delta_n)} \cdot\rho_0\right]
\end{equation}
where $\rho_0 := \ell_1 $ is the fixed point of $\mathcal{L}$, \ie \, $\mathcal{L}\cdot \rho_0=0$ normalized to $\tr[\rho_0]=1$, $R^{(\dagger)} \cdot \rho:=\rho R^\dagger $ and $R^{()}\cdot \rho = R\rho$.
This general formula gives \eg the very compact expression for the particle density:
\begin{equation}
    \langle \hpsi^\dagger(x)\hpsi(x) \rangle_{Q,R} = \tr[R\rho_0 R^\dagger] \,  ,
\end{equation}
where $\rho_0$ is a function of $Q,R$.

\subsection{Optimization}
Using the techniques of the previous section, one can get similar expressions for the expectation values of all the natural local operators that could enter in a Hamiltonian density. For example:
\begin{align}
  \langle \hpsi^{\dagger^2}\hpsi^2\rangle_{Q,R} &= \tr[R^{\dagger 2}R^2 \rho_0] \,\\
  \langle \partial_x \hpsi^\dagger \partial_x\hpsi\rangle_{Q,R} &= \tr([Q,R]^\dagger [Q,R]\rho_0) \, .
\end{align}
We thus have an explicit function (computable at cost $\propto D^3$) $\langle h \rangle_{Q,R} = f(Q,R)$ for any local Hamiltonian density. In principle, this is everything one needs. One could just put $f$ in a black-box solver and minimize it over $Q,R$ to find an approximate ground state.

For the optimization to be efficient, we need to work a bit more. We will discuss the $2$ ingredients needed (adjoint differentiation and Riemannian optimization) directly when we extend CMPS to the relativistic regime. For the plain CMPS we discussed in this section, there is fortunately a Julia package that already implement all these tricks: \texttt{CMPSKit}\footnote{\url{https://github.com/Jutho/CMPSKit.jl}}, developed by our colleagues at the university of Ghent.

\section{Relativistic difficulties}\label{sec:relativistic_difficulties}
Continuous matrix product states are an important part of the puzzle, but taken alone, they are insufficient to treat relativistic QFT variationnally. The first step is to understand why, and discuss the peculiarities introduced by relativistic theories. 

\subsection{Lattice tensor network approach to relativistic theories}
Before considering \emph{analytical} continuum limits, let us see what happens when one takes a \emph{numerical} continuum limit of relativistic model. This means taking a finer and finer lattice discretization of a relativistic field theory (say with spacing $\varepsilon$), solving it with a tensor network method, and then taking the limit of the solution numerically. 

In the non-relativistic case, \ie for Lieb-Liniger like models, this works arbitrarily well at least in theory\footnote{Practically, even in the non-relativistic case, taking $\varepsilon \rightarrow 0$ makes the energy minimization more and more difficult (and potentially wildly unstable) numerically.  But this is an essentially orthogonal problem, and does not change the fact that the state manifold remains good in principle.}, in the sense that one can keep the bond dimension $D$ fixed as $\varepsilon \rightarrow 0$ while keeping the variational part of the error approximately constant. The situation changes in the relativistic case. There, taking a finer and finer grid requires larger and larger $D$ to keep the same error budget, and it is important to understand why.

There are (primarily) two ways to discretize and solve a $1+1$ dimensional relativistic QFT:
\begin{enumerate}
    \item The ``Hamiltonian way'': keep time continuous, discretize space, solve with MPS variationally (see \eg the remarkable 2013 study of Milsted, Haegeman, and Osborne~\cite{milsted2013mpsphi4})
    \item The ``Euclidean way'': discretize space and time, and contract the Euclidean partition function with a tensor network renormalization method.
\end{enumerate}
Both approaches give the same type of behavior as the numerical continuum limit is taken. I will consider results obtained with the second method, since this is the one I initially worked on with Clément Delcamp~\cite{delcamp202giltphi4}. 

\begin{remark}[Not finite volume difficulties]\label{rk:not_IR}
Note that with the Monte Carlo method, discretizing on a finer grid requires increasing the number of lattice sites in order to keep the same total physical volume. At some point, one cannot increase the lattice size more (\eg because of memory issues) and thus discretizing finer (taming UV problems) brings finite-volume errors (IR problems). With the Monte Carlo method, there is thus always a trade-off between UV and IR errors. The situation is different for tensor networks, where increasing the system size comes almost for free. Hamiltonian MPS methods work typically directly in the thermodynamic limit, and Euclidean renormalization methods can grow the system size exponentially in the number of iterations. Thus, in both cases, the errors we see when making the lattice finer \emph{do not come} from the need to balance finite-volume errors, and have a tensor specific origin.
\end{remark}

In Euclidean space-time, one can understand the $\phi^4_{2}$ model as simply giving the probability distribution of a random scalar field $\phi$:
\begin{equation}
    P(\phi) =\frac{\exp(-S_E(\phi))}{\mathcal{Z}} \, ,
\end{equation}
with action
\begin{equation}
    S_E(\phi) = \int_{\mathbb{R}^2} \frac{1}{2}(\nabla\phi)^2 + \frac{m_0^2}{2}\, \phi^2 + g\, \phi^4
\end{equation}
and $\mathcal{Z}=\int \mathcal{D}\phi \exp(-S_E(\phi))$.
This probability measure is ill-defined without normal-ordering which we will define below. To discretizing the continuum $\mathbb{R}^2$ into a lattice, one just localizes the field on its vertices, and write
\begin{equation}
    S^\mathsf{\varepsilon}_E(\phi)=\sum_{\langle i,j \rangle} \frac{(\phi_i-\phi_j)^2}{2} + \sum_{i} \frac{\mu_0^2}{2}\, \phi_i^2 + \frac{\lambda}{4}\,\phi_i^4 \, ,
\end{equation}
where $\varepsilon$ is the lattice unit length, $\langle i,j \rangle$ denotes a sum on nearest neighbors and
\begin{equation}\label{eq:naivescaling}
    \mu_0= \mathsf{\varepsilon} m_0 ~~ \text{and} ~~ \lambda = 4 \mathsf{\varepsilon}^2 g \, .
\end{equation}
Intuitively, if there were no quartic term, taking the continuum limit would simply amount to sending $\mathsf \varepsilon \to  0$, while keeping the dimensionful parameter $m_0$ fixed. However, one has to take into account the fact that the leading effect of the quartic term is to shift the quadratic (mass) part by a term $\propto -\log(\mathsf \varepsilon)$. If we do not counter this effect, the continuum limit is a trivial model with infinite mass.

The way to keep a finite mass is to add a mass counterterm and define
\begin{equation}
    \mu_0^2 :=\mu^2 - 3\lambda A(\mu^2) \, ,
\end{equation}
where $A(\mu^2)$ corresponds to the (lattice) tadpole diagram responsible for the logarithmic divergence:
\begin{align}
    A(\mu^2)&=    \tadpole{}
    \\ \nonumber
    &= \frac{1}{N^2}\! \sum_{k_1,k_2=0}^{N-1}\frac{1}{\mu^2 + 4 \left[\sin^2(\pi k_1/N) + \sin^2(\pi k_2/N)\right]} 
    \\ \nonumber
    &\;\; \underset{N\rightarrow + \infty}{\longrightarrow} \int_{[0,1]^2} \frac{\upd k_1 \upd k_2}{\mu^2 + 4 \left[\sin^2(\pi k_1)) + \sin^2(\pi k_2)\right]} \, . 
\end{align}
It can be shown that this prescription is equivalent to the normal-ordering (\textit{i.e.} $\phi^4 \rightarrow :\phi^4:$) we have used in the operator representation~\cite{loinaz1998mc_phi4}. The continuum limit is then obtained by sending $\mathsf \varepsilon$ (or equivalently $\lambda$) to $0$, while keeping $f=\lambda/\mu^2$ fixed. 

\begin{figure}
    \centering
    \includegraphics[scale=.72]{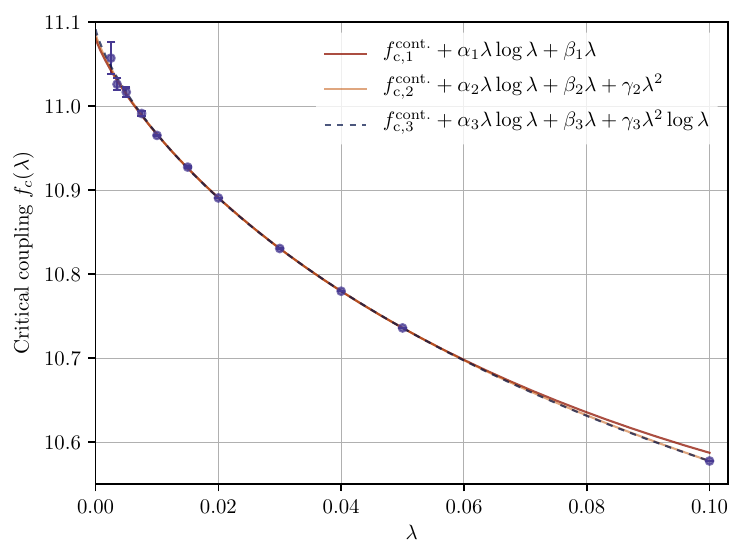}
     \includegraphics[scale=.72]{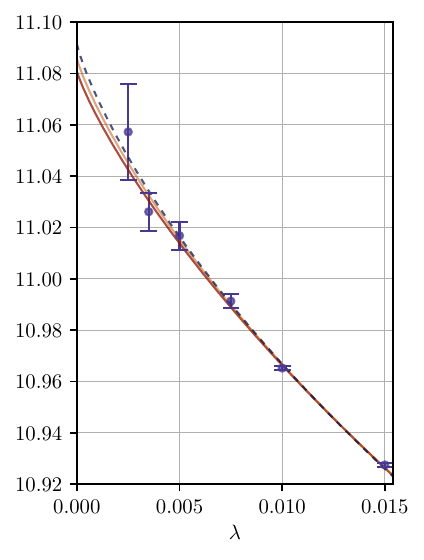}
    \caption{The critical coupling $f_c = 4 g_c$ of lattice $\phi^4$ as a function of the bare lattice coupling $\lambda$, which is a proxy for the lattice spacing ($\lambda \rightarrow 0$ is equivalent to $a\rightarrow 0$). Results from~\cite{delcamp202giltphi4}.}
    \label{fig:gilt_results}
\end{figure}

In~\cite{delcamp202giltphi4}, we had solved this lattice model with Gilt-TNR\footnote{``\textit{Graph independent local truncation - Tensor Network Renormalization}'': as its name suggests, it is a tensor network renormalization algorithm that is particularly powerful, but its specifics do not matter here.}~\cite{hauru2018gilt} with a very large bond dimension, and estimated the critical coupling for increasingly fine lattices. As often with tensor network methods, it is not easy to estimate error bars rigorously, but we could get a fairly good proxy for them by looking at our estimates for a few bond dimensions near our maximum reachable value. The critical coupling as a function of $\lambda$, which is a proxy for the lattice spacing $\varepsilon$, is shown in Fig. \ref{fig:gilt_results}. What we can see is that although tensor network methods allow to probe finer lattices than ever before, the precision does degrade dramatically when the lattice gets finer at fixed bond dimension $D$. Increasing $D$ to heroic values allowed us to get closer to the continuum limit than ever before. But ultimately, an extrapolation was needed. And extrapolations are less trivial than they seem in this context, in particular because of the $\lambda\log(\lambda)$ scaling which we conjectured to be the leading behavior.
All such plots obtained with tensor network methods, no matter the method, look fairly similar: the error explodes when one gets closer to the continuum limit.

As I argued in the previous remark \ref{rk:not_IR}, this reduction in precision does not come from the difficulty to capture IR physics.
This comes from the short-distance divergence of the entanglement entropy in relativistic theories.

\subsection{Entanglement in relativistic theories}
A relativistic quantum field theory, even a massive / gapped one, is critical at short distances. Indeed, for distances smaller than the inverse energy gap, there is no longer any scale, and a relativistic QFT is just a conformal field theory (CFT)\footnote{This is the folklore definition of a relativistic QFT: a relevant deformation of a conformal field theory. In fact, it is not fully general, and excludes gauge theories in $2$ and $3$ dimensions \cite{yin2018DH}.}. In fact, for the simple examples we consider here, the CFT describing short distance physics is just a free QFT. 

Critical theories on the lattice can still be studied with standard tensor network methods, although less accurately than gapped ones. The reason for this reduction in efficiency is best understood by looking at the bipartite entanglement entropy $S_{\frac{1}{2}}$ in the ground state of a finite system of length $L$. To define it, one just cuts the system in two equal parts and introduces $\rho_{\frac{1}{2}} = \tr_\mathrm{left}\left[\ket{0}\bra{0}\right]$:
\begin{center}
\begin{tikzpicture}

  \pgfmathsetmacro{\N}{17}            
  \pgfmathsetmacro{\spacing}{0.5}     
  \pgfmathsetmacro{\radius}{0.08}     

  \pgfmathsetmacro{\xstart}{-(\N-1)/2*\spacing}
  \pgfmathsetmacro{\xend}{(\N-1)/2*\spacing}

  \draw[dashed] (\xstart, -0.5) -- (\xstart, 0.5);
  \draw[dashed] (\xend, -0.5) -- (\xend, 0.5);

  \draw[thick] (0, -0.5) -- (0, 0.5);

  \fill[pattern=north east lines, pattern color=gray, draw=none] (\xstart,-0.1) rectangle (0,0.1);

  \foreach \i in {0,...,16} {
    \pgfmathsetmacro\x{\xstart + (\i)*\spacing}
    \filldraw[black] (\x, 0) circle (\radius);
  }

  \node[below=6pt, yshift=-2pt] at (0,-0.25) {$|0\rangle$};

  \node at (2, 0.4) {$\rho_{\frac{1}{2}}$};

  \node[below] at (\xstart, -0.5) {$-\frac{L}{2}$};
  \node[below] at (\xend, -0.5) {$\frac{L}{2}$};

  \pgfmathsetmacro{\rulerLeft}{-\spacing/2+2.25}
  \pgfmathsetmacro{\rulerRight}{\spacing/2+2.25}
  \draw[<->] (\rulerLeft, -0.25) -- (\rulerRight, -0.25);
  \node at (2.25, -0.55) {$\varepsilon$};

\end{tikzpicture}
\end{center}
The bipartite entanglement entropy is then just the Von Neumann entropy of $\rho_{\frac{1}{2}}$. In $1+1$ dimension, it is logarithmically divergent:
\begin{equation}\label{eq:entanglement_critical}
  S^\mathrm{critical}_{\frac{1}{2}}(L) = - \tr\left[\rho_{\frac{1}{2}}\log(\rho_{\frac{1}{2}})\right] \underset{L\rightarrow +\infty}{\sim}  \frac{c}{6} \log(L/\varepsilon)\, ,
\end{equation}
where $c$ is the central charge and $\varepsilon$ is some fixed scale induced by the lattice.
The bipartite entanglement entropy in a MPS is upper-bounded by the log of the bond dimensions $S^\mathrm{MPS}_{\frac{1}{2}} \leq \log(D)$. As a result, MPS introduce an effective length-scale $L_\mathrm{eff}$ in critical systems, which can be grown by increasing $D$, but is always finite. This is not so terrible: we will get more and more accurate results by increasing $D$. 

We can look at the same situation in the continuum, for a relativistic QFT:
\begin{center}
\begin{tikzpicture}

  \def\L{8}
  \def\halfL{4}
  \def\halfhalfL{3}

  \draw[dashed] (-\halfhalfL, -0.5) -- (-\halfhalfL, 0.5);
  \draw[dashed] (\halfhalfL, -0.5) -- (\halfhalfL, 0.5);

  \draw[thick] (0, -0.5) -- (0, 0.5);

  \fill[pattern=north east lines, pattern color=gray, draw=none] (-\halfL,-0.1) rectangle (0,0.1);

  \draw[thick] (-\halfL,0) -- (\halfL,0);

  \node[below=6pt,yshift=-2pt] at (0, -0.4) {$|0\rangle$};

  \node at (2, 0.4) {$\rho_{\frac{1}{2}}$};
  \node at (-\halfhalfL, -0.9) {$-M^{-1}$};
  \node at (\halfhalfL, -0.9) {$M^{-1}$};
\end{tikzpicture}
\end{center}
Relativistic models are critical ``all the way down'', \ie with infinitely small $\varepsilon$, and thus their bipartite entanglement entropy is infinite. We need to regulate it with a finite UV cutoff $\Lambda = \varepsilon^{-1}$ if we hope to get a finite answer. It is still instructive to do so, and gives:
\begin{equation}\label{eq:entanglement_relativistic}
  S^\mathrm{relativistic}_{\frac{1}{2}} (\Lambda) \underset{\Lambda \rightarrow +\infty}{\sim} \frac{c}{6}\log\left(\Lambda/M\right)\,
\end{equation}
where $M$ is the mass gap, which is the relevant IR length-scale as long as $L\gg M^{-1}$. Intuitively, this means there are more and more degrees of freedom as the UV cutoff is lifted.

If we work with a lattice discretization, the relevant ratio of length-scales $\Lambda/M$ increases as we make the discretization finer, and thus the precision degrades as we previously observed.

This behavior makes the situation even more dire in the continuum limit. Indeed, if we want to study a relativistic QFT variationally in its continuum Hilbert space, any (continuous) MPS ansatz will introduce a short distance cutoff $\Lambda_\mathrm{eff}$. However, the energy density, which is the quantity we minimize, is dominated by short distance physics. As we iterate (\eg with gradient descent), the CMPS will adjust to shorter and shorter length-scale to lower the energy. Since the expressive power of a CMPS of fixed bond dimension is limited, this means the precision at relevant length-scales (like $M^{-1}$) gradually degrades. The longer the minimization is run, the worse the results. This is exactly Feynman's third worry about the use of the variational method in relativistic QFT! 

\emph{A contrario}, there is no problem in using the variational method for critical lattice models. Indeed, the energy density is still defined at the lattice scale, and thus the energy minimization does not lead to an ever increasing length-scale for the variational ansatz. The problem really is having infinitely many degrees of freedom at arbitrarily short distance, in a scale invariant way, \emph{while having the energy sensitive to them}.

\subsection{Standard CMPS for relativistic theories}
What happens if we insist and use a standard CMPS for a relativistic model anyway? How do we concretely witness the problem? We do not even need an interacting theory, and the massive scalar field is sufficient to see what happens. In infinite volume, its Hamiltonian is:
\begin{equation}\label{eq:freeboson_original}
  H_0 = \int_\mathbb{R} h_0 := \int_\mathbb{R} \frac{\pi^2}{2} + \frac{(\nabla \phi)^2}{2} + m^2 \frac{\phi^2}{2} \, .
\end{equation}
This Hamiltonian density is infinite in the ground state, \ie $\bra{0}h_0\ket{0} = +\infty$ which is easy to see by diagonalizing $H_0$ in normal modes $a_k$
\begin{equation}
  H_0 = \frac{1}{2\pi} \int_\mathbb{R} \upd k \, \omega_k\, \frac{a_k^\dagger a_k + a_k a_k^\dagger}{2} = \frac{1}{2\pi} \int_\mathbb{R} \upd k \omega_k \left(a_k^\dagger a_k + \frac{2\pi \delta(0)}{2}\right)
\end{equation}
with $\omega_k = \sqrt{k^2  + m^2}$.
This shows that $:H_0:_{a}$, \ie $H_0$ normal-ordered with respect to the $a_k$ is well defined, with a ground state energy of $0$. In this simple free case, $h_0$ and $:h_0\!:_a$ are just related by an infinite multiple of the identity\footnote{In this free case, the initial and normal-ordered Hamiltonian thus represent the same physics, one is just technically ill-defined. This is more subtle in the interacting case where normal-ordering is also physical, and amounts to a mass counter-term.}.

The standard way to apply CMPS to this model, used \eg in~\cite{stojevic2015finiteentanglement} is to \emph{locally} relate the canonically conjugated fields verifying $[\phi(x),\pi(y)] = i\delta(x-y)$ to creation-annihilation operators verifying $[\psi(x),\psi^\dagger(y)] = \delta(x-y)$. All the possible local choices are parameterized by a new scale $\Lambda$:
\begin{align}\label{eq:local_phi_psi_link}
  \phi(x) &= \sqrt{\frac{1}{2\Lambda}}\left[\psi(x) + \psi^\dagger(x)\right]\\
  \pi(x) &= \sqrt{\frac{\Lambda}{2}} \left[\psi(x) - \psi^\dagger(x)\right]
\end{align}
With this parameterization, $H_0$ is now a local integral of $\psi,\psi^\dagger$ and thus we can use the standard CMPS toolbox to evaluate the energy density. This however introduces two types of divergences. The first is mild: $h_0$ is not normal-ordered in the $\psi,\psi^\dagger$, and thus $\langle Q,R| h_0 |Q,R\rangle$ contains $\delta(0)$ divergent terms. These problematic terms can be removed by considering $:h_0:_\psi$ instead. The second divergence is more serious. The Hamiltonian density not only contains the standard non-relativistic kinetic energy $\partial_x \psi^\dagger \partial_x \psi$ but also particle non-conserving ones $\partial_x\psi\partial_x \psi + \mathrm{h.c.}$ which are divergent when evaluated on a generic CMPS~\cite{haegeman2013calculus}, and can possibly be made negatively divergent. This is the sign that the Hamiltonian density $:h_0:_\psi$ is \emph{not} lower bounded!

One can make it lower bounded by introducing a smart regulator:
\begin{equation}\label{eq:cmps_regulator}
 :h_0^\Lambda\!:_\psi= :h_0\!:_\psi + \frac{1}{2\Lambda^2} : \partial_x \pi \partial_x \pi \!:_\psi
\end{equation}
which crucially uses the same $\Lambda$ as in eq. \eqref{eq:local_phi_psi_link}. This exactly cancels the problematic $\partial_x\psi\partial_x\psi +\mathrm{r.c.}$ term. As a result this Hamiltonian can be studied with standard CMPS: evaluating it on a given CMPS is possible, and the optimization is well behaved. The Hamiltonian regulator in \eqref{eq:cmps_regulator} acts as a UV cut-off, preventing the energy density from collapsing to $-\infty$. The price to pay is that we are no longer studying the original model, but a UV-regulated version. 

Making the ratio between the cutoff scale $\Lambda$ and the mass scale $m$ larger, one would need to increase the bond dimension $D$ of the CMPS to reach a similar precision at the scale $m$. As on the lattice, we face the difficulty that the QFT is scale invariant ``all the way down'', and thus pushing the UV regulator to lower distances makes the scale invariant window to be approximated larger. At first sight, this seems like an inevitable price to pay. 

\subsection{The \emph{correct} tensor product structure}
The root of the UV difficulties we have just described is really the fact that spatial entanglement entropy is infinite in a relativistic QFT. If we insist on writing the Hilbert space as a tensor product of factors each associated to the fields $\phi(x)$, $\pi(x)$, we face a problem.

For free QFT however, we know a different decomposition of the Hilbert space as a tensor product of factors, in which the ground state is fully disentangled. This is simply the basis of the Fourier modes $a_k$, $a_k^\dagger$, which is commonly used in Hamiltonian truncation. Of course, this only works for non-interacting QFT. But for super-renormalizable QFT as well as for asymptotically free QFT, the short distance physics is indistinguishable from that of a free theory. We can thus hope that the corrections induced by the interaction are mild, and preserve this tensor product structure at least approximately. 

It is thus tempting to work directly with this tensor product structure in Fourier space, and compress the state as a MPS directly in this basis. This was done \eg in~\cite{schmoll2023hamiltoniantruncationtensornetworks}. However, we expect matrix product states to provide efficient approximations if entanglement verifies an area law. There is no reason to expect the area law to be valid in momentum space, because interactions are wildly non-local in momentum (\eg with terms like $a_{k_1}^\dagger a_{k_2}^\dagger a_{k_3} a_{k_1 + k_2 - k_3}$). Hence, while working with this tensor product structure does fix the UV problem, directly using it with tensor networks does not solve the IR issues of Hamiltonian truncation\footnote{Such tensor network methods, just like Hamiltonian Truncation, can still be a pragmatic approach for sufficiently small systems and with careful extrapolations.}.

We can however go back to real space, because the free vacuum is unentangled in all the mode basis that annihilate it. The most natural way to go back to real-space is thus to take the inverse Fourier transform and define:
\begin{equation}\label{eq:ax_def}
  a(x) = \frac{1}{2\pi} \int_\mathbb{R}\upd k\, \e^{ikx} a_k \,
\end{equation}
which verify $\left[a(x),a^\dagger(y)\right] = \delta(x-y)$ because we have defined our $a_k$ in the convention such that $[a_k,a_{k'}] = 2\pi \delta(k-k')$ (without $\omega_k$ factor). Crucially, this canonical commutation would not hold if we had used the more natural integral measure $\frac{\upd k}{\sqrt{2\omega_k}}$ in \eqref{eq:ax_def} instead of $\upd k$ (which would then correspond to the positive frequency part of $\phi$). This annihilation operator $a(x)$ is fairly natural from a condensed matter viewpoint, and fairly suspicious from a relativistic QFT viewpoint, but we will explain later why this latter worry is unfounded.

\begin{remark}[A graphical view]
It helps to understand the change of basis we just discussed visually, assuming some discretization. The (discretized) free vacuum $\ket{0}$ is very entangled in space, but a unitary change of modes (aka a Bogoliubov transform) fully disentangles the state. Then, adding the Fourier transform on top keeps the disentangled structure. We now have a slightly different notion of spatial locality in which the free ground state is a \emph{product state}:
  \begin{center}
\begin{tikzpicture}[scale=1, thick]

\def\n{14}
\def\wiresep{0.5}
\def\circleRad{0.08}
\def\ybase{0.3}
\def\blockheight{0.6}
\def\blocksep{0.2}
\pgfmathsetmacro{\totalheight}{3*(\blockheight + \blocksep)}

\foreach \i in {0,...,13} {
  \pgfmathsetmacro\x{\i*\wiresep}
  \draw[blue] (\x, 0) -- (\x, \totalheight);
}

\foreach \j in {0,1,2} {
  \pgfmathsetmacro\y{\j*(\blockheight + \blocksep)}
  \draw[fill=white] (-0.3,\y) rectangle ({(\n-1)*\wiresep + 0.3}, {\y+\blockheight});
}

\node at ({(\n-1)*\wiresep/2}, {0.3}) {$|0\rangle$};
\node at ({(\n-1)*\wiresep/2}, {1.1}) {$U_{\mathrm{Bogoliubov}}$};
\node at ({(\n-1)*\wiresep/2}, {1.9}) {$U_{\mathrm{Fourier}}$};

\node at (8.0, 1.2) {\Large$=$};

\foreach \i in {0,...,13} {
  \pgfmathsetmacro\x{9 + \i*0.3}
  \draw[blue] (\x, {\ybase + \circleRad}) -- (\x, \totalheight);
  \draw[fill=gray, draw=black] (\x, \ybase) circle (\circleRad);
}

\end{tikzpicture}
\end{center}
If the state considered is not the free ground state $\ket{0}$ but a low energy state $\ket{\psi_0}$ with the same short distance behavior (same short distance correlation functions), we cannot expect that the state is a product state following these two steps. However, it is plausible that it is not so far from a product state, \ie weakly entangled, and thus representable by a matrix product state.
\begin{center}
  \begin{tikzpicture}[scale=1, thick]

\def\n{14}
\def\wiresep{0.5}
\def\circleRad{0.08}
\def\ybase{0.3}
\def\blockheight{0.6}
\def\blocksep{0.2}

\pgfmathsetmacro{\totalheight}{3*(\blockheight + \blocksep)}

\foreach \i in {0,...,13} {
  \pgfmathsetmacro\x{\i*\wiresep}
  \draw[blue] (\x, 0) -- (\x, \totalheight);
}

\foreach \j in {0,1,2} {
  \pgfmathsetmacro\y{\j*(\blockheight + \blocksep)}
  \draw[fill=white] (-0.3,\y) rectangle ({(\n-1)*\wiresep + 0.3}, {\y+\blockheight});
}

\node at ({(\n-1)*\wiresep/2}, {0.3}) {$|\psi_0\rangle$};
\node at ({(\n-1)*\wiresep/2}, {1.1}) {$U_{\mathrm{Bogoliubov}}$};
\node at ({(\n-1)*\wiresep/2}, {1.9}) {$U_{\mathrm{Fourier}}$};

\node at (8.0, 1.2) {\Large$=$};

\draw[orange, very thick] 
    (9, \ybase) -- ({9 + 13*0.3}, \ybase);

\foreach \i in {0,...,13} {
  \pgfmathsetmacro\x{9 + \i*0.3}
  \draw[blue] (\x, {\ybase + \circleRad}) -- (\x, \totalheight);
  \draw[fill=gray, draw=black] (\x, \ybase) circle (\circleRad);
}
\end{tikzpicture}
\end{center}

\end{remark}
This philosophy of using a Bogoliubov transform first, to disentangle as much as possible, and then represent the state as a MPS more easily, has also been used on the lattice (see \eg~\cite{krumnow2016orbitaloptim,wu2025impuritydisentangling}) to make computations \emph{more efficient}. For relativistic QFT, this is necessary to make the variational approach usable at all.

\begin{remark}[Free particle entanglement entropy] \label{rk:particleentanglement}
  It is important to note that there are two inequivalent notions of spatial locality associated to a relativistic QFT that we have successively discussed here. The first is the standard relativistic one, corresponding to the \emph{heuristic}\footnote{This decomposition is ill-defined in the continuum, but very intuitive with a discretization.} decomposition of the Hilbert space into a tensor product of factors, each associated to the local fields $\phi(x),\pi(x)$. 
With respect to this decomposition, the bipartite entanglement entropy in low energy states is UV divergent. This natural factorization cannot be used with tensor network methods without a cutoff.

The other option may be called the \emph{free particle} factorization. It decomposes the Hilbert space into a product of factors each associated to $a(x),a^\dagger(x)$. Crucially, this notion of locality \emph{does not} match the relativistic one, because $a(x)$ and $\phi(x)$ are not locally related. Instead we have:
\begin{equation}
  \phi(x) = \int_\mathbb{R}\upd y J(x-y) \left[a(y) + a^\dagger(y)\right] = [J \star (a+a^\dagger)](x)
\end{equation}
where $J$ is the inverse Fourier transform of $1/\sqrt{2\omega_k}$
\begin{equation}\label{eq:jsource_def}
  J(x) = \frac{1}{2\pi} \int_\mathbb{R} \frac{\upd k}{\sqrt{2\omega_k}} \e^{ikx} \, .
\end{equation}
This function behaves like $1/\sqrt{x}$ at short distances, and like $\e^{-m|x|}$ at large distances. As a result, the free particle factorization and the associated notion of spatial locality is not useful from an operational quantum information view point. However, it seems to be the right notion of locality to use with tensor network states, because the bipartite entanglement entropy of low energy states appears to be finite in this factorization!
\end{remark}

\section{The relativistic continuous matrix product states}
After all this work, we are now ready to define the ansatz wave-function that is the main subject of this monograph: \emph{relativistic} continuous matrix product states, or \emph{RCMPS}. This method was presented first in~\cite{tilloy2021variational,tilloy2021relativistic}, where it was applied to the $\phi^4$ model. The optimization algorithm was then refined in~\cite{tilloy2022studyquantumsinhgordonmodel}, and applied to the Sine-Gordon and Sinh-Gordon models.

\subsection{Definition}
The RCMPS is simply a CMPS in the operator basis in which we expect the entanglement entropy to be low. Concretly, this is the state where $\psi(x)$ is replaced by $a(x)$, and where the fully disentangled state $\ket{0}_\psi$ is replaced by the ground state of the free theory $\ket{0}_a$:
\begin{equation}\label{eq:rcmps_def}
  \ket{Q,R} := \tr\left[\mathcal{P}\exp\left(\int_{I} Q \otimes \mathrm{Id} + R\otimes a^\dagger(x)\right)\right] \ket{0}_a \, .
\end{equation}
As with CMPS, this path-ordered exponential \eqref{eq:rcmps_def} can be Taylor-expanded into a wave-function representation:
\begin{equation}
  \ket{Q,R} = \sum_{n=0}^{+\infty} \int_{0<x_1<x_2<\cdots < x_n<L} \hskip-2cm\upd x_1\,\upd x_2\cdots \upd x_n \; \varphi_n(x_1,\cdots,x_n)\, a^\dagger(x_1) \cdots a^\dagger(x_n) \ket{0}_\psi
\end{equation}
with
\begin{equation}\label{eq:rcmps-wavefunction}
    \varphi_n(x_1,\cdots,x_n) = \tr\left[\e^{Q x_1} R \e^{Q(x_2-x_1)} R \cdots \e^{Q(x_n-x_{n-1})} R\e^{Q (L-x_n)}\right] \, , 
\end{equation}
which could be taken as a definition of RCMPS. This representation shows that, in contrast with HT, the RCMPS state $\ket{Q,R}$ is \emph{not} sparse in the particle basis.

\subsection{Locality properties of RCMPS}

Algebraically, $\psi$ and $a$ behave in exactly the same way, and thus almost all the methods we have for RCMPS rely on earlier CMPS technology. The main change is that the Hamiltonian $H$ is no longer (the integral of) a local operator. Since $\phi = J \star (a+a^\dagger)$, the Hamiltonian density itself becomes a multi convolution with $J$. Interactions that were local in the field picture acquire a range in the $a$ picture, that fortunately decays exponentially with distance. Conceptually, this is not expected to bring issues: at least on the lattice, models with exponentially decaying interactions have the same type of physics as models with strictly finite range interactions. In particular, MPS and CMPS tend to work just as well for models with exponentially decaying interactions. Hence, from a condensed matter theory point of view, this move away from strict locality, which was motivated by quantum information considerations, is not a problem.

\begin{remark}[Is it acceptable to break Poincaré symmetry?] This operator $a(x)$ can be surprising: we are abandoning all we know about the covariance of free field theories, just for the sake of having canonical commutation relations. Could we not do better? For example, could we construct a variational manifold of states preserving more symmetries, and even ideally constructing a manifold of Poincaré invariant states? It is important to understand that there is no hope to get any such thing. The true (interacting) vacuum that we are looking for is the \emph{only} Poincaré invariant state. As a result, it is not possible to construct it explicitly (in the sense that we could compute correlation functions on this state exactly). Otherwise the model would be exactly solvable; and we do not believe this to be a generic property of interesting models.
\end{remark}

As beneficial as this change of basis is theoretically, it brings fairly substantial technical difficulties. Computing with RCMPS ends up being much more difficult than with CMPS, and a lot more work is required to build a full fledged energy minimizer. Computing with RCMPS in an efficient way is the subject of the following section.

\section{Computing with RCMPS}

Quite surprisingly, all the additional complexity introduced by the change of basis ends up bringing only a constant overhead cost compared to CMPS! Concretely, the cost of $1$ optimization step\footnote{As usual, we can only give the cost of one iteration of the optimizer. As with most non-convex minimization problems, there are no guarantees that one can reach the minimum in a small number of iterations (that remains small as $D$ is increased). In practice however, we observe that with a ``good'' optimizer, the number of iterations needed to converge grows only very mildly with $D$.} for RCMPS ultimately scales like $D^3$, as it does for CMPS\footnote{Only with a prefactor $\sim 1000$ worse for RCMPS!}. Obtaining such a scaling, which is necessary to make RCMPS practically usable, is not entirely obvious.

\subsection{Evaluating expectation values}\label{sec:expectation_values}

If we want to find the RCMPS of a given bond dimension $D$ that minimizes the energy density of a given model, we at least need to be able to evaluate it! The first step is thus to estimate the expectation values of local operators. The derivation is technical, but I believe it is important to detail it because the fact that it can be done at all is already surprising and remarkable. Indeed, in the QFT context, knowing a wave-function analytically does not imply that we can compute expectation values precisely (this requires an infinite sum that is generically not doable efficiently). The magic of RCMPS, inherited from CMPS (which inherited it from MPS), is that about everything one could want to estimate from a given state is doable efficiently. We never have to decompress the representation in the ``full'' infinite dimensional Fock space, and we can work directly in the ``compressed'' space of $Q,R$. Intuitively, the elementary operation that is most natural in this compressed space is multiplying the matrices, which costs $\propto D^3$ floating point operations. This is the lowest asymptotic cost we can hope for. Remarkably, it is also the cost we will obtain!

\subsubsection*{Naive historical evaluation}
The naive strategy, which is the one I initially used, is to leverage the fact that correlation functions of normal-ordered products of $a(x)$ are easy to compute using the CMPS formulas \eqref{eq:exact_correlations}. They give, for example, for $x_1 \leq x_2$:
\begin{equation}
  \bra{Q,R} a^\dagger (x_1) a(x_2) \ket{Q,R} = \frac{1}{\tr[\e^{L\mathbb{T}}]} \tr \left[ \e^{x_1\mathbb{T}} R^{(\delta_1)}\e^{(x_2-x_1)\mathbb{T}}\cdots R^{(\delta_n)} \e^{(L-x_n)\mathbb{T}}\right]
\end{equation}
where $\mathbb{T} = Q \otimes \mathrm{Id} + \mathrm{Id} \otimes \bar{Q} + R\otimes \bar{R}$ and $L$ is the system size. As with CMPS, this expression simplifies in the thermodynamic limit and by taking a left canonical gauge with $Q + Q^\dagger +R^\dagger R = 0$:
\begin{equation}\label{eq:twopoint_a_RCMPS_thermo}
  \bra{Q,R} a^\dagger (x_1) a(x_2) \ket{Q,R} = \tr\left[R^\dagger \e^{(x_2 - x_1)\mathcal{L}} (R\rho_0)\right] \, ,
\end{equation}
where $\mathcal{L}$ is the super-operator version of $\mathbb{T}$, \ie $\mathcal{L}(\rho) = Q\rho + \rho Q^\dagger + R\rho R^\dagger$ and $\rho_0$ is the normalized stationary state of $\mathcal{L}$, \ie $\mathcal{L}(\rho_0) = 0$ and $\tr[\rho_0]= 1$. This formula \eqref{eq:twopoint_a_RCMPS_thermo} can be evaluated efficiently, and thus it is tempting to use it to evaluate correlation functions of the field by taking the integrals with $J$ \emph{at the end}. For example, this gives:
\begin{equation}\label{eq:phi2_integralrepresentation}
  \begin{split}
    &\bra{Q,R}\! :\!\phi^2(0)\! :\!\ket{Q,R} = \int_{\mathbb{R}^2} \upd x_1 \upd x_2 \; J(x_1) J(x_2) \\
    & ~~~ \times  \left\{ 2\tr\left[R^\dagger \e^{|x_2 - x_1|\mathcal{L}} (R\rho_0)\right]+ \tr\left[R\e^{|x_2 - x_1|\mathcal{L}} (R\rho_0)\right]+\tr\left[R^\dagger \e^{|x_2 - x_1|\mathcal{L}} (\rho_0 R^\dagger)\right] \right\} \, .
  \end{split}
\end{equation}
This form is analytically instructive. For example, it allows to see that the result is finite for all RCMPS, something that was far from obvious in the first place. 

Numerically, the integrals in \eqref{eq:phi2_integralrepresentation} can be evaluated to essentially arbitrary precision using quadratures. For a double integral, this is still a viable option. But the number of integrals grows linearly as a function of the power in the field, and thus the number of quadrature points to sum over grows exponentially. Consequently, estimating $\bra{Q,R}\!\! :\!\!\phi^4(0)\!\! : \!\!\ket{Q,R}$ is already much more challenging, and getting $\bra{Q,R}\!\! :\!\! \phi^6(0)\! \! : \!\! \ket{Q,R}$ with this approach seems completely out of reach. Clearly this is the sign that something is wrong, as we do not expect the complexity of the physics to increase so stiffly as a function of the field power. And indeed, one can do better, using a differential equation approach.

\subsubsection*{Differential equation approach for \texorpdfstring{$:\!\phi^n\!:$}{phin}}

The crucial observation is that the expectation value of normal-ordered vertex operators $\langle V_\alpha(0)\rangle := \bra{Q,R} \!\! :\!\! \e^{\alpha\phi(0)} \!\! :\!\! \ket{Q,R}$ is \emph{exactly} the generating functional $\mathcal{Z}_{j',j}$ for a particular choice of source $j=j'=\alpha J$. Indeed:
\begin{align}
  \bra{Q,R} \!\! :\!\! \e^{\alpha\phi(0)} \!\! :\!\! \ket{Q,R} &= \bra{Q,R} \e^{\alpha \int_\mathbb{R}\upd x\, J(x) a^\dagger(x)} \e^{\alpha \int_\mathbb{R}\upd x \, J(x) a(x)} \ket{Q,R}\\ 
  &= \mathcal{Z}_{\alpha J,\alpha J} \, .
\end{align}
In the context of ``standard'' CMPS, the generating functional was just an intermediate tool, to be subsequently differentiated to obtain exact formulas for correlation functions of $a$. But it can also be evaluated numerically for a given source by solving an ODE. Indeed,
\begin{equation}
  \mathcal{Z}_{\alpha J,\alpha J} = \tr\left\{\mathcal{P}\exp\left[\int \mathbb{T} + \alpha J(x) (R\otimes \Id + \Id \otimes R^*)\right]\right\} \, ,
\end{equation}
which, upon going to super-operator representation and interpreting the path-ordered exponential as the solution of a linear ODE gives 
\begin{equation}
  \mathcal{Z}_{\alpha J,\alpha J} = \tr\left[\rho_{\alpha}(+\infty)\right]
\end{equation}
where $\rho_\alpha(x)$ is the solution of the ODE 
\begin{equation}\label{eq:vertex_ode}
  \frac{\upd}{\upd x} \rho_{\alpha}(x) = \mathcal{L}(\rho_\alpha(x)) + \alpha J(x) \left[R\rho_\alpha(x) + \rho_\alpha(x)R^\dagger \right]
\end{equation}
with ``initial condition'' $\rho_\alpha(-\infty)$ a\footnote{Any PSD matrix of trace $1$ will do, because $ \mathcal{L}$ projects to its stationary state anyway!} positive semi-definite (PSD) matrix of trace $1$. Because $J$ decays exponentially with $|x|$, and $\mathcal{L}$ is trace preserving, we can get a very good (numerically exact) approximation of $\rho_\alpha(+\infty)$ by solving the ODE only for a finite time. In practice, we start from $\rho_\alpha(-20) = \rho_\mathrm{ss}$ (the stationary state of $\mathcal{L}$), evolve with \eqref{eq:vertex_ode} for $\Delta x \simeq 40$, and taking $\rho_\alpha(+\infty) \simeq \rho_\alpha(+20)$. Solving the non-autonomous linear ODE for a finite time is doable efficiently (although maybe not as efficiently as we would like, see the following remark \ref{rk:non-autonomous_ODE}).

Once we know how to evaluate the expectation value of a vertex operator, field expectation values are obtained by differentiating and taking $\alpha = 0$:
\begin{equation}
  \langle :\!\phi^n(x)\!:\rangle = \frac{\partial^n}{\partial \alpha^n} \langle V_\alpha(x)\rangle \bigg|_{\alpha = 0}\,.
\end{equation}
Numerically, we could compute these derivatives by \emph{forward differentiating} through the ODE \eqref{eq:vertex_ode} using any automatic differentiation library. But we can also just as well do it explicitly. Concretely, this amounts to solving the triangular system of ODE
\begin{align}
  \forall k \geq 0, ~~ \frac{\upd}{\upd x} \rho^{(k)}(x) = \mathcal{L}[\rho^{(k)}(x)] + k J(x) [R \rho^{(k-1)}(x) + \rho^{(k-1)}(x)R^\dagger]
\end{align}
where $\rho^{(k)} := \frac{\partial^k}{\partial \alpha^k}\rho_\alpha\big|_{\alpha = 0}$, and the initial conditions $\rho^{(0)}(-\infty) = \rho_\mathrm{ss}$ and $\forall k\geq 1$, $\rho^{(k)}(-\infty) = 0$. With these notations we have $\langle :\!\phi^n(x)\!:\rangle = \tr[\rho^{(n)}(+\infty)]$. The cost of computing powers of the field thus grows only \emph{quadratically} with the power, not exponentially.

\begin{remark}[Solving non-autonomous linear differential equations] \label{rk:non-autonomous_ODE}
The main bottleneck of any RCMPS calculation is the computation of the solution of a (system of) non-autonomous \emph{linear} matrix ODE. To obtain it, we use high order (explicit) Runge-Kutta (RK) schemes that could in principle be used to solve non-linear ODEs as well. Crucially, there seems to be no simple way to exploit the linearity of the equations to obtain the solutions faster, at least for generic instances! Even at high order, RK schemes can require several thousand points to reach machine precision. Finding a better time evolution method could thus make the computation of RCMPS expectation values dramatically faster.
\end{remark}

\subsubsection*{Differential equation approach for the kinetic Hamiltonian density}
The previous technique allows to compute vertex operators and field monomials, but is not sufficient to compute operators containing spatial derivatives (or conjugate momenta), for which we need a small extension. Let us focus on a kinetic Hamiltonian of the form:
\begin{align}
  H_{0} = \int_\mathbb{R} \frac{:\!\pi^2\!:}{2} + \frac{:\!(\partial_x\phi)^2\!:}{2} + m^2 \frac{:\!\phi^2\!:}{2}= \frac{1}{2\pi}\int_\mathbb{R} \omega_k a_k^\dagger a_k
\end{align}
Introducing the positive frequency part of the field
\begin{equation}
  \varphi(x) := \int_\mathbb{R} \upd y \,  J(x-y) a(y)
\end{equation}
such that $\phi(x) = \varphi(x) + \varphi^\dagger(x)$, we have $H_0 = \int_\mathbb{R} h_0$ with 
\begin{equation} \label{eq:h0_varphi}
  h_0(x) = \partial_x \varphi^\dagger(x) \partial_x\varphi(x) + m^2 \varphi^\dagger(x) \varphi(x) \, .
\end{equation}
The second term can be evaluated quite straightforwardly by noticing that
\begin{equation}
  \langle \varphi^\dagger(0)\varphi(0)\rangle = \frac{\partial^2}{\partial \alpha \partial \beta} \mathcal{Z}_{\alpha J,\beta J}\bigg|_{\alpha = \beta = 0} \, ,
\end{equation}
where, as before, one can write $\mathcal{Z}_{\alpha J, \beta J} = \tr[\rho_{\alpha\beta}(+\infty)]$ with $\rho_{\alpha\beta}$ the solution of the ODE
\begin{equation}
  \frac{\upd}{\upd x} \rho_{\alpha\beta} = \mathcal{L}(\rho_{\alpha\beta}) + \alpha J \rho_{\alpha\beta} R^\dagger + \beta J R \rho_{\alpha\beta}
\end{equation}
with initial condition $\rho_{\alpha\beta}(-\infty) = \rho_\mathrm{ss}$.
Again, via forward differentiation, we obtain a system of ODE
\begin{align}
  \frac{\upd}{\upd x} \rho^{(1,0)} &= \mathcal{L}(\rho^{(1,0)}) + J\, \rho^{(0,0)}R^\dagger \\
  \frac{\upd}{\upd x} \rho^{(0,1)} &= \mathcal{L}(\rho^{(0,1)}) + J \, R \rho^{(0,0)} \\
  \frac{\upd}{\upd x} \rho^{(1,1)} &= \mathcal{L}(\rho^{(1,1)}) + J\, \left[ R \rho^{(1,0)} + \rho^{(0,1)} R^\dagger \right] 
\end{align}
with $\rho^{(1,0)} = \partial_\alpha \rho_{\alpha\beta} \big|_{\alpha=\beta=0}$, $\rho^{(0,1)} = \partial_\beta \rho_{\alpha\beta} \big|_{\alpha=\beta=0}$, $\rho^{(1,1)} = \partial_\alpha \partial_\beta \rho_{\alpha\beta} \big|_{\alpha=\beta=0}$,  and $\rho^{(0,0)}(x) \equiv \rho_\mathrm{ss}$. Ultimately, with these notations, $\langle \varphi^\dagger \varphi\rangle = \tr[\rho^{(1,1)}(+\infty)]$.

We are almost done, and only need a minor extension of CMPS results to compute the expectation value of the first term of \eqref{eq:h0_varphi} $\langle \partial_x\varphi^\dagger \partial_x \varphi\rangle$. It is a bit more difficult because of the derivative:
\begin{equation}\label{eq:derivative_varphi}
  \partial_x\varphi(x) = \int_\mathbb{R} \upd y\, \partial_x J (x-y) a(y) = \int_\mathbb{R}  \upd y\, J(x-y) \partial_y a(y) \, .
\end{equation}
It would be tempting to use the previous results and replace $J$ by $\partial_x J$. However the latter is not absolutely integrable in $x=0$ which makes the solution of the corresponding ODE subtle to define. It is easier to instead use the right-hand side of \eqref{eq:derivative_varphi}, and replace $a(y)$ with $\partial_y a(y)$. This works if we can be done if we are able to compute the generating functional of $\partial_y a(y)$ correlation functions:
\begin{equation}
  \mathcal{Y}_{j',j} := \bra{Q,R} \exp\left[\int \upd y\, j'(y)\partial_y a^\dagger(y)\right] \exp\left[\int \upd y\, j(y)\partial_y a(y)\right] \ket{Q,R}
\end{equation}
Fortunately, this generating functional is known as well in the CMPS context. It takes the same form as $\mathcal{Z}$ but for the substitution $R\rightarrow [Q,R]$, namely
\begin{equation}
  \mathcal{Y}_{j',j} = \tr[\varrho(+\infty)]
\end{equation}
with 
\begin{equation}
  \frac{\upd}{\upd x} \varrho = \mathcal{L} (\varrho) + j'\,  \varrho [Q,R]^\dagger + j\, [Q,R] \varrho \,.
\end{equation}
Hence, ultimately, we have $\langle \partial_x\varphi^\dagger \partial_x\varphi\rangle = \tr[\varrho^{(1,1)}(+\infty)]$ where $\varrho^{(k,\ell)}$ obeys the system of ODE:
\begin{align}
  \frac{\upd}{\upd x} \varrho^{(1,0)} &= \mathcal{L}(\varrho^{(1,0)}) + J\, \varrho^{(0,0)}[Q,R]^\dagger \\
  \frac{\upd}{\upd x} \varrho^{(0,1)} &= \mathcal{L}(\varrho^{(0,1)}) + J \, [Q,R] \varrho^{(0,0)} \\
  \frac{\upd}{\upd x} \varrho^{(1,1)} &= \mathcal{L}(\varrho^{(1,1)}) + J\, \left\{ [Q,R] \varrho^{(1,0)} + \varrho^{(0,1)}[Q,R]^\dagger \right\}\, .
\end{align}
Putting everything together, we have
\begin{equation}
  \langle h_0 \rangle = \tr[\varrho^{(1,1)}(+\infty)] + m^2 \tr[\rho^{(1,1)}(+\infty)] \, .
\end{equation}
This concludes the computation of all the local expectation values we may need for bosonic QFT, covering kinetic terms, arbitrary polynomial potentials, and exponentials. 

\begin{remark}[Is the cost of evaluating an expectation value really $D^3$?]
  The cost of applying the generator of the ODE involved is clearly always that of matrix multiplication and thus $\propto D^3$. This lead me to argue that the asymptotic cost to compute the \emph{solution} of the ODE is also $\propto D^3$ asymptotically. This is a fairly natural thing to expect, and something that one observes numerically\footnote{In fact, at least for small enough $D$, we usually observe a far \emph{better} scaling than $D^3$, because of the efficient vectorization and parallelization of linear algeabra on modern CPU.}. However, I think it is important to be clear that this is not completely rigorous. At least if we solve the ODE with a high-order explicit scheme like Runge-Kutta (the best method so far), it could be that the stiffness of the ODE increases as a function of $D$ for $Q,R$ representing approximate ground states of local Hamiltonians. This would imply that the number of steps needed to approximate the solution of the ODE to fixed precision increases as a function of $D$, which would ultimately give a worse scaling than $D^3$.
\end{remark}

\subsection{Optimizing}
Before discussing the details of RCMPS optimization, it is important to remind the reader that there is no guarantee that the minimization routine we discuss will work systematically in polynomial time. Indeed, with tensor network compression, we have replaced the convex problem of minimization on a vector space $\mathscr{H}$ by the non-convex problem of minimization on a manifold $\mathcal{M}\subset \mathscr{H}$. Generically, ground state problems are expected to be hard\footnote{Ground state problems are QMA-complete (Quantum Merlin-Arthur Complete, the quantum equivalent of NP complete), i.e. they are hard to solve even with a quantum computer.}. The best our routines can do is to gradually reduce the energy until we are close to a local minimum. In fact, this is not too dissimilar from the way Nature iteself finds approximate ground states and sometimes ignores global ones\footnote{For example, even classically, finding the lowest energy state of a disordered spin system can be exponentially expensive in the system size. But this also means the actual lowest energy state is physically irrelevant, as Nature itself does not find it...}. With any variational method to find ground states, the best one can really ask for is to find the ground state whenever Nature would find it. Practically, for most problems we look at with tensor networks, this seems to be the case. Remarkably\footnote{Having an expressive ansatz, and being able to evaluate its energy efficiently does not imply that one can efficiently find even just local minima. This is relevant for modern debates around the viability of variational quantum algorithms like QAOA or VQE: the issue is not their expressiveness (how well the manifold overlaps with the interesting corner of the Hilbert space), but their optimizability (with the infamous problem of ``barren plateaus''.).}, minimization on tensor network manifolds is practically efficient.

To minimize RCMPS efficiently, we need to follow the same philosophy that worked for uniform MPS and CMPS: Riemanian Gradient Descent (and its more recent quasi-Newton improvements). This requires understanding two problems, one technical but straightforward, the other conceptual and in my opinion more subtle:
\begin{enumerate}
  \item How to efficiently compute the gradient at a cost $\propto D^3$,
  \item How to exploit the ``natural'' geometry of the (R)CMPS manifold to not get stuck in valleys.
\end{enumerate}
It helps to start with the second one and discuss the geometry of RCMPS first.

\subsubsection*{The (R)CMPS tangent space -- redundancy and metric}
Following the standard CMPS approach, we define tangent space vectors at a point $Q,R$ ~\cite{vanderstraeten2019tangentspace}
\begin{align}\label{eq:tangent}
    \ket{V,W}_{Q,R} =\!\! \int\! \upd x \left[V_{\alpha\beta} \frac{\delta }{\delta Q_{\alpha\beta}(x)} + W_{\alpha\beta} \frac{\delta }{\delta R_{\alpha\beta}(x)}\right] \! \ket{Q,R} \,
\end{align}
(\emph{a priori}) parameterized by two complex matrices $V,W$. However, just like the parameterization of a point on the manifold was redundant (we could fix the Hermitian part of $Q$), the parameterization of the tangent space at a given point is also redundant. Using the left canonical gauge $Q = -iK -\frac{1}{2} R^\dagger R $ as before, one can fix $V=-R^\dagger W$ without losing a linearly independent direction~\cite{vanderstraeten2019tangentspace}. Hence there are only $2D^2$ independent real directions on the tangent space. 

A naive choice of metric on this tangent space is the Euclidean metric for the parameters, \ie $g^\mathrm{E}_{Q,R}(W1,W2) = \mathrm{Re}(\tr(W_1^\dagger W_2)$. This is not a good choice because it is not physical: some moves $W$ that are small for this metric change the state $\ket{Q,R}$ massively, and vice versa some large moves keep the state almost as it is. We will see it more precisely once we define the proper metric.

A better and physically more natural choice of metric on this tangent space is to take the metric induced by the Hilbert space scalar product. One can show~\cite{vanderstraeten2019tangentspace} that this overlap is simply
\begin{equation}\label{eq:overlap}
\begin{split}
  \langle W_1 | W_2 \rangle_{Q,R} &=  \tr [W_2\rho_\mathrm{ss} W_1^\dagger]
    \end{split}
\end{equation}
where $\rho_\text{ss}$ is, again, the stationary state of the Lindbladian $\mathcal{L}$. The induced metric is then just defined as the real part of this overlap \eqref{eq:overlap} 
\begin{equation}
    g(W_1,W_2)_{Q,R} = \text{Re}\left(\langle W_1 | W_2 \rangle_{Q,R}\right) \, .
\end{equation} 
Its computation is very fast, as it does not require solving an ODE, contrary to expectation values of local observables. We thus get it essentially for free. Better, the metric, which can be seen as a super-operator acting on matrices of size $D^2$ is easily invertible at cost $D^3$ (its inverse is just the right multiplication by $\rho_\mathrm{ss}^{-1}$), instead of the naive $D^6$.

In this gauge, the ``right'' metric differs from the naive Euclidean metric only by this factor $\rho_\mathrm{ss}$. The eigenvalues of this matrix correspond to the \emph{entanglement spectrum} of the state. RCMPS approximate ground states well when this entanglement spectrum decays fast. Then, truncating it to finite $D$ does not induce a large error. Hence, precisely when RCMPS are \emph{good}, this matrix $\rho_\mathrm{ss}$ has incommensurable eigenvalues, and is thus very ill-conditioned (usually by several orders of magnitude). This explains why the proper metric is so different from the Euclidean metric, and why it is absolutely crucial to take it into account.

\subsubsection*{The (R)CMPS manifold -- the choice of retraction}

The last geometric step is to understand how to move on the manifold to follow a certain tangent direction, \ie we need to define a \emph{retraction}. It is simply a function $r$ that takes an initial point $(Q,R)$ on the manifold, a tangent space vector $W$, a parameter $\alpha$, and outputs a new point obtained by moving from $(Q,R)$ in the direction $W$ for a ``time'' $\alpha$. Crucially the retraction should remain on the manifold, and ideally not destroy the gauge we chose.

In principle, the natural retraction on a manifold is given by the geodesic flow. However, geodesics are typically expensive to compute (they are the solution of a complicated non-linear matrix ODE), and it seems that flowing along them does not make optimization faster in general\footnote{I tried to optimize RCMPS using a good approximation to the geodesic retraction and approximate parallel transport. Computing the retraction then ends up dominating the optimization cost, without substantially reducing the number of iterations, at least on the examples I considered.}. Inspired by~\cite{vanderstraeten2019tangentspace}, we can instead use the ``naive'' retraction
\begin{equation}
    (K,R)\rightarrow r(K,R,W,\alpha) = \bigg(K + \frac{i}{2}\left[W^\dagger(\alpha R+\alpha^2/2 W) -  (\alpha R^\dagger+\alpha^2/2 W^\dagger) W\right] \, , \,
    R + \alpha W \bigg)\, ,
\end{equation}
which does correspond to a move in the direction of $W$ (at the leading order in $\alpha$) and does preserve the gauge along the way ($K$ remains Hermitian). Taking the ``naive'' metric is not good enough for the minimization to work (and we do need to consider the proper induced metric), but taking the naive retraction seems good enough in practice!

\subsubsection*{Gradient by backward differentiation}
Now that we know how to move on the RCMPS manifold, we need to know what is the direction that minimizes the energy. This is given by the gradient of the energy density, which we need to compute efficiently. 

We have seen in \ref{sec:expectation_values} that evaluating each term (kinetic, potential) in the energy density $\langle h \rangle$ has a cost $\propto D^3$. The naive reader may thus conclude that the cost of evaluating the gradient $\nabla_{Q,R} \langle h \rangle$ is $\propto D^5$ since we have $\propto D^2$ partial derivatives to take. This would naturally be the scaling if we were taking partial derivatives one by one with finite differences. The more educated reader will know that, thanks to \emph{backward differentiation}, it should always be possible to compute the gradient of a scalar function with the same asymptotic cost as that of computing the function itself~\cite{Griewank1989OnAD,griewankWalther2008EvaluatingDerivatives}, \ie here $\propto D^3$.

And indeed, one could in principle use automatic differentiation to backward differentiate through the various ODEs from which $\langle h\rangle$ is obtained. This differentiation at the code level is however impractical: although it would indeed give an asymptotic cost in $D^3$, it would blow the memory up even at moderate $D$. Rather, it is helpful to explicitly write down an expression for the gradient using the adjoint method, and then observe that we can compute it with a moderate amount of memory using quadratures. 

Let us illustrate the method on the expectation value of a vertex operator $\langle V_\beta\rangle_{Q,R}=\bra{Q,R} :\e^{\beta\phi}\!:\ket{Q,R}$ for $\beta$ real\footnote{For complex $\beta$, the expectation value $\langle V_\beta\rangle$ is not real, and thus $\nabla_{W} \langle V_\beta\rangle \neq g(\nabla \langle V_\beta\rangle,W) $. To get the gradient, one would first needs to split the expectation value into real and imaginary parts.}, but other local observables appearing in the Hamiltonian can be treated in the same way.

The gradient in the direction W, $\nabla_{W} \langle V_\beta\rangle = g(\nabla \langle V_\beta\rangle,W) $, is obtained via
\begin{equation}
     \langle V_\beta\rangle_{Q- \varepsilon R^\dagger W,R + \varepsilon W} =   \langle V_\beta\rangle_{Q,R} + \varepsilon \nabla_{W} \langle V_\beta\rangle_{Q,R} + O(\varepsilon^2) \,
\end{equation}
where we have parameterized the tangent space direction, \ie the allowed infinitesimal moves, using the gauge fixing discussed before. Differentiating the ODE \eqref{eq:vertex_ode} directly gives
\begin{equation}\label{eq:gradient_raw}
    \nabla_{W} \langle V_\beta\rangle=\int \upd y\, \tr\left\{ \left(\mathcal{T}\e^{\int_{y}^{+\infty} \mathcal{L}^\beta}\right) \cdot \nabla_{W}\mathcal{L}^\beta(y) \cdot\left(\mathcal{T}\e^{\int_{-\infty}^y \mathcal{L}^\beta}\right)\cdot \rho_\text{ss}\right\}.
\end{equation}
with the notation $\mathcal{L}^\beta(x)\cdot \rho = \mathcal{L} \cdot \rho + \beta J(x) (R\rho + \rho R^\dagger)$ and
\begin{equation}\label{eq:derivative_Lindblad}
\begin{split}
    \nabla_{W}\mathcal{L}^\beta(y) \cdot \rho = -R^\dagger W \rho - \rho W^\dagger R+ \frac{1}{2} \left(R\rho W^\dagger+W\rho R^\dagger\right) 
    + \beta J(y) \left(W\rho + \rho W^\dagger\right)\, ,
    \end{split}
\end{equation}
We then replace the last part of the evolution from $y$ to $+\infty$ in \eqref{eq:gradient_raw} by the adjoint evolution applied to the identity
\begin{equation}\label{eq:gradient_adjoint}
\begin{split}
    \nabla_{W} \langle V_\beta\rangle=\int \upd y\, \tr\bigg\{ \left[\left(\mathcal{T}\e^{\int_{y}^{+\infty}  \mathcal{L}^{\beta*}}\right)\cdot \mathds{1}\right]  
    \times &\nabla_{W}\mathcal{L}^\beta(y)\cdot\left[\left(\mathcal{T}\e^{\int_{-\infty}^y \mathcal{L}^\beta}\right)\cdot \rho_\text{ss}\right]\bigg\}.
\end{split}
\end{equation}
where the adjoint $\mathcal{L}^{\beta*}(y)$ of $\mathcal{L}^\beta(y)$ is defined as
\begin{equation}
    \mathcal{L}^{\beta *}(y)\cdot \mathcal{O} = Q^\dagger\mathcal{O} + \mathcal{O} Q+ \frac{1}{2} R^\dagger \mathcal{O} R + \beta J(y) \left[R^\dagger \mathcal{O} + \mathcal{O}R \right].
\end{equation}
The solution $\rho(x)=\mathcal{T}\e^{\int_{-\infty}^x \mathcal{L}^\beta}\!\cdot \rho_\text{0}$ of the forward problem and the solution  $\mathcal{O}_x=\mathcal{P}\e^{\int_{x}^{+\infty}  \mathcal{L}^{\beta *}}\!\cdot \mathds{1}$ of the backward problem can be computed by solving the corresponding ODEs. The gradient in the direction $W$ is then simply
\begin{align}
    \nabla_{W} \langle V_\beta\rangle &=\int \upd y \, \tr \left[\mathcal{O}_y \nabla_{W}\mathcal{L}^\beta (y)\cdot \rho_y\right] \\
    &=\tr\left[M_W W + M_{W}^\dagger W^\dagger\right]
\end{align}
with
\begin{equation}\label{eq:matrices_integral} 
    M_W= \int \upd y\, -\rho_y \mathcal{O}_y R^\dagger + \frac{1}{2} \rho_y R^\dagger \mathcal{O}_y + \beta J(y) \rho_y \mathcal{O}_y 
\end{equation}
The matrix $M_W$ is obtained by evaluating the integral \eqref{eq:matrices_integral} with an efficient quadrature:
\begin{equation}
  \int \upd y f(y) \underset{\text{machine precision}} {=} \sum_{k=1}^N w_k f(y_k)
  \label{eq:quadatureapproximation}
\end{equation}
Using good choices of $y_k$ and $w_k$ like the \emph{exp-sinh}~\cite{mori2001tanhsinh}  prescription\footnote{The exp-sinh quadrature is adapted for integrals over $]0,+\infty[$. Even though we want to compute an integral over $\mathbb{R}$, this quadrature  is adapted for us because we have a singularity in $y=0$. In practice, we split the integral into two integrals over $]- \infty,0[$ and $]0,+\infty[$, and use the exp-sinh quadrature, which is tolerant to boundary singularities, to compute both.}, instead of summing over all the values computed in the course of solving the ODEs, requires storing far fewer points. This is where this adjoint based method differs from computing the gradient with backpropagation at the code level.

Ultimately, the full gradient is obtained from the matrices M using $g(\nabla \langle V_\beta\rangle , W) = \nabla_W\langle V_\beta\rangle$ which gives
\begin{equation}
  \tr[ \nabla \langle V_\beta\rangle \rho_0 W + W^\dagger \rho_\text{ss} \nabla \langle V_\beta\rangle^\dagger ] = \tr[ M_W W + M_W^\dagger W^\dagger]
\end{equation}
and thus $\nabla \langle V_\beta\rangle = M_W \rho_\text{ss}^{-1}$.

The gradient of other observables can be computed with the same techniques. Hence, we have an efficient way to compute the gradient of the full energy density, and thus minimize it. Compared to the computation of expectation values themselves, we have to solve two ODEs (the forward and adjoint) instead of one, and store intermediate values (at the $\sim 1000$ quadrature points), so the computational cost is a bit more than doubled.

\subsubsection*{Refinements}
Once we have a way to compute the gradient, and a way to move on the manifold in a given descent direction (a retraction), we can simply minimize the energy using gradient descent, that is descend in the direction opposite to the gradient. This is standard Riemannian gradient descent, which is equivalent to imaginary time TDVP in the tensor network literature. This already gives a fairly efficient method to minimize the energy, but becomes slower on ``difficult'' points, typically near phase transitions.

In fact, we can pick even better descent directions using Riemannian quasi-Newton algorithms like Riemannian conjugate gradient or LBFGS (limited memory Broyden–Fletcher–Goldfarb–Shanno algorithm). To this end, it is easier to work with a Euclidean manifold, on which these quasi-Newton algorithms are reliable, and interpret the different metric as a preconditioner applied to the final descent direction. For gradient descent, Riemannian gradient descent and Euclidean gradient descent $+$ Riemannian preconditioning are equivalent. But the second interpretation allows a convenient extension to quasi-Newton. This is the strategy of Hauru, Van Damme, and Haegeman~\cite{hauru2021} which I followed in practice\footnote{I also tried fancier strategies in~\cite{tilloy2022studyquantumsinhgordonmodel}, that proved slightly faster, and which consisted in using Riemannian quasi-Newton method on the proper manifold but with a heavily regulated metric, and with a much less regulated preconditioner. I am not sure the benefits outweigh the increased complexity, especially when building extensions to more complicated models.}.

A final subtlety is related to the need for regulators. As was noted by~\cite{hauru2021} in the context of MPS, the metric we use depends on $\rho_\text{ss}$, the fixed point of $\mathcal{L}$, which can be very ill-conditioned. This is a problem since its inverse $\rho_\text{ss}^{-1}$ appears in the gradient and damages the estimate of the Hessian made by LBFGS. One thus needs to regulate the metric by redefining the matrix $\rho_\text{ss} \rightarrow \rho_\text{ss}^\varepsilon$ appearing in its definition $\rho^{\varepsilon}_\text{ss} = \rho_\text{ss} + \varepsilon \mathds{1}$. Although I have used a range of fancier methods that improve marginally the optimization, the easiest is to follow the strategy of~\cite{hauru2021}: regulate less and less as the gradient gets smaller: $\varepsilon = 10^{-2} g(\nabla e_0,\nabla e_0)$.

\subsubsection*{Implementation}
Currently, there is no general purpose package to work with RCMPS, and the codes have been mostly \emph{ad hoc} for each model that was studied. Hopefully, once more $1+1$d models are solved, it will be possible to construct a Julia package that solves essentially \emph{all} Hamiltonians within a sufficiently broad class. In the meantime, here is the strategy that is followed in the \emph{ad hoc scripts}.

To evaluate the expectations values, I simply solved the ODEs mentioned previously for $x$ from $-20$ to $20$ (which provided results indistinguishable from $]-\infty, +\infty[$). The generator of the evolution has an integrable singularity in $x=0$ which I smoothed with the change of variable $u = \e^{-x^2} x^3+(1-\e^{-x^2}) x$. I used \texttt{KrylovKit.jl} to find the stationary state $\rho_\text{ss}$ which is used in the initial condition. Then, to evolve the ODE forward, I used the package \texttt{DifferentialEquations.jl}~\cite{rackauckas2017differentialequations} with the \texttt{Vern7} solver (an explicit Runge-Kutta scheme) and a relative tolerance of $10^{-12}$. The matrix multiplications necessary for the evaluation of the ODE generator were substantially sped up using \texttt{Octavian.jl}, which relies on \texttt{LoopVectorization.jl}. For the bond dimensions I used (up to $D=32$), the only worthwhile parallelization was to compute the different expectation values appearing in the energy with different processes.

To compute the gradient of the energy, I manually implemented the adjoint method discussed previously. Taking a few hundred points was typically sufficient to make the quadrature error negligible compared to the ODE error.

For energy minimization, I have mostly used Riemannian LBFGS, with large memory parameter (\eg $m=200$), using the remarkably convenient and problem agnostic package \texttt{OptimKit.jl}. All that is needed for the optimization is to provide the function to optimize (here the energy density), its gradient, the metric, a retraction, a vector transport, and a preconditioner. 

It is important to note that thanks to these Julia packages, developed for some at the University of Ghent, the Julia implementation is much faster and cleaner than the Python version I was using in the early days. Their existence provides invaluable help.

\chapter{Solving models}

Now that we have an ansatz which we understand, and a way to optimize it, we can study particular models. However, things are not as easy as with MPS on the lattice. RCMPS cannot deal with all models out of the box, because divergences are model-specific. The construction we have so far provided still works for a wide class of models of a single scalar field with polynomial or sufficiently renormalizable exponential potential. We will thus present results for such models first. For some of the other models, a mild generalization of the ansatz is sufficient: this will be the case for coupled scalar fields. For others, like Fermionic models, an extension seems feasible but is only currently being worked out. Finally, super-renormalizable but not \emph{strongly}\footnote{I am not sure this terminology is generic, but I call a model \emph{strongly} renormalizable if it is obtained by perturbing a CFT with an operator $\mathcal{O}(x)$ of dimension $\Delta_\mathcal{O} < d/2$. This is a much stronger condition than being super-renormalizable.}-renormalizable models cannot currently be dealt with. In that case, there is no obvious fix within RCMPS theory, and it would seem another ansatz is necessary.

\section{The self-interacting scalar field}

\subsection{The model}
The $\phi^4$ model in $1+1$ dimension (sometimes written $\phi^4_2$) is the quintessential pile of dirt which we discussed in \ref{sec:pileofdirt}. This model can be defined rigorously simply by giving its Hamiltonian operator, which is a well defined, lower-bounded self-adjoint operator on Fock space:
\begin{equation}
  H = \int_\mathbb{R} \frac{:\!\pi^2\!:}{2} + \frac{:\!(\partial_x\phi)^2\!:}{2} + m^2 \frac{:\!\phi^2\!:}{2} + g :\!\phi^4 \!:
  \label{eq:phi4_definition}
\end{equation}
where we recall that the normal ordering is done with respect to the normal mode operators $a_k,a^\dagger_k$ diagonalizing the free part. This normal-ordering is equivalent to the suppression of tadpoles in a perturbative expansion in Feynman diagrams. Such a subtraction is sufficient to kill all divergences in $1+1$ dimensions, which is why this model is easier to define than its higher dimensional counterparts. 

Defining $\phi^4_2$ rigorously is one of the very first achievements of ``old-fashioned'' constructive field theory. In particular, Nelson proved in~\cite{nelson1966quartic} (see also~\cite{simon2004ed} for historical context) that the Hamiltonian $H$ of \eqref{eq:phi4_definition} has a lower-bounded density\footnote{This is a surprisingly non-trivial thing to prove. Because of normal-ordering, the Hamiltonian is not a sum of positive terms. In particular, $: \!\phi^4(x)\! :$ is \emph{not} lower-bounded. To my knowledge, no one has proved explicit (ideally numerically improvable) lower-bounds to the ground state energy density of $\phi^4_2$, that would complement the rigorous upper-bound obtained with the variational method. Standard approaches based on semi-definite programming are not immediately usable.}, a proof which was simplified\footnote{I could only have a look at the ``simplified'' proof by Federbush, the original proof of Nelson being very difficult to find online.} by Federbush~\cite{federbush1969phi4lower}. This guarantees that the variational method will work for this model: the minimum exists, and we cannot get a runaway minimization where the energy gets ever lower.

This model has simple but not fully trivial physics. Without loss of generality, we can fix $m^2=1$, which we will do from now on, by appropriately rescaling $x$. Then the model has a single parameter $g$. The model is $\mathbb{Z}_2$ symmetric for $g$ sufficiently small, but symmetry is spontaneously broken for $g > g_c \simeq 2.77$. The exact value of this critical coupling is not known analytically, but it has been estimated precisely with almost all the known numerical methods available.

Let me conclude this presentation of the model with two trivial-ish remarks aimed at dispelling some (empirically) common misunderstandings.

\begin{remark}[Not just Wilson-Fisher!]
This lovely model is sometimes unfairly belittled as being ``just'' the Wilson-Fisher fixed point, or ``just'' the Ising CFT. I risk that this is a professional deformation of people who have worked too much on conformal field theory. As trivial as this misunderstanding is, let me still attempt a (likely superfluous) clarification. Of course, it is correct that \emph{critical} $\phi^4_2$ is in the Ising CFT universality class. This means that at large distances, and when $g = g_c$, correlation functions of $\phi^4_2$ decay as power laws with the same exponents. It is also true that the phases accessible for $g\neq g_c$ match those of Ising ($\mathbb{Z}_2$ symmetry preserved or broken). But when $g\neq g_c$ nothing about the intermediate to large distance behavior of the model can be said from universality arguments. The $\phi^4_2$ model is then a massive QFT, with non-trivial scattering properties. A CFT still rules its short distance behavior, but it is the free boson\footnote{The massless free boson, without compactification radius, is strictly speaking not a well defined CFT. We use the term loosely here, to refer to the model which has as correlation functions the short distance limit of the correlation functions of the massive free boson with mass $m=1$. At the level of correlation functions, this limit is well defined.}, not the Ising CFT. Further, even when $g = g_c$, the $\phi^4_2$ model is not ``just'' the Ising CFT. It is still a theory with a scale, fixed by $m$ (the one-loop renormalized mass in path integral language). At short distances, it is, again, well approximated by the free boson CFT with central charge $c=1$, and at large distances, it is well approximated by the Ising CFT with central charge $c=1/2$. In between, for distances of order $1$ in $m^{-1}$ units, nothing can be said about the physics from universality alone, and a powerful computational method (like RCMPS) is required.
\end{remark}

\begin{remark}[Phase transition at positive renormalized mass]
It sometimes appears suspicious that the symmetry breaking happens for \emph{positive} $m^2$, as $g$ is \emph{increased}. Naively, in the path integral language, we expect symmetry breaking to occur when the bare mass is sufficiently negative compared to $g$, so that the total potential $V(\phi)$ develops minima at $\phi \neq 0$. In fact both views are compatible, and the difference comes from normal-ordering (or 1-loop renormalization of the mass). Indeed, we have $:\!\phi^4\!: = \phi^4 - (+\infty) \phi^2$, and thus increasing $g$ actually lowers the bare mass. It is thus not surprising that the symmetry breaking happens when increasing $g$, keeping positive $m^2$ fixed.
\end{remark}

\subsection{Numerical state of the art}
The $\phi^4_2$ model has been studied with almost all the numerical methods that are known for field theory. 

\subsubsection*{UV and IR cutoffs}
I think it is debatable, but not unreasonable, to claim that RCMPS is the first non-perturbative method for this \emph{exact} problem that does not require a UV or IR cutoff. Since this point is sometimes debated, I would like to argue for it here. The methods I know of are the following:
\begin{enumerate}
  \item (Borel-resummed) perturbation theory~\cite{serone2018symmetric,serone2019broken,heymans2021N8LO} \texttt{[no UV cutoff -- no IR cutoff]}
  \item Lattice Monte Carlo~\cite{loinaz1998mc_phi4} \texttt{[UV cutoff -- IR cutoff]}
  \item (Renormalized) Hamiltonian truncation~\cite{hogervorst2015, rychkov2015,eliasmiro2016,eliasmiro2017-1,eliasmiro2017-2} \texttt{[no UV cutoff -- IR cutoff]}
  \item Tensor network renormalization~\cite{delcamp202giltphi4} \texttt{[UV cutoff -- almost no IR cutoff]}
  \item (infinite) Matrix Product States~\cite{milsted2013mpsphi4,vanhecke2022scalingphi4} \texttt{[UV cutoff -- no IR cutoff]}.
  \item Light-front conformal truncation~\cite{anand2020introductionlightconeconformaltruncation} \texttt{[no UV cutoff -- no IR cutoff in a different picture]}
\end{enumerate}
Let me discuss light-front conformal truncation (LFCT) later.
Apart from LFCT, and of course perturbation theory, all those methods work with a UV or IR cutoff, \ie they deal with a slightly different model, and need to be extrapolated to apply to the true $\phi^4_2$ model. I distinguish a UV/IR \emph{cutoff}, that applies to the model itself, from whatever parameter is controlling the quality of the solution to this model. Hence, at least in my opinion, the finite loop order $n$ in perturbation theory, the finite truncation energy $E_T$ in Hamiltonian truncation, or the finite bond dimension $D$ in tensor network methods are \emph{not} UV or IR cutoffs. 

\begin{remark}[Entanglement cutoff?]
Sometimes one says that $D$ is an \emph{entanglement} cutoff, in that it limits the maximal amount of entanglement in the state. For critical models, this entanglement limitation in turn introduces a new lengthscale, which has physical effects that ressemble a hard IR cutoff. This does not imply it \emph{is} an IR cutoff.
\end{remark}

So what about LFCT? LFCT is just like Hamiltonian truncation, but for a different quantization scheme where the equal-time hypersurface $t = \text{constant}$ on which states are defined is replaced by a light-front $x=t$. I think it is fair to say that this method has no UV/IR cutoff in $1+1$ dimensions and that its truncation number is really akin to the truncation energy in ``standard'' HT. This method exploits the Lorentz invariance of the model to remove the extensive contribution of the vacuum / ground state, and provides a completely orthogonal way to deal with the thermodynamic limit.

I think it is however a bit more questionable that LFCT deals with the exact same model. Indeed, the light-front quantization scheme is difficult to relate to the standard one \cite{fitzpatrick2020}. The coupling constants in the two quantizations are related to each other non-perturbatively. As a result, the only way to connect observables computed with the light-front method to observables in the original formulation is to solve the original formulation with another method and match \emph{physical} quantities first (like the mass gap). If we understand a model in the ``condensed matter'' sense\footnote{In the high energy sense, a model is a relationship between physical properties, and thus in this sense LFCT does solve the same model as equal-time quantization methods.}, \ie as a fixed microscopic prescription from which the task is to extract physical properties, then LFCT solves a different model.

\subsubsection*{Performance of the known numerical methods}
The known numerical methods shine in different limits of the model. Perturbation theory, and its resummed version, are particularly accurate, as expected, in the small coupling limit. For very small $g$, say $g \ll 0.1$ Borel resummed perturbation theory is likely hard to beat in precision with any method. For larger coupling, the problem is usually thought to be one of convergence (for an asymptotic series), but thanks to Borel-Padé resummation, the problem is \emph{only} that high orders are factorially expensive to compute.

Hamiltonian truncation (boosted by its renormalized refinements) is likely the most accurate method if one cares about the model in a sufficiently small volume (say $L \leq m^{-1}$), as it is exact in the limit of infinitely small volume. It is also more accurate if one is interested in spectral data, which is immediately available from diagonalization of the truncated Hamiltonian.

On the contrary, if one cares more about the IR behavior of the model, methods based on a lattice discretization and extrapolated to the continuum are often the most accurate. In principle, tensor network renormalization should be slightly more efficient exactly at criticality, and give direct access to scaling dimensions, while MPS should be more efficient when one is at least slightly off criticality.

For this scalar model without sign problem, the Monte Carlo method would likely be the most accurate in more dimensions (say at least $2+1$d), in a regime where tensor network methods struggle to reach similar accuracy. But the Monte Carlo method is no longer as accurate as TNR or MPS in $1+1$d, simply because it has a harder IR cutoff. Indeed, while the cost of TNR and MPS increases with a smaller UV cutoff, their cost weakly depends on the IR cutoff (as I argued in remark \ref{rk:not_IR}) . Indeed, for TNR the IR cutoff is exponential in the number of iterations, while MPS algorithms can be run directly in the thermodynamic limit.

\begin{remark}[Personal history]
I got interested in $\phi^4_2$, and in the many ways to solve it after watching the recordings\footnote{The whole workshop was recorded and is available on Youtube: \url{www.youtube.com/playlist?list=PLx5f8IelFRgHkCsp-XyPEUn8caIUENB4j}.} 
of the workshop \emph{Hamiltonian methods in strongly coupled QFT} organized at the Institut des Hautes Études Scientifiques (IHES) in Bures-sur-Yvette by Slava Rychkov, Ami Katz, Balt Van Rees, Robert Konik. This model was clearly shown as a challenge: simple to define but non-trivial, without a clear way to cheat. This is when I got aware that it was possible to do numerics in relativistic QFT in a fully well-defined context. Being next to the office of Clément Delcamp at MPQ then, and knowing he had come up with a new numerical method (Gilt-TNR) while in Waterloo, I thought we could study $\phi^4_2$ with it as well, which had not been done yet. Having to really code something concrete to solve the model finally helped me understand renormalization and the continuum limit in the relativistic context. This planted the seed for subsequent work on RCMPS.
\end{remark}

Compared to all these methods, where should we expect RCMPS to stand? RCMPS should fully solve the UV problem, like HT, but work without an IR cutoff, and thus have an IR performance comparable to plain MPS. However, because RCMPS is more expensive than MPS at the same bond dimension, we cannot expect to push $D$ as much. Thus RCMPS should be better than lattice MPS when the extrapolation to small lattice sizes of MPS costs a lot of bond dimension. This is \emph{a priori} the case for local observables like the energy density or the field expectation value sufficiently away from criticality. Of course, near criticality, the IR starts to cost bond dimension as well, and universality starts to wash away UV physics. Thus, we expect plain MPS to still give a better accuracy than RCMPS when estimating universal IR properties (like scaling dimensions). 

\begin{remark}[Finding $g_c$]
A particularly well defined challenge for this model is to estimate its critical coupling $g_c \simeq 2.77$, \ie the coupling at which we switch from a symmetric to a symmetry-broken phase. The value of this critical coupling depends on the UV details of the model, and is not something one can compute from general arguments. But, of course, estimating it also requires being able to deal with the IR very well, otherwise one cannot know in which phase one is. This is why $g_c$ is challenging to estimate with all the known methods, because no known method deals with the UV and the critical IR at the same time and out of the box.
Currently, the best method is a double extrapolation (in UV cutoff and in entanglement) using plain MPS~\cite{vanhecke2022scalingphi4}, which gives $g_c \simeq 2.774245(8)$. A lesson is that if you are going to extrapolate anyway (since no method can do without extrapolation here), it is better to extrapolate a lot\footnote{Of course, extrapolations are only as good as the ansatz you have for the scaling behavior. This is certainly the limit of these results obtained from multiple extrapolations. Their error estimates assume that there is no systematic bias from an incorrect choice of scaling functions.}. Although I have not yet tried to push RCMPS in this direction (with a finite entanglement scaling but of course no need for UV cutoff extrapolation), I doubt RCMPS would be competitive because of the practical limits in $D$.
\end{remark}

\subsection{RCMPS results}

\subsubsection*{Optimization}
To find the RCMPS approximation of the $\phi^4_2$ ground state, I initially used the very naive method based on direct evaluation of integrals, which barely allowed me to push to $D=9$ using a cluster. I then moved on to the ODE approach, implemented in Python, that allowed to push $D$ to $25$ reliably, and get a few heroic points at $D=32$. This corresponds to the results published in~\cite{tilloy2021relativistic}. Finally, I moved on to Julia to solve the ODEs, and replaced plain Riemannian gradient descent with LBFGS using \texttt{OptimKit} which allowed to reliably get points at $D=32$ and even push to $64$. This latter code, with all its optimizations, was developed for the Sinh-Gordon model (which required far more iterations), but it can be used for $\phi^4$ as well. The following results have been recomputed by Karan Tiwana with this more recent code, and have not appeared elsewhere yet.

\subsubsection*{Energy density}
\begin{figure}
  \begin{center}
    \includegraphics[width=0.675\textwidth]{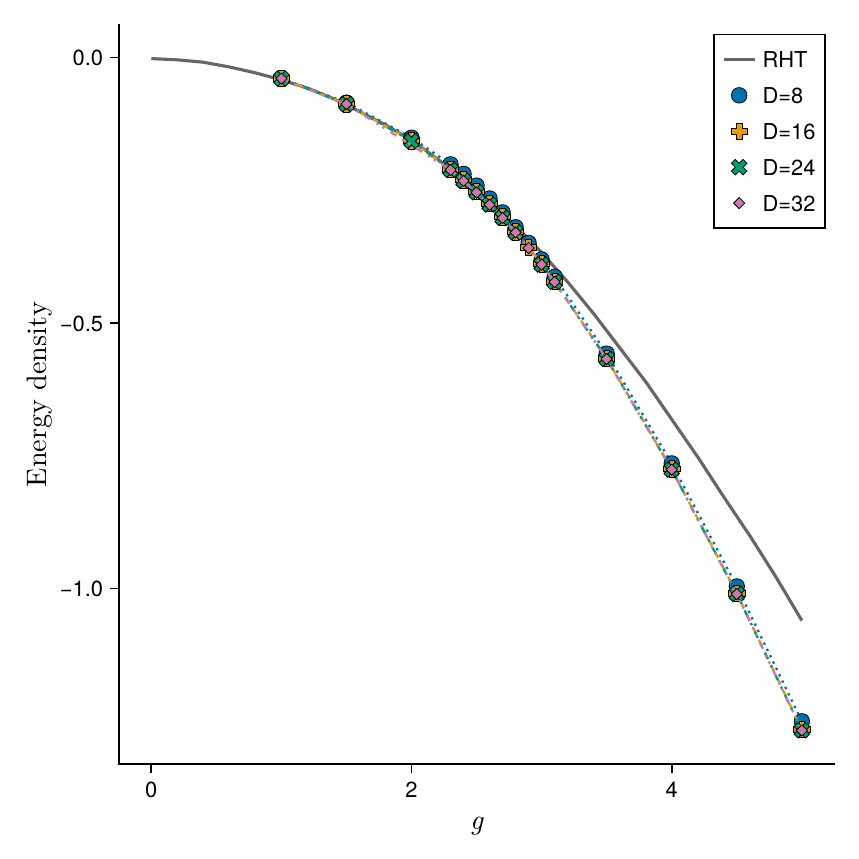}
    \includegraphics[width=.31\textwidth]{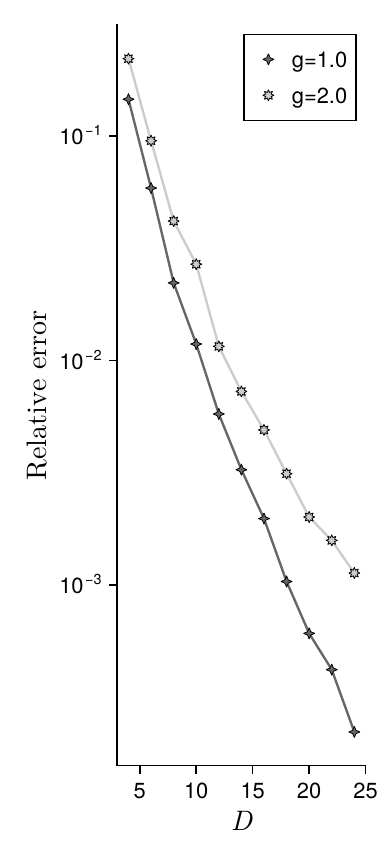}
  \end{center}
  \caption{Left: Energy density as  a function of $g$, in the thermodynamic limit, compared with renormalized Hamiltonian truncation results (RHT) obtained for a size $L=10$. Right: relative error in energy density with the method explained in the text.}
  \label{fig:e0}
\end{figure}

The energy density itself is really the observable where RCMPS shine, because it is directly the one that is optimized. The results are shown in Fig. \ref{fig:e0} 
where they are compared with the renormalized Hamiltonian truncation results of Rychkov and Vitale~\cite{rychkov2015}. Already at $D=8$, the energy density is very well approximated non-perturbatively, for all values of $g$.

Since this model is not integrable, and RCMPS is the most precise method for this observable, we estimate the error by self-comparison. Namely, we take a good approximation of the true $e0$ at very large $D$ (here $D=48$), and then look at the relative error obtained at lower $D$ (up to $D=24$). This gives an almost exponential decay of the error as a function of $D$, which is in line with theory expectations in lattice MPS~\cite{huang2015error}. This ultimately validates that for $D$ sufficiently small compared to our reference $D=48$, we have $|e0(D) - e0(48)|\gg |e0(48) - e0(+\infty)|$ and thus $|e0(D) - e0(48)|\simeq |e0(D) - e0(+\infty)|$.

Recall that the cost of the method is $\propto D^3$ and thus improving upon these results is in principle inexpensive, and one could reach machine precision on the energy density. However, in practice, getting beyond $D=64$ is difficult because of the limited precision (in \texttt{Float64}) of two crucial steps: the resolution of our large systems of ODE, and the inversion of the ill-conditioned $\rho_0$.

\subsubsection*{Magnetization}
\begin{figure}
  \begin{center}
    \includegraphics[width=0.95\textwidth]{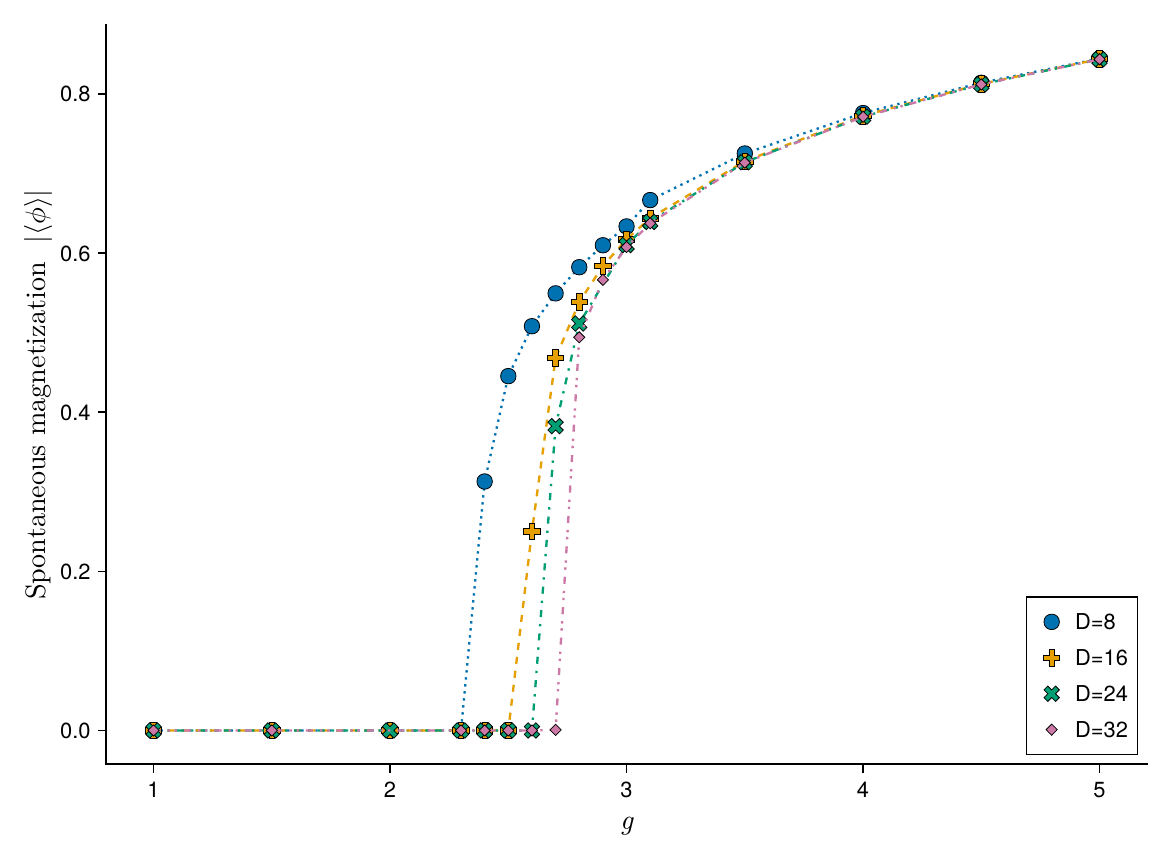}
  \end{center}
  \caption{Spontaneous magnetization $|\langle \phi \rangle|$ as a function of $g$.}\label{fig:magnetization}
\end{figure}

A natural order parameter for $\phi^4$ theory is the field average. It should be zero because of the $\mathbb{Z}_2$ symmetry of the Hamiltonian, but for sufficiently large $g$ ($g > g_c \simeq 2.77$), there is a spontaneous symmetry breaking. 

Interestingly, with RCMPS, we do not need to add a tiny external field to create this spontaneous magnetization, and the optimization over $Q,R$ breaks the symmetry on its own (with a random sign depending on the initial condition). This is quite intuitive\footnote{I am grateful to Jutho Haegeman who explained this to me when I first presented early RCMPS results.}. In the symmetry broken phase, breaking the symmetry of the ground state does not cost energy in the thermodynamic limit compared to taking the symmetric superposition of opposite magnetizations. However, the symmetric Schrödinger cat state has more entanglement. With RCMPS, we have a limited number of parameters, supporting a limited amount of entanglement: the minimization will reduce entanglement if it saves parameters to minimize the energy. This is also why the spontaneous symmetry breaking happens always slightly before ($g < g_c$) than expected, especially at lower $D$. Indeed, in the symmetric phase near the phase transition, breaking the symmetry costs only very little energy: it can be worth doing to save parameters and reduce energy elsewhere.

The precision obtained for the magnetization is not as good as for the energy density, but remains impressive, especially deep in the gapped phases. As expected, the precision gets lower in the vicinity of the second-order phase transition, where a finite entanglement scaling is in principle necessary to get unbiased results.

\begin{remark}[$\mathbb{Z}_2$ invariant RCMPS]
  It is possible to enforce the $\mathbb{Z}_2$ symmetry exactly at the RCMPS level. To this end, one remarks that $\phi \rightarrow -\phi$ translates into $a(x)\rightarrow -a(x)$ at the mode level. By choosing $K,R$ of the following block form
  \begin{equation}\label{eq:Z2_block}
    R  = \left(\begin{array}{cc}
       0  & R_1 \\
       R_2  & 0 
    \end{array}\right) ~~ \text{and} ~~
    K  = \left(\begin{array}{cc}
       K_1 & 0 \\
       0  & K_2
    \end{array}\right)\, ,
\end{equation}
one get the required invariance.
\end{remark}

\subsubsection*{Entanglement entropy}
\begin{figure}
  \begin{center}
    \includegraphics[width=0.95\textwidth]{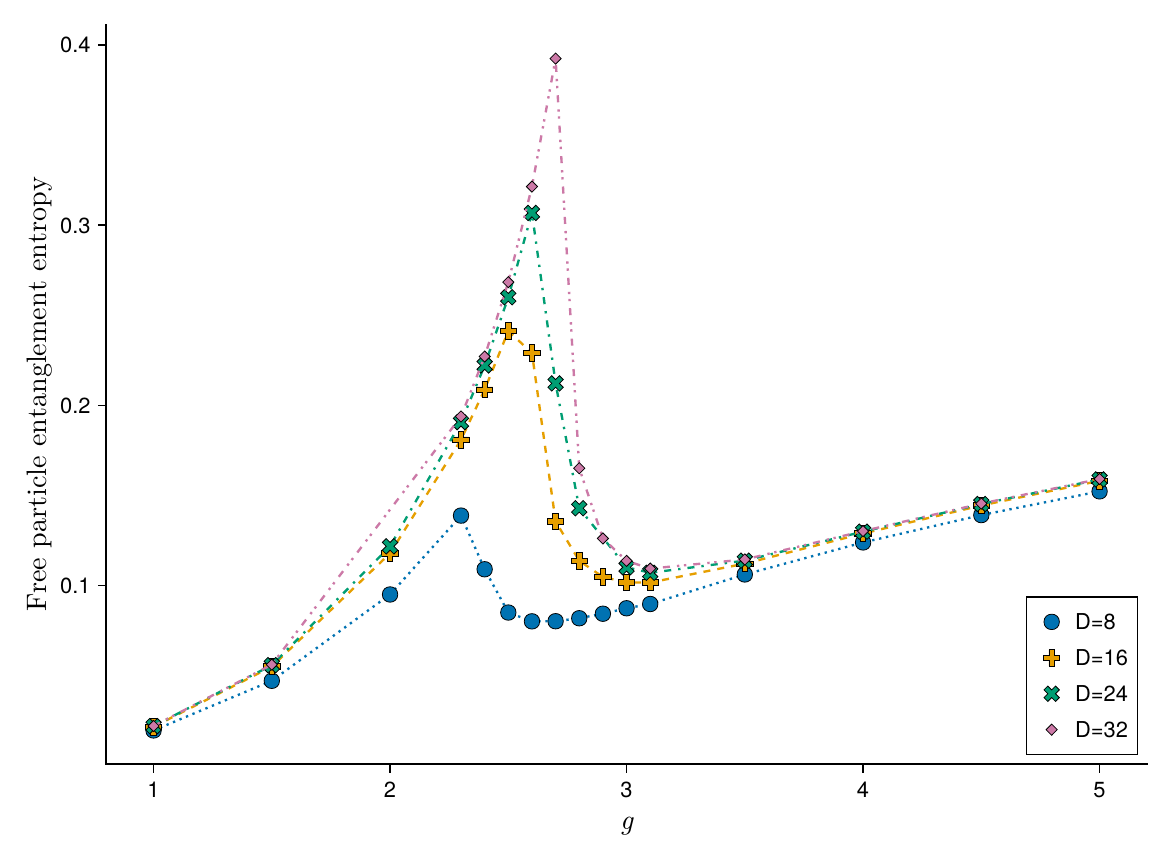}
  \end{center}
  \caption{Free particle entanglement entropy as a function of $g$.}\label{fig:entropy}
\end{figure}

We have defined a new entanglement entropy, which we called ``free particle entanglement entropy'' (for lack of a better name) in remark \ref{rk:particleentanglement}. Recall that it quantifies the amount of entanglement on top of the free vacuum, for the tensor product structure relevant to RCMPS. If our intuition is correct, it should be \emph{finite} in gapped phases, and diverge as we approach a second-order phase transition (in $\log L$ or $\log D$ depending on the cutoff used). This is indeed what we observe in Fig. \ref{fig:entropy}.
We can clearly see the divergence, regulated to order $\sim \log(D)$, near the expected transition point.

\section{The Sine-Gordon and Sinh-Gordon models}

The Sine-Gordon and Sinh-Gordon models are other scalar field models with potential $V(\phi) \propto \cos(\beta\phi)$ and $V(\phi) \propto \cosh(\beta \phi)$ respectively. These two models are interesting for many reasons. First, they are easily well defined at least for $\beta \leq \sqrt{4\pi}$ (and, we will see, also for $\beta$ larger but the situation is more subtle). They are also exactly solvable. Of course, I am not interested in trying to leverage the structure that makes them integrable, which would go against the philosophy I tried to defend in \ref{sec:pileofdirt}. But integrability provides a convenient benchmark to know the RCMPS errors \emph{exactly}, without the need for self-comparison to extrapolated values. Further, the scaling dimension of the potential of the Sine-Gordon model depends on $\beta^2$ and thus can be adjusted to make the potential more or less renormalizable. This is particularly convenient to test what happens near the domain of validity of RCMPS, where the potential goes from strongly renormalizable to not strongly renormalizable (at which point the ground state energy density collapses to $-\infty$).

Finally, the Sinh-Gordon model in particular is interesting because it should be a completely trivial model but is not. For $\beta \geq \sqrt{8\pi}$, it is manifest that the exact results from integrability, which are obtained from analytic continuation, are nonsensical (they predict a complex energy density). In fact for such large values of $\beta$, it is not clear if the model is just massless, requires an extra renormalization, or is even well defined at all.

To this day, what exactly is going on with the Sinh-Gordon model at large coupling remains unclear, despite an incredibly thorough analytic and numerical study by Konik, Lajer, and Mussardo (KLM)~\cite{konik2021}, as well as renormalization group arguments by Bernard and LeClair~\cite{bernard2022}. I was hoping to fully settle the matter using RCMPS in~\cite{tilloy2022studyquantumsinhgordonmodel}, but the results I obtained are ultimately insufficient to conclude, although they support the scenarios proposed by KLM. My hope is that future improvements in the optimizer, combined with a better analytical understanding of the model, will one day allow me to update~\cite{tilloy2022studyquantumsinhgordonmodel} and give it a real conclusion. 

The following subsections, which are dedicated to a more precise presentation of the models and its special parameter values, borrow from~\cite{tilloy2022studyquantumsinhgordonmodel} and only expand it with a few remarks.

\subsection{Two definitions of the models}
One can define the Sine-Gordon (SG) and Sinh-Gordon (ShG) models from their Hamiltonian formulation, directly in the continuum, which is the most natural for us. As I will later discuss, this definition may be inappropriate for large coupling, when new divergences occur. But for now we proceed anyway, assuming the coupling is ``small enough'' (but not perturbatively small).

The first step is to introduce the massless\footnote{Technically, this Hamiltonian is not well defined in the thermodynamic limit. But its subsequent combinations with potentials $\int_\mathbb{R} V(\phi)$ opening a gap will.} free Boson Hamiltonian (in standard, equal-time quantization):
\begin{equation}
    H_0 = \int \frac{\pi^2}{2} + \frac{(\partial_x\phi)^2}{2} \, .
\end{equation}
We insist that it does \emph{not} have a mass term. We then add a (normal-ordered) cosine or hyperbolic cosine potential
\begin{align}
    H_\text{ShG}(\beta) &=\; :\! H_0\! :_m + \int \upd x \; \frac{m^2}{\beta^2} \, :\cosh\left[\beta\, \phi(x)\right]:_m \label{eq:HShG}\\
    H_\text{SG}(\beta) &=\; : \! H_0\! :_m - \int \upd x \; \frac{m^2}{\beta^2} \, :\cos\left[\beta\, \phi(x)\right]:_m \label{eq:HSG}
\end{align}
where $\beta$ is the coupling constant that determines the physics while $m$ simply fixes the scale. Naturally the two models are related and $H_\text{ShG}(\beta) = H_\text{SG}(i\beta)$.

Note, crucially, that the normal-ordering $:\,:_m$ is done with respect to the free creation-annihilation operators of mass $m$, which are related to the field operators by
\begin{align}\label{eq:modeexpansion}
    \phi(x) &= \frac{1}{2\pi} \int \upd k \sqrt{\frac{1}{2 \, \omega_k}} \left(\e^{ikx} a_k + \e^{-ikx} a^\dagger_k \right) \\
        \pi(x) &= \frac{1}{2i\pi} \int \upd k \sqrt{\frac{\omega_k}{2}} \left(\e^{ikx} a_k - \e^{-ikx} a^\dagger_k \right) \, ,
\end{align}
where $\omega_k=\sqrt{k^2 + m^2}$ and $[a_k,a_{k'}^\dagger]=2\pi\delta(k-k')$. Those are thus the same $a_k$ we have been using throughout (\eg for the $\phi^4_2$ model).
This choice of normal-ordering completely fixes the models while leading to the simplest expressions. Other choices,  with $\tilde{m} \neq m$, lead to a simple change of scale and shift in the vacuum energy.

\begin{remark}[Massive vs massless Sin(h) Gordon model]
  There is a terminology issue that I have found creates confusions when mathematicians discuss with physicists. The models we have defined are sometimes called massless Sin(h) Gordon models, because there is no \emph{explicit} mass term in their Hamiltonian, in contrast with the \emph{massive} Sin(h)-Gordon models that would have $+\int m^2 :\!\phi^2\!:$. However, both the massive and the massless Sin(h)-Gordon models are \emph{massive} is the physicam sense, \ie they do have a mass gap. This is because $\frac{\mu^2}{\beta^2}\cosh(\beta \phi)$ (and likewise for $-\cos$) behaves to leading order in $\beta$ like a mass term.
\end{remark}

Both models are sometimes alternatively constructed in radial quantization, which is more natural from the conformal field theory (CFT) perspective. The starting point is the action of the massless free boson
\begin{equation}\label{eq:freeboson_action}
    S_0 = \int \upd^2 z \frac{1}{16\pi} \, (\nabla \varphi)^2(z) \, .
\end{equation}
The interacting models are then obtained on the plane by perturbing the dilation operator $D_0$ of the free boson by vertex operators
\begin{align}
    D_\text{ShG}(b) &= D_0 + \mu_\text{ShG} \int_C\upd z\,  \left[\mathcal{V}_b (z,z^*) + \mathcal{V}_{-b} (z,z^*)\right] \\
    D_\text{SG}(b) &=D_0 - \mu_\text{SG} \int_C\upd z\,  \left[\mathcal{V}_{ib} (z,z^*) + \mathcal{V}_{-ib} (z,z^*)\right]
\end{align}
where $C$ is the unit circle and $\mathcal{V}_b(z,z^*)=:\! \e^{b\varphi(z,z^*)}\! :$ is the vertex operator normal-ordered for the modes of the free field $\varphi$. With the convention taken for the normalization in \eqref{eq:freeboson_action}, the scaling dimension of $\mathcal{V}_b$ is $\Delta=-2 b^2$. For the Sine-Gordon model, since $\mathcal{V}_{ib}$ appears, we get $\delta = +2 b^2$.

Both constructions are equivalent up to reparameterization. Starting from the second definition, mapping the plane to a cylinder of radius $R$, and then taking the radius to infinity, one gets back the first definition~\cite{konik2021}, with the normal-ordered vertex operators identified in the following way
\begin{equation}
   :\e^{\sqrt{8\pi} a \phi(x)}\!:_m \; \longleftrightarrow \frac{m^{2a^2}\e^{2a^2 \gamma_E}}{2^{2a^2}  } \; :\e^{a\varphi(x)}\!: \;\;, \label{eq:vertex_equivalence}
\end{equation}
where $\gamma_\text{E}$ is the Euler–Mascheroni constant. The coupling constants in the two definitions thus verify
\begin{align}
    b &= \beta/\sqrt{8\pi}\\
    \mu_\text{ShG} &= \frac{m^{2+2b^2}}{2^{4+2b^2}\pi b^2} \e^{2b^2 \gamma_\text{E}} \label{eq:mushg}\\
    \mu_\text{SG} &=\frac{m^{2-2b^2}}{2^{4-2b^2}\pi b^2} \e^{-2b^2 \gamma_\text{E}} \label{eq:musg}
\end{align}
In addition, one can show (\cite{konik2021}, appendix A) that the ground energy $\varepsilon_0^{(\text{etq})}$ in the equal-time quantization approach is only a constant away from the ground energy $\varepsilon_0^{(\text{rq})}$ in the radial quantization definition 
\begin{equation}\label{eq:energy_dictionary}
    \varepsilon^{(\text{etq})}_0 = \varepsilon^{(\text{rq})}_0 - \frac{m^2}{8\pi} \, .
\end{equation}
This allows to relate the quantities computed from both approaches. Indeed, the first equal-time quantization definition is natural to use with the variational method, especially with RCMPS. This is the one we will use to get numerical results. However, most exact results we will use for comparison have been obtained from the second definition, in radial quantization.

\subsection{Remarkable values of the coupling and questions}
The Hamiltonians \eqref{eq:HShG}-\eqref{eq:HSG} we gave for the Sinh-Gordon and Sine-Gordon models define legitimate QFTs, without the need for any additional renormalization as long as  $b<1/\sqrt{2}$ (equivalently $\beta< \sqrt{4\pi}$)~\cite{froehlich1975,froehlich1977}. This is the safe regime, where normal-ordering is provably sufficient to remove all divergences in both models.

For $b\in ]1/\sqrt{2},1[$, the Sine-Gordon model can still be constructed rigorously~\cite{dimock1993}, but normal-ordering does not kill all divergences. Informally, the renormalized Hamiltonian is then the same as in \eqref{eq:HSG} up to an infinite counter term proportional to the identity. Without this divergent counterterm, the vacuum energy density of $\eqref{eq:HSG}$ is infinitely negative. This is \emph{a priori} a problem for a variational method, like RCMPS, that gives finite values of the energy density by construction. For $b>1$ (equivalently $\beta > \sqrt{8\pi}$), the scaling dimension of the cosine potential is larger than $2$ and the interaction is irrelevant. Hence, the Sine-Gordon model is no longer well defined past that point without a short distance cutoff.

The situation is less clear for the Sinh-Gordon model, even though the model \emph{a priori} looks simpler. The scaling dimension $\Delta$ of the $:\e^{\pm\beta \phi}:$ terms in the $\cosh$ is always negative, and thus the interaction should (intuitively) always be strongly relevant. However, the model was rigorously constructed by Fr\"ohlich and Park only for $b<1/\sqrt{2}$~\cite{froehlich1977}, and, as far as I know, nothing past that value is established beyond reasonable doubt. The value $\beta=\sqrt{8\pi}$, $b=1$, is remarkable because it corresponds to a \emph{formal} self-dual point ($b\rightarrow 1/b$) of the exact S-matrix. However, it is unclear if this duality is physical when the model is constructed from the definition \eqref{eq:HShG} we provided. In fact, the recent thorough analytical and numerical study of Konik, L\'ajer, and Mussardo (KLM)~\cite{konik2021} suggests that the model could be massless for $b >1$. For intermediate values, $b\in ]1/\sqrt{2},1[$, the Hamiltonian truncation (HT) data of KLM is likely not fully converged but still suggests that the Hamiltonian \eqref{eq:HShG} could exist, and that its physical properties could match those predicted by the ``exact'' solution. Quotation marks are warranted because the ``exact'' formulas for the energy density or expectation values of vertex operators are obtained from analytic continuation of formulas derived for the Sine-Gordon model. The validity of the analytic continuation is not in doubt at small coupling, but could break down past a certain threshold. For example, something could \emph{a priori} happen at $b=1/\sqrt{2}$, where the Sine-Gordon model has its first change of regime, or at $b=1$ where there the Sine-Gordon model goes through a BKT transition and ceases to exist without cutoff. KLM seem to favor the second scenario, \ie\, a transition to a massless phase for $b \geq 1$, breaking the self-duality. 
We could easily consider the massive version which has two advantages: it is not integrable, thus non-trivial, the fate of the UV and of the IR is more clearly separated.

\begin{remark}[Double flaw in Coleman's historical argument] \label{rk:coleman}
  In his landmark 1975 article~\cite{coleman1975} , Coleman used a variational argument to show that the Sine-Gordon model was defined only for $\beta \leq \sqrt{8\pi}$. This is often quoted as the original physics proof that $\beta = \sqrt{8\pi}$ is the critical value for this model (more modern arguments are just that for $\beta > \sqrt{8\pi}$, the interaction is irrelevant). Coleman's argument is variational, and uses Gaussian states. Coleman defines the state $\ket{0,m}$ as the ground state of a massive free field with mass $m \neq \mu$ ($\mu$ being the mass which appears in the Sine-Gordon Hamiltonian). Then, he evaluates the Hamiltonian density $h_\beta$ on this state: $\langle 0,m| h_\beta |0,m \rangle$· Since the state is Gaussian it can be evaluated exactly. Coleman then shows that when $\beta \geq \sqrt{8\pi}$, one can tune $m$ to make this expectation value arbitrarily negative, thus showing the model is ill-defined. In fact, the energy density $h_\beta$ is already not lower bounded for $\beta \geq \sqrt{4\pi}$, which one could in principle see with a better variational manifold than Gaussian states (we will see it numerically with RCMPS). However, this unboundedness shows nothing about whether or not the model can be defined in a reasonable sense, because it comes from a renormalization of the form $-\infty \times \Id$. In fact it is now known how to define the Sine-Gordon model rigorously up to $\beta = \sqrt{8\pi}$. So really, Coleman is ultimately correct by pure luck: his variational argument is not tight (using better states would give a stricter bound), proves nothing (one can make sense of some Hamiltonians that are not lower bounded), but miraculously hits the right threshold.
\end{remark}

\subsection{Analytical results}
For a variational method, it is particularly convenient to know the exact energy density in the ground state. Fortunately, it is known exactly for the SG and ShG models since Lukyanov and Zamolodchikov~\cite{lukyanov1997} and reads, in the KLM notations~\cite{konik2021}:
\begin{equation} \label{eq:e0_exact}
    \varepsilon^\text{rq}_0  = \frac{\pi M_\text{ShG}^2}{2 \sin\left[\pi(b +1/b)\right]} \, ,
\end{equation}
where $M_{\text{ShG}}$ is the Sinh-Gordon mass gap, which is also known exactly: 
\begin{equation}
   M_\text{ShG}(b) = \frac{4\sqrt{\pi}}{\Gamma\left(\frac{1}{2+2b^2}\right)\Gamma\left(1+\frac{b^2}{2+2b^2}\right)} \left[-\mu_\text{ShG} \pi \frac{\Gamma(1+b^2)}{\Gamma(-b^2)}\right]^\frac{1}{2+2b^2} \, ,
    \label{eq:mass}
\end{equation}
where recall that $\mu_\text{ShG} = \frac{m^{2+2b^2}}{2^{4+2b^2}\pi b^2} \e^{2b^2 \gamma_\text{E}}$. 
The mass gap $M_\text{SG}$ of the Sine-Gordon model is simply obtained by replacing $\mu_\text{ShG}$ by $\mu_\text{SG}$ [given in \eqref{eq:musg}] and replacing $b$ by $ib$ in the mass formula \eqref{eq:mass}. Historically, the result was obtained first for the Sine-Gordon model, and formula \eqref{eq:mass} is just its analytic continuation. This is important to have in mind, because for the Sine-Gordon model, $M_\text{SG}$ corresponds to the mass of the first breather, which ceases to exist when $b\rightarrow 1/\sqrt{2}$ as the model undergoes a phase transition from attractive to repulsive phase. At this point, the energy density in the SG ground state collapses to $-\infty$ (see our previous remark \ref{rk:coleman} about Coleman's argument). Indeed, for the SG model, the energy density for $b\in [1/\sqrt{2},1[$ given by formula \eqref{eq:e0_exact} is not that of the Hamiltonian $H_{SG}$, which is strictly speaking not bounded below. Rather, it corresponds to $H_{SG}$ minus an infinite multiple of the identity, a renormalized Hamiltonian which cannot be trivially written in the free Fock space without cutoff.

For the Sinh-Gordon model, the energy density given by formula \eqref{eq:e0_exact} remains lower bounded for $b\in[0,1]$ and goes to $0$ when $b\rightarrow 0$. Formula \eqref{eq:e0_exact} clearly ceases to make sense for $b\geq 1$ (it becomes complex). But the formula could break down earlier, for example already at $b=1/\sqrt{2}$.

For these integrable models, a few local expectation values are also known exactly. The most natural one is the vertex operator $G(a)=\langle:\!\e^{a\varphi}\!:\rangle$ (where $a$ can be different from $b$). This one point functions is given (in radial quantization) by the Fateev-Lukyanov-Zamolodchikov-Zamolodchikov (FLZZ) formula~\cite{fateev1998,lukyanov1997}, initially written for the Sine-Gordon model. It can be analytically continued (see KLM~\cite{konik2021}) to give, for the Sinh-Gordon model:
\begin{equation}
    \begin{split}
        G(a) = M_\text{ShG}^{-2a^2} \left[\frac{\Gamma(\frac{1}{2+2b^2}) \Gamma(\frac{2+3b^2}{2+2b^2})}{4\sqrt{\pi}}\right] \int_{\mathbb{R}^+} \!\!\frac{\upd t}{t} \left[2 a^2 \e^{-2t} - \frac{\sinh^2(2abt)}{2\sinh(b^2t)\sinh(t)\cosh(t+b^2t)}\right]\, .
    \end{split}\label{eq:flzz}
\end{equation}
This formula makes sense only up to the Seiberg bound $a \leq \frac{1}{2}\left(b + b^{-1}\right)$, after which the integral is divergent. It is tempting to trust the formula anyway, and analytically continue it to even larger values of $a$. However, as KLM showed, we then get $G(a) < 0$ which is nonsensical since $\langle:\!\e^{a\varphi}\!:\rangle \geq 0$ for $a$ real. Most likely, these vertex operators just stay at $0$ past the Seiberg bound. 

\begin{remark}[Past the Seiberg bound with RCMPS?]
  This large $a$ behavior is problematic for our variational method. At least when evaluated on a fixed RCMPS, these large $a$ vertex operators will be finite, and importantly always non-zero! The best we can hope for is that they get gradually smaller as the bond dimension is increased. Whether or not this happens is difficult to guess beforehand. As we will see, the situations is even worse, and the RCMPS result do not seem to converge even for large values \emph{within} the Seiberg bound.
\end{remark}

\subsection{Results for (massless) Sine-Gordon model}
\begin{figure}
    \centering
    \includegraphics[width=0.49\textwidth]{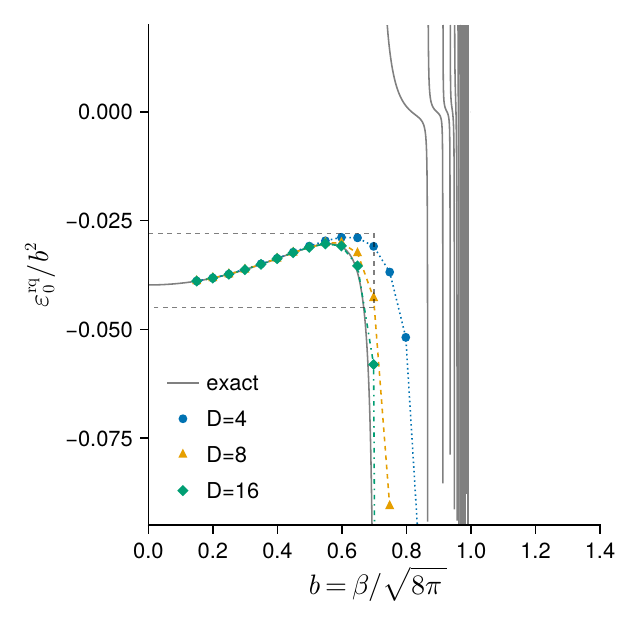}
    \includegraphics[width=0.49\textwidth]{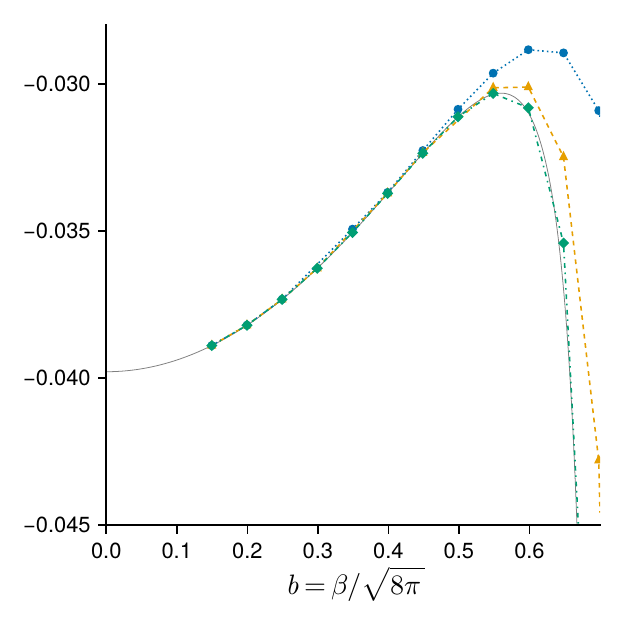}
    \caption{Rescaled ground state energy density $\varepsilon_0^\text{rq}/b^2$ of the \textbf{Sine-Gordon} model (converted to radial quantization conventions). The dashed box on the left, corresponding to the region $b\in[0,1/\sqrt{2}[$ of RCMPS approximability, is magnified on the right. Note that the exact values of the energy for $b\in]1/\sqrt{2},1[$ do not correspond to $h_\text{SG}$, but only to the resulting finite part after additional renormalization. This is why the energy density dives to $-\infty$ and then reappears from $+\infty$. Figure from~\cite{tilloy2022studyquantumsinhgordonmodel}.}
    \label{fig:energy_density_sg}
\end{figure}

We may now compute an approximate ground state of the Sine-Gordon model with RCMPS, and evaluate its energy density $\bra{Q,R} h_\text{SG} \ket{Q,R} \gtrsim \varepsilon_0^\text{etq}$. This works essentially exactly like in the $\phi^4$ model. Then, we compare with the exact formula \eqref{eq:e0_exact} using \eqref{eq:energy_dictionary} to translate the numerical (equal time) RCMPS results to radial quantization. The results are shown in Fig. \ref{fig:energy_density_sg}.

The RCMPS gives an excellent approximation to the ground state, even at moderate $D$, and at least for $b$ sufficiently away from the transition point $b_c = 1/\sqrt{2}$. For $b = 1/\sqrt{2} + \epsilon$, small values of $D$ still give a stable optimization and a very negative but finite energy density (even though the Hamiltonian density itself is no longer lower bounded). However, and as expected, I observed some runaway optimization for larger values of $D$. In this unstable case, the minimization gives lower and lower energies at every step, with a gradient that never converges to zero. This is a concrete numerical refinement of Coleman's variational argument we discussed in remark \ref{rk:coleman}.

Importantly, because of the lack of lower bound for the ``raw'' density $h_{SG}$, RCMPS cannot be used to directly compute observables in the $b \in [1/\sqrt{2},1]$ phase. One would need a different ansatz with ``pairing'', incorporating precisely the right additional UV divergence, such that energy expectation values remain finite after the infinite subtraction. Right now, the raw expectation value (normal ordered but not further renormalized) is finite by construction with RCMPS. As a result, it would be infinitely \emph{positive} after any additional renormalization, and thus infinitely far from the true ground state.

\subsection{Results for (massless) Sinh-Gordon model}
The Sinh-Gordon model is of course the more interesting model since it is where the controversy lies. For this model, the code is still essentially identical, but finding approximate ground states by energy minimization has proved far more difficult than for $\phi^4$, especially at large coupling $\beta \geq \sqrt{4\pi}$ ($b\geq 1/\sqrt{2}$). Concretely, finding ground states requires far more iterations (by a factor $10$ to $100$), and the number of iteration does seem to grow as $D$ is increased. This is the sign our optimizer is not capturing some of the ill-conditioning in the Hessian (the geometric intuition being insufficient), or that the ansatz itself does not capture crucial aspects of the physics. 

The results at $D=32$ were obtained rather heroically, after days of run on a cluster. This, added to the fact that there are very few convincing numerical results on Sinh-Gordon, is a good enough reason to show them here. Despite the slow optimization, the states at fixed $D$ are very likely converged, and their inability to approximate the ground state well at large $\beta$ is likely intrinsic to the ansatz (and something to be understood).

\subsubsection*{Energy density}
\begin{figure}
    \centering
    \includegraphics[width=0.57\textwidth]{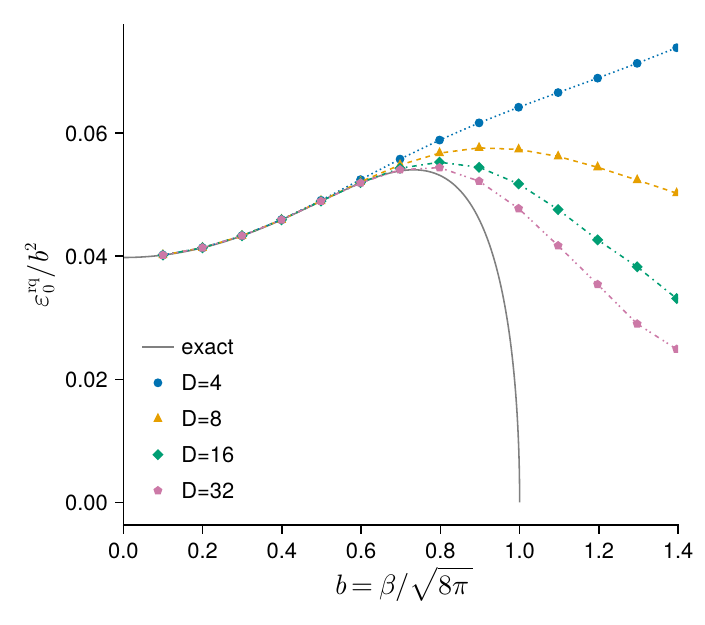}
    \includegraphics[width=0.42\textwidth]{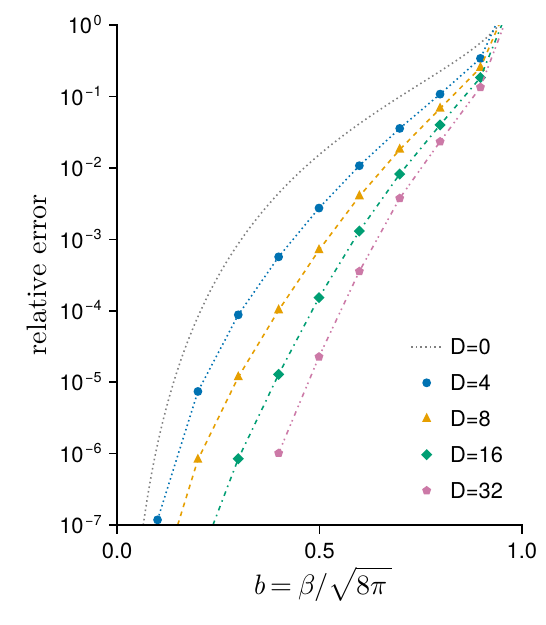}
    \caption{Rescaled ground state energy density $\varepsilon_0^\text{rq}/b^2$ of the \textbf{Sinh-Gordon} model (in radial quantization conventions). Figure from~\cite{tilloy2022studyquantumsinhgordonmodel}.}
    \label{fig:energy_density_shg}
\end{figure}

The energy minimization at fixed $D$ is well behaved for all values of the coupling, \ie\, even for $b\geq 1$, because the energy density $\varepsilon_0^\text{rq}$ is always lower bounded by zero. The results are shown in Fig. \ref{fig:energy_density_shg} for bond dimensions up to $D=32$ and coupling constants $b\in[0.1,1.4]$.

The approximation is good and improves quickly with $D$ at least until $b\approx 1/\sqrt{2}$. Then not only does the RCMPS upper bound substantially deviates from the exact value, but improvement as $D$ increases is slower. At least visually, on the energy density plot, one may suspect a transition around $1/\sqrt{2}$, where convergence gets dramatically slower in $D$. However, looking at the relative error hints at a crossover instead, with an exponential loss of precision as $b$ increases for fixed $D$.

For $b\geq 1$, the reasonable guess is that the exact value is $\varepsilon^{\text{rq}}_0 = 0$. Indeed, if formula \eqref{eq:e0_exact} is correct, $\varepsilon^{\text{rq}}_0$ reaches $0$ at $b=1$, but $\varepsilon^{\text{rq}}_0$ is also lower-bounded by $0$, and may just saturate at $0$. The numerical RCMPS results, although far from converged in $D$, are at least consistent with this hypothesis. They also provide a rigorous energy density upper bound in this controversial region.

\subsubsection*{Vertex operators}

\begin{figure}
    \centering
    \includegraphics[width=0.378\textwidth]{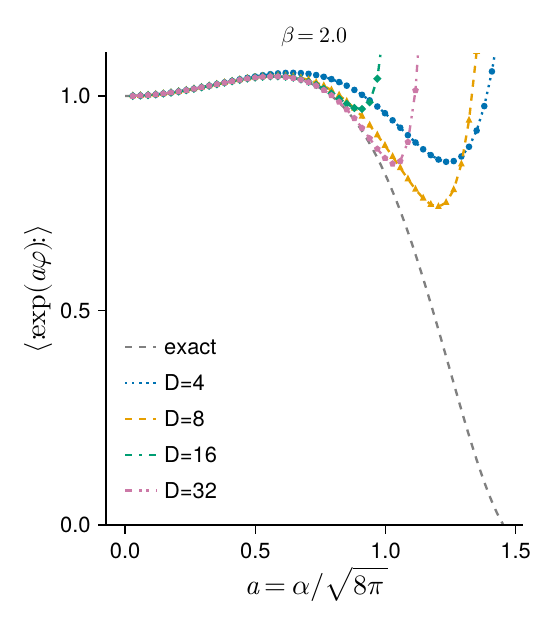}
    \includegraphics[width=0.304\textwidth]{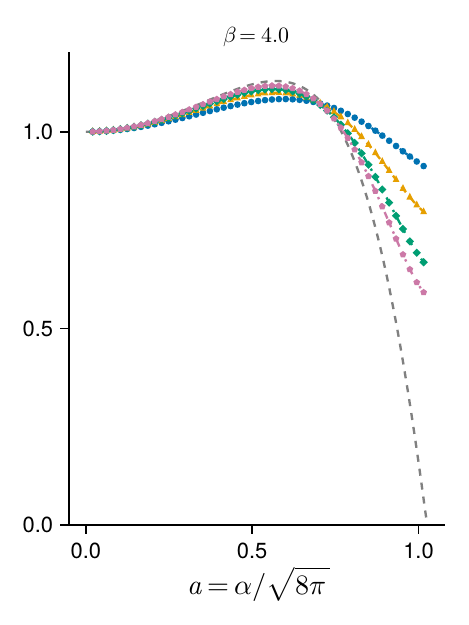}
    \includegraphics[width=0.304\textwidth]{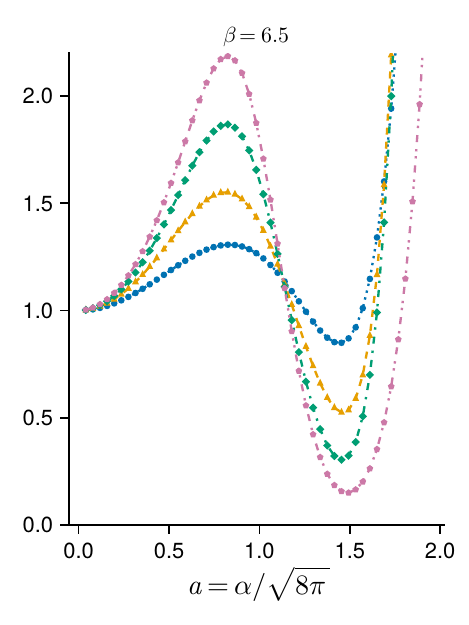}
    \caption{Vertex operator expectation values $G(a) = \langle:\!\e^{a\varphi}\!:\rangle$ for coupling constants $b=0.4, 0.8$, and $1.3$. The dashed grey line corresponds to the FLZZ formula. Figure from~\cite{tilloy2022studyquantumsinhgordonmodel}.}
    \label{fig:one_point}
\end{figure}

Once the RCMPS approximation to the ground state has been (tediously) obtained one may directly evaluate expectation values of vertex operators (in equal-time quantization conventions) $\langle :\! \e^{\alpha \phi(0)} \! :_m\rangle$ which we can then relate to $G(a)$ using \eqref{eq:vertex_equivalence}. The results for $3$ different values of the coupling ($b\simeq 0.4,0.8,1.3$) are given in Fig. \ref{fig:one_point} and correspond to the 3 main behaviors observed.

For values of the coupling in the ``safe'' region $b=0.4\in [0,1/\sqrt{2}]$, convergence as a function of the bond dimension is extremely fast for operators $:\e^{a\varphi}\! :$ with $a$ not too large (approximately $a\leq 2 b$). For larger $a$, still within the Seiberg bound, the behavior becomes extremely erratic: vertex operators do not get closer to the prediction of the FLZZ formula, and even seem to diverge. This behavior is reproducible, and occurs for independent successful energy minimizations terminated with vanishing gradient. Note that for $D=32$, the RCMPS we use to compute the vertex operators is intuitively close to the true ground state, differing in energy density only by a $10^{-6}$ relative error. It is thus possible that minimizing the energy density is insufficient to give finite values to large $a$ vertex operators in the thermodynamic limit. A possibility is that the FLZZ formula simply has a smaller domain of validity than previously believed, at least when the ShG model is defined via its Hamiltonian \eqref{eq:HShG}. Another option is that by taking the thermodynamic limit first and the large $D$ limit after, we fail to control some vacuum expectation values involving soft modes having vanishing contribution to the energy.

For values in the strong coupling region ($b=0.8$ in Fig. \ref{fig:one_point}), closer to the self dual point, convergence is slower (as expected) as a function of the bond dimension $D$, but seems more uniform in $a$, possibly even in the whole ``Seiberg allowed'' region. This is likely because minimizing the energy density controls vertex operators with $a$ of the order of the coupling $b$.

Finally, for ultra strong coupling, \ie for values beyond the self-dual point ($b=1.3$ in Fig. \ref{fig:one_point}), we enter \textit{terra incognita}. There is no reference analytical value (even speculative) to compare our numerics to, and the very existence of the model defined by its Hamiltonian $H_\text{ShG}$ is unclear. For the bond dimensions I could probe numerically, it is hard to know if the vertex operators converge very slowly to some large value, or if they do not converge at all.

\subsubsection*{Entanglement}

The convergence of the observables as a function of $D$ we obtained for the Sinh-Gordon model gets dramatically worse as the coupling constant is increased, especially past $\beta = \sqrt{4\pi}$ ($b\geq 1/\sqrt{2}$). This is to be contrasted with what happened \eg\, with the $\phi^4$ model, where convergence was fast for all values of the coupling (even $\gg 1$, deep in the symmetry broken phase), at the exception of a narrow band around the symmetry breaking point. This behavior mimics what was observed by KLM with Hamiltonian truncation techniques. It likely means that the free Fock basis becomes inefficient for approximation when the coupling constant is large, since both methods, RCMPS and HT, rely on it as a starting point. 

A good way to understand this phenomenon, at least in the RCMPS context, is to look at entanglement. As we argued before in remark \ref{rk:particleentanglement}, standard entanglement entropy is UV divergent for relativistic field theories. This is actually one of the reasons why standard tensor network techniques become inaccurate as the UV cutoff is lifted. However, we defined a notion of entanglement in the free particle basis in remark \ref{rk:particleentanglement}. This is the right notion if we care about entanglement as a measure of approximability: if it is finite, we expect RCMPS to provide a good approximation.

\begin{figure}
    \centering
    \includegraphics[width=0.49\textwidth]{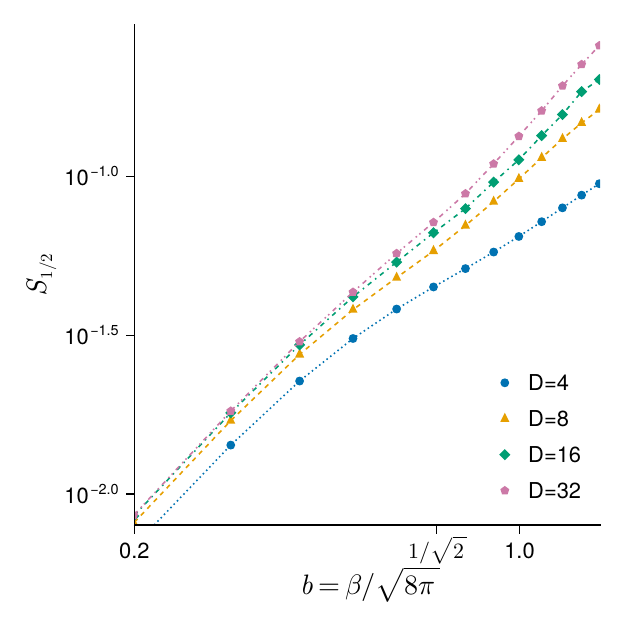}
    \includegraphics[width=0.49\textwidth]{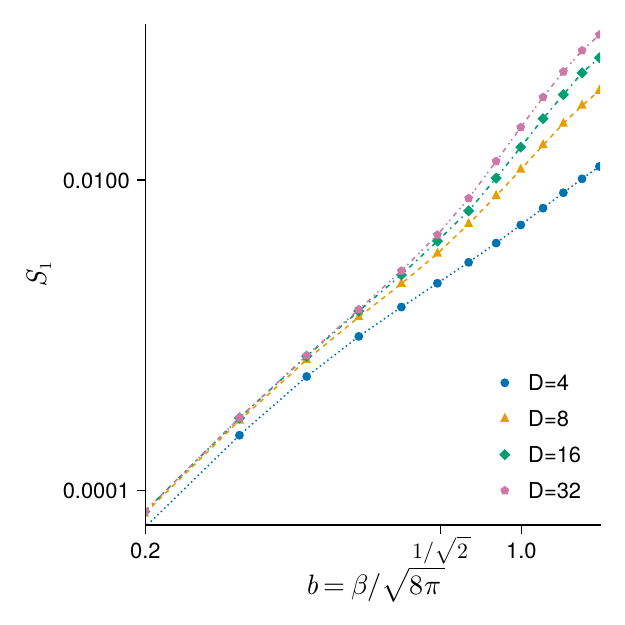}
    \caption{Half-line entanglement in the free basis for the ground state of the Sinh-Gordon model. Left: $1/2$ Renyi entropy $S_{1/2}$. Right: Von Neumann entropy $S_1$. Figure from~\cite{tilloy2022studyquantumsinhgordonmodel}. }
    \label{fig:entanglement}
\end{figure}

This exotic entanglement entropy is easy to compute in our case. We observe (see Fig. \ref{fig:entanglement}) that it increases polynomially in the coupling $b$. For $b\geq 1/\sqrt{2}$, our estimate is manifestly not converged. Either the entanglement entropy remains finite but becomes so large that much higher bond dimensions would be needed to estimate it, or there is a phase transition to infinite entanglement for some threshold coupling $b_c$ (that could plausibly be $b_c=1/\sqrt{2}$). In any case, whether it is simply a fast growth or a divergence, this increase of the entanglement entropy in the free particle basis explains why both HT and RCMPS struggle to capture accurately the properties of the ground state as the coupling is increased.

\begin{remark}[Is $\beta = \sqrt{4\pi}$ special for Sinh-Gordon?]
  So far, there is no real consensus about whether or not ``something'' happens for the Sinh-Gordon model at $\beta = \sqrt{4\pi}$. It could be that the model is defined only for $\beta$ smaller, which is however unlikely, or it could be that it departs from the integrability predictions. The latter option is not completely implausible, given that the integrability formulas rely on an analytic continuation, and that something does happen at $\sqrt{4\pi}$ for the Sine-Gordon model.
\end{remark}

\begin{remark}[The mass-perturbed Sinh-Gordon model]
  At the $\beta = \sqrt{8\pi}$ point, both the energy and mass gap of the Sinh-Gordon model go to zero, at least according to integrability predictions. Intuitively, like in the Sine-Gordon model, it could be the sign of the $\cosh$ term becoming effectively irrelevant, leaving us with the free boson CFT. Alternatively, the $\cosh$ term could be relevant, but drive the system to a peculiar interacting gapless phase. To more cleanly distinguish the potential UV problems from the standard IR criticality, it would help to consider instead the mass-perturbed Sinh-Gordon model\footnote{I thank Slava Rychkov for suggesting this option, and regret that I have not yet had the time to look more seriously into it.}, \ie the model obtained by adding an \emph{explicit} mass term in the Sinh-Gordon Hamiltonian. Such a term would close the gap everywhere, making RCMPS in principle more efficient in the IR. Any failure of convergence would then unequivocally be associated to a UV problem (either in the ansatz or in the model itself).
\end{remark}

\section[Two interacting scalar fields]{Two interacting scalar fields\protect\footnotemark}
We now consider the problem of two interacting scalar fields, which requires the first genuine extension of the ansatz beyond its original definition~\cite{tilloy2021variational,tilloy2021relativistic}. This problem turns out to be quite less trivial than expected, and dealing with multiple species is already difficult with standard CMPS. 

\footnotetext{The results in this section are preliminary and have been obtained with my PhD student Karan Tiwana. They are presented more thoroughly, and supported with extensive numerics in \cite{tiwana2025multifieldrelativisticcontinuousmatrix}.}

\subsection{The model}
Our starting point is the Hamiltonian of two relativistic scalar fields both self-interacting and interacting with one another
\begin{equation}\label{eq:H_O2}
  H(\lambda,g) = H_m^{(1)} + H_m^{(2)} + 2 \lambda \int_\mathbb{R} \phi^2_1 \phi^2_2 + g \int_\mathbb{R} \left[\phi^4_1 + \phi^4_2\right] = \int_\mathbb{R} h(\lambda,g)
\end{equation}
where
\begin{equation}
  H_m^{(i)} = \int_\mathbb{R} \frac{\pi^2_i}{2} + \frac{(\nabla\phi_i)^2}{2} + m^2 \frac{\phi_i^2}{2}
\end{equation}
To make sense of this Hamiltonian, we may diagonalize $H_m^{(1)}$ and $H_m^{(2)}$ separately into normal modes $a_k^{(1)}$ and $a_k^{(2)}$. As for $\phi^4_2$, we make the model finite\footnote{It is clear that there are no extra tadpoles to cancel, and that there are no remaining divergences, at least in perturbation theory.} by normal ordering everything with respect to the free modes, \ie we simply consider $H(\lambda,g) \rightarrow :\! H(\lambda,g) \!:$, and in practice minimize the density $:\! h(\lambda,g) \!:$.

Before studying the model variationally, let us already say a few things about it. Since we decided to take both masses and self-coupling identical (to make the number of parameters manageable), the model enjoys a fairly large discrete symmetry: the dihedral group $D_4$ (\ie the discrete symmetries of a square). Indeed, we can flip each field individually, and swap them, without changing the Hamiltonian. Additionally, when $\lambda = g$, the discrete symmetry is enhanced into a continuous $O(2)$. We expect that there is a Berezinsky-Kosterlitz-Thouless (BKT) transition for a specific coupling $\lambda_c = g_c$ along this line.

\subsection{RCMPS for two fields}
Since we now have two fields, we need a natural extension of the RCMPS ansatz, and propose to take:
\begin{equation}\label{eq:RCMPS_twofields}
\ket{Q,R_1,R_2} := \tr\left[\mathcal{P}\exp\left(\int_{I}\upd x \; Q \otimes \mathrm{Id} + R_1\otimes a^\dagger_1(x) + R_2 \otimes a^\dagger_2(x)\right)\right] \ket{0}_a \, .
\end{equation}
This manifold of states is perfectly well defined independently of the Hamiltonian \eqref{eq:H_O2}. As before, the parameterization is redundant, and we can partly fix it by taking the left canonical gauge
\begin{equation}\label{eq:leftcanonical_O2}
Q = -i K -\frac{1}{2} \left(R_1^\dagger R_1 + R_2^\dagger R_2\right) \, ,
\end{equation}
where $K$ is restricted to be self-adjoint.

As natural as it may seem, this manifold of states $\mathcal{M}_D$ is \emph{not} adapted to the QFT model we aim to solve, and is too general. Indeed, for generic $K,R_1,R_2$, the energy density $\bra{K,R_1,R_2} :\! h(\lambda,g) \!: \ket{K,R_1,R_2}$ is divergent. 
More precisely, one can show that the energy density can be written:
\begin{equation}
  \langle Q,R_1,R_2 |\!\! :\!\! h(\lambda,g) \!\!:\! \!|Q,R_1,R_2 \rangle = \langle Q,R_1,R_2 | h^\text{reg} | Q,R_1, R_2\rangle + 2 \tr([R_1,R_2]^\dagger [R_1,R_2]]\rho_\text{ss})\!\int_\mathbb{R} \!\! J^2
\end{equation}
where $\langle Q,R_1,R_2 | h^\text{reg} | Q,R_1,R_2\rangle$ is finite for all $Q,R_1,R_2$.
The culprit is the second term, which appears only for multiple species, and is indeed logarithmically UV divergent because $J^2(x)\propto 1/x$ near $0$.

To get a finite energy density, we need this problematic term to be exactly zero. This requires the additional \emph{regularity condition} $[R_1,R_2]=0$, which carves a strictly smaller subspace $\mathcal{M}_D^\text{reg} \subsetneq \mathcal{M}_D$  into the space of generic two species RCMPS. Note that this regularity condition is the same as the one encountered in the non-relativistic case~\cite{haegeman2013calculus}, although the calculation to obtain it differs slightly.

For a state $\ket{Q,R_1,R_2}$ on this regular subspace $\mathcal{M}_D^\text{reg}$, have
\begin{equation}
  \langle Q,R_1,R_2 | :\! h(\lambda,g) \!: Q,R_1,R_2 \rangle = \langle Q,R_1,R_2 | h^\text{reg} | Q,R_1, R_2\rangle + 0
\end{equation}
and thus one can optimize a finite function. To find the approximate ground state, we will thus minimize $\langle Q,R_1,R_2 | h^\text{reg} | Q,R_1, R_2\rangle$ \emph{under the constraint}\footnote{Note, of course, that $h^\text{reg}$ taken alone, can have an expectation far lower than the ground state energy if we consider a generic state not belonging to the regular manifold!} $[R_1,R_2]=0$.

\subsection{Expectation values}
One we are on the regular subspace and do not have a divergent term, the expectation values are computed essentially like with the $1$ species RCMPS. One can use almost the same ODE coming from the same generating functionals. There are only two novelties.

The first is that one needs to use a new transfer operator $\mathcal{L}$ is
\begin{equation}\label{eq:lindblad_O2}
  \begin{split}
    \mathcal{L}(\rho) :&= Q\rho + \rho Q^\dagger + R_1\rho R_1^\dagger + R_2\rho R_2^\dagger\\
                       &= - i [K,\rho] + \mathcal{D}[R_1](\rho) + \mathcal{D}[R_2](\rho)
\end{split}
\end{equation}
with $\mathcal{D}[R](\rho) = R\rho R^\dagger - \frac{1}{2}\{ R^\dagger R,\rho \} $. This is the natural generalization of the transfer operator we had before. The second novelty is that we have a new cross term to compute: $\langle Q,R_1,R_2 |\!:\!\!\phi_1^2\!\!: :\!\!\phi_2^2\!\!:\!|Q,R_1,R_2\rangle$. This is however straightforward with the same generating functional techniques as before, and one obtains an ODE with the same complexity as the others.

\subsection{Optimizing on the regular submanifold}
To find the ground state energy (and the associated ground state), our minimization problem is
\begin{equation}
  \begin{split}
  \min ~ &~ \langle K,R_1,R_2 | h^\text{reg} | K,R_1, R_2\rangle\\
  \text{under} ~&~ [R_1,R_2] = 0 ~\text{(regularity condition)} \, ,
\end{split}
\end{equation}
where we use the left-canonical gauge parameterization $Q=-iK - R_1^\dagger R_1/2 - R_2^\dagger R_2 /2$.

To solve this minimization problem, we have in principle many options. We choose to stay with the Riemannian optimization approach, which was successful before. We thus see the subspace of regular RCMPS as a submanifold\footnote{Technically, as Karan reminded me, it is more of a ``physicist'' submanifold, with many singular points where the constraint is degenerate. Yet in practice, gradient descent trajectories do not fall into these isolated points.} of general RCMPS, and need to define its tangent space, natural metric, and retraction(s).

\subsubsection*{Raw tangent vectors}
For a RCMPS with two species of particles (so far not regular), tangent vectors are defined analogously to the standard case:
\begin{align}\label{eq:tangent_O2}
    \ket{V,W_1,W_2}_{Q,R_1,R_2} =\!\! \int\! \upd x \left[V_{\alpha\beta} \frac{\delta }{\delta Q_{\alpha\beta}(x)} + W_{1,\alpha\beta} \frac{\delta }{\delta R_{1,\alpha\beta}(x)}+ W_{2,\alpha\beta} \frac{\delta }{\delta R_{2,\alpha\beta}(x)}\right] \! \ket{Q,R_1,R_2} \, .
\end{align}
As before, we may fix the gauge condition on the tangent space:
\begin{equation}\label{eq:tangentgaugeO2}
  V= -R_1^\dagger W_1 - R_2^\dagger W_2 \, ,
\end{equation}
in order to give the overlap of tangent vectors a particularly simple (local) form:
\begin{equation} \label{eq:tangent_overlap_O2}
  \bra{V,W_1,W_2} \tilde{V}, \tilde{W}_1, \tilde{W}_2\rangle = \tr\left[\left(W_1^\dagger \tilde{W}_1 + W_2^\dagger \tilde{W}_2\right) \rho_\text{ss}\right],
\end{equation}
where again $\rho_\text{ss}$ is the trace $1$ stationary state of the transfer superoperator $\mathcal{L}$ \eqref{eq:lindblad_O2}.

\subsubsection*{Regular tangent space projection}
Our tangent vectors, parameterized by arbitrary matrices $W_1, W_2$, are ``raw'' tangent vectors: they live in the tangent space of the full $2$-species RCMPS manifold $\mathcal{M}_D$. We need to understand how to project any such raw tangent vector on the ``regular'' tangent space of the RCMPS submanifold $\mathcal{M}_D^\text{reg}$ with $[R_1,R_2] = 0$. To this end, we need to preserve this condition to leading order, namely ask
\begin{equation}
  [R_1 + \varepsilon W_1\, ,\, R_2 + \varepsilon W_2] = o(\varepsilon^2) \;\; \implies \;\; [R_1,W_2] = [R_2,W_1] \,.
\end{equation}
This differential condition characterizes the regular tangent space fully.

We also need a natural notion of orthogonal projection onto this regular tangent space. Namely, if we are given $W_1,W_2$ on the raw tangent space, we can find the closest regular tangent vector parameterized by $\xi_1,\xi_2$ simply solving:
\begin{equation}
  \xi_1,\xi_2 = \underset{[R_1,\xi_2] = [R_2,\xi_1] }{\argmin} \big\| \, \ket{W_1,W_2}_{Q,R} - \ket{\xi_1,\xi_2}_{Q,R}\, \big\|^2\,.
\end{equation}
Using the explicit form of the tangent space overlap \eqref{eq:tangent_overlap_O2} and neglecting the constant term we have:
\begin{equation}
  \xi_1,\xi_2 = \underset{[R_1,\xi_2] = [R_2,\xi_1] }{\argmin} \tr\left[\left(\xi_1^\dagger\xi_1 + \xi_2^\dagger\xi_2\right)\rho_\text{ss}\right] - 2 \mathrm{Re}\bigg\{\tr\left[\left(\xi_1^\dagger W_1 + \xi_2^\dagger W_2\right)\rho_\text{ss}\right]\bigg\} \, .
\end{equation}

This is a convex minimization problem with quadratic objective and linear constraints, hence it is guaranteed to be efficiently solvable. We may use black-box solvers (like \texttt{IPOPT}), but it helps to write the solution explicitly and exploit the fact that we act on matrices by matrix multiplication only\footnote{This is a general subtlety in MPS algorithms: the operators acting on matrices are always acting via matrix multiplication, which means they cost $\propto D^3$ to apply and not $D^4$ as would be the case for a generic linear operator on $D\times D$ matrices.}. There are various strategies to solve it. A natural option is to parameterize explicitly the regular tangent space (\ie solve the constraint first). Another approach is to use a Lagrangian method instead, which appears to be a bit easier here. For this minimization problem, the Lagrangian $\mathscr{L}(\xi_1,\xi_2,\Lambda)$ reads:
\begin{equation}\label{eq:lagrangian}
\mathscr{L}(\xi_1,\xi_2,\Lambda) = \tr\left[\left(\xi_1^\dagger\xi_1 + \xi_2^\dagger\xi_2\right)\rho_\text{ss}\right] - 2 \mathrm{Re}\Big\{\tr\left[\left(\xi_1^\dagger W_1 + \xi_2^\dagger W_2\right)\rho_\text{ss} - \Lambda^\dagger \left([R_1,\xi_2] - [R_2,\xi_1]\right)\right]\Big\} \,
\end{equation}
where $\Lambda$ is a $D\times D$ complex matrix of Lagrange multipliers. The minimum is found where the Lagrangian is stationary. We could in principle take derivatives with respect to real and imaginary parts but it is easier to take complex Wirtinger derivatives. One then notices that differentiating with respect to the variable or its complex conjugate gives the same constraints, and thus the stationary conditions reduce to:
\begin{equation}\label{eq:lagrangian_stationary}
  \frac{\partial \mathscr{L}}{\partial \Lambda^\dagger} = \frac{\partial \mathscr{L}}{\partial \xi_1^\dagger} = \frac{\partial \mathscr{L}}{\partial \xi_2^\dagger} = 0
\end{equation}
This implies the following system
\begin{align}
  \frac{\partial \mathscr{L}}{\partial \Lambda^\dagger}=0 \implies  0 &= [R_1,\xi_2] - [R_2,\xi_1] \label{eq:partial_Lambda}\\
  \frac{\partial \mathscr{L}}{\partial \xi_1^\dagger}=0 \implies 0 &= (\xi_1 - W_1) \rho_\text{ss} - [R_2^\dagger,\Lambda]\label{eq:partial_xi1} \\
  \frac{\partial \mathscr{L}}{\partial \xi_2^\dagger}=0 \implies 0 &= (\xi_2 - W_2) \rho_\text{ss} + [R_1^\dagger,\Lambda] \label{eq:partial_xi2}
\end{align}
We can now use \eqref{eq:partial_xi1} and \eqref{eq:partial_xi2} to express $\xi_1$ and $\xi_2$ as a function of $\Lambda$, 
\begin{align}
  \xi_1 & = W_1 + [R_2^\dagger,\Lambda] \rho_\text{ss}^{-1}\\
  \xi_2 & = W_2 - [R_1^\dagger,\Lambda]\rho_\text{ss}^{-1}
\end{align}
and then inject it into \eqref{eq:partial_Lambda} to find an equation for $\Lambda$:
\begin{equation}
\underset{:= \Delta}{\underbrace{[R_1,W_2] - [R_2,W_1]}} = \underset{:= \mathcal{A}(\Lambda)}{\underbrace{ [R_1,[R_1^\dagger,\Lambda]\rho_\text{ss}^{-1}] + [R_2,[R_2^\dagger,\Lambda]\rho_\text{ss}^{-1}]}} \, .
\end{equation}
This is a linear system for $\Lambda$, that we may solve exactly or iteratively using \eg the generalized minimal residual method (GMRES)\footnote{Again we use the \texttt{KrylovKit} implementation.}. Writing formally $\mathcal{A}^{-1}(\Delta)$ the solution of this linear system we finally have:
\begin{align}\label{eq:regulartangentprojectionfinal}
  \xi_1 & = W_1 + [R_2^\dagger,\mathcal{A}^{-1}(\Delta)] \rho_\text{ss}^{-1}\\
  \xi_2 & = W_2 - [R_1^\dagger,\mathcal{A}^{-1}(\Delta)]\rho_\text{ss}^{-1} \,.
\end{align}
We thus have an efficient way to (orthogonally) project any tangent vector (and in particular the gradient of the energy density) onto the appropriate regular subspace.

\subsubsection*{Retraction}
The previous derivation only tells us how to project tangent vectors onto directions that keep the state on the regular manifold for infinitesimal moves. Let us for conciseness write $u = \{Q,R_1,R_2\}$ a point on the regular submanifold, and $\xi = \{\xi_1,\xi_2\}$ a direction on regular tangent space. What we have shown is that $u + \varepsilon \xi$ is on the regular submanifold up to terms of order $\varepsilon^2$.

We now need a \emph{retraction} $r(\alpha,u,\xi)$ that gives us a full trajectory that stays on the manifold no matter the value of $\alpha$. Again, a natural choice would be the geodesic retraction but it is difficult to compute for the highly non-trivial submanifold at hand. For optimization purposes, any retraction that is cheap to compute is in principle acceptable. The requirement to be a retraction is just that $r(\alpha,u,\xi)\in \mathcal{M}_\text{reg}$ and
\begin{equation}
  \frac{\upd}{\upd \alpha} r(\alpha,x,\xi)\bigg|_{\alpha=0} = \xi \, .
\end{equation}
There are many ways to construct such a retraction, which primarily differ in numerical stability. Let us present one family options which can be combined. In what follows, note that there is no need to do anything subtle for the retracted $K$, which has no impact on the regularity of the state. We may just take the first order move
\begin{equation}
  K \rightarrow \tilde{K} = K + \alpha \frac{i}{2}(\xi_1^\dagger R_1 + \xi_2^\dagger R_2 - \text{h.c.})
\end{equation}
as retracted $K$, which is indeed self-adjoint. The real difficulty lies in making $R$ matrices commute post retraction.

The simplest option is to note that $\tilde{R}_1 = R_1+ \alpha \xi_1$ and $\tilde{R}_2 =R_2 + \alpha \xi_2$ should commute to order $\alpha^2$ for $(\xi_1,\xi_2)$ on the regular tangent space. The matrices $\tilde{R}_1$ and $\tilde{R}_2$ are thus co-diagonalizable to order $\alpha$ (but not $\alpha^2$). The solution is to \eg modify $\tilde{R}_2$ mildly so that it is diagonalizable in the same basis as $\tilde{R}_1$. Concretely, diagonalize $\tilde{R}_1$: $R_1 = V_1 \mathsf{D} V_1^{-1}$, and then cut the off-diagonal part of $\tilde{R}_2$ in this basis
\begin{equation}
  \tilde{R}_2 \longrightarrow \tilde{\tilde{R}}_2 =  V_1\; \text{diag} \big[V_1^{-1} \tilde{R}_2 V_1\big] \; V^{-1} \, .
\end{equation}
This retraction $r(\alpha,u,\xi) = (\tilde{K},\tilde{R}_1,\tilde{\tilde{R}}_2)$ works well numerically, except when $\tilde{R}_1$ is very close to a non-diagonalizable matrix (with non-trivial Jordan blocks), or a matrix with degenerate eigenvalues. In that case, we may consider another natural retraction, obtained by swapping $\tilde{R}_1$ and $\tilde{R}_2$ in the previous procedure. In practice, we may just test both retractions and pick the one which seems the best conditioned.

\subsubsection*{Beyond gradient descent}
With the geometric ingredients we have given so far, we have all that is needed for Riemannian gradient descent. We could in principle extend to quasi-Newton methods, as in the single species case. The difficulty is that we do not have a natural base Euclidean manifold. That is, even if we take the Euclidean metric instead of the Hilbert induced metric on the tangent space, our base manifold is not Euclidean. Even with the Euclidean metric, our retraction is not geodesic, and we have no natural isometric vector transport. Quasi-Newton algorithm could still work, but taking natural choices of vector transport and metric gave a fairly unstable LBFGS algorithm, working only in the immediate vicinity of the optimum. Currently, staying with plain Riemannian gradient descent has proved more reliable.

\subsection{Results}
We can finally look at very preliminary numerical results based on this geometric optimization approach. 

\subsubsection*{Comparison with perturbation theory}

\begin{figure}
  \begin{center}
    \includegraphics[width=0.48\textwidth]{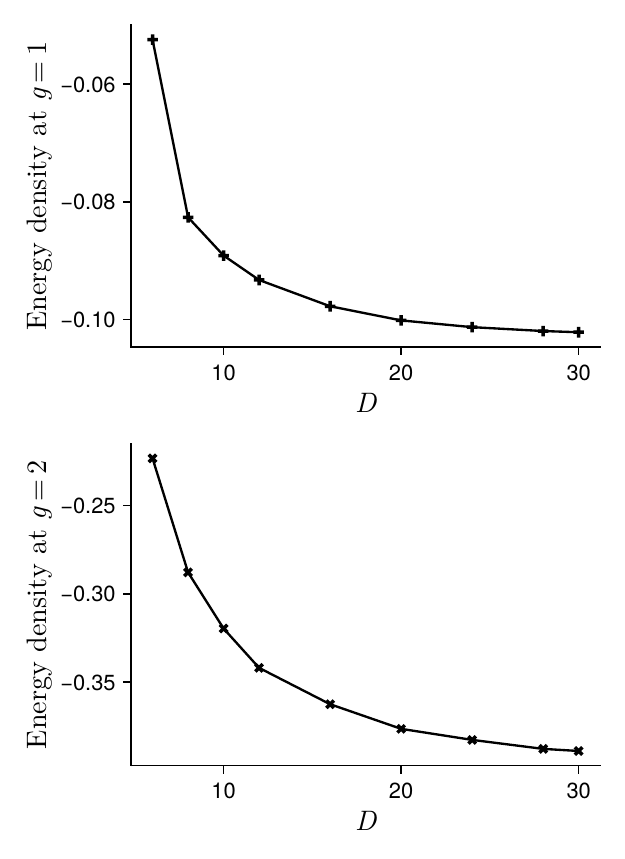}~
    \includegraphics[width=0.48\textwidth]{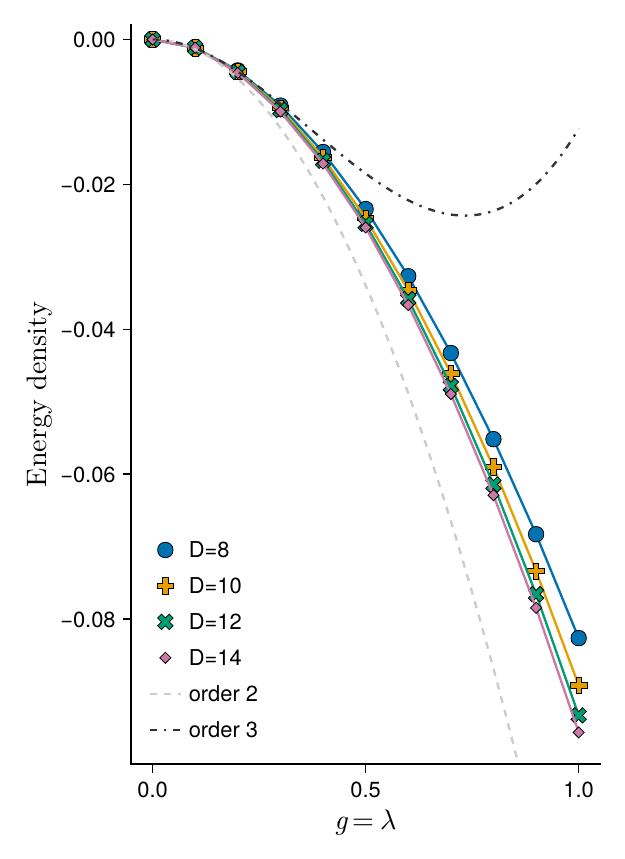}
  \end{center}
  \caption{Left: energy density as a function of $D$ for $g=\lambda=1$ (top) and $g=\lambda=2$ (bottom). Right: energy density for small $g=\lambda$ compared with perturbation theory at order $g^2$ and $g^3$.}\label{fig:o2_energy}
\end{figure}

The very first step is to minimize the energy density for increasing $D$ and see if the results converge. This is indeed the case. Although this was retrospectively not the simplest choice, we looked at two points $g=1.0$ and $g=2.0$ on the $O(2)$ invariant line $g=\lambda$. As shown in Fig. \ref{fig:o2_energy}, the values $e_0(D)$ do converge, although slower than for $\phi^4$.

The second sanity check is a comparison with another method. To my knowledge, there are no non-perturbative numerical results available in the literature for this model. To make sure we did not make mistakes, the easiest is thus to compare RCMPs values of the energy density against second and third order perturbation theory. The corresponding diagrams are the same as the ones in~\cite{serone2018symmetric}, only their multiplicity changes. The results are shown in Fig. \ref{fig:o2_energy} and demonstrate 1) that the RCMPS results agree for very small $g$ with perturbation theory, even at very moderate $D$, 2) that the RCMPS results are much more accurate already for $g\geq0.3$ (at least if we accept the spread between the two largest bond dimensions as a crude proxy for the error).

\subsubsection*{A simple observable on a simple line}

\begin{figure}
  \begin{center}
    \includegraphics[width=0.48\textwidth]{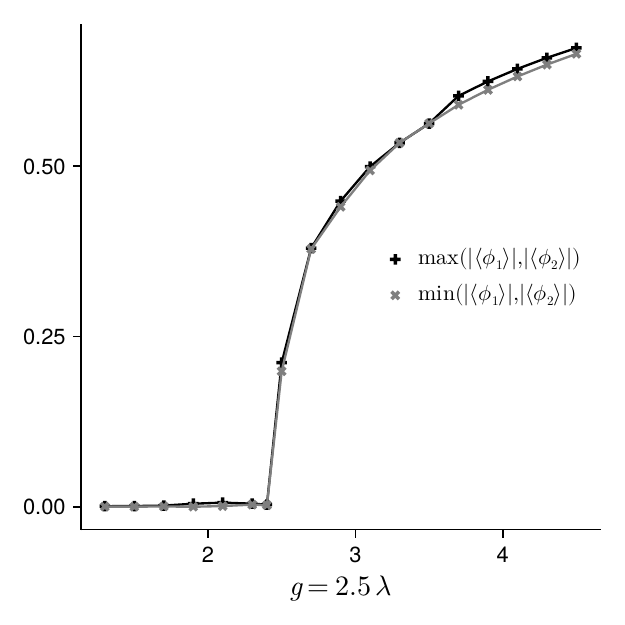}~
    \includegraphics[width=0.48\textwidth]{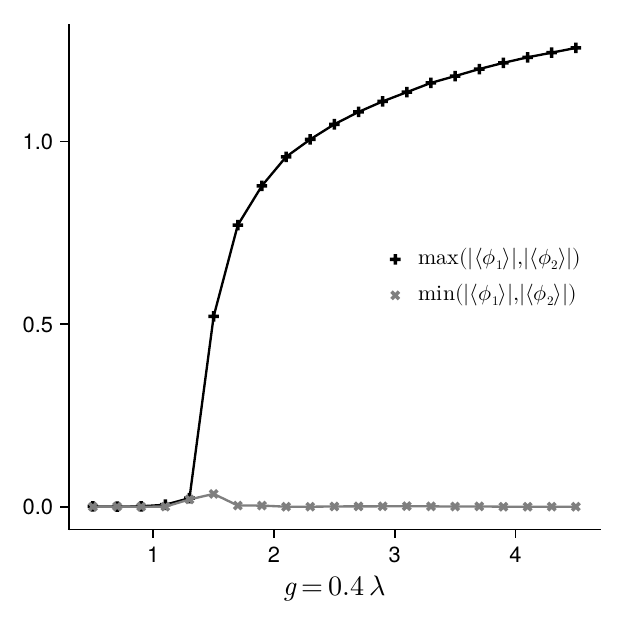}
  \end{center}
  \caption{Minimum and maximum absolute expectation values of the fields $\phi_1$ and $\phi_2$ for $D=18$ and $g=2.5\lambda$ on the left, and $g=0.4\lambda$ on the right.}\label{fig:o2_observables}
\end{figure}

The phase diagram of this model is quite subtle because the group of invariance is $D_4$, which admits $10$ subgroups. With the limited objective of showing that the method works, we may consider two cuts in $g,\lambda$ space displaying different types of phase transitions.

The simplest setting is when $g \gg \lambda$. In that case, the model behaves like two weakly coupled $\phi^4$ models. As the coupling is increased, each half goes from a symmetric to a symmetry broken phase, via an Ising second-order phase transition. Preliminary results for $g=2.5 \lambda$ and $D=18$ are shown in Fig. \ref{fig:o2_observables} and we indeed observe the expected behavior. The data is a bit noisier than for $\phi^4$, but we suspect this is because our bond dimension is too low, convergence being a bit slower with two fields.

A more complex behavior, involving the two bosons in a genuine way, can be obtained when $g\ll \lambda$. In that case the cross term $:\!\phi^2_1\!: :\!\phi^2_2\!:$ dominates. For small enough coupling we are, again, in a fully symmetric phase. However, as the coupling is increased, the potential develops minima where one field is $0$ and the other is not. At that point, it becomes advantageous to break the $\mathbb{Z}_2$ symmetry of only one of the fields. Preliminary results for $g=0.4$ and $D=18$ are shown in Fig. \ref{fig:o2_observables} and display the expected behavior in a very clean way.

In analogy with the Ashkin-Teller model, we expect that this model has other more subtle phases (\eg Baxter, nematic, ...), that cannot be seen from a simple mean-field analysis \cite{Baxter1982ExactlySolved}. We already see hints of them in numerics, but larger bond dimensions are required to get clean phase boundaries. Finally, on the $g=\lambda$ line, which is $O(2)$ invariant, we expect a BKT transition from a gapped to a gapless phase. We see clear signs of this transition in the behavior of the free particle entropy, and hope to study it more precisely in the near future. This will also require larger bond dimensions, and ideally an explicit inclusion of $O(2)$ symmetry into the block structure of the $Q,R_1,R_2$ matrices.

\chapter{Computing observables beyond correlation functions}

At least for $1+1$ dimensional bosonic QFT, RCMPS provide (equal-time) expectation values of local operators to impressive precision. Apart from extending the technique to a broader set of models, an important orthogonal question is to access more observables. In this chapter, I discuss two methods we have developed to access extended operators (defects) and spectral data.

\section[Extended operators]{Extended operators\protect\footnotemark}

\footnotetext{This section is primarily a compressed version of~\cite{tiwana2025}, an article written with Karan Tiwana and Edoardo Lauria.}

\subsection{Defects, impurities, extended operators}

\subsubsection*{General defects}
In a $d$-dimensional space-time, a $p$-dimensional defect is a modification of the $d$-dimensional action (the ``bulk theory'') that has support on $p$-dimensional submanifold $B$ of $\mathbb{R}^d$. More explicitly, but at a (so far) non-rigorous level, we consider ``bulk'' QFTs with an action of the form
\begin{equation}
    S_\text{bulk}(\phi) = \int_{\mathbb{R}^d} \frac{(\nabla \phi)^2}{2} + V(\phi)\,,
\end{equation}
Since we are working with RCMPS, we will quickly fix $d=2$. The ``defect'' QFT is specified by the action
\begin{equation}
    S(\phi) = S_\text{bulk}(\phi) + S_\text{defect}(\phi) = S_\text{bulk}(\phi) + \int_B f(\phi)\,.
\end{equation}
Expectation values can be computed with this modified action in the same way as for a standard QFT. For example:
\begin{equation}
    \langle \phi(x_1)\cdots \phi(x_n)\rangle_\text{defect} = \frac{\int \mathcal{D}\phi \; \phi(x_1)\cdots \phi(x_n) \; \e^{-S(\phi)} }{\int \mathcal{D}\phi  \; \e^{-S(\phi)} }= \frac{\langle \phi(x_1)\cdots \phi(x_n) \, \e^{-\int_B f(\phi)}\,\rangle_\text{bulk}}{\langle \e^{-\int_B f(\phi)}\,\rangle_\text{bulk}}\,.
\end{equation}
Physically, defects can be used to model boundaries, interfaces, and impurities which are very common in realistic low-energy systems (\eg in the Kondo effect~\cite{Wilson:1974mb,KONDO1970183}). Considering defects is also useful in the formal study of QFT, and \eg allows for a characterization of symmetries that are not captured by local QFT operators~\cite{Gaiotto:2014kfa} (see~\cite{Cordova:2022ruw} for a recent review).

%
\subsubsection*{From impurities to extended operators}

In $d=1+1$ space-time dimension, we can consider line defects (\ie $p=1$): a one dimensional deformation of the two dimensional action, which we take to be a straight line. In the Hamiltonian picture, such a defect is naturally interpreted as an impurity, namely a perturbation of the Hamiltonian localized at one point $x_0$ in space, and persistent in time:
\begin{equation}
    H = H_\text{bulk} + H_\text{defect} = H_\text{bulk} + f\left[\hphi(x_0)\right]\,.
\end{equation}
If the defect is only a finite or semi-infinite line, this corresponds to a \emph{local quench}, that is an impurity that is turned on only for some time. 

For generic quantum many-body systems, such impurities/quenches give rise to rich physics but introduce challenges with the variational method. First, translation invariance in space is lost. Second, in the local quench case, the quantum state needed to evaluate expectation values becomes time dependent, and thus needs to be evolved \eg with the time dependent variational principle (TDVP)~\cite{haegeman2011_original_tdvpMPS,vanderstraeten2019tangentspace}.

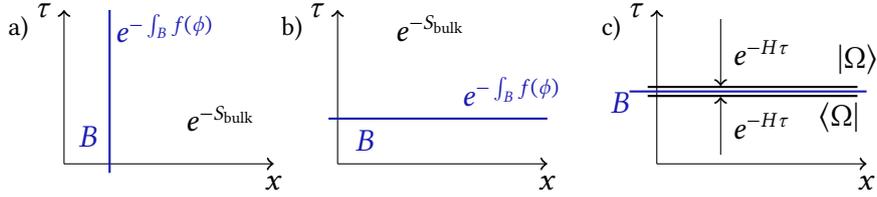
\begin{figure}[htp]
    \centering
    \begin{tikzpicture}[scale=1.2]

    \tikzset{line width=0.8pt}
    \definecolor{bulkblue}{rgb}{0.1, 0.1, 0.7}
    
    \begin{scope}
      \node at (-0.5,1.5) {\footnotesize a)};
      
      \draw[->] (0,0) -- (0,1.7) node[left] {$\tau$};
      \draw[->] (0,0) -- (2.3,0) node[below] {$x$};
      
      \draw[bulkblue, thick] (0.5,-0.1) -- (0.5,1.7);
      \node[bulkblue, left] at (0.5,0.3) {$B$};
      \node[bulkblue] at (1.1,1.5) {$e^{-\int_B f(\phi)}$};
      
      \node[right] at (1.2,0.5) {$e^{-S_{\text{bulk}}}$};
    \end{scope}

    \begin{scope}[xshift=3cm]
      \node at (-0.5,1.5) {\footnotesize b)};
      
      \draw[->] (0,0) -- (0,1.7) node[left] {$\tau$};
      \draw[->] (0,0) -- (2.3,0) node[below] {$x$};
      
      \draw[bulkblue, thick] (-0.1,0.5) -- (2.3,0.5);
      \node[bulkblue, below] at (0.3,0.5) {$B$};
      
      \node[bulkblue] at (1.9,0.8) {$e^{-\int_B f(\phi)}$};
      
      \node[right] at (0.5,1.5) {$e^{-S_{\text{bulk}}}$};
    \end{scope}
    
    \begin{scope}[xshift=6.5cm]
      \node at (-0.5,1.5) {\footnotesize c)};
      
      \draw[->] (0,0) -- (0,1.7) node[left] {$\tau$};
      \draw[->] (0,0) -- (2.3,0) node[below] {$x$};
      
      \draw[bulkblue, thick] (-0.3,0.8) -- (2.3,0.8);
      \node[bulkblue] at (-0.4,0.68) {$B$};
      
    
      \draw[->] (0.7,1.6) -- (0.7,0.86) node[midway, right] {$e^{-H\tau}$};
      \draw[->] (0.7,0.1) -- (0.7,0.74) node[midway, right] {$e^{-H\tau}$};
      
      \draw[thick] (-0.1,0.85) -- (2.2,0.85);
      \node[below] at (2.0,0.80) {$\bra{\Omega}$};
      
      \draw[thick] (-0.1,0.75) -- (2.2,0.75);
      \node[above] at (2.2,0.86) {$\ket{\Omega}$};
    \end{scope}
    \end{tikzpicture}
    \caption{a) original defect, interpreted as an impurity, b) equivalent rotated defect, interpreted as an extended operator c) representation of the defect expectation value in the Hamiltonian formalism: the bulk dynamics from $+\infty$ and $-\infty$  is equivalent to a projection of the state at fixed $\tau$ to the bulk ground state $\ket{\Omega}$. Figure from~\cite{tiwana2025}}
    \label{fig:defect_rotation}
\end{figure}

Fortunately, a simple observation makes a plain variational study possible for large subset of defects and operator insertions. For a relativistic QFT in imaginary time, \ie a Euclidean invariant theory, we can rotate the quantization direction, exchange the role of space and time, and make the defect spacelike instead of timelike (see Fig. \ref{fig:defect_rotation}). In this new picture, time evolution is generated by the bulk Hamiltonian, and expectation values are simply evaluated by applying an extended operator to the translation-invariant bulk vacuum, $\ket{\Omega}$:
\begin{equation}\label{eq:rotated_defect_expectations}
   \langle \phi(x_1) \cdots \phi(x_n)\rangle_\text{defect} = \frac{\bra{\Omega} \, \hphi(x_1) \cdots \hphi(x_n)\,  \e^{-\int_B f(\hphi)} \ket{\Omega}}{\bra{\Omega} \e^{-\int_B f(\hphi)} \ket{\Omega}}\,.
\end{equation}

This representation is convenient because $\ket{\Omega}$ can be replaced by its variational approximation as a RCMPS $\ket{\psi_0}$, simply computed from the bulk theory. The last ingredient is technical, non-trivial, and specific to RCMPS: if the operator insertions and defect are aligned (same imaginary time), and if $f$ is \emph{linear} (\ie a so called magnetic defect), the right-hand side of \eqref{eq:rotated_defect_expectations} is efficiently computable at the same asymptotic cost with the defect than in the vanilla no defect case. All the complex optimization toolbox required to find RCMPS ground states for translation invariant models can thus be reused without modification. All that is needed is a particular post-processing of this ground state, which is comparatively cheap.

\subsection{Magnetic defects with RCMPS}\label{sec:RCMPS_defects}

\subsubsection*{Defect expectation value}
We now discuss the last technical RCMPS ingredient. Let us consider the expectation value of the extended operator $\Dmu=\e^{-\mu \int_{-L}^0 \hphi}$. While the defect has a finite size $L$, we take the thermodynamic limit for the bulk. Given the RCMPS ground state $\ket{Q,R}$ of the $\phi^4$ model, the expectation value of $\Dmu$ is given by:
\begin{align}
	\langle 0,g |\Dmu |0,g\rangle \simeq \, \bra{Q,R} e^{-\mu \int_{-L}^0  \hphi}\ket{Q,R}\,.
\end{align}
Here, $\ket{Q,R} \simeq \ket{0,g}$ is the RCMPS approximation of a given bond dimension $D$ to the true ground state of the $\phi^4$ model at coupling $g$.  Using the Baker-Campbell-Hausdorff formula to normal order $\mathcal{D}_L$  (see appendix \ref{DefectExp} ) we have
\begin{align}\label{eq:defRCMPS}
	\begin{split}
		\bra{Q,R}\Dmu\ket{Q,R} = \mathcal{Z}_{-\mu G, -\mu G}\,\exp\left[\frac{\mu^{2}}{2}\int_{\mathbb{R} }\upd x\; G(x)^2\right]\,,
	\end{split}
\end{align}
with the modified ``source''
\begin{align}
	G(x) \equiv \int_{-L}^{0} \upd y \, J(x-y)\,,
\end{align}
and $J(x)$ being the source defined in eq.~\eqref{eq:jsource_def}. As before, the path ordered exponential in $ \mathcal{Z}_{-\mu G, -\mu G}$ is the solution to an ODE \eqref{eq:vertex_ode} now involving the modified source $G(x)$. The additional exponential factor in $\braket{\mathcal{D}_{L}}_{Q,R}$ can be incorporated into the ODE as a term proportional to the identity to get (in superoperator form)
\begin{equation}\label{eq:denODE}
	\frac{\upd}{\upd x} \rho(x) = \mathcal{L}\cdot \rho(x) - \mu G(x) \left[ R\rho(x) + \rho(x) R^\dagger \right] + \frac{\mu^{2}}{2} G^{2}(x) \rho(x)\,,
\end{equation}
which gives
\begin{align}
	\bra{Q,R}\Dmu\ket{Q,R} = \lim_{x \to \infty} \, \mathrm{tr}\,[\rho(x)]~, ~~ \text{for the initial condition} \qquad \lim_{x\to -\infty} \rho(x)= \rho_{0}\,.
\end{align}
Hence, computing the expectation value of the defect operator, which is extended, is not more difficult (asymptotically in $D$) than computing expectation values of local operators, as we advertised.

\subsubsection*{Defect correlation functions}
Correlation functions of local operators in the defect theory can be computed in a similar manner. We define the expectation value of a vertex operator in the full defect theory as
\begin{align}\label{eq:Defect_vertex}
    \langle V_b (x) \rangle_{\text{defect}} = \frac{\bra{0,g} :e^{b\hat{\phi(x)}}: \text{e}^{-\mu \int_{-L}^{0} \hat{\phi}}\ket{0,g}}{\bra{0,g}\text{e}^{-\mu \int_{-L}^{0} \hat{\phi}}\ket{0,g}}~\,,
\end{align}
where $\ket{0,g}$ is the exact ground state of the Hamiltonian at coupling $g$.
This expression, upon iterated differentiation with respect to $b$, allows one to compute expectation value of $:\phi^k(x):$ in the defect theory. We just computed the denominator, and the numerator can be evaluated in a similar way (see appendix \ref{sec:DefectVertex} for more details). Ultimately, we obtain 
\begin{align}
	\begin{split}
		\frac{\bra{Q,R} V_{b}(x) \Dmu \ket{Q,R}}{\bra{Q,R} \Dmu \ket{Q,R}}
		& = \frac{\mathcal{Z}_{s_x,s_x}}{\mathcal{Z}_{-\mu G, -\mu G}} \; \times\exp\Big{(}-b\,\mu\int_{\mathbb{R}} \upd y \int_{-L}^{0} \upd z \:J(x-y)J(y-z)\Big{)}\,,
	\end{split} \label{eq:vertex_defect_solution}
\end{align} where
\begin{equation}
	s_x(y) = b\,J(x-y) - \mu\int_{-L}^{0} \upd z \: J(z-y)\,.
\end{equation}
Again, this means that the expectation value of the vertex operator in the defect theory can be computed efficiently as the ratio of the trace of the solution of matrix ODEs. Numerically, the only new difficulty is that r.h.s. of \eqref{eq:vertex_defect_solution} is the ratio of two terms that grow exponentially in $L$, the defect size (they are exponentials of an extensive quantity). This can however be alleviated with simple normalization strategies outlined in appendix \ref{app:normalization_tricks}, which allow one to take the large defect limit $L\rightarrow \infty$ numerically.

\subsection{Some results}

All is left then is to compute defect expectation values on good ground state approximations. In~\cite{tiwana2025}, we have considered $4$ relevant coupling regimes: $g \ll 1$, \ie perturbative, $g = 2$, \ie  strongly coupled, $g\simeq 2.77$, \ie critical, and $g=4$, symmetry broken. The first allows a comparison with perturbation theory, the second allows the method to shine, the third shows its limits, and finally the fourth shows interesting physics. Let us summarize the results we obtained here.

\subsubsection*{\texorpdfstring{Weak coupling $g=0.1$}{Weak coupling g=0.1}}

\begin{figure}[htp]
	\centering
	\subfloat[]{
		\includegraphics[width=0.48\textwidth]{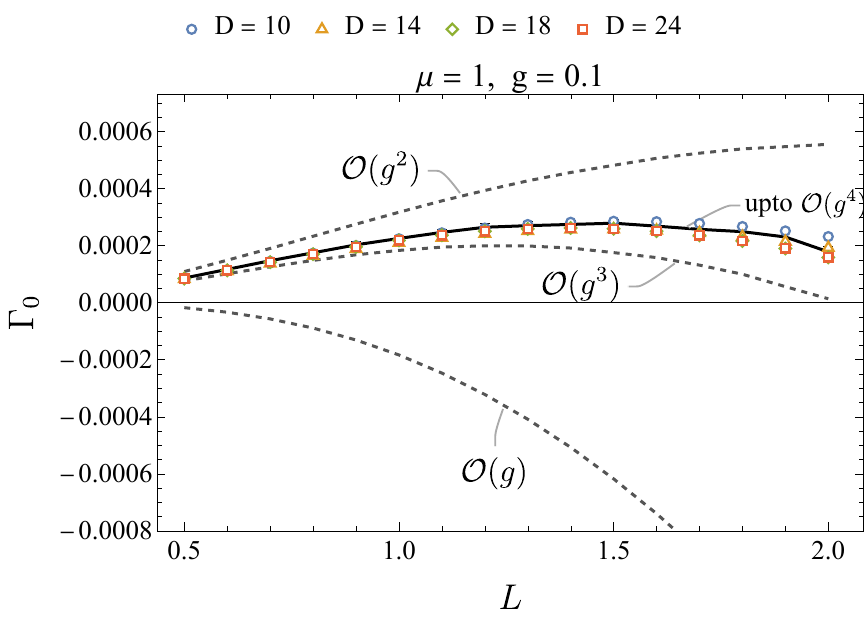}}
	\subfloat[]{
		\includegraphics[width=0.48\textwidth]{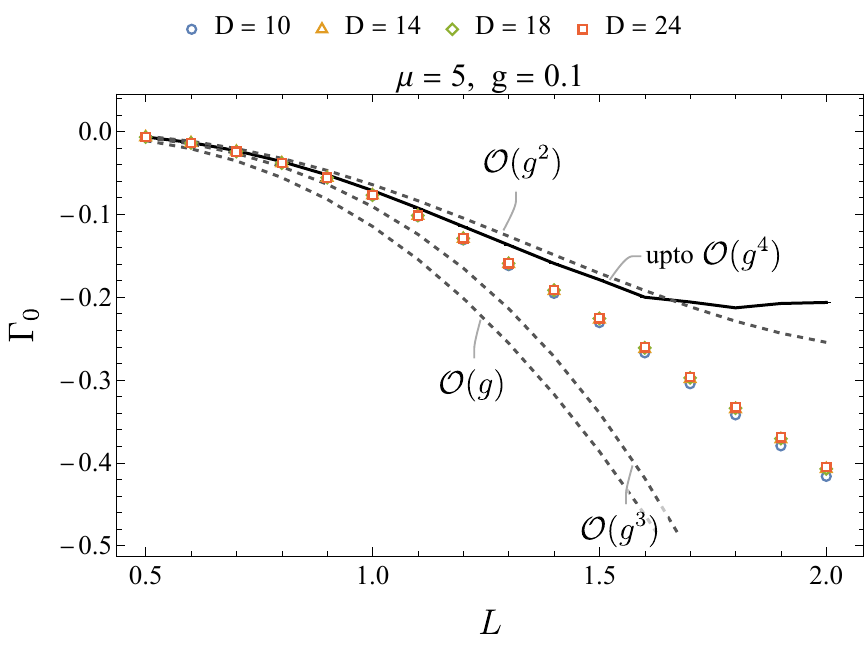}}
    \caption{Perturbative prediction for $\Gamma_0 \equiv  \frac{\bra{0,g} \Dmu \ket{0,g}}{\bra{0,0} \Dmu \ket{0,0}} -1$ up to $\mathcal{O}(g^4)$ vs. RCMPS (colored markers). The error bars are obtained from the Monte Carlo errors in the evaluation of the diagrams and are smaller than the size of the data points. Figure from~\cite{tiwana2025}}
	\label{fig:0pt_vs_L_pert}
\end{figure}

We compare the perturbative calculation for the defect expectation value and RCMPS in Fig.~\ref{fig:0pt_vs_L_pert}. We took $g=0.1$, and $\mu=1$ and $\mu=5$ for the defect coupling. A small bond dimension, $D\geq 10$ is sufficient to get well-converged RCMPS results, whereas we have to go to a fairly high order, $g^4$, in the perturbation expansion to get similar values even just at $\mu=1$. For larger $\mu$, the defect is no longer a small perturbation as $L$ grows, and the perturbation theory results are inaccurate while RCMPS results remain just as well converged.

\begin{remark}[Perturbative expansion in $g$ and $\mu$]
From the fact that the $g$-expansion is \emph{exact} in $\mu$, which is a consequence of working with a defect that is linear in $\phi$, one might have expected perturbation theory to be accurate for fixed $g=0.1$ and large $\mu$. However, as is clear from these numerical results, this is not the case. Indeed, Feynman diagrams of order $n$ scale like $g^n \mu^{2n+2}$, and thus become large when $\mu$ increases as $g$ is kept small but fixed.
\end{remark}

\subsubsection*{\texorpdfstring{Strong coupling $g=2.0$}{Strong coupling g=2.0}}

We now explore the strong-coupling (but still symmetric) regime $g=2$. This time, we can  consider a one-point function $|\langle \phi(x)\rangle_{\text{defect}}|$ for finite values of $g$, $\mu$ and $L$, as a function of $x$. 

We take $g=2$, $\mu =1, 4$, and $L=6$.  The results are intuitive. First $|\langle \phi(x)\rangle_{\text{defect}}|$ decays exponentially as we move away from the edge of the defect into the bulk, as expected. Second, increasing the defect coupling $\mu$ (keeping all other parameters fixed), $|\langle \phi(x)\rangle_{\text{defect}}|$ increases, as it should since the bulk field tends to align with the defect. Third, the field expectation value remains continuous at $x=0$, \ie as we cross the defect. For $x<0$ it measures a \emph{defect magnetization} and stabilizes to a constant value in the middle of the defect.

\begin{figure}[htp]
	\centering
	\subfloat[]{
		\includegraphics[width=0.48\textwidth]{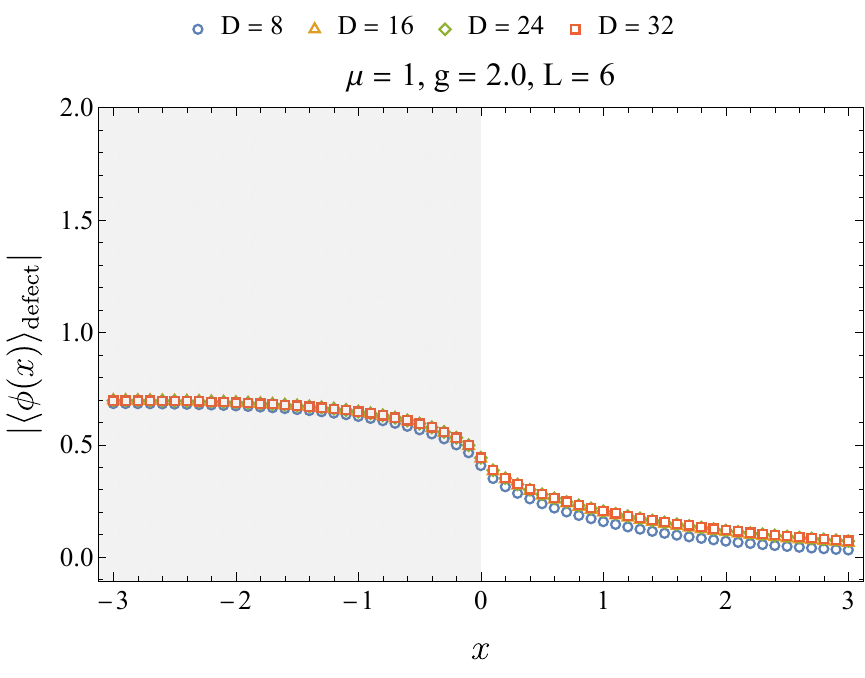}
	}
	\subfloat[]{
		\includegraphics[width=0.48\textwidth]{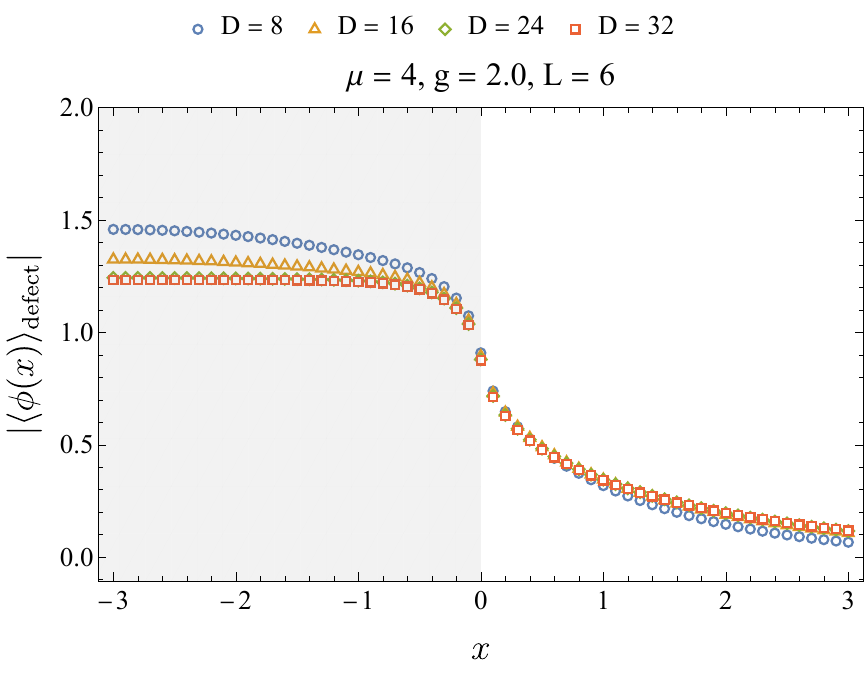}
	}	
  \caption{Bulk one-point functions from RCMPS computation as a function of the distance from the defect (shaded region) in the strong-coupling, symmetric regime, for $\mu=1$ (left) and $\mu=4$ (right). Figure from~\cite{tiwana2025}}
	\label{fig:1pt_vs_tau_strong}
\end{figure}

At large $\mu$, the RCMPS results, although accurate at $D\geq 24$, are arguably slower to converge, especially inside the defect. We believe this is fairly intuitive, and a property of all variational methods. Our RCMPS states are approximate ground states of the bulk model, obtained by minimizing an energy density. From the ground state, one can define a probability distribution for the spatial average of the field in any finite size interval. The RCMPS approximates this probability very well overall, but is inevitably less accurate in the tails, for very small probabilities that almost do not contribute to the energy density. As we increase $\mu$ we shift the center of the distribution, and thus get more sensitive to its tail. As a result, we need a larger bond dimension to preserve the accuracy. This shift of the distribution is stronger closer to the defect, and even more so inside it, and thus we need larger bond dimensions inside the defect. Note however that even deep in the defect and at large $\mu$ we can easily reach bond dimensions such that the results are visually converged in Fig.~\ref{fig:1pt_vs_tau_strong}. We do not know of any other method that could reach a similar accuracy for such large perturbations both in $g$ and $\mu$ in the continuum.

\subsection{\texorpdfstring{Critical coupling $g\simeq 2.771525$}{Critical coupling g = 2.771525, comparison with scaling expectations}}
We then fine tune the model at criticality $g_c\simeq 2.771525$~\cite{delcamp202giltphi4,vanhecke2022scalingphi4}. At this specific coupling, and as far as IR physics is concerned, the $\phi^4$ model is in the Ising universality class, which is well understood. 

However, this universal regime is also where the accuracy of RCMPS degrades because of their finite entanglement entropy. To obtain accurate results, one needs to do a finite entanglement scaling analysis~\cite{pollmann2009_finiteentanglementscaling}. Here, we stick with a plain variational study, which also helps to demonstrate the limitations of RCMPS.

We looked at $|\langle \phi(x)\rangle_{\text{defect}}|$ with $\mu=1, 4$ and $L=30$. For $1\ll x \ll L$ we should be sensitive to the critical (scaling) behavior of the model. The scaling behavior of the one-point function was computed in~\cite{Cuomo:2021kfm} using Boundary Conformal Field Theory techniques (see~\cite{DiFrancesco:1997nk,Cardy:2004hm} for a review): 
	\begin{align}\label{critical1pt}
		\langle \phi(x) \rangle_\text{defect}= \frac{2^{1/8}\sqrt{C_\phi}}{|x|^{1/8}}\,,
	\end{align}
where $C_\phi$ is a non-universal normalization constant which is obtained from the two-point function of the field (without defect) at large distance
\begin{equation}\label{eq:1pt_scaling_prediction}
    \bra{0,g_c} \hphi(x_1) \hphi(x_2) \ket{0,g_c} \sim \frac{C_\phi}{|x_1-x_2|^{1/4}} \,.
\end{equation}
Our results are reasonably well converged for small enough distance, but our bond dimensions remain manifestly too low, at least without finite entanglement scaling, to reliably fit such an exponent.

\begin{figure}[htp]
	\centering
	\subfloat[]{
		\includegraphics[width=0.48\textwidth]{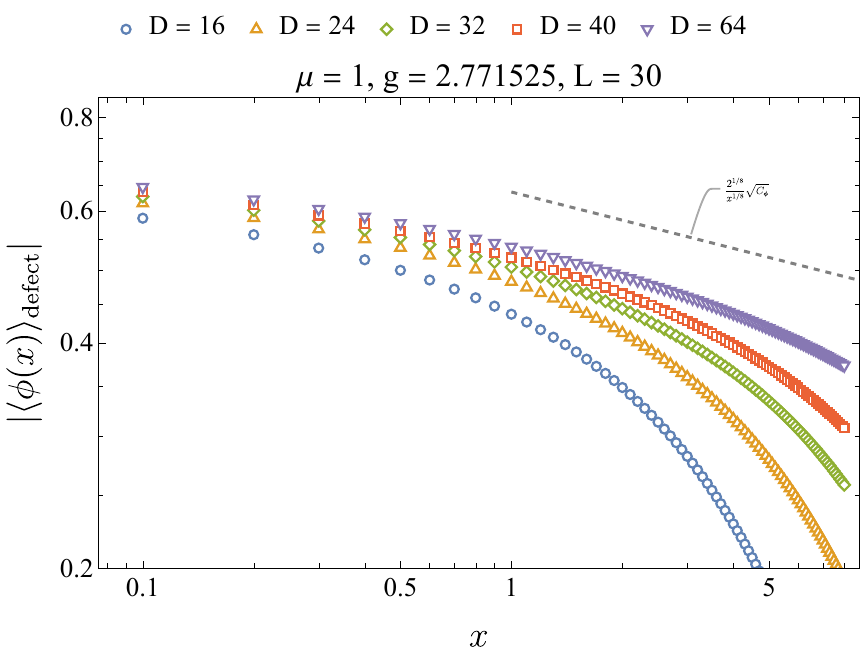}
	}
	\subfloat[]{
		\includegraphics[width=0.48\textwidth]{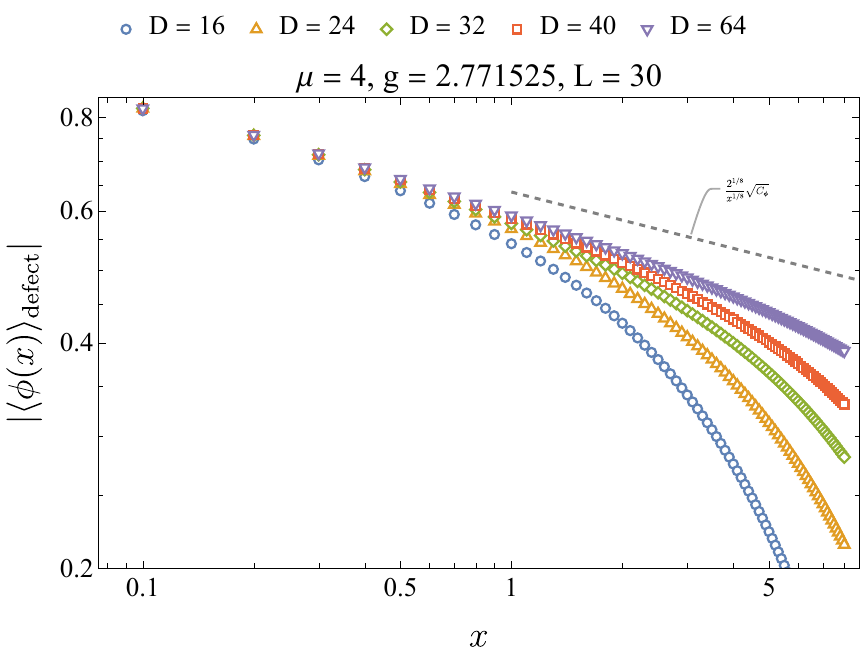}
	}	
  \caption{One-point functions obtained from RCMPS computation as a function of the distance from the defect at critical coupling, for $\mu=1$ and $\mu=4$. The dashed line is the expected exact scaling. Figure from~\cite{tiwana2025}}
	\label{fig:1pt_vs_tau_critical}
\end{figure}

\subsection{\texorpdfstring{Symmetry broken phase, coupling $g=4$}{Symmetry broken phase, coupling g=4}}
We now go deep in the symmetry broken phase. There are two options for symmetry breaking, corresponding to very different physical situations:
\begin{itemize}
  \item The bulk can be symmetry broken in the direction favored by the defect, \[ \text{sign}(\bra{0,g} \hat{\phi}(x)\ket{0,g})= - \text{sign} (\mu) .\] This is what would happen spontaneously in a statistical mechanical system with a defect, initialized in the symmetric phase but then slowly cooled down. The bulk would then pick the average field value favored by the defect.
  \item The bulk can be symmetry broken ``against'' the defect, \ie \[ \text{sign}(\bra{0,g} \hat{\phi}(x)\ket{0,g})= + \text{sign} (\mu).\] Physically, this would occur if symmetry breaking happens first, yielding positive $\bra{0,g} \hat{\phi}(x)\ket{0,g}$, and only then the defect coupling $\mu$ is progressively sent to positive values. The bulk field would then remain stuck ``against'' the defect.
\end{itemize}

\begin{figure}[htp]
	\centering
	\subfloat[]{
	\includegraphics[width=0.48\textwidth]{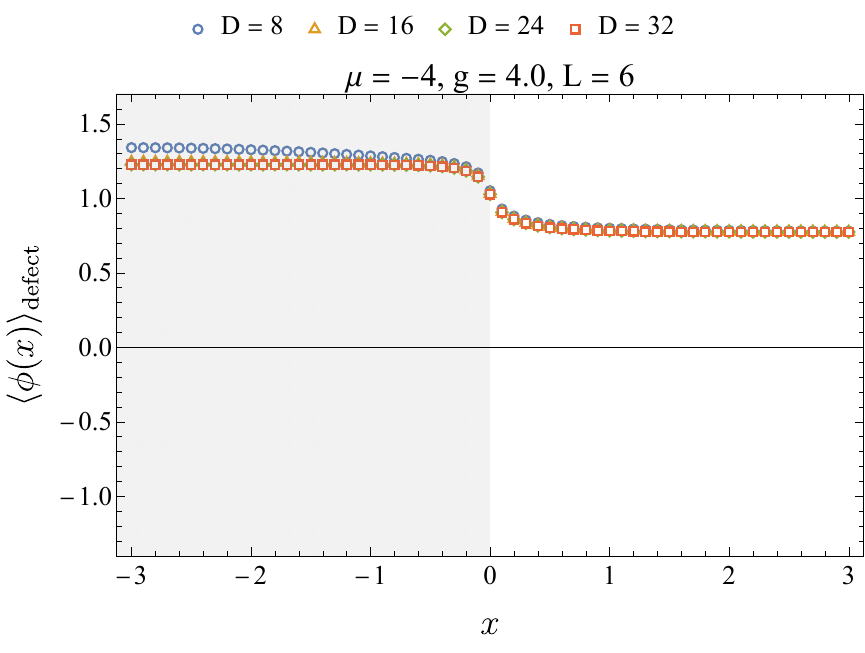}
	}
	\subfloat[]{
		\includegraphics[width=0.48\textwidth]{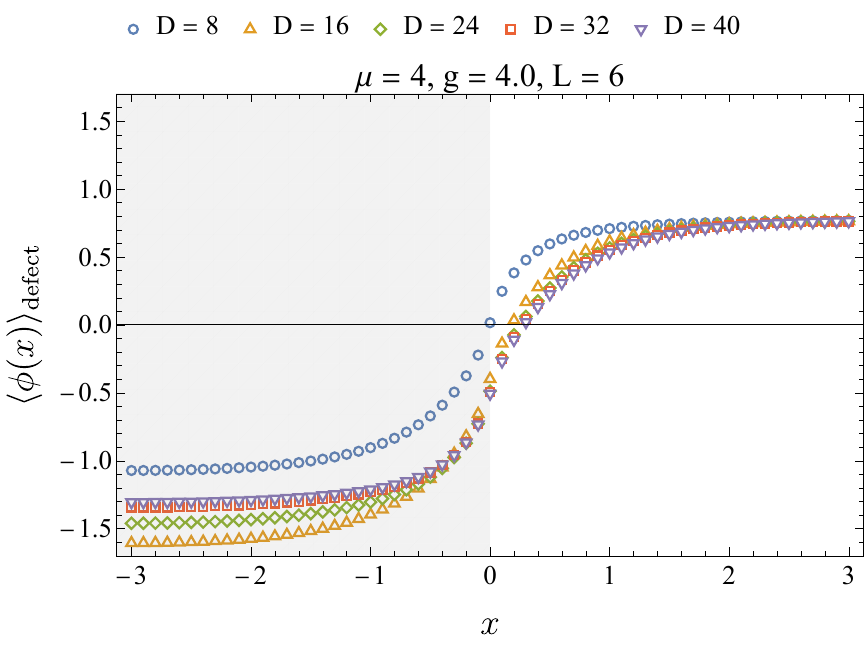}
	}
  \caption{One-point functions from RCMPS computation as a function of the distance from the defect (shaded region) in the symmetry-broken regime. In (a) symmetry breaking along the defect, in (b) symmetry breaking against the defect. In both plots $\bra{Q,R}\hphi(x)\ket{Q,R}>0$. Figure from~\cite{tiwana2025}}
	\label{fig:1pt_vs_tau_supcritical}
\end{figure}

We consider the $1$-point function in the presence of a defect of fixed size $L=6$ for these two types of symmetry breaking in Fig. \ref{fig:1pt_vs_tau_supcritical}, where we took RMCPS states with $\bra{Q,R}\hphi(x)\ket{Q,R}>0$.
These results show the expected sign change in the field when moving from the defect to the bulk when the bulk is symmetry broken against the defect. Clearly, in this latter situation, the RCMPS results are slower to converge, but still provide accurate results at large $D$. Even setting aside the UV difficulties, such calculations would be very difficult to carry on the lattice with the Monte Carlo method.

\section[Spectral and real-time data]{Spectral and real-time data \texorpdfstring{\protect\footnotemark}{}}

\footnotetext{This section presents first results of a promising research line explored with Sophie Mutzel. A more thorough discussion appeared (after the first version of this manuscript was written) in \cite{mutzel2025extractingquantumfieldtheory}.}

We saw that RCMPS allow to get exceptionally precise local expectation values in the vacuum, without UV or IR cutoff. However, we do not have a direct way to estimate useful spectral data like the mass.

As we will discuss later in \ref{sec:excitation}, there exists tensor network methods to compute the excitation spectrum directly. But they are difficult to setup in our case. Further it seems that the relativistic case we consider is for once far easier! Indeed, because of Lorentz/Euclidean invariance, we can in principle directly estimate the spectrum from the decay of \emph{vacuum} correlation functions in \emph{space}. And since these correlation functions are very precise with RCMPS, perhaps we can just use them to get precise spectral data directly!

This seducing option is not as direct as one might expect. In what follows, we first show how a direct estimation of the mass from the asymptotic decay of RCMPS correlation functions is possible but imprecise. We then introduce a linear programming approach, greatly inspired by the recent work of Scott Lawrence~\cite{lawrence2024spectral}, to get more precise estimates. This latter approach can also be greatly extended beyond the computation of the gap, to give access to far more observables (including real-time data). 

\subsection{Some analytical results on RCMPS correlators}

Let us first understand RCMPS two-point functions, and more specifically $C_\text{Q,R}(x) := \langle\phi(x)\phi(0)\rangle_{Q,R}$. Because of the mild non-locality and convolutions with $J$, RCMPS correlation functions do not behave as simple sums of exponentials. 

Instead, it is possible to show:
\begin{equation}\label{eq:analytic_RCMPS}
  C_\text{Q,R}(x)  =\frac{1}{2\pi}K_{0}{\left(m|x|\right)}-\frac{1}{\pi}\sum_{i}\int\frac{c_{i}}{2\sqrt{p^2+m^2}}\frac{\lambda_{i}}{\lambda_{i}^{2}+p^{2}}e^{-i p x}\upd p \, ,
\end{equation}
where the first term corresponds the free propagator contribution, $\lambda_i$ are the eigenvalues of the transfer operator $\mathbb{T}$, and the coefficients $c_i$ are expressed in terms of the corresponding left $\ket{l_i}$ and right $\bra{r_i}$ eigenvectors
\begin{equation}
  \begin{split}
    c_i = &\, \langle l_{0}|(R\otimes1)|r_{i}\rangle\langle l_{i}|(R\otimes1)\,|r_{0}\rangle+\langle l_{0}|(R\otimes1)\,|r_{i}\rangle\langle l_{i}|(1\otimes R^{*})\,|r_{0}\rangle \\ 
          &+\langle l_{0}|(1\otimes R^{\ast})\:|r_{i}\rangle\langle l_{i}|(R\otimes1)\:|r_{0}\rangle+\langle l_{0}|(1\otimes R^{\ast})\:|r_{i}\rangle \,.
  \end{split}
\end{equation}
The integral in \eqref{eq:analytic_RCMPS} does not have an exact expression, but one can compute its asymptotic behavior when $x\rightarrow +\infty$. Preliminary notes from Sophie Mutzel demonstrate that, as expected, it depends only $\lambda_1$, the eigenvalue of $\mathbb{T}$ with smallest real part. At least when $\lambda_1$ is real and $\lambda_1 \leq m$ one has:
\begin{equation}
  C_{Q,R}(x) \underset{x\rightarrow \infty}{\sim} -\frac{c_1 \lambda_1}{\pi \sqrt{m^2 - \lambda_1^2}} \exp(-\lambda_1 |x|)\, .
\end{equation}
This strict exponential decay is slightly different from the one we expect for a relativistic QFT. Indeed for a massive free field of mass $M$, we have
\begin{equation}
  \langle\phi(x)\phi(0)\rangle_\text{free} = \frac{1}{2\pi} K_0(M|x|) \underset{x\rightarrow +\infty}{\sim} \frac{\exp(-M|x|)}{\sqrt{4\pi M |x|}} \, .
\end{equation}
From the K\"all\'en-Lehmann spectral representation (which we will soon say more about), we expect this asymptotic decay to be generic, even for an interacting massive model.

This analysis shows that RCMPS correlation functions cannot match the asymptotic decay of true QFT correlation functions, at least in the strong sense that their ratios remain close to $1$. They can only match in the weaker sense that their logs are equivalent.

\begin{figure}
  \begin{center}
    \includegraphics[width=0.99\textwidth]{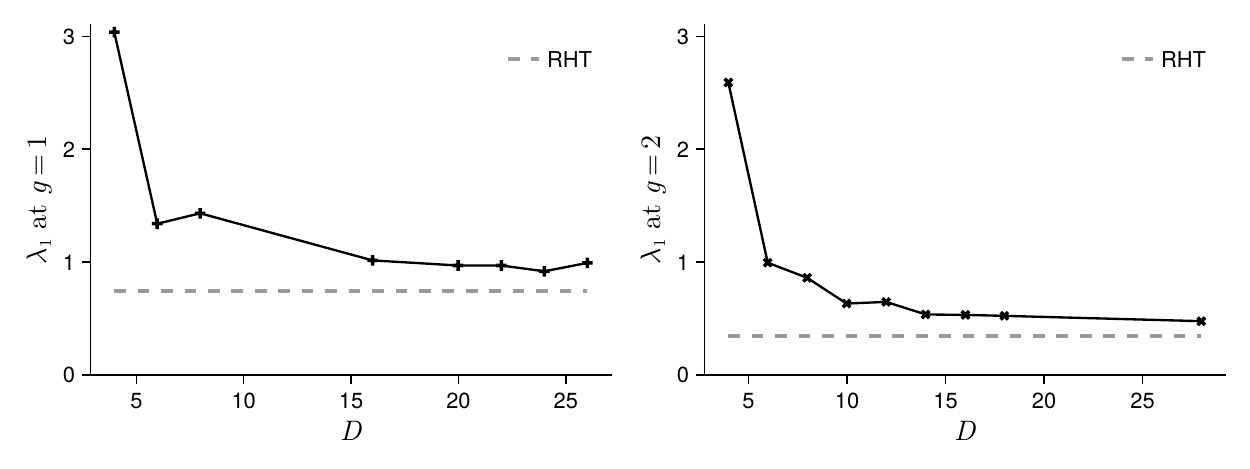}
  \end{center}
  \caption{Eigenvalue of $\mathbb{T}$ with smallest real part as a function of $D$, compared with the mass gap estimated from renormalized Hamitonian truncation. Left, for $g=1$. Right, $g=2$.}\label{fig:mass_fit}
\end{figure}

Nonetheless, we may still take $\lambda_1(D)$ as an estimate of the mass gap $M$, which should converge when $D\rightarrow +\infty$. The results are shown in Fig. \ref{fig:mass_fit} and compared with the precise renormalized Hamiltonian truncation results of Elias-Miro \textit{et. al.}~\cite{eliasmiro2017-2}, which is a much more direct approach. They are arguably underwhelming, and one would presumably need much larger $D$ to get $\lambda_1$ closer to $M$. 

What is \emph{a priori} surprising is that the correlation functions themselves are essentially indistinguishable by eye for $D\geq 16$, at least in linear scale and for $x \in [0,5]$. By focusing only on the asymptotic properties, which are not exactly captured by RCMPS, we seem to not benefit from the precision we have at all.

\subsection{Spectral function from linear programming}
We move on to a better method, which relies on the spectral representation.

\subsubsection*{K\"all\'en-Lehmann representation}
One can write the two-point function of a $1+1$ dimensional QFT in the following K\"all\'en-Lehmann spectral representation:
\begin{equation}\label{eq:kl_representation}
  \langle \phi(x)\phi(0)\rangle = \int_{\mathbb{R}^+} \upd \mu^2 \rho(\mu^2)\, \frac{K_0(\mu |x|)}{2\pi}
\end{equation}
where the \emph{spectral density} $\rho(\mu^2)$ is positive and $\int \rho(\mu^2)\upd \mu^2 = 1$.

Importantly, the analytic structure of $\rho(\mu^2)$ is well understood. For a massive scalar theory with repulsive interactions, it should be of the form 
\begin{equation}\label{eq:rho_analytic}
  \rho(s) = Z \delta(s - M^2) + \rho^\text{reg}(s) \theta(s-4 M^2)
\end{equation}
where $M$ is the mass gap and $\theta$ is the Heaviside function (the threshold $4M^2$ is changed to $9M^2$ if the theory is $\mathbb{Z}^2$ symmetric).  The problem thus becomes: can we invert the transform in equation \eqref{eq:kl_representation} to obtain $\rho$ from an accurate RCMPS estimate $\langle \phi(x) \phi(0)\rangle_{Q,R}$ of the two-point function? This would in turn trivially allow to estimate the mass.

\subsubsection*{Reconstructing the smeared spectral density}
Fitting the spectral density from a real-space correlation function is a notoriously difficult problem, akin to the inversion of a Laplace transform.
The approach we follow instead is an extension to the continuum of the recent proposal of Lawrence~\cite{lawrence2024spectral}, and which consists in turning the fit into a linear program. 

First, we consider a smeared spectral density:
\begin{equation}
\rho^\sigma(s) = \frac{1}{\sqrt{2\pi}\sigma} \int_\mathbb{R^+} \rho(u)\, \e^{- (s-u)^2/(2\sigma^2)} \, \upd u \, ,
\end{equation}
which is now a smooth function. We would like to know how large or small it can be at a fixed $s$, given our numerical RCMPS data. Let us first try to maximize it. This requires solving over functions $\rho$ the following problem
\begin{align}
  \max_\rho & ~~~~~ \rho^\sigma(s) \label{eq:objective}\\ 
  \text{under}  &~~~ \bullet ~ \forall u\geq 0,  \rho(u)\geq 0 \label{eq:psd} \\ 
                &~~~ \bullet ~ \int \upd u\, \rho(u) = 1\label{eq:normalized}\\ 
                  & ~~~ \bullet ~ \forall x \in \mathbb{R}, \langle \phi(x)\phi(0)\rangle_\text{Q,R} = \int_{\mathbb{R}^+} \upd u\, \rho(u)\, \frac{K_0(\sqrt{u}|x|)}{2\pi}\label{eq:x_constraint}
  \end{align}
  This problem is infinite in two ways: it has infinitely many variables $\rho(u)$, and infinitely many constraints (at all the points $x$ in \eqref{eq:x_constraint}). However, a nice feature is that it is a linear program: we optimize a linear cost function \eqref{eq:objective} under the linear inequalities \eqref{eq:psd} and under the linear equalities \eqref{eq:normalized} and \eqref{eq:x_constraint}. 

The next step is to make this linear program finite. We start by reducing the number of constraints and choose a finite number of points $x_k$ where we enforce it. This \emph{relaxation} of the constraints turns the solution into an upper bound to the true value.

We then discretize $u$ and thus $\rho(u)$ into a finite number of variables $u_l$, $\rho_l$. This, however, is no longer a relaxation, and we should make sure the error associated to this step is much smaller than other errors.

Finally, we should accept that our RCMPS result are not exact, and in fact have a small systematic bias $\delta$ that depends on $D$ and that we do not quite know. We should thus give this slack $\delta$ to the (now discretized) constraint \eqref{eq:x_constraint}.

With these two discretizations and extra slack, the problem now reads:
\begin{align}
  \max_\rho & ~~~~~ \frac{1}{\sqrt{2\pi}\sigma} \sum_l \rho_l \, \e^{- (s-u_l)^2/(2\sigma^2)} \Delta_l \label{eq:objective_disc}\\ 
  \text{under}  &~~~ \bullet ~ \forall l,  \rho_l\geq 0 \label{eq:psd_disc} \\ 
                &~~~ \bullet ~ \sum_l \rho_l \Delta_l = 1\label{eq:normalized_disc}\\ 
                  & ~~~ \bullet ~ \forall k, \, \langle \phi(x_k)\phi(0)\rangle_\text{Q,R} -\delta  \leq \sum_l  \rho_l\, \frac{K_0(\sqrt{u_l}|x_k|)}{2\pi} \Delta_l\leq \langle \phi(x_k)\phi(0)\rangle_\text{Q,R} +\delta \label{eq:x_constraint_disc}\,
  \end{align}
  where $\Delta_l = u_{l+1} - u_l$. The last equality constraint is now a pair of linear inequalities because of the slack $\delta$ we have introduced. This is still a linear program, which is now finite, and that we can thus solve with the simplex method for thousands of variables $\rho_l$ and hundreds of constraints. 

We may of course solve the exact same problem for the $\min$, which gives us a lower bound for the smeared spectral density. If we knew the error $\delta$ of our RCMPS results (and, again, if our discretization in $\rho$ were arbitrarily fine) these lower and upper bound on the smeared spectral density would be fully rigorous! But we have no way to know our error exactly with RCMPS. 

Instead we proceed differently. We start from a large slack $\delta$, and reduce it until the lower and upper bound obtained with the simplex method coincide. Below this error threshold, the constraints in the linear program \eqref{eq:objective_disc} are not satisfiable\footnote{This shows that there is no relativisitc QFT Hamiltonian of which our RCMPS is the exact ground state.}. Exactly at the threshold, we obtain our best guess for the smeared spectral density as well as a good retrospective approximation of our RCMPS error\footnote{This is not exactly the estimate of the distance between our RCMPS correlators and the true two-point function, which would be the natural definition of the RCMPS error. Rather, it is the distance between our RCMPS two point function and the closest \emph{physical} two-point function.}.

We have tried this approach at various couplings and bond dimensions. Although this is still preliminary, we could reconstruct quite well the Dirac contribution to the smeared spectral density. Estimating $\rho^\sigma(s)$ at larger $s$, to get access to $\rho^\text{reg}$, proved more difficult to do precisely\footnote{This is likely intrinsic: we verified that we could construct synthetic spectral densities differing wildly for $s$ sufficiently large, but with correlation functions almost numerically indistinguishable.}.

\subsubsection*{Boostrapping the mass}
\begin{figure}
  \begin{center}
    \includegraphics[width=0.99\textwidth]{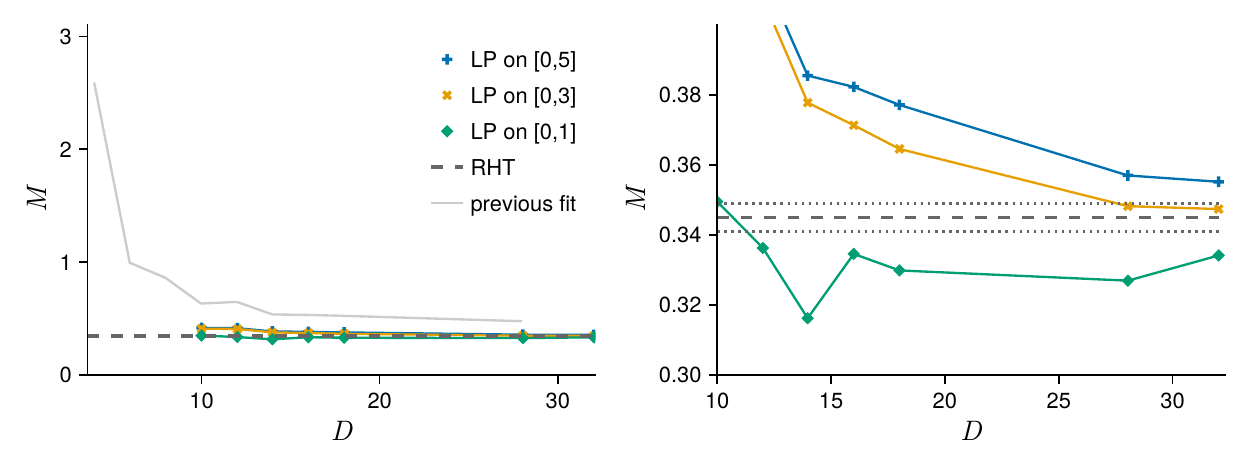}
  \end{center}
  \caption{Mass gap estimated with the bootstrap approach for $g=2$ using the RCMPS value of the correlator in the domains $x\in[0,5]$, $x\in[0,3]$, and $x\in[0,1]$. On the left, the convergence is compared with the asymptotic fit approach. On the right, we zoom in and compare more precisely with the carefully extrapolated Renormalized Hamiltonian Truncation results of~\cite{eliasmiro2017-2}. The thin dotted lines correspond to a $2\sigma$ distance from the mean, obtained from the estimated systematics reported in~\cite{eliasmiro2017-2}.}\label{fig:mass_bootstrap}
\end{figure}

If we care primarily about the mass gap, instead of particular values of the (smeared) spectral density, we can use a more direct (and precise) \emph{bootstrap} approach. 

To this end, we can change of variables in the linear program. Using \eqref{eq:rho_analytic}
\begin{equation}
\rho(s) = Z \delta(s - M^2) + \rho^\text{reg}(s) \theta(s-4 M^2)
\end{equation}
it is tempting to optimize over $Z$, $M$, and $\rho^\text{reg}(s)$ instead of $\rho(s)$. However, the dependence in $M$ is not linear, and we would lose the ability to use an efficient solver. Instead, we \emph{fix} $M$, and solve the discretized linear program \eqref{eq:objective_disc}, replacing $\rho$ with its expression in terms of $Z$ and $\rho^\text{reg}$. We do not care about the value at the optimum, but just want to see if the constraints are satisfiable.

Starting from a large RCMPS slack $\delta$, we find a range of $M$ where the constraints can be satisfied. Then, as we reduce $\delta$, this allowed range shrinks, and finally reaches a single point $M^\text{opt}$. We find this threshold easily by dichotomy, and take the resulting mass as our best guess for the gap.

In this procedure, it is important to choose the constraint points $x_k$ in a domain where the RCMPS are accurate. Using results at various $D$ as a proxy of our systematic error, we observed that the error of $C_{Q,R}(x)$ is more or less constant as a function of $x$, for values of $x$ of the order of the inverse mass gap. Then, for larger values, even though the two point function decays, we observe that the absolute error increases slightly! This is another reason why estimating the mass from the asymptotic decay is bad with RCMPS. Hence, for the constant absolute error slack $\delta$ we allow ourselves, it is better to use constraints only at points $x_k$ in the interval \footnote{The profile of the error may also depend on $D$, which we ignore here for simplicity} $[0,\text{a few} \, M^{-1}]$. 

In practice, we increase the number of constraints and variables until the values of the mass do not change. The numerical results we obtain for $g=2$ are shown in \ref{fig:mass_bootstrap} for various natural choices of constraint intervals. In all cases, the convergence compared to the asymptotic fit approach is impressive. Especially for an interval $\sim [0,M^{-1}]$, our results get very close to the renormalized Hamiltonian Truncation values (which, we insist, are obtained from careful extrapolations, and directly have access to this quantity).

In the end, we believe this bootstrap method is a promising and original approach to estimate the mass. With some fine-tuning, it should allow to get the mass itself with a precision not much below that of correlation functions. 

\subsection{Extensions}
There are natural extensions of this linear programming / bootstrap approach. 

Technically, we could consider the two-point functions of multiple operators, like $\langle \mathcal{O}_i(x)\mathcal{O}_j(0)\rangle$. This would give us access to a (positive) spectral density matrix, and the linear program would become a semi-definite program. We could also try to remove the need for a discretization of $\rho$, using the (convex) dual formulation of the linear problem (an approach followed also by Lawrence~\cite{lawrence2024spectral}). 

Physically, the K\"all\'en-Lehmann representation also relates the spectral density to the real-time correlation functions. The latter are just linear forms on $\rho$ and can thus be bounded with the same linear program we have used so far. Preliminary calculations show excellent convergence for short time, and as expected, noisier estimates for larger times. It would be interested to see in which cases this approach is more or less efficient than dedicated time-evolution methods for tensor networks.

\chapter{Difficulties and open problems}

The variational method we have presented is still in its early stages. As a result, it suffers from limitations, and naturally points to open problems. The objective of this chapter is to discuss them as honestly as possible, and sketch a tentative program of improvements.

\section{Beyond toy models}
The most obvious limitation of the RCMPS method so far is that it can be used only for a small range of models (that could arguably be called \emph{toy models}). Indeed, at this stage, the method has been applied only to bosonic models in $1+1$ dimension (with only the recent addition of multiple species of bosons). Certainly, if this remains so, the method will be but a curiosity, and the philosophy that underpins it will never graduate from the \emph{pile of dirt}. It is thus the first priority to increase its range of applicability. Let us first discuss the most difficult extension, beyond $1+1$ dimensions, before gradually coming back to easier extensions that could reasonably expected in the near future.

\subsection{Higher dimensions}

\subsubsection*{Genesis and definition of continuous tensor networks states}
Just after CMPS were introduced by Cirac and Verstraete in 2010~\cite{verstraete2010}, people started wondering if and how the construction could be extended to higher dimensions. However, while CMPS were a straightforward continuum limit of MPS, no such straightforward continuum limit seemed to be obtainable from PEPS (Projected entangled pair states). First attempts were made by Jennings and collaborators~\cite{jennings2015ctns}, but I think it is fair to say that it remained unclear how to obtain a Euclidean invariant ansatz in the continuum from the PEPS construction.

\begin{figure}[!h]
\floatbox[{\capbeside\thisfloatsetup{capbesideposition={right,top},capbesidewidth=0.55\textwidth}}]{figure}[0.42\textwidth]
{\caption{\small Coarse-graining a MPS (on the far left), the number of physical degrees of freedom (blue legs) increases while the bond dimension (orange legs) stays fixed. Coarse-graining a PEPS (on the left), both dimensions increase, meaning both need to become fields in the continuum limt. An intuition for the resulting state is shown below. Images from~\cite{tilloy2019ctns}\label{fig:scaling} \\~~ \\ ~~~~~~\includegraphics[width=0.5\textwidth]{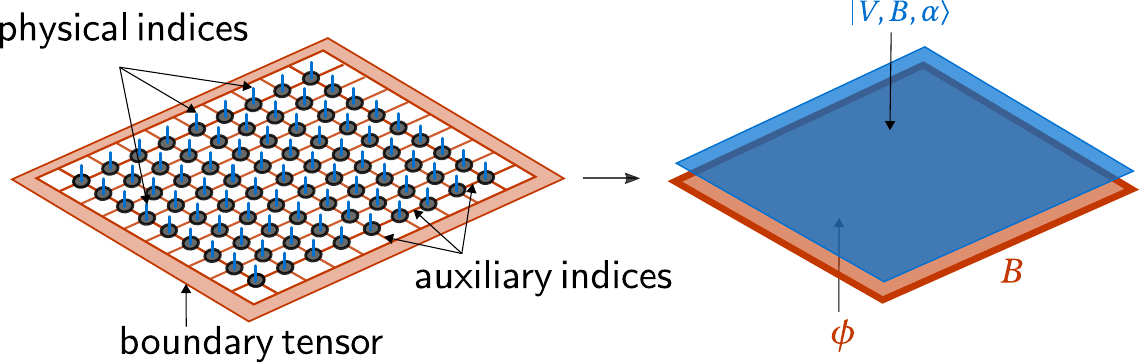}}}
{\includegraphics[width=0.4\textwidth]{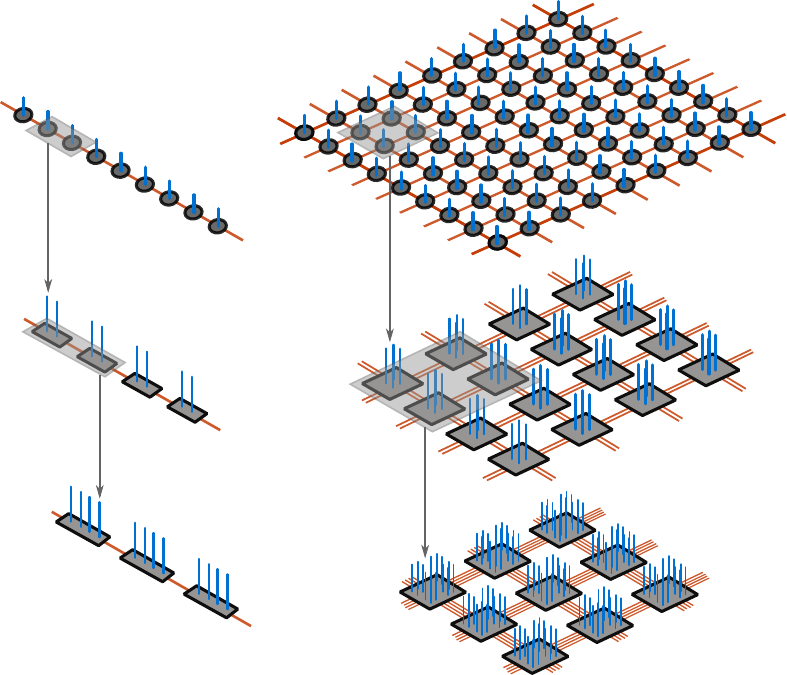}}
\end{figure}

I think we understood with Cirac that the only way forward was to accept drastic changes. Most importantly, we believed (and still do) that it was absolutely necessary to abandon a parameterization of the ansatz in terms of finite dimensional tensors. The reason for it is the coarse/fine graining argument depicted in Fig. \ref{fig:scaling}. If we have a continuum version of PEPS, that we assume to be stable by fine-graining, then necessarily the bond degrees of freedom need to be infinite dimensional (fields instead of discrete indices).

Accepting this move from indices to fields, and taking one natural continuum limit, we proposed the following definition in~\cite{tilloy2019ctns}. In $\db$ space dimensions, for a (non-relativistic) QFT living in $\Omega\subset \mathbb{R}^\db$, a continuous tensor network state\footnote{It might have been a better choice to call this ansatz a continuous PEPS or CPEPS. But we had the feeling our construction was more general and could in principle be used to build a continuous MERA. However, we never made this latter idea concrete.} (CTNS) is a state parameterized by two functions $V$ and $\alpha$ (and a boundary functional $B$) and given by
\begin{equation}\label{eq:ctns_def}
\ket{V,B,\alpha}=\int \mathcal{D} \phi\,  B(\phi_{\partial\Omega})\exp\left\{ - \int_{\Omega \subset \mathbb{R}^\db}\!\!\!\!\!\!\! \upd^\db x \, \frac{ [\nabla\cdot\phi(x)]^2}{2} + V[\phi(x)] - \alpha[\phi(x)]\, \hpsi(x) \right\} \ket{0}_\psi\, .
\end{equation}
The field $\phi$ here is an auxiliary scalar field (or, more generally, a vector of scalar fields), and the functional integral means that it is given by a Euclidean QFT in $\db$ space-time dimensions. Importantly, this is one dimension \emph{less} than the \emph{physical} non-relativistic theory in $\db+1$ space-time that we aim to solve variationally. This is natural, and similar to what happens with tensor networks on the lattice: \emph{we trade a space dimension for a variational optimization}. 

Even if bond degrees of freedom $\phi$ are now fields (in one dimension less), their action can be typically be specified with a finite number of parameters, \eg by expanding the functions $V$ (analog of $Q$) and $\alpha$ (analog of $R$) as polynomials (or simple exponentials) in $\phi$. Of course, if the resulting functional integral is not Gaussian, contracting the state is \emph{a priori} still difficult. While the problem is reduced from $2+1$ dimensions to $2$ (or equivalently $1+1$), the latter is still a many body problem. Understanding how one can approximately contract the state by variationally solving the boundary theory is crucial both for practical calculations, and because it is ultimately the best justification that we have defined the ``correct'' CTNS ansatz. Before explaining how this boundary picture is obtained, let me add 3 general remarks that also suggest the CTNS ansatz is right.

\begin{remark}[Insights from the Gaussian case]\label{rk:gctns}
When $V$ is at most quadratic and $\alpha$ at most linear in the field, the CTNS expectation values can be computed exactly, as the corresponding functional integral is Gaussian. We considered this restriction (Gaussian CTNS, or GCTNS) in~\cite{karanikolaou2021gctns} with Teresa Karanikolaou and Patrick Emonts, and applied it to simple quadratic bosonic models in $2+1$ dimensions. Of course, such models are exactly solvable, but we pretented we did not know, and tried to find their ground state variationally, by gradient descent. Importantly, in $2+1$ dimensions, non-relativistic energy densities are infinitely negative, which should be a problem for variational optimization. However, we observed that GCTNS diverge in exactly the same way as the true ground state, which allows to carry the variational minimization on the renormalized finite part of the energy density. This is a non-trivial test that CTNS have the right short distance behavior! At least for super-renormalizable auxiliary fields, the short distance behavior is the same for GCTNS and ``interacting'' CTNS, and thus this result extends beyond the Gaussian case.
\end{remark}

\begin{remark}[Intriguing link between complex $\phi^4_2$ and interacting bosons] \label{rk:ctns_knowles}
Another hint that CTNS capture the physics of non-relativistic bosons comes from a recent mathematical physics result by Fr\"ohlich, Knowles, Schlein, and Sohinger~\cite{frohlich2024euclidean}. In this paper, the authors demonstrate that for an interacting Bose gas in $2+1$ dimensions, in the limit where the density of the gas is large and the range of the interaction is small, (thermal) particle densities of the gas are given by the correlation functions of Euclidean $\phi^4$ model in $1+1$ dimensions. Colloquially,
\begin{equation}
\tr\left[ \psi^\dagger(x_1) \cdots  \psi^\dagger(x_n)  \psi(x_1) \cdots \psi(x_n) \e^{-\beta H}\right] \propto \int \, \mathcal{D} \phi \; \phi^*(x_1) \cdots \phi^*(x_n) \phi(x_1) \phi(x_n) \e^{-S(\phi)}\, ,
\end{equation}
where $H$ is the non-relativistic Hamiltonian of interacting bosons in $2$ space dimensions and $S$ is the action of Euclidean $\phi^4_2$. The parameters of $H$ can be matched to the parameters in $S$. This form is very reminiscent of what one would obtain assuming that the expectation value is taken over a CTNS, except that the right-hand side would be taken as an expectation value over a pair of fields. It suggests one should be able to describe the interacting bose gas, at least in some limit, \emph{exactly} as a CTNS with only few fields and a very simple potential $V$. 
\end{remark}

\begin{remark}[Link with Hall effect ansatz wavefunctions]\label{rk:ctns_hall}
If we replace the auxiliary scalar fields appearing in the CTNS path integral by \emph{chiral} fields --for which path integrals are not really well defined but the final result may still exist--, we can obtain \emph{exactly} the many ansatz wavefunctions that have been used to describe fractional quantum Hall states. In fact, such a CTNS was foreseen already in 2012 by Dubail, Read, and Rezayi in~\cite{dubail2012} in the particular case of an auxiliary chiral fermion. This is yet another hint that CTNS have the right structure, since they (or a minor extension) could describe (chiral) topological phases.
\end{remark}

\subsubsection*{Operator representation}
While the functional integral definition \eqref{eq:ctns_def} of CTNS is convenient to immediately see Euclidean invariance, an operator picture is more adapted to understand the contraction strategy. It also helps make it clearer that it is the natural extension of CMPS.

\begin{figure}[!h]
\floatbox[{\capbeside\thisfloatsetup{capbesideposition={right,top},capbesidewidth=0.45\textwidth}}]{figure}[0.52\textwidth]
{\caption{\small Illustration of the dimensional reduction allowed by discrete and continuous tensor networks in  a $\mathbf{d}=2$ cylinder. One of the space direction of the physical theory becomes an imaginary time direction of the auxiliary (bond) relativistic field theory. In correlation functions, one gets two copies of the auxiliary theory, coupled by the connection of physical indices.  Image from~\cite{tilloy2019ctns}}\label{fig:boundary}}
{\includegraphics[width=0.5\textwidth]{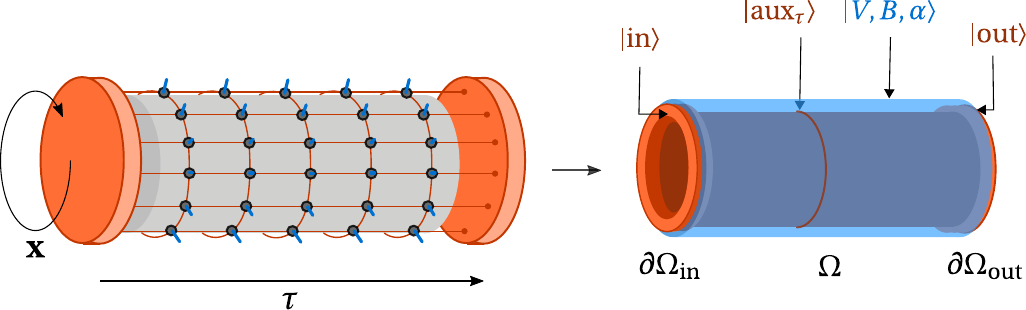}}
\end{figure}

This operator representation can be obtained for a domain $\Omega$ that can be written as a cartesian product $\Omega=[-T/2,T/2] \times S$. In that case, we write $x=(\tau,\xb)$ where $\tau \in [-T/2,T/2]$ and $\xb \in S$. Following the standard textbook recipe to go from functional integral to operator representation, we translate \eqref{eq:ctns_def} into: 
\begin{equation}\label{eq:operator}
\begin{split}
  \ket{V,B,\alpha}=\tr \bigg[\hat{B} \mathcal{T} \exp\bigg(\!\!-\!\!\!\! \int\limits_{-T/2}^{T/2} \!\!\!\upd \tau & \!\int_S \!\upd\xb  \;\sum_{k=1}^D \frac{\left[\hat{\pi}_k(\xb)\right]^2 }{2} \\
&+  \frac{[\nabla \hat{\phi}_k(\xb)]^2}{2}  + V[\hat{\phi}(\xb)] - \alpha[\hat{\phi}(\xb)]\,\psi^\dagger(\tau,\xb) \bigg)\bigg]\ket{0}
\end{split}
\end{equation}
where $\mathcal{T}$ is the $\tau$-ordering operator, $\hat{\phi}_k(\xb)$ and $\hat{\pi}_k(\xb)$ are $k$ canonically conjugated pairs of (auxiliary) field \emph{operators}: $[\hat{\phi}_k(\xb),\hat{\phi}_l(\yb)]=0$, $[\hat{\pi}(\xb)_k,\hat{\pi}_l(\yb)]=0$, and $[\hat{\phi}_k(\xb),\hat{\pi}_l(\yb)]=i \delta_{k,l}\,\delta^{d-1}(\xb-\yb)$ (the first $\delta$ is Kronecker, the second is Dirac).  These operators act on $\mathscr{H}_\text{aux} =\mathcal{F}[L^2(S)]^D $, \ie $D$ copies of a bosonic Fock space on a $d-1$ dimensional space. The trace is taken over this auxiliary Hilbert space. This operator representation is illustrated in Fig. \ref{fig:boundary} where it is compared with its equivalent in the discrete.

\subsubsection*{Boundary contraction with (R)CMPS}
To compute correlation function with CTNS, the most natural way is to follow exactly the same strategy as with CMPS and introduce the generating functional $\mathcal{Z}_{j',j}$:
\begin{equation}
  \mathcal{Z}_{j',j}=\frac{\bra{V,B,\alpha} \exp\left(\int_\Omega j'\cdot \psi^\dagger \right) \exp\left(\int_\Omega j\cdot \psi  \right) \ket{V,B,\alpha}}{\langle V,B,\alpha|V,B,\alpha\rangle}
\end{equation}
Exploiting the operator definition of the CTNS \eqref{eq:operator}, expanding the $\tau$-ordered exponential into an infinite product of infinitesimal exponentials, we get:
\begin{equation}\label{eq:operator-ordered}
    {\mathcal{Z}}_{j',j} = \tr\left[B\otimes B^* \, \mathcal{T}\exp\left(\int_{-T/2}^{T/2} \mathds{T}_{j'j}\right)\right]
\end{equation}
with the \emph{transfer matrix} (with sources):
\begin{equation}
  \mathds{T}_{j'j} = \int_S -\mathcal{H}\otimes \mathds{1} -\mathds{1}\otimes \mathcal{H}^* + \alpha[\hat{\phi}]  \otimes \alpha[\hat{\phi}]^*  + j\, \alpha[\hat{\phi}]\otimes \mathds{1} + j'\, \mathds{1} \otimes \alpha[\hat{\phi}]^*
\end{equation}

The functional derivatives can then be carried explicitly and one obtains \eg , for $-T/2<\tau_2 <\tau_1<T/2$:
\begin{equation}
\begin{split}
  \langle \psi^\dagger(x_1) \psi(x_2)\rangle_{V, \alpha} = \tr &\big\{B\otimes B^* \cdot \mathcal{M}_{T/2,\tau_1} \cdot  [\mathds{1} \otimes \hat{\alpha}^*(x_1)] 
    \!\! \cdot \mathcal{M}_{\tau_1,\tau_2} \cdot [\hat{\alpha}(x_2)\otimes \mathds{1}] \cdot \mathcal{M}_{\tau_2,-T/2}  \big\}
\end{split}
\end{equation}
with  the map $\mathcal{M}_{u,v}= \mathcal{T}\exp[\int_v^u \mathds{T}]$, the \emph{transfer operator} $\mathds{T}:=\mathds{T}_{00}$ and the compact notation $\alpha[\hat{\phi}(x)] =  \alpha(x)$. More generally, as in the CMPS case, correlation functions are given by the trace of a succession of propagators $\mathcal{M}$ followed by operator insertions of $\alpha \otimes \mathds{1}$ (respectively $\mathds{1}\otimes \alpha^*$) in the positions corresponding to $\psi$ (respectively $\psi^\dagger$).

When we take the thermodynamic limit, we see that $\mathcal{M}_{-\infty, \tau_1}$ projects onto the left eigenstate $\bra{\ell_0}$ with largest real eigenvalue of $\mathbb{T}$, and likewise $\mathcal{M}_{\tau_2,+\infty}$ projects onto the right eigenstate $\ket{r_0}$ with largest real eigenvalue of $\mathbb{T}$. Hence, as long as we are interested in computing operators in the original model that depend locally on the fields $\psi,\psi^\dagger$ (as does the Hamiltonian density for example), their expectation value can be written $\bra{\ell_0} f(\phi,\phi^*) \ket{r_0}$ in the auxiliary (lower dimensional) space. 

If the original model is in $2+1$ dimensions, $\ket{r_0}$ is the state of a \emph{relativistic} QFT in $1$ dimension less, hence $1+1$. To compute it, and we can simply try to find the stationary state of $\mathbb{T}$ in RCMPS form. If $\mathbb{T}$ is self-adjoint (which is not generic but helps build intuition), then we can even find this RCMPS variationally as the ground state of $\mathbb{T}$. Otherwise we may simply use an evolution algorithm like TDVP to find the stationary state of a non-Hermitian $\mathbb{T}$. All the progress we made with RCMPS thus allows, in principle, the contraction of arbitrary CTNS in $2$ dimensions!

\begin{remark}[The right ansatz]
This boundary contraction method with (R)CMPS is, I believe, the ultimate test that we indeed constructed the continuum counterpart to PEPS. Indeed, a PEPS is a 2 dimensional tensor network that can be contracted using boundary MPS. Likewise, a CTNS is an ansatz for theories in 2 space dimensions that can be contracted with a (R)CMPS. What seemed like a bold choice then --dropping a finite bond dimension-- seems fairly obvious from the boundary perspective.
\end{remark}

\begin{remark}[Need for RCMPS]
In the 2019 paper with Ignacio Cirac~\cite{tilloy2019ctns}, which appeared before the definition of RCMPS, we claimed that one could contract the boundary with standard CMPS. This was the right intuition, but not fully accurate if taken literally. Since $\mathbb{T}$ is the time translation generator of a Euclidean invariant theory, we expect its stationary state to have the same UV-divergent entanglement entropy that only RCMPS can capture. I did not understand this subtlety then. 
\end{remark}

\begin{remark}[Need for at least 2 species]
  At least if we stick with a standard path integral and do not construct a CTNS from chiral fields (as in remark \ref{rk:ctns_hall}), $\mathbb{T}$ acts on a Hilbert space containing $2 \times D$ fields, \ie at least $2$ fields. Consequently, to contract even the simplest non-Gaussian CTNS, one needs RCMPS for $2$ species of relativistic bosons. 
\end{remark}
These remarks make it clear, I hope, that the recent refinements of CMPS into RCMPS, and the extension to multiple species were non-trivial prerequisites to contract CTNS. Finally, I think we have everything needed in principle to start contracting and optimizing simple non-Gaussian CTNS.

\subsubsection*{Relativistic QFT in \texorpdfstring{2+1}{$2+1$} dimensions}
Continuous tensor networks can be extended from $1$ to $2$ space dimensions, but the move from non-relativistic to relativistic models currently seems more difficult.

The first observation is that the simplest model in $2+1$ dimensions, $\phi^4_3$, is \emph{not} made finite by simply suppressing tadpole graphs (or equivalently, normal ordering the Hamiltonian). Indeed, there is a subleading divergence, the sunset diagram, visible at second order in perturbation theory:
\begin{equation}
\begin{aligned}
\begin{tikzpicture}[baseline=-0.5ex, scale=0.6, line width=1pt]
  \draw (-2,0)--(-1,0);
  \draw (1,0)--(2,0);
  
  \filldraw (-1,0) circle(0.08);
  \filldraw (1,0) circle(0.08);

  \draw (0,0) circle(1);
  \draw (-1,0)--(1,0);
\end{tikzpicture}
\quad &=\quad
\frac{\lambda^2}{6}\int^\Lambda \frac{d^3k}{(2\pi)^3}\frac{d^3q}{(2\pi)^3}\, \frac{1}{k^2 + m^2}\frac{1}{q^2 + m^2}\frac{1}{(p - k - q)^2 + m^2}\\ 
      &=\quad \frac{\lambda^2}{384\pi^2}\log\left(\frac{\Lambda}{m}\right) + \text{finite terms} \; .
\end{aligned}
\end{equation}
This is not a problem to define the model rigorously. This is the only new divergence, and the model remains super-renormalizable. Intuitively, one simply needs to add a subleading counterterm to the mass that precisely cancels this UV divergence. Mathematical physicists have shown\footnote{The first complete proof of this result by Feldman and Osterwalder~\cite{feldman1976phi43} (building on earlier work by Glimm and Jaffe~\cite{glimm1971phi43_lowerbound}) is one of the main achievements of ``modern'' constructive field theory based on the Osterwalder-Schrader axioms (i.e. Wick rotation and discretization).}, now with many different methods~\cite{hairer2014phi43,jagannath2023phi43}, that this recipe indeed leads to a well defined model verifying all the Wightman axioms.

However, Glimm and Jaffe also showed~\cite{glimm1971phi43_lowerbound} that one cannot define the (renormalized) Hamiltonian of the $\phi^4_3$ model as an operator acting in the original free Fock space, which makes a direct translation of the RCMPS approach to the $2+1$ dimensional case impossible to follow. Finding an alternative route to build a variational ansatz seems like a very difficult problem.

\subsection{Beyond the renormalizability gap}
One can separate the problem of extra UV divergences from the other problems of $2+1$ dimensions. To this end, one should look at $1+1$ dimensional relativistic QFT that have an interaction term given by an operator $\mathcal{O}$ with scaling dimension $d\geq \Delta_\mathcal{O}\geq d/2$ with $d=2$ (\ie super-renormalizable, but not \emph{strongly} renormalizable).

Such models already have divergences that are not cured by normal-ordering, and thus display a similar UV problem as that of their $2+1$ dimensional counterpart. This problem has been well identified in the Hamiltonian truncation, with proposed solutions. In the context, the authors typically consider perturbations of generic interacting CFT that naturally have weakly relevant perturbations.

We could even avoid this complication of going to generic CFT in the UV first, and stick to perturbations of the free boson. The easiest example is perhaps the Sine-Gordon model (which we discussed before) beyond the free Fermion point, \ie for $\beta > \sqrt{4\pi}$. In this regime, the vertex operators $:\!\e^{\pm i\beta \phi}\!:$ are no longer strongly relevant, and make the vacuum energy infinitely negative. This example remains very simple because only the vacuum energy is renormalized. 

By construction RCMPS can only give finite expectation values for normal ordered operators, and there is thus no way to match this divergent behavior for fixed $Q,R$. What happens in that case is precisely what Feynman feared would happen in general: the energy minimization algorithm gives a sequence of states $\ket{Q_n,R_n}$ with lower and lower energy, but observables $\langle Q_n R_n | \mathcal{O} |Q_n,R_n \rangle$ do not improve. This is because these states $\ket{Q_n,R_n}$ remain infinitely far from the ground state. Without modifying the ansatz, the only way forward is to add a regulator, penalizing UV overfitting. The interest of the approach would be diminished, but it might still compare favorably with other methods requiring more violent regulators.

A strategy without regulator could be to construct an ansatz that is formally defined in Fock space like RCMPS, with parameters than can be tuned to make it divergent in just the right way, while keeping a subset of parameters adjustable. It is not completely implausible that this can be done, because this is exactly what happend with Gaussian CTNS in $2+1$ dimensions (see remark \ref{rk:gctns}). But I have no idea how to even begin the construction of such an ansatz.

\subsection{Perturbation of interacting CFT}
Even if we stick with strongly relevant perturbations, we could consider starting from of generic QFT, \ie models with a Hamiltonian of the form
\begin{equation}
  H = H_\mathrm{CFT} + \lambda \int \upd x \mathcal{O}(x)
\end{equation}
with $\Delta_\mathcal{O} \leq 1$.

To generalize RCMPS in this context, I think there are two questions to answer:
\begin{enumerate}
  \item What state (capturing the UV) should one start from? In the plain RCMPS case it was the \emph{massive} free vacuum $\ket{0}_m$.
  \item What tensor product structure should one use to build the (C)MPS on top? In the plain RCMPS case it was the factorization induced by $a(x),a^\dagger(x)$, unusual operators obtained from the Fourier transform of the normal modes $a_k,a_k^\dagger$.
\end{enumerate}

I think a very good candidate for the first are the regularized CFT boundary states which have already been used variationally (but without CMPS on top) by Cardy~\cite{cardy2017boundary} and further expanded recently by Konechny~\cite{konechny2023boundary}. In the functional integral representation, one can naturally associate a quantum state to a boundary. The idea is to take a boundary that preserves conformal invariance (such boundary states are classified), and then evolve the corresponding state $\ket{B}$ with the CFT Hamiltonian $H_\mathrm{CFT}$ for a time (or distance) $\ell$:
\begin{equation}
  \ket{\psi_\ell} := \e^{-\ell H_\mathrm{CFT}}\ket{B}
\end{equation}
In the path integral representation, this corresponds to defining the state at a distance $\ell$ from the boundary. This state $\ket{\psi_\ell}$ has the right properties. Because it is ``cooled down'' by the CFT Hamiltonian, its UV behavior is indistinguishable from the CFT vacuum. This is clear from the path integral representation: if I zoom in on a point at a fixed distance from the boundary, and the model is scale invariant, then the boundary has no effect. On the other hand, for distances larger than $\ell$, this state has a scale and is thus well behaved in the IR. At least superficially, it does behave like the vacuum of a free massive boson with $m\simeq \ell^{-1}$ which we had used for plain RCMPS.

It is less clear at this stage which tensor factorization to use, or equivalently which (local-ish) operators to choose to replace $a(x)$ and build a tensor network on top of $\ket{\psi_\ell}$. Recently, Vardian has proposed using operators constructed from linear combinations of the stress energy tensor~\cite{vardian2024}. This choice reproduces some features of $a(x)$, in that $\ket{\psi_\ell}$ is indeed annihilated. However, the resulting CMPS is (at least superficially) divergent, and the operators are not associated to a tensor product structure.

It is possible that there will not be exact analogs of $a(x)$ and an associated continuous tensor product structure. However, we expect at the very least that we should be able to cut space into discrete (possibly overlapping) chunks, to which one could associate a Hilbert space by adding excitations on top of $\ket{\psi_\ell}$. From Hamiltonian truncation intuition, if the chunks are very small, the effective Hilbert space should be small. Then, one could build a MPS in the factorization induced by this cut. It would be a discrete \emph{variational} ansatz, for a fully continuous theory. A promising approach to get a split of space into orthogonal chunks would be to use a basis of wavelets, an approach we are currently exploring with Molly Kaplan in the much easier context of non-relativistic field theories.

\subsection{Fermionic models}
Fermionic models are a particulary interesting target because they are typically ill-suited for the Monte-Carlo method, are subtle to discretize (\eg with the Fermion doubling problem) and yet the standard CMPS formalism works \emph{a priori} without drastic modifications. Let us explain briefly the strategy, and the remaining challenges.

\subsubsection*{Standard definition}
Following Haegeman and collaborators~\cite{haegeman2013calculus}, one can define a (standard) CMPS for fermions in just the same way as for bosons:
\begin{equation}
  \ket{Q,R} = \tr\left\{\mathcal{P}\exp\left[\int\upd x Q + R_\alpha\psi^\dagger_\alpha(x)\right]\right\}\ket{0}_\psi
\end{equation}
where $\{\psi_\alpha(x),\psi^\dagger_\beta(y)\}=\delta_{\alpha,\beta}\delta(x-y)$ and $\{\psi_\alpha(x),\psi_\beta(y)\}=0$. 

Correlation functions can be computed in the same way as in the bosonic case, although with slight complications to keep track of the signs coming from the anti-commutation relations~\cite{haegeman2013calculus}. The regularity conditions for typical fermionic kinetic terms impose that $\{R_\alpha,R_\beta\} = 0$, and in particular $R_\alpha$ has to be nilpotent, even for a single fermion.

\subsubsection*{Relativistic extension}
For fermionic RCMPS, we should just follow the strategy that worked for bosons, and define the state by acting on the vacuum $\ket{0}$ of a free massive fermion with the Fourier transform $b(x)$ of the normal-mode creation operator $b_k$. For a Dirac fermion, the latter are related to the field via:
\begin{equation}\label{eq:fermionicmodes}
  \psi_{\alpha}(x) = \frac{1}{2\pi}\int \frac{\upd k} {\sqrt{2\omega_k}} \left[b_{1,k} u_{\alpha}(k) e^{-ik x} + b^{\dagger}_{2,k} v_{\alpha}(p) e^{ik x}\right],
\end{equation}
with
\begin{equation}
  u(k) = \begin{pmatrix} \sqrt{\omega_k-k} \\ 
                \sqrt{\omega_k+k} \end{pmatrix}
\end{equation}
and
\begin{equation}
v(k) = \begin{pmatrix} \sqrt{\omega_k-k} \\ 
                -\sqrt{\omega_k+k} \end{pmatrix}\, .
\end{equation}
With this notation, the massive free Dirac Hamiltonian is
\begin{equation}
:H_D: = :\int \overline{\psi}\left[-i\gamma_{1} \partial_{1} + m\right] \psi(x) \upd x: =
\int \frac{d k}{2\pi} \omega_k  \left(b_{1,k}^{\dagger}b_{1,k} + b^{\dagger}_{2,k}b_{2,k}\right)
\end{equation}
It is then natural to define the fermionic RCMPS as:
\begin{equation}\label{eq:fermionicRCMPS}
  \ket{Q,R} = \tr\left\{\mathcal{P}\exp\left[\int\upd x Q + R_\alpha b^\dagger_\alpha(x)\right]\right\}\ket{0}  \, ,
\end{equation}
with $b_\alpha(x)$ the inverse Fourier transform of $b_{\alpha,k}$.
Following a similar strategy as for bosonic RCMPS,it is possible to compute local expectation values. Crucially, one can verify that as long as the regularity condition $\{R_\alpha,R_\beta \}$ holds, the Dirac Hamiltonian density is finite. This makes us confident the starting point is sound.

\subsubsection*{Models and strong renormalizability}
So far, the story has remained fairly close to the bosonic case. It may thus come as a surprise that no Fermionic model has been solved with the method yet. This is because we have not yet discussed the main way in which fermions are different. 

In the bosonic case, the most natural models one can write are all strongly renormalizable. This is because the field itself, $\phi$, has a scaling dimension $0$, and thus all normal ordered powers $:\!\phi^n\!:$ are also dimension zero and thus well defined. Vertex operators $:\!\e^{i\beta \phi}\!:$ do have a non-zero scaling dimension $\Delta_\beta = \beta^2 / 4\pi$. As a result, the only not-strongly-renormalizable model we encountered was the Sine-Gordon model for $\beta \geq \sqrt{4\pi}$. 

However, in the fermionic case, most models one can write are not strongly renormalizable, simply because the fermionic field $\psi$ has scaling dimension $\Delta_\psi = 1/2$. Consequently the mass term $m \bar{\psi}\psi$ is already right on the threshold of "strong" renormalizability, while the most natural interacting term, $(\bar{\psi}\gamma^\mu\psi)(\bar{\psi}\gamma_\mu\psi)$ is even marginal! Consequently, the most natural non-trivial model, the Thirring model, cannot be dealt with variationally, at least in the way we have discussed so far.

Because the mass term is already on the threshold of strong renormalizability, one may suspect that no interacting fermionic theory can be considered at all. This, however, would be excessively pessimistic. In $1+1$ dimensions, there are still $2$ fermionic theories one can think of, that have a strongly relevant interaction. The trick is that in both cases, the interaction term is not local as a function of $\psi$

The first example is the Ising Field Theory. One way to understand it is as the most general relevant deformation of the Ising CFT. In the fermionic picture, it can be written as a massive Majorana fermion perturbed with a magnetic term. Writing it in terms of a single ``standard'' fermion $\psi$, one starts from the free Hamiltonian
\[H_0=\int \upd x\;\Big[ -\,i\,\psi^\dagger\partial_x\psi \;+\; m\,:\!\psi^\dagger\psi\!: \Big],
\]
(which is already a massive perturbation of the Ising CFT) and perturbs it additionally with a magnetic term:
\[
H \;=\; H_0 \;-\; h \int \upd x\,\sigma(x) \, .
\]
In this equation $\sigma$ is the spin, which, in the discrete, is related non-locally to the fermionic degrees of freedom by a Jordan-Wigner transformation. Its scaling dimension is $\Delta_\sigma = 1/8$ which is strongly relevant, and thus in principle adapted to our variational method. It is natural to work in the fermionic picture, as it describes well the UV, and deal with the $\sigma$ term variationally. In principle, its non-local nature is no problem. But in practice, the missing piece right now is an ability to compute $\langle Q,R|\sigma(x)|Q,R\rangle$ for massive fermions. The different sources of non-locality (of $\sigma(x)$ and $b(x)$ in terms of $\psi(x)$) do not play nicely together, and I could not yet find an explicit ODE for $\rho$ giving $\tr[\rho]=\langle Q,R |\sigma | Q,R \rangle$ (which was the strategy so far). I suspect this is just a technicality, and that the Ising Field Theory can ultimately be solved with RCMPS, but this will require more work.

The second example is the Schwinger model, which we are currently studying in Paris. There are various subtleties, but no obvious roadblock. The Schwinger model is a low dimensional toy model of QED. In $1+1$ dimensions, the electromagnetic field can be integrated out. In Coulomb gauge, what remains is an interaction between fermions that grows linearly with distance. The resulting Hamiltonian is
\[
H \;= :\!H_D\!: + \; \frac{e^2}{2}\int \upd x\,\upd y \;\rho(x)\,\big[(-\partial_x^2)^{-1}\big]_{x,y}\,\rho(y),
\]
where $\rho = \psi^\dagger\psi$ is the density. This interaction term has scaling dimension $0$. This is consistent with bosonization, where the interaction becomes $\propto \phi^2$. Why not study the bosonized model directly then? Because the resulting massive Sine-Gordon model has an interaction term exactly at the threshold of strong renormalizability (and the threshold point is already divergent, if only logarithmically). In the fermionic picture however, this problematic interaction is just a mass term, that can be dealt with via a Bogoliubov transform. We can thus integrate this in our starting free vacuum. 

The RCMPS calculations for the Schwinger model remain non-trivial. There are new ODEs to derive for expectation values, and the optimization manifold is even more subtle than for multiple species of scalar fields because of extra regularity conditions. But I hope that we can have some results in the coming year.

To summarize, in contrast with the bosonic setup with infinitely many models, there are only $3$ models with fermions, which all have difficulties:
\begin{center}
\begin{tabular}{|l|c|c|c|}
\hline
 & Computable & Strongly renormalizable & Local (in Fermions) \\
\hline
Thirring model     & \ding{51} & \ding{55} & \ding{51} \\ \hline
Ising Field Theory & \ding{55} & \ding{51} & \ding{55} \\ \hline
Schwinger model    & \ding{51} & \ding{51} & \ding{55} \\
\hline
\end{tabular}
\end{center}
Right now, it seems only the Schwinger model can be tackled with RCMPS without extra ingredients or tricks. But solving \emph{any} fermionic model would be a major advance.

\section{Getting more out of RCMPS in theory}

On the lattice, tensor network states are far from \emph{just} being a numerical method. While it all started this way, with White's DMRG, tensor networks have grown as a powerful theoretical tool. For example, they have provided a very explicit understanding of plain topological and symmetry protected topological phases of matter~\cite{schuch2011classifying}. 

Intuitively, this is because TNS have allowed a change of paradigm. Since TNS are ground states of local Hamiltonians and all local Hamiltonian ground states are TNS (approximately, and with caveats), one can understand theory space by studying TNS directly. This is easier than considering a space of Hamiltonians first, and then trying to characterize their ground state physics.

This type of formal progress has not had an equivalent in the continuum yet. I think it could be useful to mention what the open problems are.

\subsection{Understanding convergence}

In my opinion, if we want to mimic the  historical development of the understanding on the lattice, an important step is to understand how well (R)CMPS converge to ground states.

In $1+1$ dimensions on the lattice, there are essentially two steps in this argument:
\begin{itemize}
  \item Showing that a state verifying an area law is well approximated by an MPS (the Cirac \& Verstraete argument~\cite{cirac2006mps})
  \item Showing that ground states of local gapped Hamiltonians do verify a similar area law (Hastings' proof\footnote{Technically, Hastings' proof directly demonstrates that ground states of local gapped Hamiltonians are well approximated by MPS. This is because Hastings only proved the edge case of area law for Von Neumann entropy, which on its own is not sufficient to use the Cirac-Verstraete argument requiring $\alpha$-Renyi entropy with $\alpha <1$.}~\cite{hastings2007}).
\end{itemize}

In the (relativistic) continuum, we have seen that standard entanglement entropy verifies an area law but with an infinite prefactor. However, we also saw that by considering an exotic entanglement entropy (see remark \ref{rk:particleentanglement}), adapted to the RCMPS tensor factorization, we could get a finite result. This was just numerics but a powerful \emph{a posteriori} justification of the method. A first step would be to prove that this is indeed the case. More precisely, assuming a given state verifies the area law (for some $\alpha$-Renyi entropy), is it possible to show that it can be well approximated by RCMPS? This seems fairly natural, and a direct extension of the Cirac-Verstraete argument.

The second step would be to show that this exotic entanglement entropy is indeed a relevant quantity for the low energy states of gapped relativistic Hamiltonians. It seems more difficult in the continuum but relativistic QFT could paradoxically make things easier. The most natural way in which it could help is that space and time behave in the same way. The gap is thus immediately related to the decay of spatial correlations, something which is much less trivial on the lattice. In fact, this was precisely Hastings' intuition for (gap + locality) implying short range spatial correlations!

Even if this connection cannot be made as precise and sharp as on the lattice, it would be nice to see if it can be related to the precision of the method. In particular, the dramatic reduction in accuracy observed for the Sinh-Gordon model could plausibly be associated to a transition in entanglement entropy, dominated by the UV (for example, with the emergence of a divergent part for large enough coupling).

\begin{remark}[Proving the optimizer converges is too difficult]
  In these formal developments, the question of expressiveness of the ansatz is orthogonal the problem of its optimization. In general, there is no hope to prove that minimization over RCMPS can be done in polynomial time (as a function of $D$). The best we can probably have is an algorithm running in polynomial time  \emph{per iteration} (usually $D^3$), with a number of iterations that does not seem to grow exponentially with system size.
  The situation is analogous on the lattice. We know MPS can in principle approximate ground states well, we know DMRG runs in $D^3$ iterations, but it is not known that the number of DMRG iterations remains small as a function of $D$. It \emph{typically} works, but is not true in the worst case~\cite{schuch2008mpsoptimhardness}.
\end{remark}

\begin{remark}[A weaker result -- density]
  A much weaker result that seems reachable in the short term is to show that translation invariant RCMPS are dense in the appropriate Fock space. To my knowledge, this is expected beyond reasonable doubt, but not proved. What is known is that if we allow space dependent $Q(x), R(x)$, then CMPS are dense in Fock space because they can reproduce sums of coherent states. But even for a translation invariant target, the approximate CMPS are not translation invariant with this construction. It would be nice to prove directly that a translation invariant RCMPS can approximate arbitrarily \emph{well} all translation invariant Fock states. Of course this approximation will be inefficient in general, but this would be a first step towards showing efficient approximation of a subset of the Fock space.
\end{remark}

\subsection{Theory classification}

MPS approximate well the ground states of local gapped Hamiltonian, but reciprocally, for a given MPS, one can construct a local gapped Hamiltonian of which it is the ground state. The existence of such \emph{parent} Hamiltonian demonstrates that studying the space of MPS is indeed like studying the ground space of (a class of) Hamiltonians. As a consequence, one can understand phases of matter (like symmetry protected topological phases) at the state level directly.

In the continuum, there is no such strong result. The problem of finding parent Hamiltonians for CMPS has been attacked by many students and postdocs in the field. Very little has been published because everyone arrives more or less at the same conclusion: such a parent Hamiltonian would have to be fairly non-natural, and involve derivatives of arbitrary high orders. This is where the continuum starkly differs from the lattice: in the continuum, we naturally demand ``standard'' kinetic terms, to accept a parent Hamiltonian as reasonable.

In the relativistic context, Hamiltonians take an even more constrained form. A natural extension of the parent Hamiltonian problem is then: given a RCMPS, is there a \emph{relativistic} Hamiltonian of which it is the unique gapped ground state? We now have accumulated a series convincing (albeit not fully rigorous) arguments, that RCMPS \emph{never} have such parent Hamiltonians (apart from the trivial RCMPS with $Q,R=0$).

Using linear programming, we could show numerically that there exists no strictly positive spectral density $\rho_{\mu^2}$ associated to the two point function $\langle Q,R | \phi(x) \phi(0)\ket{Q,R}$ for specific $Q,R$ that we chose (which were approximations of the $\phi^4$ ground state). So at least there exists some $R$ and $Q$ for which the resulting RCMPS has no parent Hamiltonian. But the result seems more general. For example, it seems impossible to match the asymptotic decay of correlators predicted by the K\"all\'en-Lehmann representation theorem, with that of RCMPS, at least if we assume $\rho_{\mu^2}$ does not have support near $\mu=0$ which is the case for a gapped Hamiltonian. Making this argument sharper would definitely be worthwhile, even if it is ultimately to prove a no-go theorem.

\section{Getting more out of RCMPS in practice}

On the lattice, one can use MPS to find ground states of gapped Hamiltonians, and then evaluate local expectations on them, to essentially machine precision. But one can also do much more. For example, one can approximate thermal states, simulate real-time evolution, calculate the excitation spectrum, and scale entanglement to large values to probe gapless systems. This expanded toolbox, beyond simple ground state optimization, is not available in the continuum. I now discuss possible steps to improve the situation and related open problems.

\subsection{Improved optimization}

The first step to get closer to the situation on the lattice is to improve the variational optimization, which is currently very brittle. Indeed, the Riemannian algorithm (in its gradient descent of LBFGS form) currently has several hyper-parameters, which need to be subtly fine-tuned to get a decently fast and stable minimization. On top of this, each iteration can be as slow as a few minutes for large $D$. Finally, for some models, the random initial point $Q,R$ we start from also requires some form of fine tuning of its distribution. We are clearly far from the stability and reliability of an algorithm like VUMPS.

\subsubsection*{Improved descent with mixed-canonical form}

On the lattice, a lot of the extra robustness comes from having a mixed canonical form, which one can obtain only through well-conditioned operations (no explicit inversion of an ill conditioned matrix like $\rho_0$). Finding a way to work efficiently with such a mixed form in the continuum would ideally allow us to remove the metric regulator and thus converge to the ground state much faster.

Other improvements depend on such a mixed-canonical form. First, it is the starting point to construct bond expansion methods. Currently, to find the ground state approximation at a large $D$, the best (and in fact \emph{only}) strategy is to take random $Q,R$ of size $D\times D$, and run the minimization from there. One starts from a huge energy, and it takes many iterations even just to reach an energy one could have already achieve with much lower bond dimension. Only the very final steps refine the energy further. On the lattice, with bond-expansion techniques, one can start from a state at lower $D$, already giving a good approximation, and gradually increase its bond dimensions. The difficulty lies in having this bond expansion step keep the energy as it is, while providing a well conditioned state of higher bond dimension that can be optimized. For example, simply padding a lower dimension $Q,R$ with zeros would immediately break the algorithm (for example $\rho_0$ would not be invertible). All the bond expansion strategies on the lattice rely on an efficient mixed-canonical form.

\subsubsection*{Improved evaluation of energy and gradient}

In addition to reducing the number of iterations, one can try to make each iteration faster. The main bottleneck is the computation of the energy expectation value and its gradient, obtained by the resolution of a system of linear matrix ODEs. Recall that the archetypal such ODE (needed to compute vertex operators) is:
\begin{equation}\label{eq:basic_ODE}
  \frac{\upd}{\upd x} \rho(x) = \mathcal{L} (\rho(x)) + \alpha J(x) \left[R\rho(x)+\rho(x) R^\dagger\right]
\end{equation}
where $\mathcal{L} (\rho) = Q\rho + \rho Q^\dagger + R\rho R^\dagger$ is of the Lindblad form when $Q = -iK -\frac{1}{2}R^\dagger R$ with $K=K^\dagger$ (left-canonical gauge condition). The function $J$, introduced in \eqref{eq:jsource_def}, is the inverse Fourier transform of $1/\sqrt{2\omega_k}$:
\begin{equation}\label{eq:jsource_bis}
  J(x) = \frac{1}{2\pi} \int_\mathbb{R} \frac{\upd k}{\sqrt{2}\; (k^2 + m^2)^{1/4}} \e^{ikx} \, ,
\end{equation}
which has the asymptotic behavior $J(x) \underset{x\rightarrow 0}{\sim} \frac{1}{|x|^{1/2}}$ (an integrable singularity) and $J(x) \underset{x\rightarrow + \infty}{\sim} \e^{-m|x|}$. The objective is to compute $\rho(+\infty)$ starting from any initial condition of trace $1$. The other ODEs required to compute the energy density share exactly the same structure.

Currently, as we discussed in remark \ref{rk:non-autonomous_ODE}, the fastest method is simply to solve the ODE as if it were a completely generic non-linear non-autonomous ODE, with an explicit high-order Runge-Kutta scheme (\eg \texttt{DP5} in \texttt{DifferentialEquations.jl}). This does not use the linearity of the equation, nor the special form of $J$, and as a result requires thousands of evaluations, for a final precision markedly below that of \texttt{Float64}.

Interestingly, there is no general purpose algorithm that is more efficient than Runge-Kutta for linear ODEs that are not autonomous and for which computing commutators of the generator is expensive. Short of developing such a new algorithm, one could think of exploiting the special form of $J$ better. Indeed because of its asymptotic behavior:
\begin{itemize}
  \item The $J$ term in \eqref{eq:basic_ODE} is \emph{dominant} near $x=0$ and one can do a perturbative expansion in $\mathcal{L}$ in this regime,
  \item The $J$ term is \emph{negligible} when $x\gg 1/m$ and one can do a perturbative expansion in $\alpha J$.
\end{itemize}
When only one term is dominant, the ODE becomes time-independent, and can be solved with an efficient Krylov exponentiation. One could thus use two very accurate (numerically exact) perturbative expansions with Krylov exponentiaiton, connected by a Runge-Kutta solver in the intermediate regime where both terms are of the same magnitude. This could provide a small gain in evaluation time, but most importantly bring us closer to machine precision. This is of course only one strategy, but I suspect there are many ways to do better than plain Runge-Kutta.

\begin{remark}[Can one brute-force it with a GPU farm?]
  It is tempting, and not unreasonable, to think of making iterations faster simply by throwing more compute at it, or implementing the same algorithm better. However, there seems to be little gain to find there. Solving an ODE is a very sequential problem, which is notoriously difficult to parallelize. Further, each function evaluation in the ODE is essentially a sequence of matrix multiplications of intermediate size (say $32\times 32$ or at most right now $64\times 64$). In this regime of small matrices, vector units in a single CPU are faster than multiple CPUs or GPUs. Provided one uses the vector units efficiently (\eg with \texttt{MKL} or \texttt{Octavian.jl}), there is not much more one can do. One might think of \emph{batching} some operations, that is to solve many ODEs in parallel at the same time, to benefit from the power of GPUs. But it is not so clear how this could benefit the optimization, besides perhaps making the line search slightly faster.
\end{remark}

\subsubsection*{Including symmetries}

Global symmetries on the lattice constrain the elementary tensors of matrix product states to have a certain block structure. Taking this structure into account has been a very powerful strategy in the discrete, drastically reducing the numerical cost for models with continuous symmetry groups like $U(1)$ or $SU(2)$. This is so well understood on the lattice that it is now included in standard MPS codes like \texttt{MPSKit} or \texttt{ITensor}.

A similar construction can be followed for RCMPS, and we plan to present the resulting block structure soon. For example, for two coupled scalars, it allows to construct states with exact $O(2)$ invariance. There are still two difficulties: optimizing the state for a fixed block structure, and finding the size of the blocks. The first is a standard problem in Riemannian optimization, on a more constrained sub-manifold respecting the block structure. In principle, there is no obstacle, one just needs to code it. The second problem is \emph{a priori} a difficult one. One certainly does not want to try all the possible block sizes, which would be combinatorially expensive.

On the lattice, one can choose the block sizes progressively, using the very bond expansion techniques we still lack in the continuum. This is another reason why being able to grow $D$ is important for RCMPS. Temporarily, before we develop a good bond expansion method for RCMPS, one could try a greedy search, restarting the full optimization at each change of the block structure.

\subsection{Excitation ansatz} \label{sec:excitation}

Finding an approximate MPS ground state $\ket{A}$ gives access to much more than ground state physics. Indeed, one can look for excited states in the tangent space of the approximate vacuum. This has been used for standard CMPS already~\cite{vanderstraeten2019tangentspace}, with the same tangent vectors \eqref{eq:tangent} that we used before. Since the tangent space is a vector space, one can just diagonalize the Hamiltonian to find excitations.

With RCMPS, the non-locality of the Hamiltonian in the $a(x),a^\dagger(x)$ makes everything more tedious and difficult, but it is likely not out of reach fundamentally. This would provide an interesting complement to the approach to estimate the mass we presented in the previous chapter, based on linear programming.

\subsection{Real-time evolution}

The Riemannian optimization picture we have presented is equivalent to the time dependent variational principle in imaginary time. In principle, we have everything needed to switch to real-time evolution. Apart from a few $i$ here and there, the main difference is that we would have to take small time steps $\upd t$ instead of potentially large increments $\alpha$.

Of course, the usual tensor network caveats apply. For a local quench, that is if we inject only a finite amount of energy into the system, one expects the RCMPS evolution to be accurate for a long time. However, the RCMPS will no longer be translation invariant, which is not something our current codebase can deal with\footnote{It is not easy to optimize even standard CMPS without translation invariance, despite recent progress by Tuybens and colleagues~\cite{tuybens2022cmps}.}.

For a global quench (extensive injection of energy, obtained \eg by changing a coupling uniformly), only short times are reachable in principle, because entanglement entropy grows linearly with time. However, this is something that one could simulate with only minor modifications of existing codes.

\section{Conclusion}
We started this journey by arguing for a pragmatic approach to quantum field theory, starting from simple models (our pile of dirt), and growing progressively to the more complex theories of the real world. Admittedly, we made only minor progress, and understood only the very first models. 

For those simple field theories, in $1+1$ dimensions, we managed to do what Feynman deemed unlikely. We made the variational approach work in the strongest sense, without a discretization or a finite volume. We did so by combining the insights of Hamiltonian truncation and of (continuous) matrix product states. 

We now have a powerful method --energy minimization over the space of relativistic matrix product states-- which allows to obtain a precise compressed representation of the vacuum. It can then be used to compute local expectation values, and some extended operators. The precision can be refined systematically, and the asymptotic scaling is favorable for gapped models (plausibly polynomial cost with super-polynomial decay of the error). Extending this method to other models in $1+1$ dimensions is non-trivial but is either already in progress or seems feasible. Extending from vacuum correlation function to spectral and real-time data is also within reach. I plan to keep on working on these extensions, as I believe they are important and likely to ultimately be successful.

We are still far from the models used to describe the real world at a fundamental level. To get there, one will need to understand if and how the variational method can be applied for QFT with less relevant interaction potentials (from ``strongly renormalizable'' to generic ``super-renormalizable''). It will also be necessary to make continuous tensor networks in $d+1$ dimensions numerically efficient. These are more difficult problems, and either of them could very well be insurmountable. If so, RCMPS will perhaps remain at least an interesting curiosity. And even then, one may hope that some of the lessons we learned along the way will remain valid more broadly.

\appendix 
\addcontentsline{toc}{chapter}{Appendices} \addtocontents{toc}{\protect\setcounter{tocdepth}{0}}

\chapter{Technicalities}\label{appendix:technicalities}

\section{Detailed CMPS calculations}

\subsection{Proofs of expectation value computations} \label{app:cmps_expect}

Our objective is to compute the expression for the norm of $\ket{Q,R}$ from eq. \eqref{eq:normCMPS} which we recall here:
\begin{equation}
    \bra{Q,R}Q,R\rangle=\tr_{\mathbb{C}^D\otimes \mathbb{C}^D}\left[\exp (L \mathbb{T})\right]
\end{equation}
with the transfer operator $\mathbb{T}$
\begin{equation}
    \mathbb{T} = Q\otimes \mathds{1} + \mathds{1} \otimes Q^* + R\otimes R^*
\end{equation}
There are many proofs of this result, one that relies on the discretization but is quite transparent, one that works directly in the continuum but is less transparent, and finally one that is both directly in the continuum and transparent but requires more advanced mathematics (quantum It\^o calculus).

\begin{proof}[Proof strategy 1: discretizing]
From \eqref{eq:cmps_bj} we have
\begin{equation}
    \ket{Q,R}\simeq \tr\left\{\prod_{j=1}^{L/\varepsilon} \exp\left[\varepsilon Q + \sqrt{\varepsilon} R\, b^\dagger_j \right] \right\}\ket{0}_\psi 
\end{equation}
hence
\begin{align}
    \bra{Q,R} Q,R\rangle &=\bra{0} \tr \left\{\prod_{j=1}^{L/\varepsilon} \exp\left[\varepsilon Q^* + \sqrt{\varepsilon} R^*\, b_j \right]\right\}\tr\left\{ \prod_{j=1}^{L/\varepsilon}\exp\left[\varepsilon Q+ \sqrt{\varepsilon} R\, b^\dagger_j \right] \right\}\ket{0} \\
    &=  \tr\left\{\prod_{j=1}^{L/\varepsilon} \bra{0}\exp\left[\varepsilon \mathds{1}\otimes Q^* + \sqrt{\varepsilon} \mathds{1}\otimes R^*\, b_j \right]\exp\left[\varepsilon Q\otimes \mathds{1} + \sqrt{\varepsilon} R\otimes \mathds{1}\, b^\dagger_j \right] \ket{0}_{b_j} \right\}
\end{align}
We could put everything into the same trace using that $\tr[A\otimes B] = \tr[A]\tr[B]$. We also used that the bosonic vacuum is simply a tensor product of states $\ket{0}_{b_j}$ annihilated by the $b_j$. Finally, expanding the exponentials to order $O(\varepsilon)$ 
\begin{align}
     \bra{Q,R} Q,R\rangle &\simeq \tr\left\{\prod_{j=1}^{L/\varepsilon} \bra{0} 1 + \varepsilon (\mathds{1}\otimes Q^*+ Q \otimes\mathds{1} + R \otimes R^* b_j b_j^\dagger) + \sqrt{\varepsilon} (\mathds{1}\otimes R^*\, b_j +R\otimes \mathds{1}\, b^\dagger_j)   \ket{0}_{b_j} \right\}\\
     &\simeq \tr\left\{\prod_{j=1}^{L/\varepsilon} 1 + \varepsilon \mathbb{T}  \right\}\simeq \tr\left\{\prod_{j=1}^{L/\varepsilon} \exp\left(\varepsilon \mathbb{T}\right)  \right\} \simeq \tr \left[\exp L \mathbb{T}\right]
\end{align}
\end{proof}
\begin{remark}[Direct MPS version]
In fact, although it is written in a slightly more cumbersome way as if we knew nothing about tensor networks, this first proof is the direct translation of the standard one for MPS. $\Phi=\exp[\varepsilon \mathbb{T}]$
\begin{align}
    \Phi:&= \sum_{i\geq 0} A_i \otimes A_i^* \\
    &= \mathds{1} + \varepsilon \left( Q\otimes \mathds{1} + \mathds{1} \otimes Q^* +  R\otimes R^*\right) + \varepsilon^2 (Q \otimes Q^*)+ \cdots\\
    &=\exp(\varepsilon \mathbb{T})+ o(\varepsilon)
\end{align}
this allows to understand the role of the factor $\sqrt{\varepsilon}$ 
\end{remark}

\begin{proof}[Proof strategy 2: using the continuum wave-function]
We can evaluate the norm directly in the continuum using the wavefunction expression of the CMPS \eqref{eq:cmps-wavefunction}:
\begin{align}
   \langle Q,R \ket{Q,R} &= \sum_{n=0}^{+\infty} \int_{0<x_1<x_2<\cdots < x_n<L} \hskip-2cm\upd x_1\,\upd x_2\cdots \upd x_n \; \varphi_n(x_1,\cdots,x_n)\varphi_n^*(x_1,\cdots,x_n)
\end{align}
For $0<x_1<x_2<\cdots < x_n<L$:
\begin{align}
    \varphi_n\varphi_n^*&=\tr\left[\e^{Q x_1} R \e^{Q(x_2-x_1)} \cdots R\e^{Q (L-x_n)}\right] \tr\left[\e^{Q^* x_1} R^* \e^{Q^*(x_2-x_1)} \cdots R^*\e^{Q^* (L-x_n)}\right]\\
    &= \tr\left[\left(\e^{Q x_1} \otimes \e^{Q^* x_1} \right) \left(R\otimes R^*\right) \left(\e^{Q(x_2-x_1)}\otimes \e^{Q^*(x_2-x_1)} \right)\cdots  \left(R\otimes R^*\right)\left(\e^{Q(L-x_n)}\otimes \e^{Q^*(L-x_n)}\right)\right]\\
    &= \tr\left[\e^{(Q\otimes \mathds{1} + \mathds{1}\otimes Q^*) x_1}  \left(R\otimes R^*\right) \e^{(Q\otimes \mathds{1} + \mathds{1}\otimes Q^*) (x_2-x_1)}\cdots  \left(R\otimes R^*\right)\e^{(Q\otimes \mathds{1} + \mathds{1}\otimes Q^*) (L-x_n)}\right]
\end{align}
We can now re-exponentiate this expression using the same technique as for the derivation of the wavefunction representation of CMPS, with the substitution $Q\rightarrow Q\otimes \mathds{1} + \mathds{1}\otimes Q^*$, $R\rightarrow R\otimes R^*$, and $\psi^\dagger(x)\rightarrow \mathds{1}$.
\end{proof}
This second proof has the advantage that it is directly in the continuum, but the fact that it works may seem a bit miraculous at first sight, and at least not obviously guessable from standard MPS techniques. There is, finally, a neat continuum proof that uses the formalism of quantum noise.

\begin{proof}[Proof strategy 3: quantum It\^o calculus]
Since the final result is the trace of an exponential, the easiest derivation would be by differentiation, to extract the generator $\mathbb{T}$ directly. The need for the formalism of quantum noise (or quantum It\^o calculus) comes because differentiating expressions containing $\upd\xi^\dagger:=\psi^\dagger(x) \upd x$ is non-trivial. Indeed, we have $\bra{0}\upd \xi \upd \xi^\dagger\ket{0} = \upd x$, which means that $\upd \xi$ behaves more like the root of a differential rather than differential (this is the quantum equivalent of a white noise). Hence, when expressions depend on $\upd \xi$, we need to push to second order to differentiate them, otherwise we miss squares of $\upd \xi$ that are of order $\upd x$ (this is the quantum equivalent of It\^o's lemma). This is just the continuum manifestation of the factor $\sqrt{\varepsilon}$ instead of $\varepsilon$ in front of $R$ that we have seen over and over again.

Let us write down again the overlap
\begin{align}
    \langle Q,R \ket{Q,R} &= \bra{0}\tr \bigg\{\underset{\tilde{U}_{0,L}}{\underbrace{\mathcal{P} \exp\left[ \int_I \upd x \, Q^* + R^* \hpsi(x)\right]}}\bigg\} \tr \bigg\{\underset{U_{0,L}}{\underbrace{\mathcal{P} \exp\left[ \int_I \upd x \, Q  + R \hpsi^\dagger(x)\right]}}\bigg\} \ket{0}\\
    &= \bra{0} \tr[\tilde{U}_{0,L}] \tr[U_{0,L}] \ket{0}\\
    &= \tr[ \bra{0} U_{0,L} \otimes  \tilde{U}_{0,L} \ket{0}] \, .
\end{align}
To obtain $\mathbb{T}$ we should now differentiate $\bra{0}U_{0,x} \otimes \tilde{U}_{0,x}\ket{0} $ with respect to the endpoint $x$, going to second order in $\upd \xi_x = \hpsi(x)\upd x$ (which is ultimately what quantum stochastic calculus boils down to):
\begin{align}
    \upd \bra{0}  U_{0,x} \otimes \tilde{U}_{0,x}\ket{0} &= \bra{0}  U_{0,x} (Q\upd x + R \upd \xi) \otimes U^\dagger_{0,x}(Q^*\upd x + R^* \upd \xi^\dagger_x) \ket{0}\\
    &=\bra{0} \left( U_{0,x} \otimes U^\dagger_{0,x}\right) \left( Q\otimes \mathds{1} \upd x + \mathds{1} \otimes Q^* \upd x + R\otimes R^* \, \upd \xi_x \upd \xi^\dagger_x\right)\ket{0} \label{eq:ito_beforeaverage}\\
    &=\bra{0}  U_{0,x} \otimes \tilde{U}_{0,x} \ket{0} \, \mathbb{T} \, \upd x \label{eq:ito_afteraverage}
\end{align}
hence finally $\langle Q,R \ket{Q,R} = \tr [\exp(L\mathbb{T})]$ as advertised. 
\end{proof}

\begin{remark}[Subtlety: It\^o vs Stratonovich]
To go for \eqref{eq:ito_beforeaverage} to \eqref{eq:ito_afteraverage} one needs to use the fact that $\upd \xi_x$ acts infinitesimally after $U_{0,x}$, \ie\, $U_{0,x}$ contains $\upd \xi^\dagger$ only up to $x-\upd x$. As a result the vacuum $\ket{0_x}$  is not modified, and thus we can take the expectation value directly without worrying about $U$. This corresponds to the quantum It\^o convention. In contrast taking the Stratonovich formulation amounts to having $\upd \xi_x$ centered around the last $x$ in $U_{0,x}$, which gives standard differentiation rules (no need to go to second order), but makes taking expectation values less trivial (this second order term has to be recovered somehow). 
\end{remark}
The reader should choose whatever proof they are comfortable with, but it is important to keep in mind that discretizing is not necessary.

\begin{remark}[Overlap of two different CMPS] \label{rk:overlap}
The $3$ previous proofs generalize immediately to the computation of arbitrary overlaps of CMPS $\bra{Q_1,R_1} Q_2,R_2\rangle$. One simply needs to formally substitute $Q^*,R^* \rightarrow Q_1,R_1$ and $Q,R\rightarrow Q_2,R_2$.
\end{remark}

\section{Derivation of RCMPS formulae for defects}
\label{app:tensors}

In this appendix, we present the full derivation of the RCMPS formulae given in section \ref{sec:RCMPS_defects}. Below $\ket{Q,R}$ is a RCMPS approximation to the true (interacting) ground state $\ket{0,g}$ of the $\phi^4$ model at coupling $g$.

\subsection{Expectation value of the defect operator}\label{DefectExp}

We start with the expectation value of the defect operator $\Dmu$:
\begin{align}
	\begin{split}
		\langle 0,g|\Dmu |0,g\rangle = \frac{\int   [D\phi ]\, \Dmu e^{-S_B} }{\int [D\phi ] e^{-S_B } }\,.
	\end{split}
\end{align}
Plugging in the definition of $\hphi(x)$
\begin{equation}
	\hphi(x) = \int_\mathbb{R} \upd y \, J(x-y)[\hat{a}(y) + \hat{a}^{\dagger}(y)]~, \quad J(x) = \frac{1}{2\pi} \int_\mathbb{R}\frac{\upd k}{\sqrt{2\omega_k}}e^{-i k x}\,.
\end{equation}
we can use the Baker-Campbell-Hausdorff (BCH) formula to put $\mathcal{D}_L$ in normal-ordered form
\begin{align}\label{eq:denominator}
	\begin{split}
		\Dmu   &= \; :\Dmu:\times \exp\left[\frac{\mu^{2}}{2}\int_{[-L,0]^2} \upd x \, \upd x \,^{\prime}\int_{\mathbb{R}} \upd y \; J(x-y)J(x^{\prime}-y)\right]\\
		&= \;  :\Dmu: \times \langle 0,0|\Dmu |0,0\rangle\,,
	\end{split}
\end{align}
where $\ket{0,0}$ is the ground state at coupling $0$, hence the Fock vacuum.

So far, everything was exact. Replacing $\ket{0,g}$ by its RCMPS approximation we get 
\begin{equation} 
\langle 0,g|:\Dmu :|0,g\rangle \simeq \bra{Q,R}:\Dmu : \ket{Q,R} = \mathcal{Z}_{-\mu G,-\mu G}\,,
\end{equation} and thus
\eqref{eq:denominator} gives:
\begin{equation}
    \langle 0,g|\Dmu |0,g\rangle \simeq \mathcal{Z}_{-\mu G, -\mu G}\times \exp\left[\frac{\mu^{2}}{2}\int_{\mathbb{R}} \upd y \; G(y)^2\right]\,,
\end{equation}
where $\mathcal{Z}_{j^{'},j}$ is the RCMPS generating functional, and $G$ is the modified ``source'', \ie 
\begin{align}
	&\mathcal{Z}_{-\mu G, -\mu G} = \tr\Big{[}\mathcal{P} \exp \int_{\mathbb{R}} \upd x \, \: \mathbb{T} - \mu G(x) (R\otimes\id + \id\otimes R^*)\Big{]}~~ \text{with}\quad G(x) := \int_{-L}^0 \upd y J(x-y)\,.
\end{align}
This path ordered exponential can be written as the solution to an ordinary differential equation (ODE). Putting the ODE in superoperator form as in the main text, we finally get
\begin{align}
	\bra{0,g}\Dmu\ket{0,g} \simeq \lim_{x \to \infty} \, \mathrm{tr}\,[\rho(x)]~, ~~ \text{for the initial condition}~~~ \qquad \lim_{x\to -\infty} \rho(x)= \rho_{0}\,.
\end{align}
and 
\begin{equation}\label{eq:denODE_bis}
	\frac{\upd}{\upd x} \rho(x) = \mathcal{L}\cdot \rho(x) - \mu G(x) \left[ R\rho(x) + \rho(x) R^\dagger \right] + \frac{\mu^{2}}{2} G^{2}(x) \rho(x)\,,
\end{equation}
where we recall that
\begin{align}
	\mathcal{L} \cdot \rho = Q\, \rho + \rho \, Q^{\dagger} + R \, \rho \, R^{\dagger}\,.
\end{align}

\subsection{One-point functions in the defect model}

\subsubsection{Vertex operators in the defect model}\label{sec:DefectVertex}

Next, we consider the vertex operator in the full defect theory \eqref{eq:Defect_vertex}, which we differentiate in \ref{app:monomial} to obtain expectation values of one point functions in the $\phi^4$ model in presence of the defect. It is defined as
\begin{align}\label{eq:Defect_vertex_bis}
    \langle V_b (x) \rangle_{\text{defect}} = \frac{\bra{0,g} :e^{b\hat{\phi}(x)}: \text{e}^{-\mu \int_{-L}^{0} \hat{\phi}}\ket{0,g}}{\bra{0,g}\text{e}^{-\mu \int_{-L}^{0} \hat{\phi}}\ket{0,g}}\,.
\end{align}
We first focus on the numerator, as the denominator has just been computed.

The first step, again exact, is to normal-order the operator being evaluated using the Baker-Campbell-Hausdorff formula:
\begin{equation}
\begin{split}
    : e^{\, b \, \hat{\phi}(x)}: e^{-\mu \int_{-L}^0 \upd y \, \: \hphi(y)} = &\,: e^{ b \, \hphi(x)-\mu \int_{-L}^0 \upd y \, \: \hphi(y) }: \; \bra{0,0}\Dmu \ket{0,0}\\
    & \times   \exp\left[- b\,\mu\,\int_{\mathbb{R}} \upd y  \:J(x-y)G(y)\right]\,,
    \end{split}
\end{equation}
where we have explicitly separated the $\bra{0,0}\Dmu \ket{0,0}$ for convenience as it also appears in the denominator and thus ultimately cancels out.

Replacing the true ground state by its variational approximation yields
\begin{equation}
\begin{split}
    \bra{0,g} :e^{b\hat{\phi}(x)}: \text{e}^{-\mu \int_{-L}^{0} \hat{\phi}}\ket{0,g} \simeq &\,\mathcal{Z}_{s_x,s_x} \; \bra{0,0}\Dmu \ket{0,0}\\
    & \times   \exp\left[- b\,\mu\,\int_{\mathbb{R}} \upd y \:J(x-y)G(y)\right]\,,
\end{split}
\end{equation}
where 
\begin{equation}{\label{eq:smearedsource}}
	s_x(y) := b\,J(x-y) - \mu\int_{-L}^0 \upd z \: J(z-y) = b\, J(x-y) - \mu \,  G(y) \,.
\end{equation}
Again, we can write $\mathcal{Z}_{s_x,s_x}$ as the solution of an ODE to get
\begin{equation}\label{eq:vertsolution}
	\bra{Q,R} : e^{\, b \, \hphi(x)}: \Dmu \ket{Q,R} = \langle 0,0|\Dmu |0,0\rangle \, \lim_{y\to +\infty}\tr[\rho(y)]\,,
\end{equation}
with
\begin{align}\label{eq:defectvertexODE}
	\frac{\upd}{\upd y} \rho(y) &= \mathcal{L}\cdot \rho(x) + s_x(y)  \left[R\rho(x) + \rho(x) R^\dagger\right]- b \, \mu \, G(y)J(x-y)\rho(y)\,,
\end{align}
and with the initial condition $\lim_{y\rightarrow-\infty}\rho(y) = \rho_{0}$ with $\tr[\rho_0]=1$. Finally, we can put numerator and denominator together, which cancels the $\bra{0,0} \Dmu\ket{0,0}$ term and we get
\begin{equation}\label{eq:vertex_as_trace}
    \langle V_b (x) \rangle_{\text{defect}} \simeq \lim_{y\rightarrow \infty}  \frac{\tr[\rho(y)]}{\tr[\tilde{\rho}(y)]}\,,
\end{equation}
where $\tilde{\rho}(y)$ is the solution of \eqref{eq:defectvertexODE} with $b=0$ (or equivalently of \eqref{eq:denODE_bis} without $G^2$ term).

\subsubsection{Field monomials in the defect model} \label{app:monomial}

With explicit expressions for defect vertex operators, we may now compute expectation values of field monomials in the full defect theory. The latter are defined as:
\begin{equation}\label{eq:npointdef}
	\langle : \phi^{n}(x): \rangle_{\text{defect}} = \frac{ \langle 0,g | : \hphi^{n}(x): e^{-\mu \int_{-L}^0 \upd x \, \: \hphi(x)} |0,g\rangle}{ \langle 0,g | e^{-\mu \int_{-L}^0 \upd x \, \: \hphi(x)}|0,g \rangle}\,.
\end{equation}
Again, we already have the denominator from \eqref{DefectExp}. For the numerator, we can differentiate the defect vertex operator w.r.t $b$,
\begin{align}\label{eq:npointdef2}
	\begin{split}
		\langle 0,g| : \hphi^{n}(x): e^{-\mu \int_{-L}^0d x \, \: \hphi(x)} |0,g\rangle  
		& = 	\frac{\partial
			^{n}}{\partial b^{n}}\langle 0,g| : e^{\, b \, \hphi(x)}: \Dmu |0,g\rangle \Big|_{b=0}\,.
	\end{split}
\end{align}
The strategy is now to forward differentiate the ordinary differential equation giving us $\rho(y)$ appearing in \eqref{eq:vertex_as_trace}
\begin{align}
	\bra{0,g} : \hphi^{n}(x): e^{-\mu \int_{-L}^0 \upd x \, \: \hphi(x)} \ket{0,g} \, &\simeq \, \bra{Q,R} :\hphi^{n}(x): e^{-\mu \int_{-L}^0 \upd x \, \: \hphi(x)} \ket{Q,R} \\
    \, &= \, \langle 0,0|\Dmu |0,0\rangle \times \, \lim_{y\to +\infty}\tr\left[\rho^{(n)}(y)\right]\,, \label{eq:monomial_from_rho_l2}
\end{align}
where $\rho^{(n)}(y):= \partial_b^n \rho(y) |_{b=0}$ and $\rho$ obeys the ODE \eqref{eq:defectvertexODE}. The matrices $\rho^{(k)}(y)$ for $0\leq k\leq n$ obey the triangular system of matrix ODEs
\begin{align}\label{eq:numODE}
	\frac{\upd}{\upd y} \rho^{(k)}(y) &= \mathcal{L} \cdot \rho^{(k)}(y) + k J(x-y)\left[R\rho^{(k-1)}(y) + \rho^{(k-1)}(y) R^\dagger\right] \\
    &- \mu G(y)\left[R\rho^{(k)}(y) + \rho^{(k)}(y) R^\dagger\right] - k\, \mu \, J(x-y)G(y)\rho^{(k-1)}(y)\,,
\end{align}
with the initial conditions, $ \lim_{y\rightarrow-\infty}\rho^{(0)}(y) = \rho_{0}$ and $\lim_{y\rightarrow-\infty}\rho^{(n)}(y) = 0$ for $k > 0$. Here, again the factor $\langle 0,0|\Dmu |0,0\rangle$ arises both in the numerator \eqref{eq:monomial_from_rho_l2} and the denominator \eqref{eq:denominator} and hence drops out in the computation of monomials \eqref{eq:npointdef}. Putting all together, we find
\begin{equation}\label{eq:monomial_final_as_trace}
	\langle : \phi^{n}(x): \rangle_{\text{defect}} = \lim_{y\to +\infty} \frac{\tr[\rho^{(n)}(y)]}{\tr[\rho^{(0)}(y)]}\,.
\end{equation}

\subsubsection{Normalization tricks} \label{app:normalization_tricks}
The vertex operator and field monomial expectation values in the presence of the defect given in \eqref{eq:vertex_as_trace} and \eqref{eq:monomial_final_as_trace} are ratios of traces, each of which is typically growing exponentially with the size $L$ of the defect. Thus, computing each term independently and taking the ratio quickly gives large numerical errors in double precision, especially when numerically taking the semi-infinite defect line limit. This should be avoided.

A simple way to make things better behaved is to periodically normalize both the matrix in the numerator and the matrix in the denominator by the trace of the denominator as the ODE is being solved. For example, in the case of the computation of the field expectation
\begin{equation}\label{eq:field_as_trace}
	\langle : \phi(x): \rangle_{\text{defect}} = \lim_{y\to +\infty} \frac{\tr[\rho^{(1)}(y)]}{\tr[\rho^{(0)}(y)]}\,,
\end{equation}
we can pick a number of checkpoints $y_k$ (either a priori, or each time $\tr\left[\rho^{(0)}(y)\right]$ gets too large or too small) and renormalize both $\rho^{(0)}(y_k)$ and $\rho^{(1)}(y_k)$:
\begin{align}
    \rho^{(0)}(y_k^+) &= \frac{\rho^{(0)}(y_k)}{\tr\left[\rho^{(0)}(y_k)\right]}\,, \\
    \rho^{(1)}(y_k^+) &= \frac{\rho^{(1)}(y_k)}{\tr\left[\rho^{(0)}(y_k)\right]} \,.
\end{align}
Inserting these normalizations do not change the value of the expectation value \eqref{eq:field_as_trace}.

Alternatively, one could define the continuously normalized matrices
\begin{align}
    \sigma^{(0)}(y) &= \frac{\rho^{(0)}(y)}{\tr\left[\rho^{(0)}(y)\right]} \\
    \sigma^{(1)}(y) &= \frac{\rho^{(1)}(y)}{\tr\left[\rho^{(0)}(y)\right]} \,,
\end{align}
which indeed obey a system of closed ODEs. However these ODEs are non-linear, and we found it numerically preferable to use the previous periodic normalization strategy. No matter which strategy is used, one can compute expectation values for large or even numerically semi-infinite defects without suffering from instabilities of precision loss.

\subsection{Strategy for more general expectation values} \label{app:RCMPS_2point}
Following the same techniques, one can compute expectation values of arbitrary strings of vertex operators or normal-ordered monomials in the presence of the defect. The steps are the same as before, and we only outline the general strategy here. One first considers expectation values of products of vertex operators
\begin{align}\label{eq:Defect_vertex_multiple}
    \langle V_{b_1} (x_1) V_{b_2}(x_2) \dots V_{b_n}(x_n) \rangle_{\text{defect}} = \frac{\bra{0,g} :e^{b_1\hat{\phi}(x_1)}\!: \, :e^{b_2\hat{\phi}(x_2)}\!: \, \dots \, :e^{b_n\hat{\phi}(x_n)}\! : \; \text{e}^{-\mu \int_{-L}^{0} \hat{\phi}}\ket{0,g}}{\bra{0,g}\text{e}^{-\mu \int_{-L}^{0} \hat{\phi}}\ket{0,g}}~ \,\,,
\end{align}
for $x_1  <x_2 <\cdots < x_n$. Then, focusing on the numerator, we use the Baker-Campbell-Hausdorff formula iteratively to put all the fields under the same exponential
\begin{equation}
\begin{split}
    :e^{b_1\hat{\phi}(x_1)}\!: \, :e^{b_2\hat{\phi}(x_2)}\!: \, \dots \, :e^{b_n\hat{\phi}(x_n)}\! : \text{e}^{-\mu \int_{-L}^{0}\hat{\phi}} &=\\ \mathcal{N}(b_1,b_2,\dots,b_n,x_1,x_2,\cdots,x_n) \; \times \; &:e^{b_1\hat{\phi}(x_1) + b_2\hat{\phi}(x_2) + \dots + b_n\hat{\phi}(x_n) -\mu \int_{-L}^{0}\hat{\phi}} : \,,
\end{split}
\end{equation}
where $\mathcal{N}$ is an explicit scalar contribution coming from the normal ordering. The expectation value of the remaining normal-ordered exponential on a RCMPS is then just $\mathcal{Z}_{s,s}$ with
\begin{equation}
    s(y) := \sum_{j=1}^n b_j\, J(x_j-y) - \mu \,  G(y) \,,
\end{equation}
and can thus be obtained by solving a simple linear ODE like before. 

Then, to compute expectation values of normal-ordered field monomials at different points, one may just differentiate with respect to the $b$'s
\begin{equation}
    \langle :{\phi}^{k_1}(x_1): \, :{\phi}^{k_2}(x_2): \, \dots \, :{\phi}^{k_n}(x_n): \rangle_{\text{defect}}\, =\partial_{b_1}^{k_1} \partial_{b_2}^{k_2} \dots \partial_{b_n}^{k_n}  \langle V_{b_1} (x_1) V_{b_2}(x_2) \dots V_{b_n}(x_n) \rangle_{\text{defect}} \bigg|_{b=0} \,. 
\end{equation}

As an illustration, following these steps one gets:
\begin{equation}
    \langle\phi(x_1)\phi(x_2)\rangle_\text{defect} = \lim_{y\rightarrow + \infty}\frac{\tr[\rho^{(1,1)}(y)]}{\tr[\rho^{(0,0)}(y)]}\,,
\end{equation}
with
\begin{align}
	\begin{split}
		\frac{\upd \rho^{(0,0)}(y)}{\upd y} &= [\mathcal{L}- \mu G(y)\mathcal{R}]\cdot \rho^{(0,0)}(y)\\
		\frac{\upd \rho^{(1,0)}(y)}{\upd y}&= [\mathcal{L}- \mu G(y)\mathcal{R}]\cdot \rho^{(1,0)}(y) + [J(x_1-y)(\mathcal{R}-\mu \, G(y)\id)]\cdot \rho^{(0,0)}(y)\\
		\frac{\upd \rho^{(0,1)}(y)}{\partial y}&= [\mathcal{L}- \mu \, G(y)\mathcal{R}]\cdot \rho^{(0,1)}(y) + [J(x_2-y)(\mathcal{R}-\mu G(y)\id)]\cdot \rho^{(0,0)}(y)\\
		\frac{\upd \rho^{(1,1)}(y)}{\upd y} &= [\mathcal{L}- \mu \, G(y)\mathcal{R}]\cdot \rho^{(1,1)}(y) + [J(x_2-y)(\mathcal{R}-\mu \, G(y)\id)]\cdot \rho^{(1,0)}(y)\\
		& \quad + [J(x_1-y)(\mathcal{R}-\mu \, G(y)\id)]\cdot \rho^{(0,1)}(y) + J(x_1-y)J(x_2-y)\rho^{(0,0)}(y)\,,
	\end{split}
\end{align}
with $\mathcal{R}\cdot \rho = R \rho + \rho R^\dagger$ and the initial conditions $\lim_{y\to -\infty} \rho^{(0,0)}(y)= \rho_{0}$ and $\lim_{y\to -\infty} \rho^{(i,j)}(y)= 0$. 

While we do not show the result of two-point functions with defects in the main text, we verified numerically that they converged approximately as fast as the defect one-point functions as a function of $D$ and matched perturbation theory at sufficiently small coupling.
\backmatter
\cleardoublepage
\phantomsection

\cleardoublepage
\phantomsection

\cleardoublepage
\phantomsection

\addcontentsline{toc}{chapter}{References}
\bibliography{thesis}

@article{stojevic2015finiteentanglement,
  title = {Conformal data from finite entanglement scaling},
  author = {Stojevic, Vid and Haegeman, Jutho and McCulloch, I. P. and
            Tagliacozzo, Luca and Verstraete, Frank},
  journal = {Phys. Rev. B},
  volume = {91},
  issue = {3},
  pages = {035120},
  numpages = {16},
  year = {2015},
  month = {Jan},
  publisher = {American Physical Society},
  doi = {10.1103/PhysRevB.91.035120},
  url = {https://link.aps.org/doi/10.1103/PhysRevB.91.035120},
}

@article{haegeman2013calculus,
  title = {Calculus of continuous matrix product states},
  author = {Haegeman, Jutho and Cirac, J. Ignacio and Osborne, Tobias J. and
            Verstraete, Frank},
  journal = {Phys. Rev. B},
  volume = {88},
  issue = {8},
  pages = {085118},
  numpages = {21},
  year = {2013},
  month = {Aug},
  publisher = {American Physical Society},
  doi = {10.1103/PhysRevB.88.085118},
  url = {https://link.aps.org/doi/10.1103/PhysRevB.88.085118},
}

@article{haegeman2010relativistic,
  title = {Applying the Variational Principle to ($1+1$)-Dimensional Quantum
           Field Theories},
  author = {Haegeman, Jutho and Cirac, J. Ignacio and Osborne, Tobias J. and
            Verschelde, Henri and Verstraete, Frank},
  journal = {Phys. Rev. Lett.},
  volume = {105},
  issue = {25},
  pages = {251601},
  numpages = {4},
  year = {2010},
  month = {Dec},
  publisher = {American Physical Society},
  doi = {10.1103/PhysRevLett.105.251601},
  url = {https://link.aps.org/doi/10.1103/PhysRevLett.105.251601},
}

@misc{tilloy2022studyquantumsinhgordonmodel,
  title = {A study of the quantum Sinh-Gordon model with relativistic continuous
           matrix product states},
  author = {Antoine Tilloy},
  year = {2022},
  eprint = {2209.05341},
  archivePrefix = {arXiv},
  primaryClass = {hep-th},
  url = {https://arxiv.org/abs/2209.05341},
}

@article{krumnow2016orbitaloptim,
  title = {Fermionic Orbital Optimization in Tensor Network States},
  author = {Krumnow, C. and Veis, L. and Legeza, \"O. and Eisert, J.},
  journal = {Phys. Rev. Lett.},
  volume = {117},
  issue = {21},
  pages = {210402},
  numpages = {6},
  year = {2016},
  month = {Nov},
  publisher = {American Physical Society},
  doi = {10.1103/PhysRevLett.117.210402},
  url = {https://link.aps.org/doi/10.1103/PhysRevLett.117.210402},
}

@article{wu2025impuritydisentangling,
  title = {Disentangling interacting systems with fermionic Gaussian circuits:
           Application to quantum impurity models},
  author = {Wu, Ang-Kun and Kloss, Benedikt and Krinitsin, Wladislaw and Fishman
            , Matthew T. and Pixley, J. H. and Stoudenmire, E. M.},
  journal = {Phys. Rev. B},
  volume = {111},
  issue = {3},
  pages = {035119},
  numpages = {18},
  year = {2025},
  month = {Jan},
  publisher = {American Physical Society},
  doi = {10.1103/PhysRevB.111.035119},
  url = {https://link.aps.org/doi/10.1103/PhysRevB.111.035119},
}

@misc{schmoll2023hamiltoniantruncationtensornetworks,
  title = {Hamiltonian truncation tensor networks for quantum field theories},
  author = {Philipp Schmoll and Jan Naumann and Alexander Nietner and Jens
            Eisert and Spyros Sotiriadis},
  year = {2023},
  eprint = {2312.12506},
  archivePrefix = {arXiv},
  primaryClass = {quant-ph},
  url = {https://arxiv.org/abs/2312.12506},
}

@article{milsted2013mpsphi4,
  title = {Matrix product states and variational methods applied to critical
           quantum field theory},
  author = {Milsted, Ashley and Haegeman, Jutho and Osborne, Tobias J.},
  journal = {Phys. Rev. D},
  volume = {88},
  issue = {8},
  pages = {085030},
  numpages = {23},
  year = {2013},
  month = {Oct},
  publisher = {American Physical Society},
  doi = {10.1103/PhysRevD.88.085030},
  url = {https://link.aps.org/doi/10.1103/PhysRevD.88.085030},
}

@article{delcamp202giltphi4,
  title = {Computing the renormalization group flow of two-dimensional ${
           \ensuremath{\phi}}^{4}$ theory with tensor networks},
  author = {Delcamp, Clement and Tilloy, Antoine},
  journal = {Phys. Rev. Res.},
  volume = {2},
  issue = {3},
  pages = {033278},
  numpages = {15},
  year = {2020},
  month = {Aug},
  publisher = {American Physical Society},
  doi = {10.1103/PhysRevResearch.2.033278},
  url = {https://link.aps.org/doi/10.1103/PhysRevResearch.2.033278},
}

@article{hauru2018gilt,
  title = {Renormalization of tensor networks using graph-independent local
           truncations},
  author = {Hauru, Markus and Delcamp, Clement and Mizera, Sebastian},
  journal = {Phys. Rev. B},
  volume = {97},
  issue = {4},
  pages = {045111},
  numpages = {18},
  year = {2018},
  month = {Jan},
  publisher = {American Physical Society},
  doi = {10.1103/PhysRevB.97.045111},
  url = {https://link.aps.org/doi/10.1103/PhysRevB.97.045111},
}

@article{loinaz1998mc_phi4,
  title = {Monte Carlo simulation calculation of the critical coupling constant
           for two-dimensional continuum ${\ensuremath{\varphi}}^{4}$ theory},
  author = {Loinaz, Will and Willey, R. S.},
  journal = {Phys. Rev. D},
  volume = {58},
  issue = {7},
  pages = {076003},
  numpages = {5},
  year = {1998},
  month = {Sep},
  publisher = {American Physical Society},
  doi = {10.1103/PhysRevD.58.076003},
}

@article{tilloy2021relativistic,
  title = {Relativistic continuous matrix product states for quantum fields
           without cutoff},
  author = {Tilloy, Antoine},
  journal = {Phys. Rev. D},
  volume = {104},
  issue = {9},
  pages = {096007},
  numpages = {14},
  year = {2021},
  month = {Nov},
  publisher = {American Physical Society},
  doi = {10.1103/PhysRevD.104.096007},
  url = {https://link.aps.org/doi/10.1103/PhysRevD.104.096007},
}

@article{tilloy2021variational,
  title = {Variational method in relativistic quantum field theory without
           cutoff},
  author = {Tilloy, Antoine},
  journal = {Phys. Rev. D},
  volume = {104},
  issue = {9},
  pages = {L091904},
  numpages = {5},
  year = {2021},
  month = {Nov},
  publisher = {American Physical Society},
  doi = {10.1103/PhysRevD.104.L091904},
  url = {https://link.aps.org/doi/10.1103/PhysRevD.104.L091904},
}

@article{heymans2021N8LO,
  author = {Heymans, Gustavo O. and Pinto, Marcus Benghi},
  title = {Critical behavior of the 2d scalar theory: resumming the N8LO
           perturbative mass gap},
  journal = {Journal of High Energy Physics},
  year = {2021},
  month = {Jul},
  day = {22},
  volume = {2021},
  number = {7},
  pages = {163},
  issn = {1029-8479},
  doi = {10.1007/JHEP07(2021)163},
  url = {https://doi.org/10.1007/JHEP07(2021)163},
}

@article{serone2019broken,
  author = {Serone, Marco and Spada, Gabriele and Villadoro, Giovanni},
  title = {$\lambda \phi^4$ theory --- Part II. the broken phase beyond
           NNNN(NNNN)LO},
  journal = {Journal of High Energy Physics},
  year = {2019},
  month = {May},
  day = {08},
  volume = {2019},
  number = {5},
  pages = {47},
  issn = {1029-8479},
  doi = {10.1007/JHEP05(2019)047},
  url = {https://doi.org/10.1007/JHEP05(2019)047},
}

@article{serone2018symmetric,
  author = {Serone, Marco and Spada, Gabriele and Villadoro, Giovanni},
  title = {$\lambda\phi^4$ theory --- Part I. The symmetric phase beyond
           NNNNNNNNLO},
  journal = {Journal of High Energy Physics},
  year = {2018},
  month = {Aug},
  day = {23},
  volume = {2018},
  number = {8},
  pages = {148},
  issn = {1029-8479},
  doi = {10.1007/JHEP08(2018)148},
  url = {https://doi.org/10.1007/JHEP08(2018)148},
}

@inbook{feynman1988,
  author = { Richard P. Feynman },
  title = {Difficulties in Applying the Variational Principle to Quantum Field
           Theories},
  booktitle = {Variational Calculations in Quantum Field Theory},
  chapter = {},
  year = {1987},
  pages = {28-40},
  doi = {10.1142/9789814390187_0003},
  publisher = {World Scientific Publishing, Singapore},
}

@article{vanderstraeten2019tangentspace,
  title = {{Tangent-space methods for uniform matrix product states}},
  author = {Laurens Vanderstraeten and Jutho Haegeman and Frank Verstraete},
  journal = {SciPost Phys. Lect. Notes},
  pages = {7},
  year = {2019},
  publisher = {SciPost},
  doi = {10.21468/SciPostPhysLectNotes.7},
}

@article{vanderstraeten2019spectrum,
  title = {Simulating excitation spectra with projected entangled-pair states},
  author = {Vanderstraeten, Laurens and Haegeman, Jutho and Verstraete, Frank},
  journal = {Phys. Rev. B},
  volume = {99},
  issue = {16},
  pages = {165121},
  numpages = {5},
  year = {2019},
  month = {Apr},
  publisher = {American Physical Society},
  doi = {10.1103/PhysRevB.99.165121},
  url = {https://link.aps.org/doi/10.1103/PhysRevB.99.165121},
}

@article{verstraete2004thermal,
  title = {Matrix Product Density Operators: Simulation of Finite-Temperature
           and Dissipative Systems},
  author = {Verstraete, F. and Garc\'{\i}a-Ripoll, J. J. and Cirac, J. I.},
  journal = {Phys. Rev. Lett.},
  volume = {93},
  issue = {20},
  pages = {207204},
  numpages = {4},
  year = {2004},
  month = {Nov},
  publisher = {American Physical Society},
  doi = {10.1103/PhysRevLett.93.207204},
  url = {https://link.aps.org/doi/10.1103/PhysRevLett.93.207204},
}

@article{hackl2020,
  title = {{Geometry of variational methods: dynamics of closed quantum systems}
           },
  author = {Lucas Hackl and Tommaso Guaita and Tao Shi and Jutho Haegeman and
            Eugene Demler and J. Ignacio Cirac},
  journal = {SciPost Phys.},
  volume = {9},
  issue = {4},
  pages = {48},
  year = {2020},
  publisher = {SciPost},
  doi = {10.21468/SciPostPhys.9.4.048},
  url = {https://scipost.org/10.21468/SciPostPhys.9.4.048},
}

@article{kshetrimayum2019thermal,
  title = {Tensor Network Annealing Algorithm for Two-Dimensional Thermal States
           },
  author = {Kshetrimayum, A. and Rizzi, M. and Eisert, J. and Or\'us, R.},
  journal = {Phys. Rev. Lett.},
  volume = {122},
  issue = {7},
  pages = {070502},
  numpages = {6},
  year = {2019},
  month = {Feb},
  publisher = {American Physical Society},
  doi = {10.1103/PhysRevLett.122.070502},
  url = {https://link.aps.org/doi/10.1103/PhysRevLett.122.070502},
}

@article{hogervorst2015,
  title = {Truncated conformal space approach in $d$ dimensions: A cheap
           alternative to lattice field theory?},
  author = {Hogervorst, Matthijs and Rychkov, Slava and van Rees, Balt C.},
  journal = {Phys. Rev. D},
  volume = {91},
  issue = {2},
  pages = {025005},
  numpages = {35},
  year = {2015},
  month = {Jan},
  publisher = {American Physical Society},
  doi = {10.1103/PhysRevD.91.025005},
  url = {https://link.aps.org/doi/10.1103/PhysRevD.91.025005},
}

@article{eliasmiro2016,
  title = {The renormalized Hamiltonian truncation method in the large E T
           expansion},
  volume = {2016},
  ISSN = {1029-8479},
  url = {http://dx.doi.org/10.1007/JHEP04(2016)144},
  DOI = {10.1007/jhep04(2016)144},
  number = {4},
  journal = {Journal of High Energy Physics},
  publisher = {Springer Science and Business Media LLC},
  author = {Elias-Miró, J. and Montull, M. and Riembau, M.},
  year = {2016},
  month = {Apr},
  pages = {1–34},
}

@article{rychkov2015,
  title = {Hamiltonian truncation study of the ${\ensuremath{\varphi}}^{4}$
           theory in two dimensions},
  author = {Rychkov, Slava and Vitale, Lorenzo G.},
  journal = {Phys. Rev. D},
  volume = {91},
  issue = {8},
  pages = {085011},
  numpages = {26},
  year = {2015},
  month = {Apr},
  publisher = {American Physical Society},
  doi = {10.1103/PhysRevD.91.085011},
}

@article{james2018hamiltonian_truncation,
  doi = {10.1088/1361-6633/aa91ea},
  year = 2018,
  month = {feb},
  publisher = {{IOP} Publishing},
  volume = {81},
  number = {4},
  pages = {046002},
  author = {Andrew J A James and Robert M Konik and Philippe Lecheminant and
            Neil J Robinson and Alexei M Tsvelik},
  title = {Non-perturbative methodologies for low-dimensional
           strongly-correlated systems: From non-Abelian bosonization to
           truncated spectrum methods},
  journal = {Reports on Progress in Physics},
}

@article{eliasmiro2017-1,
  title = {NLO renormalization in the Hamiltonian truncation},
  author = {Elias-Mir\'o, Joan and Rychkov, Slava and Vitale, Lorenzo G.},
  journal = {Phys. Rev. D},
  volume = {96},
  issue = {6},
  pages = {065024},
  numpages = {43},
  year = {2017},
  month = {Sep},
  publisher = {American Physical Society},
  doi = {10.1103/PhysRevD.96.065024},
  url = {https://link.aps.org/doi/10.1103/PhysRevD.96.065024},
}

@article{eliasmiro2017-2,
  author = {Elias-Mir{\'o}, Joan and Rychkov, Slava and Vitale, Lorenzo G.},
  title = {High-precision calculations in strongly coupled quantum field theory
           with next-to-leading-order renormalized Hamiltonian Truncation},
  journal = {JHEP},
  year = {2017},
  volume = {2017},
  number = {10},
  pages = {213},
  issn = {1029-8479},
  doi = {10.1007/JHEP10(2017)213},
  url = {https://doi.org/10.1007/JHEP10(2017)213},
}

@article{carleo2017neural,
  author = {Carleo, Giuseppe and Troyer, Matthias},
  title = {Solving the quantum many-body problem with artificial neural networks
           },
  volume = {355},
  number = {6325},
  pages = {602--606},
  year = {2017},
  doi = {10.1126/science.aag2302},
  publisher = {American Association for the Advancement of Science},
  issn = {0036-8075},
  journal = {Science},
}

@article{choo2020fermionic,
  title = {Fermionic neural-network states for ab-initio electronic structure},
  author = {Choo, Kenny and Mezzacapo, Antonio and Carleo, Giuseppe},
  journal = {Nature communications},
  volume = {11},
  number = {1},
  pages = {1--7},
  year = {2020},
  publisher = {Nature Publishing Group},
}

@article{srednicki1993,
  title = {Entropy and area},
  author = {Srednicki, M.},
  journal = {Phys. Rev. Lett.},
  volume = {71},
  issue = {5},
  pages = {666--669},
  numpages = {0},
  year = {1993},
  month = {Aug},
  publisher = {American Physical Society},
  doi = {10.1103/PhysRevLett.71.666},
  url = {https://link.aps.org/doi/10.1103/PhysRevLett.71.666},
}

@article{eisert2010,
  title = {Colloquium: Area laws for the entanglement entropy},
  author = {Eisert, J. and Cramer, M. and Plenio, M. B.},
  journal = {Rev. Mod. Phys.},
  volume = {82},
  issue = {1},
  pages = {277--306},
  numpages = {0},
  year = {2010},
  month = {Feb},
  publisher = {American Physical Society},
  doi = {10.1103/RevModPhys.82.277},
  url = {https://link.aps.org/doi/10.1103/RevModPhys.82.277},
}

@article{wolf2008,
  title = {Area Laws in Quantum Systems: Mutual Information and Correlations},
  author = {Wolf, M. M. and sredVerstraete, F. and Hastings, M. B. and Cirac, J.
            I.},
  journal = {Phys. Rev. Lett.},
  volume = {100},
  issue = {7},
  pages = {070502},
  numpages = {4},
  year = {2008},
  month = {Feb},
  publisher = {American Physical Society},
  doi = {10.1103/PhysRevLett.100.070502},
  url = {https://link.aps.org/doi/10.1103/PhysRevLett.100.070502},
}

@inbook{Ran2020,
  author = "Ran, Shi-Ju and Tirrito, Emanuele and Peng, Cheng and Chen, Xi and
            Tagliacozzo, Luca and Su, Gang and Lewenstein, Maciej",
  title = "Two-Dimensional Tensor Networks and Contraction Algorithms",
  bookTitle = "Tensor Network Contractions: Methods and Applications to Quantum
               Many-Body Systems",
  year = "2020",
  publisher = "Springer International Publishing",
  address = "Cham",
  pages = "63--86",
  isbn = "978-3-030-34489-4",
  doi = "10.1007/978-3-030-34489-4_3",
  url = "https://doi.org/10.1007/978-3-030-34489-4_3",
}

@article{page1993volume,
  title = {Average entropy of a subsystem},
  author = {Page, Don N.},
  journal = {Phys. Rev. Lett.},
  volume = {71},
  issue = {9},
  pages = {1291--1294},
  numpages = {0},
  year = {1993},
  month = {Aug},
  publisher = {American Physical Society},
  doi = {10.1103/PhysRevLett.71.1291},
  url = {https://link.aps.org/doi/10.1103/PhysRevLett.71.1291},
}

@article{dahlsten2014volume,
  doi = {10.1088/1751-8113/47/36/363001},
  year = 2014,
  month = {aug},
  publisher = {{IOP} Publishing},
  volume = {47},
  number = {36},
  pages = {363001},
  author = {Oscar C O Dahlsten and Cosmo Lupo and Stefano Mancini and Alessio
            Serafini},
  title = {Entanglement typicality},
  journal = {Journal of Physics A: Mathematical and Theoretical},
}

@article{bridgeman2017review,
  doi = {10.1088/1751-8121/aa6dc3},
  year = 2017,
  month = {may},
  publisher = {{IOP} Publishing},
  volume = {50},
  number = {22},
  pages = {223001},
  author = {Jacob C Bridgeman and Christopher T Chubb},
  title = {Hand-waving and interpretive dance: an introductory course on tensor
           networks},
  journal = {Journal of Physics A: Mathematical and Theoretical},
}

@article{fannes1992,
  author = "Fannes, M. and Nachtergaele, B. and Werner, R. F.",
  title = "Finitely correlated states on quantum spin chains",
  journal = "Commun. Math. Phys.",
  year = "1992",
  month = "Mar",
  day = "01",
  volume = "144",
  number = "3",
  pages = "443--490",
  issn = "1432-0916",
  doi = "10.1007/BF02099178",
  url = "https://doi.org/10.1007/BF02099178",
}

@article{verstraete2006groundstates,
  title = {Matrix product states represent ground states faithfully},
  author = {Verstraete, F. and Cirac, J. I.},
  journal = {Phys. Rev. B},
  volume = {73},
  issue = {9},
  pages = {094423},
  numpages = {8},
  year = {2006},
  month = {Mar},
  publisher = {American Physical Society},
  doi = {10.1103/PhysRevB.73.094423},
  url = {https://link.aps.org/doi/10.1103/PhysRevB.73.094423},
}

@article{montangero2009,
  title = {Critical exponents with a multiscale entanglement renormalization
           Ansatz channel},
  author = {Montangero, S. and Rizzi, M. and Giovannetti, V. and Fazio, Rosario},
  journal = {Phys. Rev. B},
  volume = {80},
  issue = {11},
  pages = {113103},
  numpages = {4},
  year = {2009},
  month = {Sep},
  publisher = {American Physical Society},
  doi = {10.1103/PhysRevB.80.113103},
  url = {https://link.aps.org/doi/10.1103/PhysRevB.80.113103},
}

@incollection{evenbly2013,
  author = {Evenbly, G. and Vidal, G.},
  title = {Quantum Criticality with the Multi-scale Entanglement Renormalization
           Ansatz},
  booktitle = {Strongly Correlated Systems. Numerical Methods},
  editor = {A. Avella and F. Mancini },
  year = {2013},
  chapter = {4},
  isbn = {978-3-642-35105-1},
  pages = {99--130},
  numpages = {32},
  doi = {10.1007/978-3-642-35106-8},
  publisher = {Springer-Verlag},
  address = {Berlin DE},
}

@article{haferkamp2020pepscontraction,
  title = {Contracting projected entangled pair states is average-case hard},
  author = {Haferkamp, Jonas and Hangleiter, Dominik and Eisert, Jens and Gluza,
            Marek},
  journal = {Phys. Rev. Research},
  volume = {2},
  issue = {1},
  pages = {013010},
  numpages = {9},
  year = {2020},
  month = {Jan},
  publisher = {American Physical Society},
  doi = {10.1103/PhysRevResearch.2.013010},
  url = {https://link.aps.org/doi/10.1103/PhysRevResearch.2.013010},
}

@misc{verstraete2004renormalization,
  title = {Renormalization algorithms for Quantum-Many Body Systems in two and
           higher dimensions},
  author = {F. Verstraete and J. I. Cirac},
  year = {2004},
  eprint = {cond-mat/0407066},
  archivePrefix = {arXiv},
  primaryClass = {cond-mat.str-el},
}

@article{vidal2007,
  title = {Entanglement Renormalization},
  author = {Vidal, G.},
  journal = {Phys. Rev. Lett.},
  volume = {99},
  issue = {22},
  pages = {220405},
  numpages = {4},
  year = {2007},
  month = {Nov},
  publisher = {American Physical Society},
  doi = {10.1103/PhysRevLett.99.220405},
  url = {https://link.aps.org/doi/10.1103/PhysRevLett.99.220405},
}

@misc{huang2015computing,
  title = {Computing energy density in one dimension},
  author = {Yichen Huang},
  year = {2015},
  eprint = {1505.00772},
  archivePrefix = {arXiv},
  primaryClass = {cond-mat.str-el},
}

@article{white1993,
  title = {Density-matrix algorithms for quantum renormalization groups},
  author = {White, S. R.},
  journal = {Phys. Rev. B},
  volume = {48},
  issue = {14},
  pages = {10345--10356},
  numpages = {0},
  year = {1993},
  month = {Oct},
  publisher = {American Physical Society},
  doi = {10.1103/PhysRevB.48.10345},
  url = {https://link.aps.org/doi/10.1103/PhysRevB.48.10345},
}

@article{verstraete2010,
  title = {Continuous Matrix Product States for Quantum Fields},
  author = {Verstraete, F. and Cirac, J. I.},
  journal = {Phys. Rev. Lett.},
  volume = {104},
  issue = {19},
  pages = {190405},
  numpages = {4},
  year = {2010},
  month = {May},
  publisher = {American Physical Society},
  doi = {10.1103/PhysRevLett.104.190405},
  url = {https://link.aps.org/doi/10.1103/PhysRevLett.104.190405},
}

@misc{tilloy2017interactingquantumfieldtheories,
  title = {Interacting quantum field theories as relativistic statistical field
           theories of local beables},
  author = {Antoine Tilloy},
  year = {2017},
  eprint = {1702.06325},
  archivePrefix = {arXiv},
  primaryClass = {quant-ph},
  url = {https://arxiv.org/abs/1702.06325},
}

@article{mori2001tanhsinh,
  title = {The double-exponential transformation in numerical analysis},
  journal = {Journal of Computational and Applied Mathematics},
  volume = {127},
  number = {1},
  pages = {287-296},
  year = {2001},
  note = {Numerical Analysis 2000. Vol. V: Quadrature and Orthogonal Polynomials
          },
  issn = {0377-0427},
  doi = {https://doi.org/10.1016/S0377-0427(00)00501-X},
  url = {https://www.sciencedirect.com/science/article/pii/S037704270000501X},
  author = {Masatake Mori and Masaaki Sugihara},
}

@article{hauru2021,
  title = {{Riemannian optimization of isometric tensor networks}},
  author = {Markus Hauru and Maarten Van Damme and Jutho Haegeman},
  journal = {SciPost Phys.},
  volume = {10},
  issue = {2},
  pages = {40},
  year = {2021},
  publisher = {SciPost},
  doi = {10.21468/SciPostPhys.10.2.040},
  url = {https://scipost.org/10.21468/SciPostPhys.10.2.040},
}

@article{rackauckas2017differentialequations,
  title = {Differentialequations.jl--a performant and feature-rich ecosystem for
           solving differential equations in julia},
  author = {Rackauckas, Christopher and Nie, Qing},
  journal = {Journal of Open Research Software},
  volume = {5},
  number = {1},
  year = {2017},
  publisher = {Ubiquity Press},
}

@misc{nelson1966quartic,
  title = {A quartic interaction in two dimensions, in" Mathematical theory of
           elementary particles", eds. R. Goodman and I. Segal},
  author = {Nelson, E},
  year = {1966},
  publisher = {MIT Press, Cambridge, Mass},
}

@article{simon2004ed,
  title = {Ed Nelson’s work in quantum theory},
  author = {Simon, Barry},
  journal = {Diffusion, Quantum Theory, and Radically Elementary Mathematics (WG
             Faris, ed.)},
  volume = {47},
  pages = {75--93},
  year = {2004},
  doi = {10.1515/9781400865253.75},
}

@article{federbush1969phi4lower,
  author = {Federbush, Paul},
  title = {Partially Alternate Derivation of a Result of Nelson},
  journal = {Journal of Mathematical Physics},
  volume = {10},
  number = {1},
  pages = {50-52},
  year = {1969},
  month = {01},
  issn = {0022-2488},
  doi = {10.1063/1.1664760},
  url = {https://doi.org/10.1063/1.1664760},
  eprint = {https://pubs.aip.org/aip/jmp/article-pdf/10/1/50/19242126/50\_1 \_
            online.pdf},
}

@article{vanhecke2022scalingphi4,
  title = {Entanglement scaling for $\ensuremath{\lambda}{\ensuremath{\phi}}_{2}
           ^{4}$},
  author = {Vanhecke, Bram and Verstraete, Frank and Van Acoleyen, Karel},
  journal = {Phys. Rev. D},
  volume = {106},
  issue = {7},
  pages = {L071501},
  numpages = {6},
  year = {2022},
  month = {Oct},
  publisher = {American Physical Society},
  doi = {10.1103/PhysRevD.106.L071501},
  url = {https://link.aps.org/doi/10.1103/PhysRevD.106.L071501},
}

@article{coleman1975,
  title = {Quantum sine-Gordon equation as the massive Thirring model},
  author = {Coleman, Sidney},
  journal = {Phys. Rev. D},
  volume = {11},
  issue = {8},
  pages = {2088--2097},
  numpages = {0},
  year = {1975},
  month = {Apr},
  publisher = {American Physical Society},
  doi = {10.1103/PhysRevD.11.2088},
  url = {https://link.aps.org/doi/10.1103/PhysRevD.11.2088},
}

@article{konik2021,
  title = {Approaching the self-dual point of the sinh-Gordon model},
  author = {Konik, Robert and L{\'a}jer, M{\'a}rton and Mussardo, Giuseppe},
  journal = {JHEP},
  volume = {2021},
  number = {1},
  pages = {1--85},
  year = {2021},
  publisher = {Springer},
  doi = {10.1007/JHEP01(2021)014},
}

@article{froehlich1975,
  title = {Quantized "Sine-Gordon" Equation with a Nonvanishing Mass Term in Two
           Space-Time Dimensions},
  author = {Fr\"ohlich, J\"urg},
  journal = {Phys. Rev. Lett.},
  volume = {34},
  issue = {13},
  pages = {833--836},
  numpages = {0},
  year = {1975},
  month = {Mar},
  publisher = {American Physical Society},
  doi = {10.1103/PhysRevLett.34.833},
  url = {https://link.aps.org/doi/10.1103/PhysRevLett.34.833},
}

@article{froehlich1977,
  title = {Remarks on exponential interactions and the quantum sine-Gordon
           equation in two space-time dimensions},
  author = {Fr{\"o}hlich, J{\"u}rg and Park, Yong Moon},
  journal = {Helvetica Physica Acta},
  volume = {50},
  number = {3},
  pages = {315--329},
  year = {1977},
  url = {https://www.e-periodica.ch/cntmng?pid=hpa-001\%3A1977\%3A50\%3A\%3A982},
}

@article{dimock1993,
  title = {Construction of the two-dimensional Sine-Gordon model for $\beta < 8
           \pi$},
  author = {Dimock, J and Hurd, TR},
  journal = {Communications in mathematical physics},
  volume = {156},
  number = {3},
  pages = {547--580},
  year = {1993},
  publisher = {Springer},
  doi = {10.1007/BF02096863},
}

@misc{anand2020introductionlightconeconformaltruncation,
  title = {Introduction to Lightcone Conformal Truncation: QFT Dynamics from CFT
           Data},
  author = {Nikhil Anand and A. Liam Fitzpatrick and Emanuel Katz and Zuhair U.
            Khandker and Matthew T. Walters and Yuan Xin},
  year = {2020},
  eprint = {2005.13544},
  archivePrefix = {arXiv},
  primaryClass = {hep-th},
  url = {https://arxiv.org/abs/2005.13544},
}

@article{bernard2022,
  author = {Bernard, Denis and LeClair, Andr{\'e}},
  title = {The sinh-Gordon model beyond the self dual point and the freezing
           transition in disordered systems},
  journal = {Journal of High Energy Physics},
  year = {2022},
  month = {May},
  day = {04},
  volume = {2022},
  number = {5},
  pages = {22},
  issn = {1029-8479},
  doi = {10.1007/JHEP05(2022)022},
  url = {https://doi.org/10.1007/JHEP05(2022)022},
}

@article{lukyanov1997,
  title = {Exact expectation values of local fields in the quantum sine-Gordon
           model},
  journal = {Nucl. Phys. B},
  volume = {493},
  number = {3},
  pages = {571-587},
  year = {1997},
  issn = {0550-3213},
  doi = {https://doi.org/10.1016/S0550-3213(97)00123-5},
  author = {Sergei Lukyanov and Alexander Zamolodchikov},
}

@article{fateev1998,
  title = {Expectation values of local fields in the Bullough-Dodd model and
           integrable perturbed conformal field theories},
  journal = {Nucl. Phys. B},
  volume = {516},
  number = {3},
  pages = {652-674},
  year = {1998},
  issn = {0550-3213},
  doi = {https://doi.org/10.1016/S0550-3213(98)00002-9},
  author = {Vladimir Fateev and Sergei Lukyanov and Alexander Zamolodchikov and
            Alexei Zamolodchikov},
}

@article{tiwana2025,
  author = {Tiwana, Karan and Lauria, Edoardo and Tilloy, Antoine},
  title = {A relativistic continuous matrix product state study of field
           theories with defects},
  journal = {Journal of High Energy Physics},
  year = {2025},
  month = {May},
  day = {13},
  volume = {2025},
  number = {5},
  pages = {97},
  issn = {1029-8479},
  doi = {10.1007/JHEP05(2025)097},
  url = {https://doi.org/10.1007/JHEP05(2025)097},
}

@article{tilloy2019ctns,
  title = {Continuous Tensor Network States for Quantum Fields},
  author = {Tilloy, Antoine and Cirac, J. Ignacio},
  journal = {Phys. Rev. X},
  volume = {9},
  issue = {2},
  pages = {021040},
  numpages = {17},
  year = {2019},
  month = {May},
  publisher = {American Physical Society},
  doi = {10.1103/PhysRevX.9.021040},
  url = {https://link.aps.org/doi/10.1103/PhysRevX.9.021040},
}

@article{jennings2015ctns,
  doi = {10.1088/1367-2630/17/6/063039},
  url = {https://dx.doi.org/10.1088/1367-2630/17/6/063039},
  year = {2015},
  month = {jun},
  publisher = {IOP Publishing},
  volume = {17},
  number = {6},
  pages = {063039},
  author = {Jennings, David and Brockt, Christoph and Haegeman, Jutho and
            Osborne, Tobias J and Verstraete, Frank},
  title = {Continuum tensor network field states, path integral representations
           and spatial symmetries},
  journal = {New Journal of Physics},
}

@article{dubail2012,
  title = {Edge-state inner products and real-space entanglement spectrum of
           trial quantum Hall states},
  author = {Dubail, J. and Read, N. and Rezayi, E. H.},
  journal = {Phys. Rev. B},
  volume = {86},
  issue = {24},
  pages = {245310},
  numpages = {32},
  year = {2012},
  month = {Dec},
  publisher = {American Physical Society},
  doi = {10.1103/PhysRevB.86.245310},
  url = {https://link.aps.org/doi/10.1103/PhysRevB.86.245310},
}

@article{karanikolaou2021gctns,
  title = {Gaussian continuous tensor network states for simple bosonic field
           theories},
  author = {Karanikolaou, Teresa D. and Emonts, Patrick and Tilloy, Antoine},
  journal = {Phys. Rev. Res.},
  volume = {3},
  issue = {2},
  pages = {023059},
  numpages = {11},
  year = {2021},
  month = {Apr},
  publisher = {American Physical Society},
  doi = {10.1103/PhysRevResearch.3.023059},
  url = {https://link.aps.org/doi/10.1103/PhysRevResearch.3.023059},
}

@article{frohlich2024euclidean,
  title = {The Euclidean $\phi^4_2$ theory as a limit of an interacting Bose gas
           },
  author = {Fr{\"o}hlich, J{\"u}rg and Knowles, Antti and Schlein, Benjamin and
            Sohinger, Vedran},
  journal = {Journal of the European Mathematical Society},
  year = {2024},
  doi = {10.4171/jems/1454},
}

@article{feldman1976phi43,
  title = {The Wightman axioms and the mass gap for weakly coupled $\phi^4_3$
           quantum field theories},
  journal = {Annals of Physics},
  volume = {97},
  number = {1},
  pages = {80-135},
  year = {1976},
  issn = {0003-4916},
  doi = {https://doi.org/10.1016/0003-4916(76)90223-2},
  url = {https://www.sciencedirect.com/science/article/pii/0003491676902232},
  author = {Joel S Feldman and Konrad Osterwalder},
}

@article{glimm1971phi43_lowerbound,
  author = {Glimm, James and Jaffe, Arthur},
  title = {Positivity of the $\varphi^4_3$ Hamiltonian},
  journal = {Fortschritte der Physik},
  volume = {21},
  number = {7},
  pages = {327-376},
  doi = {https://doi.org/10.1002/prop.19730210702},
  url = {https://onlinelibrary.wiley.com/doi/abs/10.1002/prop.19730210702},
  eprint = {https://onlinelibrary.wiley.com/doi/pdf/10.1002/prop.19730210702},
  year = {1973},
}

@article{jagannath2023phi43,
  title = {A simple construction of the dynamical $\phi^4_3$ model},
  volume = {376},
  ISSN = {1088-6850},
  url = {http://dx.doi.org/10.1090/tran/8724},
  DOI = {10.1090/tran/8724},
  number = {3},
  journal = {Transactions of the American Mathematical Society},
  publisher = {American Mathematical Society (AMS)},
  author = {Jagannath, Aukosh and Perkowski, Nicolas},
  year = {2023},
  month = jan,
  pages = {1507–1522},
}

@article{hairer2014phi43,
  author = "Hairer, Martin",
  title = "{Regularity structures and the dynamical $\Phi^4_3$ model}",
  eprint = "1508.05261",
  archivePrefix = "arXiv",
  primaryClass = "math.PR",
  doi = "10.4310/cdm.2014.v2014.n1.a1",
  journal = "Curr. Dev. Math.",
  volume = "2014",
  number = "1",
  pages = "1--49",
  year = "2014",
}

@article{cardy2017boundary,
  title = {{Bulk Renormalization Group Flows and Boundary States in Conformal
           Field Theories}},
  author = {John Cardy},
  journal = {SciPost Phys.},
  volume = {3},
  pages = {011},
  year = {2017},
  publisher = {SciPost},
  doi = {10.21468/SciPostPhys.3.2.011},
  url = {https://scipost.org/10.21468/SciPostPhys.3.2.011},
}

@misc{konechny2023boundary,
  title = {RG boundaries and Cardy's variational ansatz for multiple
           perturbations},
  author = {Anatoly Konechny},
  year = {2023},
  eprint = {2306.13719},
  archivePrefix = {arXiv},
  primaryClass = {hep-th},
  url = {https://arxiv.org/abs/2306.13719},
}

@article{vardian2024,
  author = {Vardian, Niloofar},
  title = {A CFT dual for evaporating black holes: boundary continuous matrix
           product states},
  journal = {Journal of High Energy Physics},
  year = {2024},
  month = {Dec},
  day = {03},
  volume = {2024},
  number = {12},
  pages = {22},
  issn = {1029-8479},
  doi = {10.1007/JHEP12(2024)022},
  url = {https://doi.org/10.1007/JHEP12(2024)022},
}

@article{schuch2008mpsoptimhardness,
  title = {Computational Difficulty of Finding Matrix Product Ground States},
  author = {Schuch, Norbert and Cirac, Ignacio and Verstraete, Frank},
  journal = {Phys. Rev. Lett.},
  volume = {100},
  issue = {25},
  pages = {250501},
  numpages = {4},
  year = {2008},
  month = {Jun},
  publisher = {American Physical Society},
  doi = {10.1103/PhysRevLett.100.250501},
  url = {https://link.aps.org/doi/10.1103/PhysRevLett.100.250501},
}

@misc{luscher1998advancedlatticeqcd,
  title = {Advanced Lattice QCD},
  author = {Martin Lüscher},
  year = {1998},
  eprint = {hep-lat/9802029},
  archivePrefix = {arXiv},
  primaryClass = {hep-lat},
  url = {https://arxiv.org/abs/hep-lat/9802029},
}

@article{petreczky2019qcdfinelattice,
  title = {Strong coupling constant and heavy quark masses in ($2+1$)-flavor QCD
           },
  author = {Petreczky, P. and Weber, J. H.},
  journal = {Phys. Rev. D},
  volume = {100},
  issue = {3},
  pages = {034519},
  numpages = {17},
  year = {2019},
  month = {Aug},
  publisher = {American Physical Society},
  doi = {10.1103/PhysRevD.100.034519},
  url = {https://link.aps.org/doi/10.1103/PhysRevD.100.034519},
}

@article{Wilson:1974mb,
  author = "Wilson, Kenneth G.",
  title = "{The Renormalization Group: Critical Phenomena and the Kondo Problem}
           ",
  reportNumber = "CLNS-296",
  doi = "10.1103/RevModPhys.47.773",
  journal = "Rev. Mod. Phys.",
  volume = "47",
  pages = "773",
  year = "1975",
}

@book{KONDO1970183,
  title = {Theory of Dilute Magnetic Alloys},
  editor = {Frederick Seitz and David Turnbull and Henry Ehrenreich},
  series = {Solid State Physics},
  publisher = {Academic Press},
  volume = {23},
  pages = {183-281},
  year = {1970},
  issn = {0081-1947},
  doi = {https://doi.org/10.1016/S0081-1947(08)60616-5},
  url = {https://www.sciencedirect.com/science/article/pii/S0081194708606165},
  author = {J. Kondo},
}

@inproceedings{Cordova:2022ruw,
  author = "Cordova, Clay and Dumitrescu, Thomas T. and Intriligator, Kenneth
            and Shao, Shu-Heng",
  title = "{Snowmass White Paper: Generalized Symmetries in Quantum Field Theory
           and Beyond}",
  booktitle = "{Snowmass 2021}",
  eprint = "2205.09545",
  archivePrefix = "arXiv",
  primaryClass = "hep-th",
  month = "5",
  year = "2022",
}

@article{Gaiotto:2014kfa,
  author = "Gaiotto, Davide and Kapustin, Anton and Seiberg, Nathan and Willett,
            Brian",
  title = "{Generalized Global Symmetries}",
  eprint = "1412.5148",
  archivePrefix = "arXiv",
  primaryClass = "hep-th",
  doi = "10.1007/JHEP02(2015)172",
  journal = "JHEP",
  volume = "02",
  pages = "172",
  year = "2015",
}

@article{haegeman2011_original_tdvpMPS,
  title = {Time-Dependent Variational Principle for Quantum Lattices},
  author = {Haegeman, Jutho and Cirac, J. Ignacio and Osborne, Tobias J. and Pi
            \ifmmode \check{z}\else \v{z}\fi{}orn, Iztok and Verschelde, Henri
            and Verstraete, Frank},
  journal = {Phys. Rev. Lett.},
  volume = {107},
  issue = {7},
  pages = {070601},
  numpages = {5},
  year = {2011},
  month = {Aug},
  publisher = {American Physical Society},
  doi = {10.1103/PhysRevLett.107.070601},
  url = {https://link.aps.org/doi/10.1103/PhysRevLett.107.070601},
}

@article{pollmann2009_finiteentanglementscaling,
  title = {Theory of Finite-Entanglement Scaling at One-Dimensional Quantum
           Critical Points},
  author = {Pollmann, Frank and Mukerjee, Subroto and Turner, Ari M. and Moore,
            Joel E.},
  journal = {Phys. Rev. Lett.},
  volume = {102},
  issue = {25},
  pages = {255701},
  numpages = {4},
  year = {2009},
  month = {Jun},
  publisher = {American Physical Society},
  doi = {10.1103/PhysRevLett.102.255701},
  url = {https://link.aps.org/doi/10.1103/PhysRevLett.102.255701},
}

@article{borinsky2023tropical,
  title = {Tropical Monte Carlo quadrature for Feynman integrals},
  author = {Borinsky, Michael},
  journal = {Annales de l’Institut Henri Poincar{\'e} D},
  volume = {10},
  number = {4},
  pages = {635--685},
  year = {2023},
  doi = {10.4171/AIHPD/158},
}

@article{rossi2017diagrammatic,
  title = {Determinant Diagrammatic Monte Carlo Algorithm in the Thermodynamic
           Limit},
  author = {Rossi, Riccardo},
  journal = {Phys. Rev. Lett.},
  volume = {119},
  issue = {4},
  pages = {045701},
  numpages = {5},
  year = {2017},
  month = {Jul},
  publisher = {American Physical Society},
  doi = {10.1103/PhysRevLett.119.045701},
  url = {https://link.aps.org/doi/10.1103/PhysRevLett.119.045701},
}

@article{fernandez2022tci,
  title = {Learning Feynman Diagrams with Tensor Trains},
  author = {N\'u\~nez Fern\'andez, Yuriel and Jeannin, Matthieu and Dumitrescu,
            Philipp T. and Kloss, Thomas and Kaye, Jason and Parcollet, Olivier
            and Waintal, Xavier},
  journal = {Phys. Rev. X},
  volume = {12},
  issue = {4},
  pages = {041018},
  numpages = {30},
  year = {2022},
  month = {Nov},
  publisher = {American Physical Society},
  doi = {10.1103/PhysRevX.12.041018},
  url = {https://link.aps.org/doi/10.1103/PhysRevX.12.041018},
}

@article{yurov1990truncated,
  author = {YUROV, V. P. and ZAMOLODCHIKOV, AL. B.},
  title = {TRUNCATED COMFORMAL SPACE APPROACH TO SCALING LEE-YANG MODEL},
  journal = {International Journal of Modern Physics A},
  volume = {05},
  number = {16},
  pages = {3221-3245},
  year = {1990},
  doi = {10.1142/S0217751X9000218X},
  URL = { https://doi.org/10.1142/S0217751X9000218X },
  eprint = { https://doi.org/10.1142/S0217751X9000218X },
}

@misc{huang2015error,
  title = {Computing energy density in one dimension},
  author = {Yichen Huang},
  year = {2015},
  eprint = {1505.00772},
  archivePrefix = {arXiv},
  primaryClass = {cond-mat.str-el},
  url = {https://arxiv.org/abs/1505.00772},
}

@article{Cuomo:2021kfm,
  author = "Cuomo, Gabriel and Komargodski, Zohar and Mezei, M\'ark",
  title = "{Localized magnetic field in the O(N) model}",
  eprint = "2112.10634",
  archivePrefix = "arXiv",
  primaryClass = "hep-th",
  doi = "10.1007/JHEP02(2022)134",
  journal = "JHEP",
  volume = "02",
  pages = "134",
  year = "2022",
}

@book{DiFrancesco:1997nk,
  author = "Di Francesco, P. and Mathieu, P. and Senechal, D.",
  title = "{Conformal Field Theory}",
  doi = "10.1007/978-1-4612-2256-9",
  isbn = "978-0-387-94785-3, 978-1-4612-7475-9",
  publisher = "Springer-Verlag",
  address = "New York",
  series = "Graduate Texts in Contemporary Physics",
  year = "1997",
}

@article{Cardy:2004hm,
  author = "Cardy, John L.",
  title = "{Boundary conformal field theory}",
  eprint = "hep-th/0411189",
  journal = {arXiv e-prints},
  archivePrefix = "arXiv",
  month = "11",
  year = "2004",
}

@article{tuybens2022cmps,
  title = {Variational Optimization of Continuous Matrix Product States},
  author = {Tuybens, Beno\^{\i}t and De Nardis, Jacopo and Haegeman, Jutho and
            Verstraete, Frank},
  journal = {Phys. Rev. Lett.},
  volume = {128},
  issue = {2},
  pages = {020501},
  numpages = {6},
  year = {2022},
  month = {Jan},
  publisher = {American Physical Society},
  doi = {10.1103/PhysRevLett.128.020501},
  url = {https://link.aps.org/doi/10.1103/PhysRevLett.128.020501},
}

@misc{lawrence2024spectral,
  title = {Model-free spectral reconstruction via Lagrange duality},
  author = {Scott Lawrence},
  year = {2024},
  eprint = {2408.11766},
  archivePrefix = {arXiv},
  primaryClass = {hep-lat},
  url = {https://arxiv.org/abs/2408.11766},
}

@article{schuch2011classifying,
  title = {Classifying quantum phases using matrix product states and projected
           entangled pair states},
  author = {Schuch, Norbert and P\'erez-Garc\'{\i}a, David and Cirac, Ignacio},
  journal = {Phys. Rev. B},
  volume = {84},
  issue = {16},
  pages = {165139},
  numpages = {21},
  year = {2011},
  month = {Oct},
  publisher = {American Physical Society},
  doi = {10.1103/PhysRevB.84.165139},
  url = {https://link.aps.org/doi/10.1103/PhysRevB.84.165139},
}

@article{hastings2007,
  doi = {10.1088/1742-5468/2007/08/P08024},
  url = {https://dx.doi.org/10.1088/1742-5468/2007/08/P08024},
  year = {2007},
  month = {aug},
  publisher = {},
  volume = {2007},
  number = {08},
  pages = {P08024},
  author = {Hastings, M B},
  title = {An area law for one-dimensional quantum systems},
  journal = {Journal of Statistical Mechanics: Theory and Experiment},
}

@article{cirac2006mps,
  title = {Matrix product states represent ground states faithfully},
  author = {Verstraete, F. and Cirac, J. I.},
  journal = {Phys. Rev. B},
  volume = {73},
  issue = {9},
  pages = {094423},
  numpages = {8},
  year = {2006},
  month = {Mar},
  publisher = {American Physical Society},
  doi = {10.1103/PhysRevB.73.094423},
  url = {https://link.aps.org/doi/10.1103/PhysRevB.73.094423},
}

@Article{brooks1984,
author={Brooks, E. D.
and Frautschi, S. C.},
title={Scalars coupled to fermions in 1+1 dimensions},
journal={Zeitschrift f{\"u}r Physik C Particles and Fields},
year={1984},
month={Sep},
day={01},
volume={23},
number={3},
pages={263-273},
issn={1431-5858},
doi={10.1007/BF01546194},
url={https://doi.org/10.1007/BF01546194}
}

@article{yin2018DH,
  author = "Yin, Xi",
  title = "{Aspects of Two-Dimensional Conformal Field Theories}",
  doi = "10.22323/1.305.0003",
  journal = "PoS",
  year = 2018,
  volume = "TASI2017",
  pages = "003"
}

@book{griewankWalther2008EvaluatingDerivatives,
  author    = {Griewank, Andreas and Walther, Andrea},
  title     = {Evaluating Derivatives: Principles and Techniques of Algorithmic Differentiation},
  edition   = {2},
  publisher = {SIAM},
  year      = {2008},
  doi       = {10.1137/1.9780898717761},
}

@incollection{Griewank1989OnAD,
  author    = {Griewank, Andreas},
  title     = {On Automatic Differentiation},
  booktitle = {Mathematical Programming: Recent Developments and Applications},
  editor    = {Iri, Masao and Tanabe, Kunio},
  pages     = {83--108},
  year      = {1989},
  publisher = {Kluwer Academic Publishers},
  address   = {Dordrecht},
url = {https://softlib.rice.edu/pub/CRPC-TRs/reports/CRPC-TR89003.pdf}
}

@Article{fitzpatrick2020,
author={Fitzpatrick, A. Liam
and Katz, Emanuel
and Walters, Matthew T.},
title={Nonperturbative matching between equal-time and lightcone quantization},
journal={Journal of High Energy Physics},
year={2020},
month={Oct},
day={14},
volume={2020},
number={10},
pages={92},
issn={1029-8479},
doi={10.1007/JHEP10(2020)092},
url={https://doi.org/10.1007/JHEP10(2020)092}
}

@book{Baxter1982ExactlySolved,
  author    = {Baxter, Rodney J.},
  title     = {Exactly Solved Models in Statistical Mechanics},
  publisher = {Academic Press},
  address   = {London},
  year      = {1982},
  isbn      = {978-0-12-083180-7}
}

@misc{mutzel2025extractingquantumfieldtheory,
      title={Extracting quantum field theory dynamics from an approximate ground state}, 
      author={Sophie Mutzel and Antoine Tilloy},
      year={2025},
      eprint={2512.19594},
      archivePrefix={arXiv},
      primaryClass={quant-ph},
      url={https://arxiv.org/abs/2512.19594}, 
}

@misc{tiwana2025multifieldrelativisticcontinuousmatrix,
      title={Multi-Field Relativistic Continuous Matrix Product States}, 
      author={Karan Tiwana and Antoine Tilloy},
      year={2025},
      eprint={2511.20762},
      archivePrefix={arXiv},
      primaryClass={quant-ph},
      url={https://arxiv.org/abs/2511.20762}, 
}

@Article{cheng2026normalizingflows,
	title={{Lecture notes on normalizing flows for lattice quantum field theories}},
	author={Miranda C. N. Cheng and Niki Stratikopoulou},
	journal={SciPost Phys. Lect. Notes},
	pages={110},
	year={2026},
	publisher={SciPost},
	doi={10.21468/SciPostPhysLectNotes.110},
	url={https://scipost.org/10.21468/SciPostPhysLectNotes.110},
}

\end{document}